\documentclass[preprint,3p,times,sort&compress]{elsarticle}
\usepackage{multirow,setspace,times,amssymb,amsmath,graphicx,color,rotating,subfigure,url}
\usepackage{lineno}
\usepackage{natbib}
\usepackage{booktabs}%
\usepackage{longtable}%
\usepackage{soul}
\usepackage{subfigure}
\usepackage[table]{xcolor}
\usepackage{tabularx}
\usepackage{graphicx} 
\usepackage{epstopdf}
\usepackage{mathrsfs}
\usepackage{makecell}
\usepackage[bookmarks=true,colorlinks,linkcolor=blue,anchorcolor=blue,citecolor=blue,unicode]{hyperref}
\usepackage{bookmark}
\usepackage{appendix}

\journal{Chaos, Solitons and Fractals}
\begin{document}

\begin{frontmatter}
\newpage
\title{Quantifying the status of economies in international crop trade networks: An correlation structure analysis of various node-ranking metrics}

\author[SB,RCE]{Yin-Ting Zhang}
\author[SB,RCE,Math]{Wei-Xing Zhou\corref{cor1}}
\ead{wxzhou@ecust.edu.cn} 
\cortext[cor1]{Corresponding author.}
\address[SB]{School of Business, East China University of Science and Technology, Shanghai 200237, China}
\address[RCE]{Research Center for Econophysics, East China University of Science and Technology, Shanghai 200237, China}
\address[Math]{School of Mathematics, East China University of Science and Technology, Shanghai 200237, China}
\begin{abstract}
International food trade is a growing complement to gaps in domestic food supply and demand, but it is vulnerable to disruptions due to some unforeseen shocks. This paper assembles the international crop trade networks using maize, rice, soybean, and wheat trade data sets from 1986 to 2020. We assess the importance of economies using multidimensional node importance metrics. We analyze the correlation structure of different node important metrics based on the random matrix theory and incorporate 20 metrics into a single metric. We find that some metrics have many similarities and dissimilarities, especially for metrics based on the same trade flow directions. We also find that European economies have a significant impact on the iCTNs. Additionally, economies with poor crop production play a major role in import trade, whereas economies with higher food production or smaller populations are crucial to export trade. Our findings have practical implications for identifying key economies in the international crop trade networks, preventing severe damage to the food trade system caused by trade disruptions in some economies, maintaining the stability of the food supply, and ensuring food security.

\end{abstract}

\begin{keyword} 
Complex network, node importance, food security, food trade, key economies

\end{keyword}

\end{frontmatter}


\section{Introduction}
\label{S1:Introduction}

Global food security, defined as the state where people are able to get sufficient, available, and nutritious food \cite{Barrett-2010-Science,Teeuwen-Meyer-Dou-Nelson-2022-NatFood}, is always a worldwide concern \cite{Rosegrant-Cline-2003-Science,Barrett-2010-Science}. It is a challenge to support the demand of the growing population \cite{Edreira-Andrade-Cassman-vanIttersum-vanLoon-Grassini-2021-NatFood}. The latest State of Food Security and Nutrition in the World report, promulgated by five United Nations agencies, announces that food insecurity is escalating\footnote{See \url{https://www.fao.org/3/cc0639en/cc0639en.pdf}.}. Climate change, conflict, and epidemics expose millions of people to hunger and malnutrition risks \cite{Janssens-Havlik-Krisztin-Baker-Frank-Hasegawa-Leclere-Ohrel-Ragnauth-Schmid-Valin-VanLipzig-Maertens-2020-NatClimChang}. Extreme weather poses one of the most serious threats to food production \cite{Zurek-Hebinck-Selomane-2022-Science}, shifting regional consumption and trade patterns \cite{Janssens-Havlik-Krisztin-Baker-Frank-Hasegawa-Leclere-Ohrel-Ragnauth-Schmid-Valin-VanLipzig-Maertens-2020-NatClimChang}. Conflicts like the current Russia-Ukraine crisis have the potential to exacerbate global food insecurity\footnote{Assessing Food Insecurity in 2022/23 at National and Sub-National Levels in 50 Countries Vulnerable to the Effects of the Ukraine-Russia Crisis (2022/08/07). See \url{https://www.fao.org/3/cb9447en/cb9447en.pdf}.}. The epidemic of diseases also has a negative impact on the food system through labor losses and trade restrictions \cite{Laborde-Martin-Swinnen-Vos-2020-Science}. Under the shadow of the COVID-19 pandemic, global hunger worsened in 2020. Due to conflicts and the COVID-19 pandemic, export restrictions imposed by some economies, cause disruptions to staple food flows in the international market. Because these restrictions reduce external food supply, they have the potential to cause a spiral rise in global food prices \cite{Bouet-Debucquet-2012-RevWorldEcon}, spreading supply instability to trading partners \cite{Jagermeyr-Robock-Elliott-Muller-Xia-Khabarov-Folberth-Schmid-Liu-Zabel-Rabin-Puma-Heslin-Franke-Foster-Asseng-Bardeen-Toon-Rosenzweig-2020-ProcNatlAcadSciUSA}.

Food availability in an economy relies on domestic production, reserves, and trade. Food production reductions are compensated first by obtaining reserves and using food imports, and then by diminishing both domestic use and food exports \cite{Jagermeyr-Robock-Elliott-Muller-Xia-Khabarov-Folberth-Schmid-Liu-Zabel-Rabin-Puma-Heslin-Franke-Foster-Asseng-Bardeen-Toon-Rosenzweig-2020-ProcNatlAcadSciUSA}. It is now clearer that trade plays a crucial role in narrowing the gap between domestic consumption and supply. The international food trade provides access to approximately 24\% of the food consumed by the global population \cite{DOdorico-Carr-Laio-Ridolfi-Vandoni-2014-EarthFuture}. If extreme events hinder food trade, the instability of the world market will exacerbate \cite{Anderson-Nelgen-2012-OxfRevEconPolicy}, thereby festering food shortages in resource-poor economies with food deficits \cite{Jagermeyr-Robock-Elliott-Muller-Xia-Khabarov-Folberth-Schmid-Liu-Zabel-Rabin-Puma-Heslin-Franke-Foster-Asseng-Bardeen-Toon-Rosenzweig-2020-ProcNatlAcadSciUSA}. Food trade within or between economies diversifies food supplies and rationalizes resource allocation, which assists in reducing the vulnerability of food systems to various types of shocks \cite{Laborde-Martin-Swinnen-Vos-2020-Science}. However, the formation of extensive and close trade relationships between economies may cause shocks to propagate to importers.

When an economy with a significant impact on the global food market restricts food exporting or importing as a result of an external disturbance, other economies and even the global food trade system are impacted. One then wonders how to identify critical economies and evaluate their influence on the trade system. To that end, we use complex networks as a tool to address this problem. We construct international crop trade networks (iCTNs) and investigate critical nodes in topology structures using international trade data sets from four staple crops. Our analysis is rooted in the core concept that the importance of a node is equal to its destructive effect on the network after it is deleted. The most common node importance metrics include degree scales, eigenvector scales, path scales, and so on. Different metrics depict various characteristics and demonstrate the influence rankings of nodes. Here, we summarize four types of indicators, including global metrics, local metrics, modular metrics, and metrics based on information theory. It helps us measure the influence of economies on the international crop trade networks in a holistic way. Despite the fact that numerous studies have used a specific metric based on a specific network property to identify key nodes, there is a critical knowledge gap regarding the correlation structure between different node importance metrics. Random matrix theory (RMT) is a successful approach to recognizing correlations among objects \cite{Garas-Argyrakis-2007-PhysicaA}. An open question is how to construct a comprehensive metric to measure the influence of an economy. We heuristically apply the random matrix theory for constructing optimal portfolios in the stock market to construct a composite node importance indicator.

It is widely assumed that measuring the trade influence of economies in the iCTNs is critical and advantageous in order to avoid massive trade blockages when shocks occur. In this paper, we apply various node importance metrics of four dimensions for each iCTN to identify critical economies in the food trade system. We specifically employ random matrix theory to examine the correlation structure of various node importance metrics. We find that the trade influence of economies in the iCTNs differs across node metric scales. However, some metrics that are based on the same direction of trade flows show special correlations. This new knowledge can inform a new mindset to obtain a composite node importance index. In this regard, we fictionalize the optimal eigenportfolio of eigenvalues to estimate an economy's composite importance. Our approach does not distinguish between different types of shocks and simplifies the form of trade constraints. But our results highlight the importance of some economies in the global trade system and therefore point to the need to ensure unimpeded trade between critical economies.

The subsequent sections of this article are organized as follows: In Section~\ref{S2:DataMethod}, we provide an overview of the sources of our data and our research methods. Section~\ref{S3:Results} presents the findings of our analysis, which includes an examination of the correlation structure among different node importance metrics using random matrix theory, and the integration of 20 metrics into a single metric. In the final section, we summarize the key contributions of our paper, as well as its limitations and potential future directions.

\section{Data and method}
\label{S2:DataMethod}

\subsection{Data and iCTN construction}

Most of the calories consumed by the world's population come from four staple crops (maize, rice, soybean, and wheat) \cite{DOdorico-Carr-Laio-Ridolfi-Vandoni-2014-EarthFuture}. We collect bilateral food trade data from 1986 to 2020, provided by the FAOSTAT data sets on international trade flows from the Food and Agriculture Organization (FAO, http://www.fao.org). Note that the quantities and prices of crops vary across different crops, making it difficult to compare them on an equal footing. To solve this problem, we calculate caloric trade flows by multiplying calories per 1000 t and volumes.

We assemble the international crop trade networks, where economies are portrayed as nodes and trade relationships as links from importers to exporters. That is to say, we define the iCTN for a given crop $\mathrm{crop}$ as: $F^{crop}(t) = \left(\mathscr{V}^{crop}(t), W^{crop}(t) \right)$, where ${\mathscr{V}}^{crop}(t)$ is the set of all nodes (economies) and $W^{crop}(t)$ is the weight of links (i.e. the caloric trade flows). Having constructed the international maize, rice, soybean, and wheat trade networks from 1986 to 2020, we investigate and compare the characteristics and evolution of these different networks.

\begin{figure}[h!]
      \centering
     \subfigbottomskip=-1pt
     \subfigcapskip=-5pt
     \centering
     \subfigure[]{\includegraphics[width=0.45\linewidth]{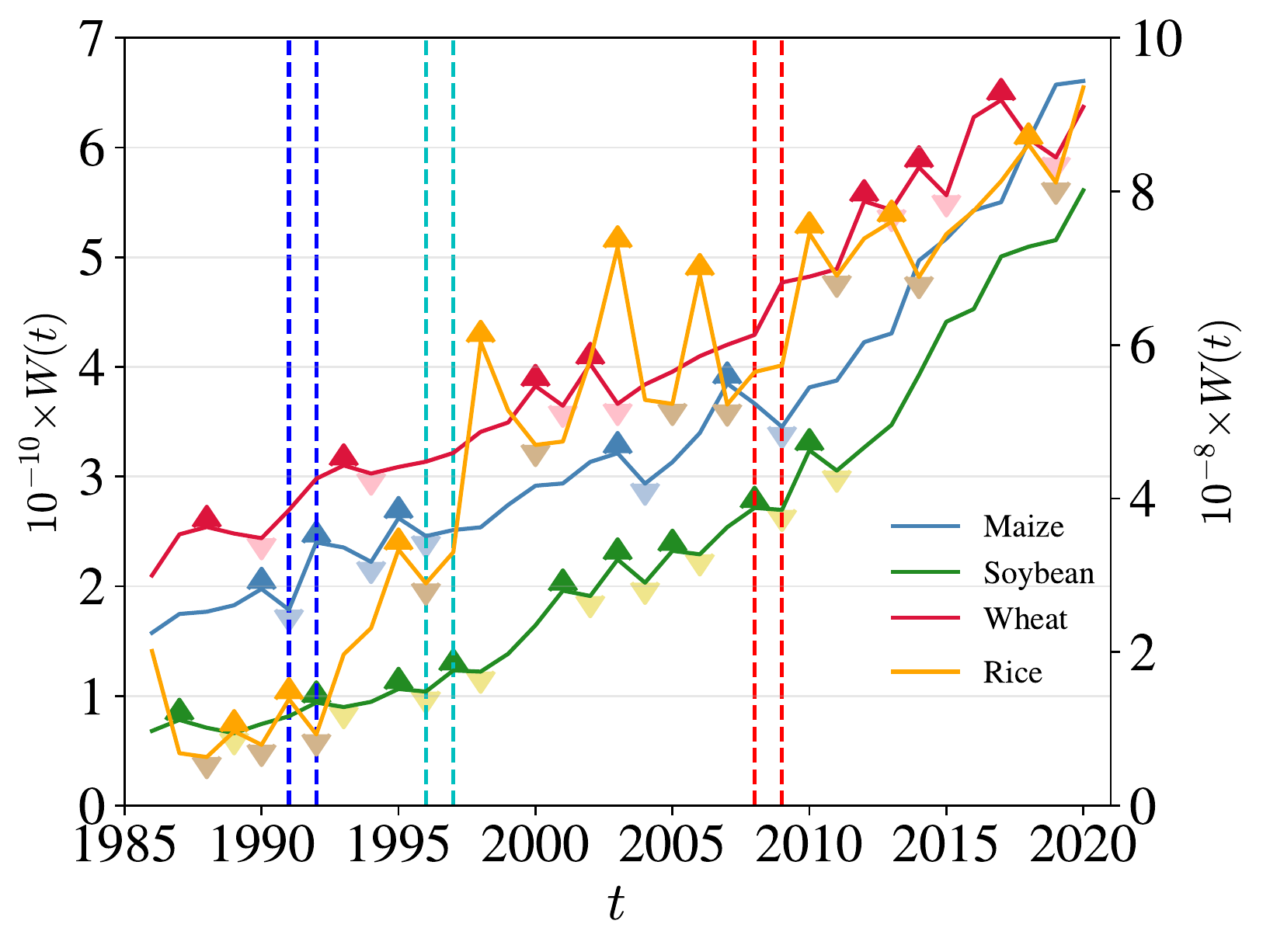}}
     \subfigure[]{\includegraphics[width=0.475\linewidth]{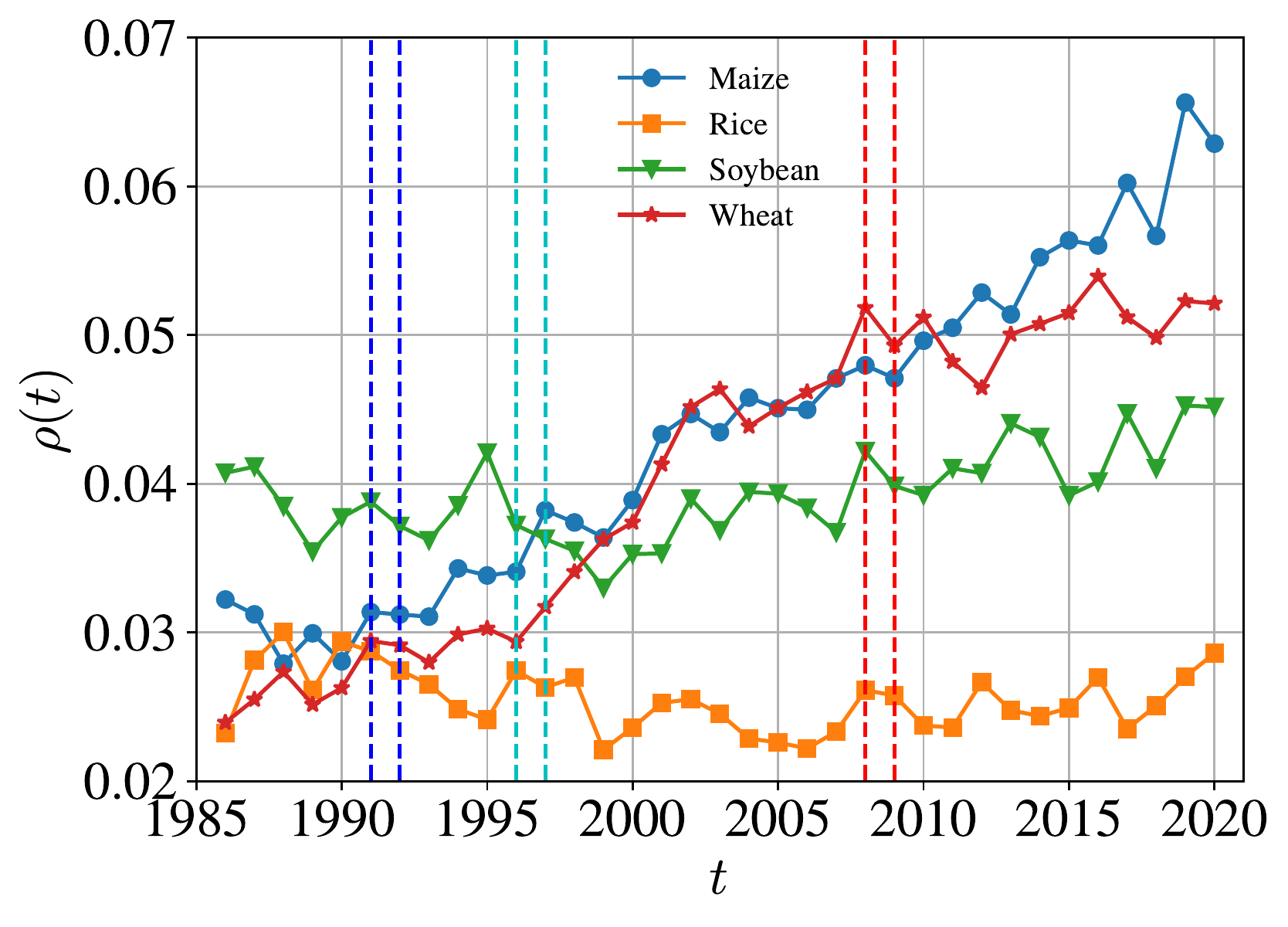}}
      \caption{Evolution of the summary statistics of the four iCTNs from 1986 to 2020. Curves in different colors are corresponding to different crops. The events corresponding to the dotted purple, blue and red lines are the dissolution of the Soviet Union (1991/1992), the Asian financial crisis (1996/1997) and the food price crisis (2008/2009).}
      \label{Fig:iCTN:sumW:density}
\end{figure}

To track the dynamics of the iCTNs, we displaying the weights of links. In this paper, the total weight $W(t)$ is defined as the caloric values of imports and exports across all economies, that is $W(t)=s_i^{\mathrm{in}}(t)+s_i^{\mathrm{out}}(t)$. The network density $\rho$ depicts overall tightness. In the iCTNs, the closer the density is to 1, the more frequent the trade activities among economies are. Figure~\ref{Fig:iCTN:sumW:density} shows the evolution of the weights and the density of four iCTNs from 1986 to 2020. Different colors stand for various types of crops. The up and down arrows indicate that the upward (downward) trend changes into the downward (upward) trend. Dotted lines with different colors depict three food-related crises. On the whole, the weights show an upward trend but experience small fluctuations, as presented in Fig.~\ref{Fig:iCTN:sumW:density}(a). We can observe that the dynamics of the weights vary across different crops. Notably, prior to 2018, wheat consistently held the top spot as the primary crop in the global food trade. In 2018-2020, maize's weight overtook wheat's. It is worth noting that the extent to which special events affect the weights is different in different iCTNs. The collapse of the Soviet Union in 1991 and the Asian financial crisis in 1996/1997 led to a decline in the trade volume of maize, soybeans, and rice, except for wheat. When looking at the evolution of network density, it is obvious that the links between economies were generally closer in the global food market from 1986 to 2020. Following the Southeast Asian financial crisis, the density of the international maize trade network has surpassed that of the international soybean trading network. Furthermore, the density increased significantly and ranked first following the 2008 food price crisis. {\color{red}{The evolutionary properties of degree, strength, clustering coefficients, et al. have been investigated \cite{Zhang-Zhou-2022-FrontPhysics}. We have also studied the degree distributions and found that the in-degree and out-degree distributions have different forms in different iCTNs \cite{Zhang-Zhou-2021-Entropy}. Since the import and export data of the international crop trade provided by FAO have been updated to 2020, we focus on the iCTNs in 2020, which are the latest trade networks.}}

\begin{figure}[h!]
    \centering
    \includegraphics[width=0.24\linewidth]{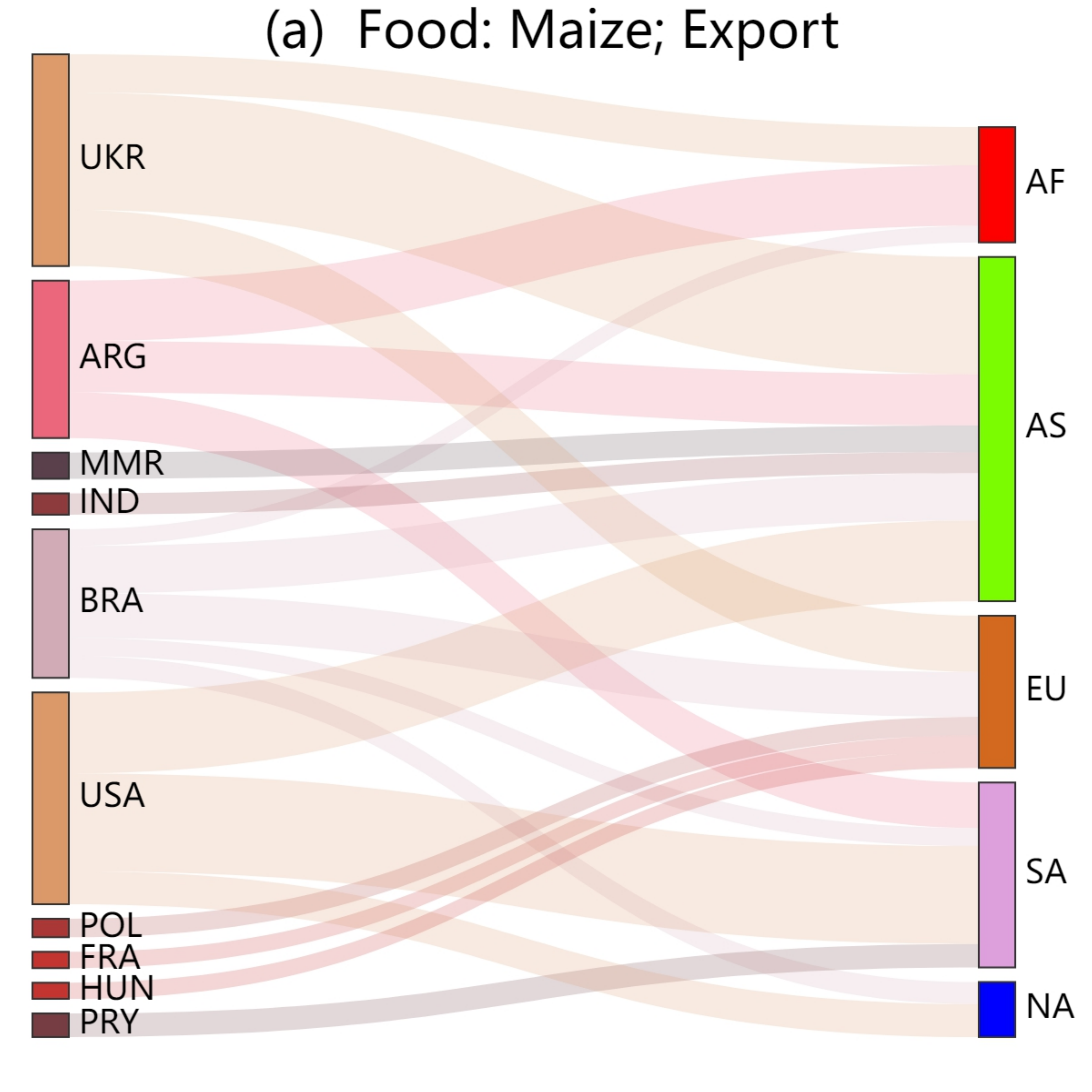}
    \includegraphics[width=0.24\linewidth]{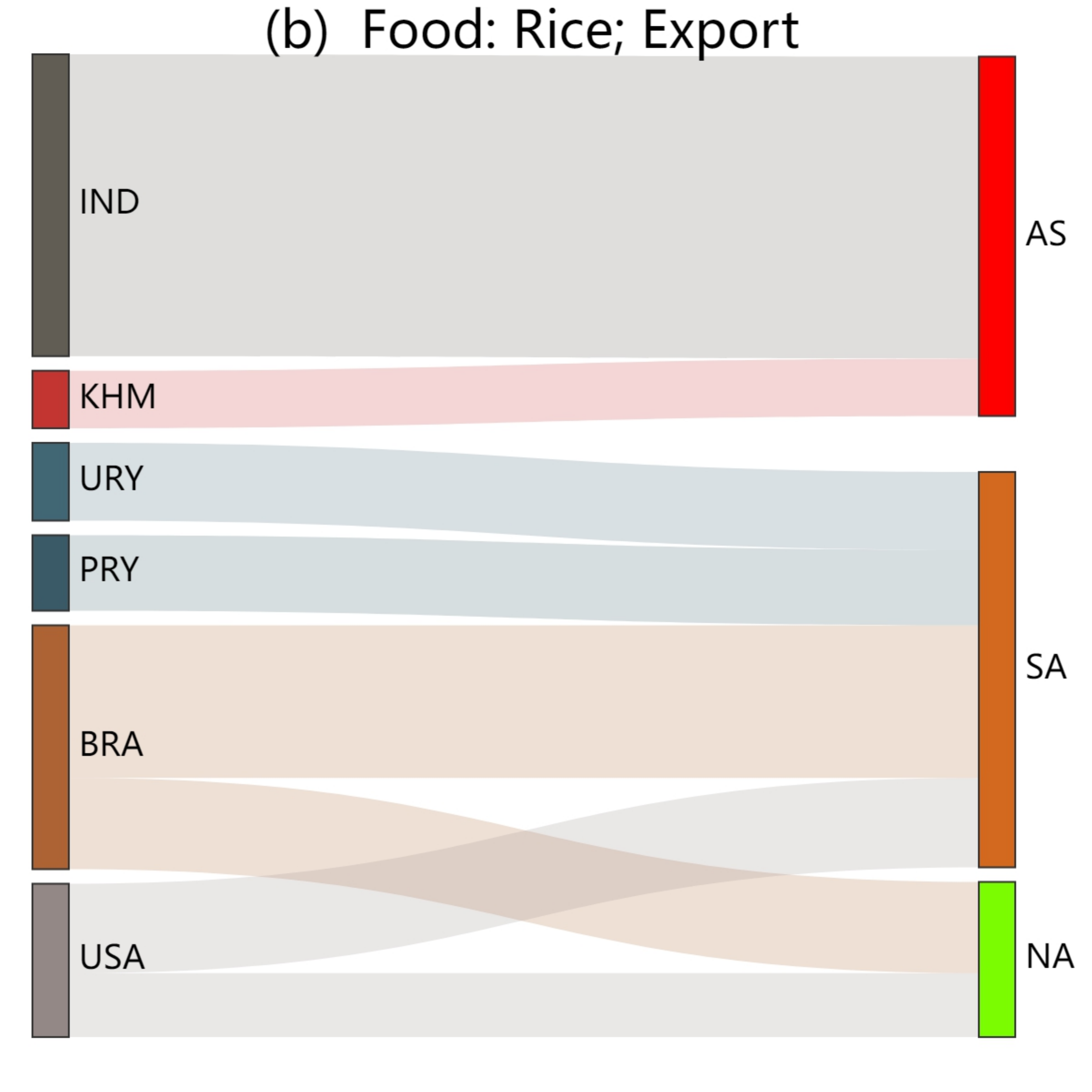}
    \includegraphics[width=0.24\linewidth]{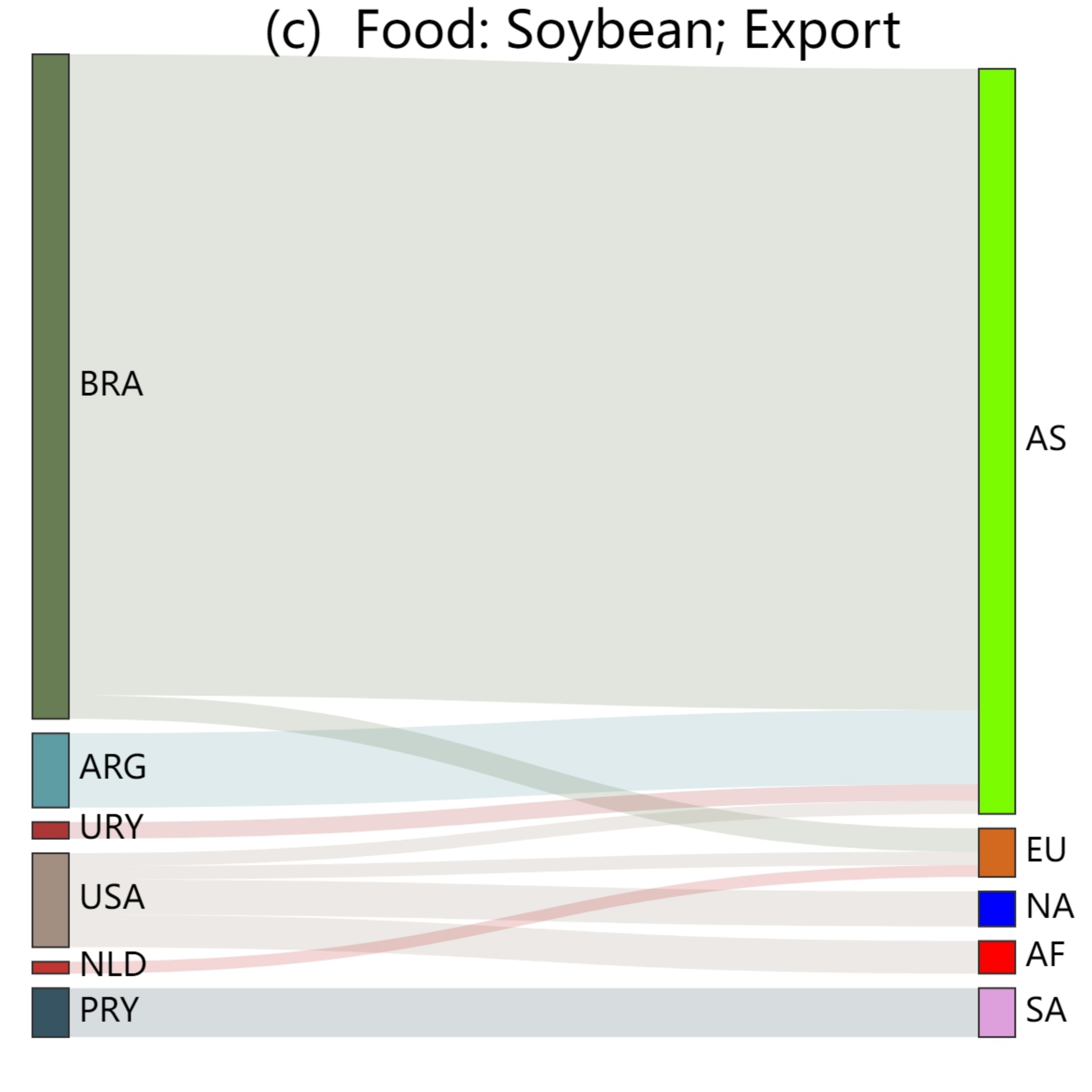}
    \includegraphics[width=0.24\linewidth]{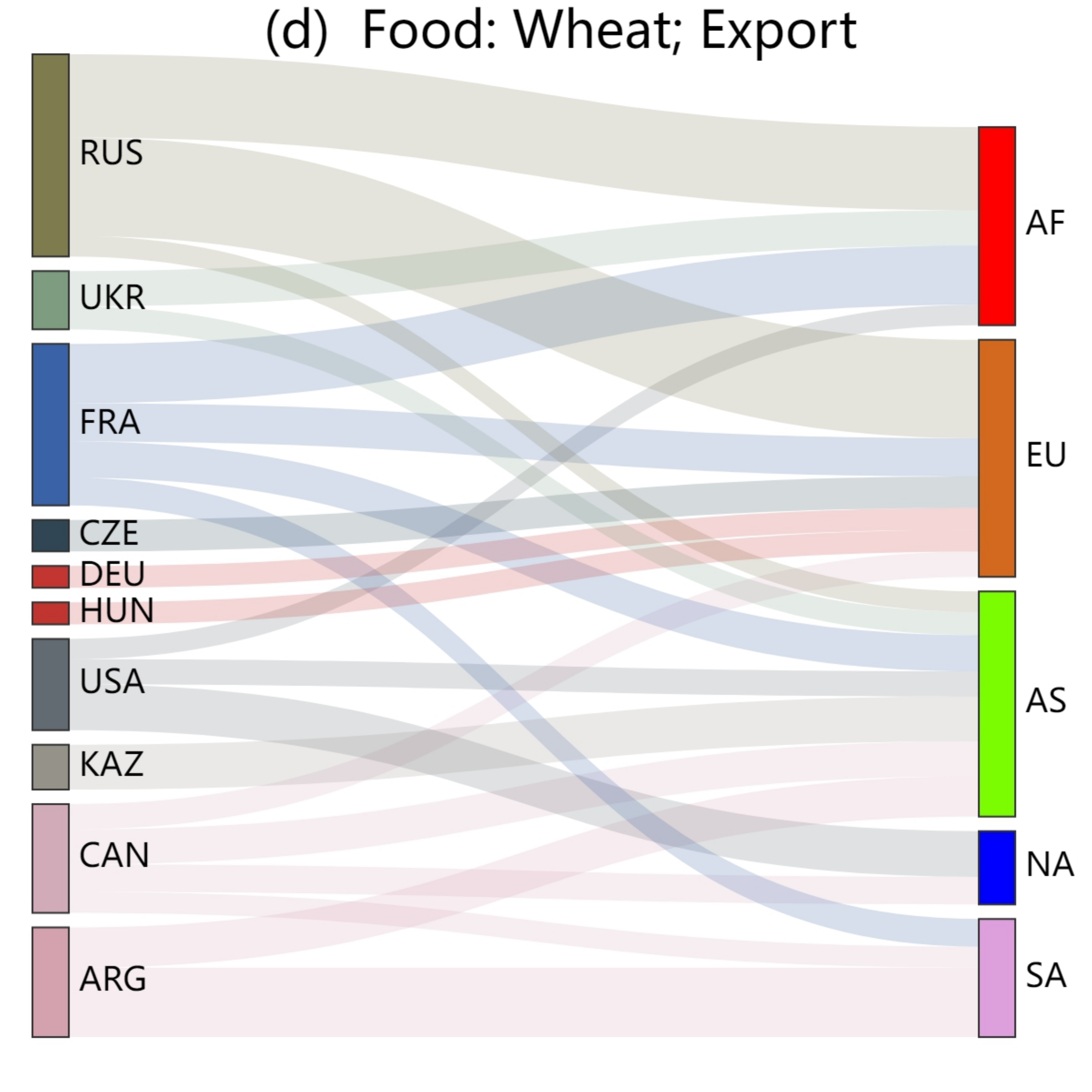}\\
    \includegraphics[width=0.24\linewidth]{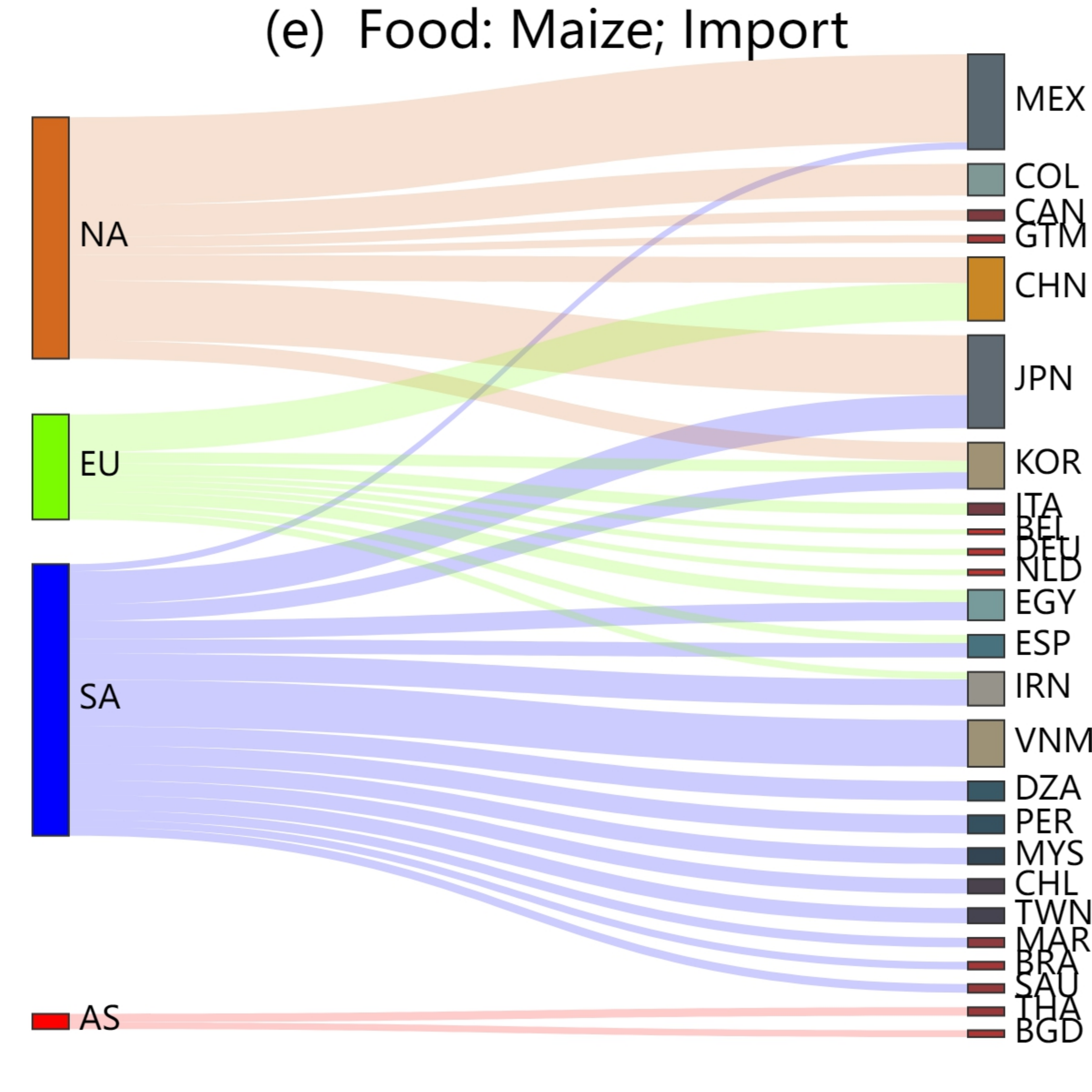}
    \includegraphics[width=0.24\linewidth]{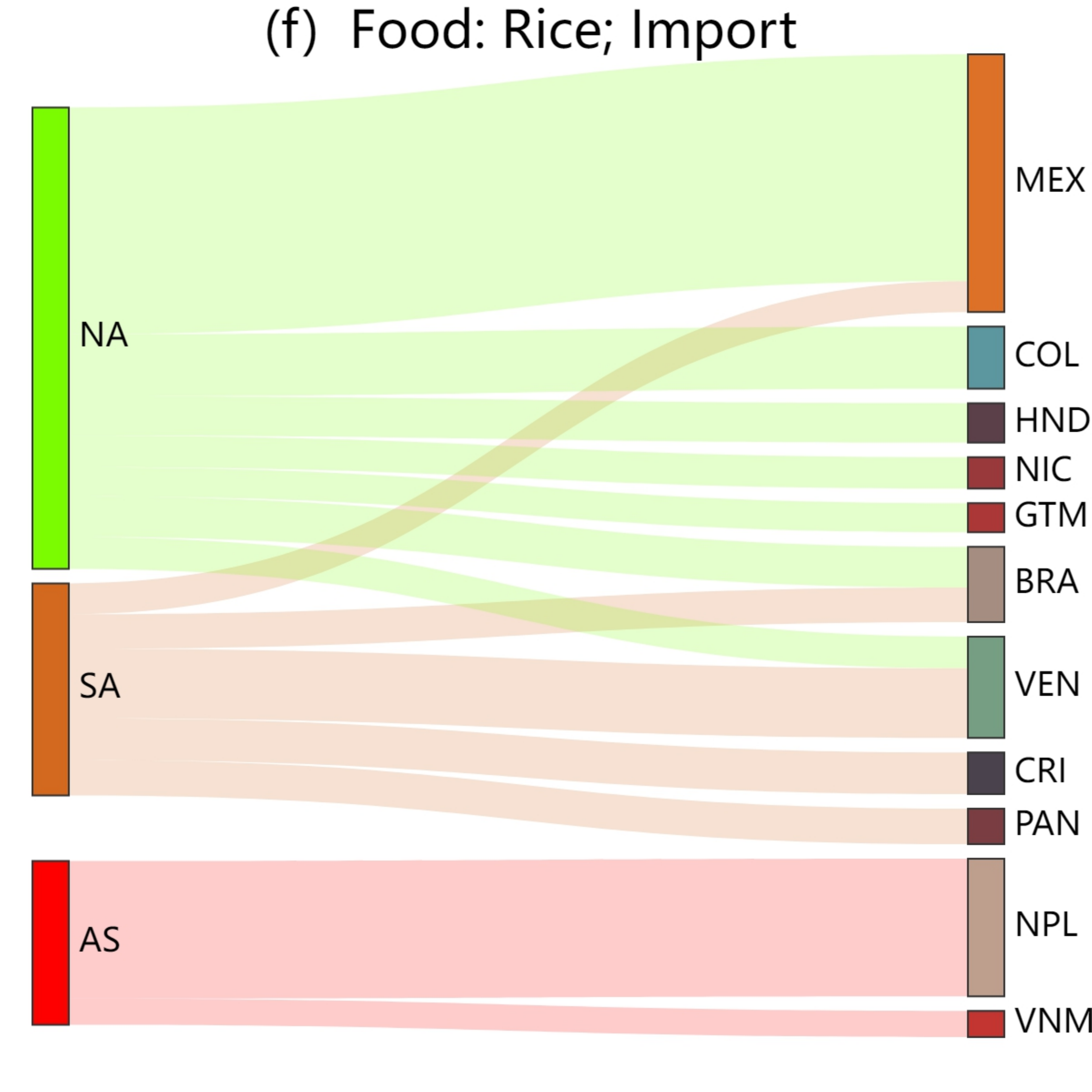}
    \includegraphics[width=0.24\linewidth]{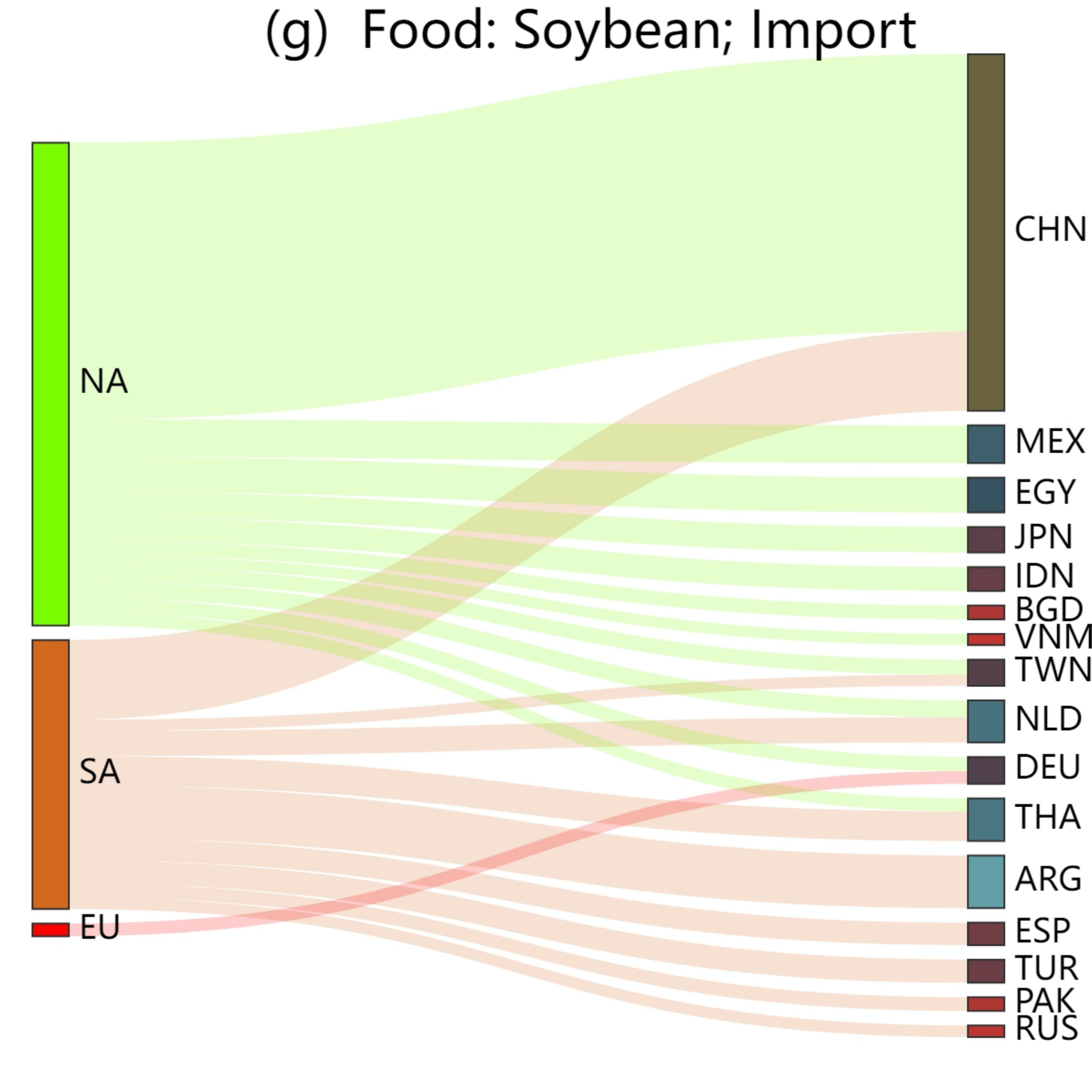}
    \includegraphics[width=0.24\linewidth]{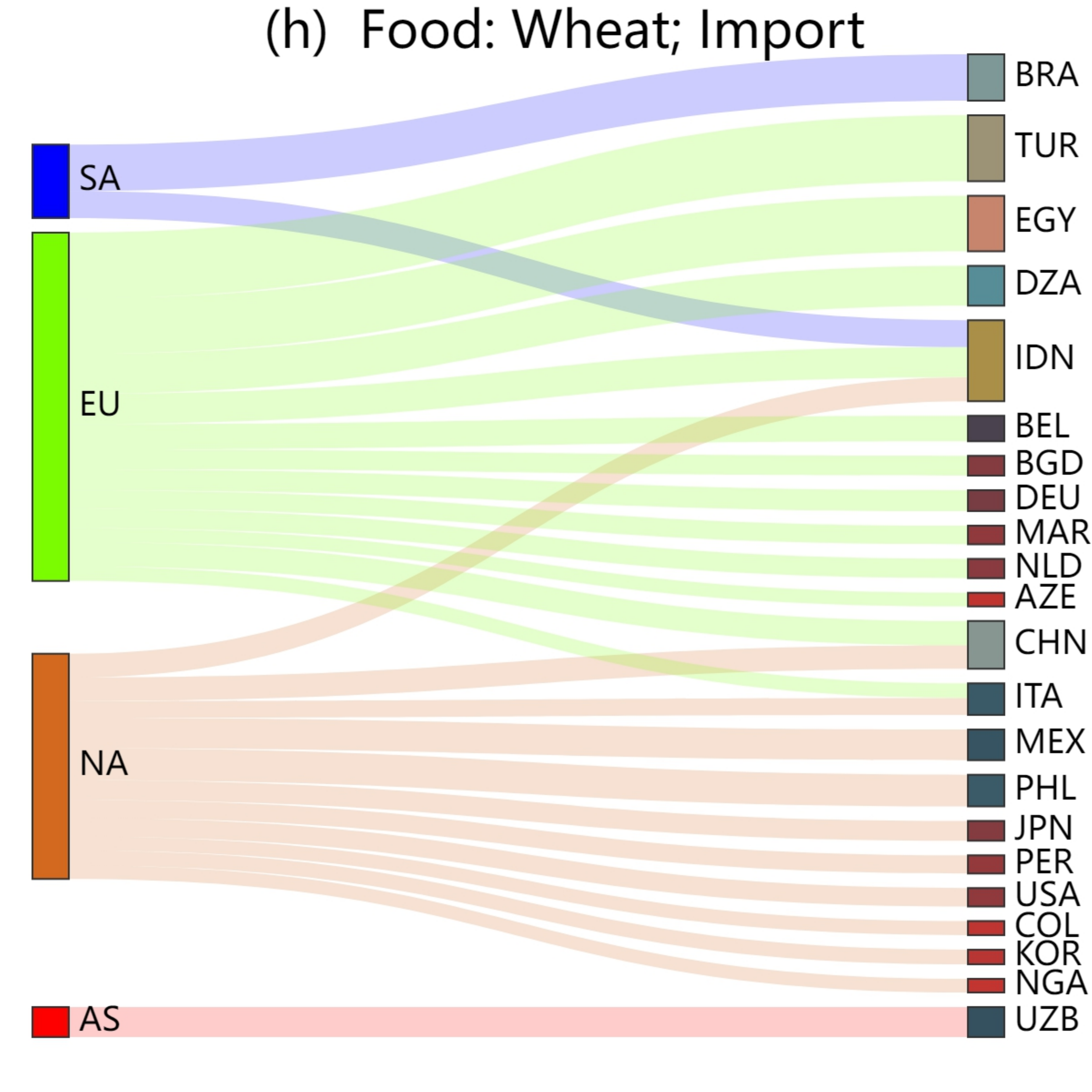}
    \caption{International crop trade flows in 2020. For clarity, we have shown only the links with high trade values, accounting for 2\% of the total number of links. The columns from left to right respectively describe maize, rice, soybean and wheat. The top row shows export trade calories from economies to corresponding regions. The bottom row shows import trade calories from regions to corresponding economies. The size of symbols means the values of trade calories. Outgoing links from an economy are shown with the same color as the origin region.Different colors of symbols correspond to different regions. Note that AF refers to Africa, AS refers to Asia, EU refers to Europe, NA refers to North America, and SA refers to South America.}
    \label{Fig:iCTN:TradeFlow:Sankey}
\end{figure}

Figure~\ref{Fig:iCTN:TradeFlow:Sankey} is the Sankey diagram of the regional distribution of imports and exports of the four iCTNs in 2020. Figure~\ref{Fig:iCTN:TradeFlow:Sankey}(a-d) and (e-h) respectively depict major economies' food export flows and import sources. Asia is the main food-importing region, while North America plays an important role in food exports. The regional patterns of food trade vary across crop sectors. The U.S. is an important exporter of the four crops, while Saudi Arabia has a significant impact on the import of maize and rice. Besides, China is a major maize and soybean importer.
 
Overall, the international food trade shows a booming trend with increasingly strong trade links. However, different crops have different trade patterns, and different economies have different impacts on the iCTNs. This motivates our focus on the trade influences of economies in different food trade networks in the subsequent study.

\subsection{Economic importance metrics}

\subsubsection{Economic importance metrics based on the local structure of the iCTNs}

There are many metrics to measure the importance of nodes in a network \cite{Hao-Manuel-Matus-Zhang-Zhou-2009-PhysRep}. The larger the degree indicates that the node is more important in the network. However, the degree of a node in a large network is often too large to measure the node importance intuitively. \textbf{Degree centrality} is used as a direct measure. The greater the degree centrality of a node, the more neighbors it has in the network and the greater its influence \cite{Garlaschelli-Loffredo-2005-PhysicaA,Ou-Guo-Liu-2022}. It is simply defined as
\begin{equation}
   DC_i=\frac{k_i}{N_\mathscr{V}-1}.
   \label{Eq:centrality:degree}
\end{equation}
Since the iCTNs we consider are directed networks, we use $\textbf{in-degree}$ and $\textbf{out-degree centrality}$ taking into account the directions of links \cite{Du-Tang-Qi-Wang-Han-Yang-2020-PhysicaA}:
\begin{equation}
   DC_i^{\mathrm{in}}=\frac{k_i^{\mathrm{in}}}{N_\mathscr{V}-1},
   \label{Eq:centrality:degree:in}
\end{equation}
and
\begin{equation}
   DC_i^{\mathrm{out}}=\frac{k_i^{\mathrm{out}}}{N_\mathscr{V}-1}.
   \label{Eq:centrality:degree:out}
\end{equation}

Degree centrality lays emphasis on the node itself. It considers that the more neighbors a node connects to, the greater the node's role in risk propagation, but neglects the node's neighborhood and the topology of the network. $\textbf{Semi-local centrality}$, on the other hand, considers that the influence of a node is also related to the influence of its neighbors \cite{Chen-Lu-Shang-Zhang-Zhou-2012-PhysicaA}. It is defined as,
\begin{equation}
    LC_i=\sum_{u \in \tau(i)} \sum_ {v \in \tau(u)} k_v,
    \label{Eq:centrality:Semi_local}
\end{equation}
where $\tau(i)$ denotes the set of nearest neighbor nodes of node $i$ and $k_v$ denotes the number of neighbors of node $v$. In directed networks, it is possible to construct $\textbf{in-semi-local}$ and $\textbf{out-semi-local centrality}$, 
\begin{equation}
    IL_i=\sum_{u \in \tau^{\mathrm{in}}(i)} \sum_ {v \in \tau^{\mathrm{in}}(u)}k_v^{\mathrm{in}},
    \label{Eq:centrality:Semi_local:in}
\end{equation}
\begin{equation}
    OL_i=\sum_{u \in \tau^{\mathrm{out}} (i)} \sum_ {v \in \tau^{\mathrm{out}}(u)}k_v^{\mathrm{out}},
    \label{Eq:centrality:Semi_local:out}
\end{equation}
where $\tau^{\mathrm{in}}(\cdot)$ and $\tau^{\mathrm{out}} (\cdot)$ respectively calculate the set of economies which food import from and which food export to for a given economy. $k_v^{\mathrm{in}}$ and $k_v^{\mathrm{out}}$ are respectively the in-degree and out-degree of node $v$.

The importance of a node depends on both the number and the importance of its neighboring nodes, so we use $\textbf{eigenvector centrality}$ to measure the influence of a node, where the eigenvector is the eigenvector corresponding to the maximum eigenvalue of the network adjacency matrix \cite{Oehlers-Fabian-2021-Mathematics}. It means,
\begin{equation}
\begin{aligned}
\centering
\mathbf{A} \cdot \mathbf{e} &=\lambda \mathbf{e},
\end{aligned}
\label{Eq:eigenvector}
\end{equation}
The eigenvector of maximum eigenvalues is given by solutions to the recursive equation,
\begin{equation}
\begin{aligned}
e_i &=\max _{r} \frac{\sum_{s \in \mathscr{V}_{i}} e_{s}^{r-1}}{\sum_{k \in \mathscr{V}} e_{k}^{r}}, \quad \text { with } \quad e_{i}^{0}=1,
\label{Eq:final:eigenvector}
\end{aligned}
\end{equation}
where $A=\{a_{ij}\}$ is the network adjacency matrix, $\mathbf{e}$ is the eigenvector corresponding to the largest eigenvalue, $e_{i}$ is the $i^{th}$ element in the vector $\mathbf{e}$, $\mathscr{V}_{i}$ is the set of neighboring nodes of node $i$, and $r$ is the number of iterations, thus we get the node $i$'s Eigenvector centrality. That is,
\begin{equation}
EC_i=e_i,
\label{Eq:Centrality:Eigenvector}
\end{equation}

$\textbf{Katz centrality}$ is another metric for measuring network centrality and is commonly used in biological fields, for example, to assess the potential neural activity of neurons. It not only takes into account the centrality of the node itself, but also the centrality of its neighboring nodes \cite{1953-Katz-Psycho}. Katz centrality measures the importance of a node $i$ by counting the number of neighboring nodes of the node $i$ and the number of all nodes connected to the node $i$ that pass through these neighboring nodes \cite{Peter2014-ProcRSocA-MathPhysEngSci,Li-Chu-Qiao-2020-NeuralNetw},
\begin{equation}\label{Eq:centrality:katz}
KC_{i}=\sum_{k=1}^{\infty} \sum_{j=1}^{N_\mathscr{V}} \alpha^{k} a_{ij}^{k},
\end{equation}
where $a_{ij}^{k}$ is the number of nodes passed from node $i$ to node $j$ along a path of length $k$, and $\alpha$ is the decay factor satisfying $0 <1/\lambda_{\max}<1$ ($\lambda_{\max}$ is the maximum eigenvalue of the network adjacency matrix), which is set to 0.01 here. For node $j$, the farther it is from node $i$, the smaller its contribution to the Katz centrality of node $i$.

$\textbf{PageRank}$ (PG) algorithm was proposed as an algorithm to calculate the importance of Internet pages, and then it was widely used to identify nodes in a directed network that have an important influence on the network structure. In the international food trade networks, the larger the value of a economy's PageRank indicates that the more food import trade partners the economy has and the more food it exports to other economies. It suggests that the economy play a crucial role in the trade network. PageRank can be calculated by a simple iterative algorithm and corresponds to the principal eigenvector of the normalized network connectivity matrix \cite{Brin-Page-1998-ComputNetwISDNSyst}.

Two metrics are introduces in the Hyperlink-Induced Topic Search (HITs) algorithm: $\textbf{Authorities}$ (AU) and $\textbf{Hubs}$ (HU). Similar to PageRank, the HITs algorithm was once used to rank web pages \cite{Kleinberg-1999-JACM}. {\color{red}{A node with a high authority value is pointed to by numerous other nodes with high hub values, while a node with a high hub value is pointed to numerous other nodes with high authority values. This approach has been utilized in a number of economic sectors as a tool for characterizing the network structure of web pages. The eigenvector for the matrix $A^TA$ in a network is the HITS authority vector, while the eigenvector for the matrix $AA^T$ is the HITS hub vector. Each economy in the iCTNs has an authority value and a hub value, 
which are related to the directed edges of the economy and other economies.}} Therefore, by calculating the Authority and Hub, a node's importance can be determined.

The $\textbf{clustering coefficient}$ (CC) is used to describe the aggregation of nodes in a network. The higher the value, the more important the node is in the network. A node's clustering coefficient reflects the likelihood of connectivity with its neighbors. The clustering coefficient of a node in a directed weighted network can be determined in a number of different methods. In this paper, it is described as the ratio of the total number of directed triangles that node $i$ has actually formed to the total number of directed triangles that node $i$ could form \cite{Fagiolo-2007-PhysRevE}, i.e.
\begin{equation}
    CC_i=\frac{1}{k_i(k_i-1)-2k_i^{R}} \sum_{sr}(\hat{w}_{is} \hat{w}_{sr} \hat{w}_{ri})^{1/3},
    \label{Eq:DiGraph:Clustering:Node}
\end{equation}
where $k_i$ is the total degree of node $i$ ($k_i=k_i^{\mathrm{in}}+k_i^\mathrm{out}$), node $s$ and $r$ are neighbors of node $i$, and $\hat{w}_{sr}= \frac {w_{sr}}{\max(w)}$, $\max(w)$ means the maximum of network weights.

\subsubsection{Economic importance metrics based on the global structure of the iCTNs}

The global features of the node are taken into account in the evaluation since the significance of a node in a network is tied to the network's overall structure.

$\textbf{Betweenness centrality}$ is the ratio of all shortest pathways between two nodes that travel through a specific node to all other shortest paths \cite{Freeman-1977-Sociometry}. It gauges an economy's capacity to function as an intermediary in the food trade networks, that is as a trade conduit between different countries or regions. An economy's importance in the global food trade networks increases with its betweenness centrality. It is calculated as
\begin{equation}
BC_i=\sum_{p,q \in \mathscr{V}, p \neq q \neq i }\frac{h_{pq}^i}{g_{pq}},
\label{Eq:centrality:betweenness}
\end{equation}
where $h_{pq}^i$ is the number of shortest paths from $p$ to $q$ through $i$, and $g_{pq}$ is the total number of all shortest paths from $p$ to $q$.

$\textbf{Closeness centrality}$ is used to measure the proximity between a node and other nodes \cite{Qi-Fuller-Wu-Wu-Zhang-2012-InfSci}. If a node is closer to other nodes, then the less it needs to rely on others when disseminating information. In the food trade networks, the greater the closeness centrality of an economy indicates that the economy is less likely to be constrained by other. We use the average shortest distance from node $i$ to all other nodes to represent the Closeness centrality. In a directed network, we can define $\textbf{in-closeness centrality}$ (IC) and $\textbf{out-closeness centrality}$ (OC) as
\begin{equation}
IC_i=\frac{D_i^{\mathrm{in}}}{N_\mathscr{V}-1}\frac{D_i^{\mathrm{in}}}{\sum _{j=1}^{D_i^{\mathrm{in}}}d_{ji}},
\label{Eq:centrality:closeness:in}
\end{equation}
\begin{equation}
OC_i=\frac{D_i^\mathrm{out}}{N_\mathscr{V}-1}\frac{D_i^\mathrm{out}}{\sum _{j=1}^{D_i^\mathrm{out}}d_{ij}},
\label{Eq:centrality:closeness:out}
\end{equation}
where $D_i^{\mathrm{in}}$ and $D_i^\mathrm{out}$ denote respectively the number of nodes that a node can reach $i$ and the number of nodes that a node $i$ can reach, and $d_{ji}$ and $d_{ij}$ denote the shortest path length from node $j$ to node $i$ or from node $i$ to node $j$, respectively.

The most influential nodes in a network with a community structure should have the following characteristics: serve as the community center and play a ``bridging'' role in each community. Therefore, the focus of the key node ranking problem should not be limited to the core nodes in the network, but also the nodes in the Structural Hole. Burt proposed the $\textbf{constraint index}$ (CI) in 1992, which represents the ratio of the contribution of node $i$ to its neighbors to all its contributions \cite{ Burt-2004-AmJSociol}. In the food trade networks, the higher constraint coefficient of a node, the greater the bridging role of the node in the community structure.

\subsubsection{Economic importance metrics based on the modular structure of the iCTN}

Community structure appears to be common in a social network \cite{Girvan-Newman-2002-ProcNatlAcadSciUSA}. The gathering nodes in a community show that there is a higher density of links between them than other nodes outside the community \cite{Clauset-Newman-Moore-2004-PhysRevE}. Modularity is an important feature of a network and a part of agglomeration algorithms for detecting community structure. It is plausible that nodes with the same role should perform analogous topological characters \cite{Guimera-Amaral-2005-Nature}. Thus we can extract the significant information about the role of nodes by considering properties of nodes within module. More often than not not, the role of a node in a network can be measured by its $\textbf{within-module degree}$ (IM) and $\textbf{participation coefficient}$ (PC) \cite{Guimera-Amaral-2005-Nature}. 

$\textbf{Within-module degree}$ (IM) measures the extent of connection of a node to other nodes in their module \cite{Guimera-Amaral-2005-Nature}, which is suitable for portraying the relationship between economies and gauging an economy's influence within the module. For a given network, within-module degree is:
\begin{equation}
IM_{i}=\frac{\kappa_{i,m_i}-\overline{\kappa}_{m_i}}{\sigma_{\kappa_{m_i}}},
\label{Eq:module:within}
\end{equation}
where $\kappa_{i,m_i}$ is the number of links of node $i$ to other nodes in its module $m_i$, $\overline{\kappa}_{m_i}$ is the average of $\kappa_i$ over all nodes in $m_i$, and $\sigma_{\kappa_{m_i}}$ is the standard deviation of $\kappa_{i}$ in $m_i$. The value of $IM_{i}$ increases with the number of trade links of the economy $i$ to other economies in its module $m_i$. It indicates that the economy plays a pivotal role in the modular structure.

Occasionally, nodes with the same $IM_{i}$ may arise different roles. Participation coefficient of node $i$ is used to measure the uniformly distributed of its links among all the modules \cite{Guimera-Amaral-2005-Nature},
\begin{equation}
PT_{i}=1-\sum_{m=1}^{N_{M}}\left(\frac{\kappa_{i,m_i}}{k_{i}}\right)^{2},
\label{Eq:module:participant}
\end{equation}
where $N_{M}$ is the number of modules involving node $i$.

To measure how well-influence of nodes outside the module to node $i$, we apply $\textbf{outside-module degree}$ (OM) \cite{Xu-Pan-Muscoloni-Xia-Cannistraci-2020-NatCommun},
\begin{equation}
OM_{i}=\frac{\kappa_{i,m_i}^{'}-\overline{\kappa}^{'}_{m_i}}{\sigma_{\kappa_{m_i}}},
\label{Eq:module:outside}
\end{equation} 
similarly, where $\kappa_{i,m_i}^{'}$ is the number of links of node $i$ to other economies outside its module $m_i$, $\overline{\kappa}_{m_i}$ is the average of $\kappa_i$ over all nodes outside $m_i$.

\subsubsection{Economic importance metrics based on mutual information metric}

The importance of a node depends on not only the network topological structure, but also the strength of the connection between nodes. Therefore in this paper we consider $\textbf{mutual information}$ metric that is an essential point in Information theory to access the ability of nodes to transit information. Shannon argues that information is something useful obtained by message receiver through passing messages , thus eliminating the ``uncertainty'' that existed before the communication \cite{Borst-Theunissen-1999-NatNeurosci}. It is widely believed that the iCTNs are complex communication systems. We can measure the importance of nodes based on information theory. It is plausible to surmise that the measure of a node's importance in a network can be fulfilled according to the amount of information contained in the node \cite{Liu-Jin-Zhang-Xu-2014-JSupercomput}.

For a undirected and unweighted network, mutual information of a node depends on node degree,
\begin{equation}
I_{ij}= \begin{cases}\ln k_{i}-\ln k_{j}, &j
\in \mathscr{V}_i \\ 0, & \text { others }\end{cases}.
\label{Eq:mutual:information:node}
\end{equation}
Information of node $i$ is the sum of mutual information between it and other node, that is,
\begin{equation}
MI_i=\sum_{j=1}^{N} I_{ij},
\label{Eq:mutual:information:undirected}
\end{equation}

Since the iCTNS are directed and weighted, the directions and weights of links are crucial. Based on the above method, we consider node information of a directed and weighted network. A node may be both a transmitting and a receiving node at the same time. Therefore, we define the probability of out-links and in-links of node $i$. Since $w_{ij}$ is the weight of the link from node $i$ to $j$, the probability of out-links is the ratio of $w_{ij}$ to all weighted links transiting information from node $i$ that is equal to out-strength of node $i$. We calculate the probability of out-links as follows,
\begin{equation}
    p^\mathrm{out}_{i}= \frac{w_{ij}}{s^\mathrm{out}_i}.
\end{equation}
Similarly, we obtain the probability of in-links as follows,
\begin{equation}
    p^{\mathrm{in}}_{j}= \frac{w_{ij}}{s^{\mathrm{in}}_j}.
\end{equation}
Then the mutual information $I_{ij}^w$ from node $i$ to $j$ is,
\begin{equation}
   MI_{ij}= \begin{cases}\ln \frac{s^\mathrm{out}_i}{w_{ij}}-\ln \frac{s^{\mathrm{in}}_j}{w_{ij}}, &j
   \in \mathscr{V}_i^\mathrm{out} \\ 0, & \text {others}
\end{cases}.
\label{Eq:mutual:information:weighted:node}
\end{equation}
So, the information embraced in node $i$ is the sum of the mutual information from node $i$ to all $i$-pointing nodes minus the sum of the mutual information from all nodes pointing to $i$,
\begin{equation}
  MI_i=\sum_{j \in \mathscr{V}^\mathrm{out}_i} I_{ij}-\sum_{s \in \mathscr{V}^{\mathrm{in}}_i} I_{si}.
\label{Eq:mutual:information:weighted}
\end{equation}
It goes without saying that nodes possessing more information are more important in a network.  {\color{red}{In the iCTNs, mutual information is transmitted between economies $i$ and $j$ if there is a link between them. If there is no link, their mutual information is 0. As a result, the quantity of reciprocal information both reflects the connection and importance of economies in the trade network.}}

\subsection{Correlation structure of node importance metrics}

Random matrix theory is often used to study financial markets \cite{Dai-Xiong-Zhou-2021-FinancResLett}. Here, we use it to analyze the correlation structure between different node importance metrics. Based on the above 20 node importance metrics, we obtained the rankings of trade influence of selected economies in four iCTNs from 1986-2020. We set the rankings of an economy $i$ for a given iCTN in time $t$ as $NI_i^{crop}(t)$ and analyze the correlation structure of different ranking based on each importance metric. Our approach works for all iCTNs in any time $t$, so time $t$ and crop type $\mathrm{C}$ are ignored to predigest the formula.

We obtain a matrix of node importance rankings for each year of the four iCTNs from 1986-2020 and calculate the correlation coefficients of importance rankings among different nodes,
\begin{equation}
\label{Eq:C:correlation}
c_{pq}(i)=\left\langle r_{p}(i) r_{q}(i)\right\rangle,
\end{equation}
where $r_{p}(i)$ is the importance ranking of $i$-th economy based on the $p$-th node importance index.

By resolving the following equation, we procure the eigenvalues and related eigenvectors of the correlation coefficient matrix $\mathbf{C}$ for various grains across various years,
\begin{equation}
    \mathbf{C}=\mathbf{U} \Lambda \mathbf{U}^{\mathrm{T}},
    \label{Eq:C:eigenvalue}
\end{equation}
where $\mathbf{U}$ represents matrix of the eigenvectors $u_i$, $\Lambda$ represents the diagonal matrix of the eigenvalues $\lambda_i$, $\mathbf{U}^{\mathrm{T}}$ is the transpose of $\mathbf{U}$. The probability density function $\Lambda$ is  
\begin{equation}
  f_{\mathbf{C}}(\lambda)=\frac{1}{M} \frac{\mathrm{d} n(\lambda)}{\mathrm{d} \lambda},
  \label{Eq:C:PDF:eigenvalue}
\end{equation}
where $n(\lambda)$ means the number of eigenvalues which are smaller than $\Lambda$ of the matrix $C$. Thus, we construe a random matrix as, 
\begin{equation}
    \mathbf{R}=\frac{1}{T} \mathbf{X} \mathbf{X}^{\mathrm{T}},
    \label{Eq:RMT:define}
\end{equation}
where ${\mathbf{X}}$ is an $M \times N$ matrix that incorporates $M$ nodes and $N$ random elements with zero mean and unit variance. Here $N=20$, and $M$ is the number of nodes in the network each year. When $M \rightarrow \infty$ and $N \rightarrow \infty$, if $Q=N / M>1$, the probability density function $f_{\mathrm{RMT}}(\lambda)$ of $\lambda$ of $R$ is given by \cite{Edelman-1988-SIAMjmaa,Laloux-Cizeau-Bouchaud-Potters-1999-PhysRevLett,Sengupta-Mitra-1999-PhysRevE},
\begin{equation}
    f_{\mathrm{RMT}}(\lambda)=\frac{Q}{2 \pi} \frac{\sqrt{\left(\lambda_{\max }^{\mathrm{RMT}}-\lambda\right)\left(\lambda-\lambda_{\min }^{\mathrm{RMT}}\right)}}{\lambda},
    \label{Eq:RMT:PDF:eigenvalue}
\end{equation}
where $\lambda \in\left[\lambda_{\min }^{\mathrm{RMT}}, \lambda_{\max }^{\mathrm{RMT}}\right]$, and $\lambda_{\min }^{\mathrm{RMT}}$ and $\lambda_{\max }^{\mathrm{RMT}}$ are respectively the minimum and maximum eigenvalues of $\mathbf{R}$, which can be calculated as follows,
\begin{equation}
    \lambda_{\min }^{\mathrm{RMT}}=1+\frac{1}{Q}-2 \sqrt{\frac{1}{Q}}~,
    \label{Eq:RMT:eigenvalue:min}
\end{equation}
and
\begin{equation}
\label{Eq:RMT:eigenvalue:max}
\lambda_{\max}^{\mathrm{RMT}}=1+\frac{1}{Q}+2\sqrt{\frac{1}{Q}}~.
\end{equation}

\subsection{Comprehensive node importance metric}

The inverse participation ratio (IPR) is used to describe the importance of elements of the matrix \cite{Fyodorov-Mirlin-1992-PhysRevLett,Plerou-Gopikrishnan-Rosenow-Amaral-Stanley-1999-PhysRevLett}, which measures the contribution of the $l$th element of $u^{k}$ to the eigenvectors. Likewise, IPR has also been used to measure the degree to which the eigenvector deviates from the RMT results. The deviation of the eigenvector is inversely correlated with IPR, meaning that the lower the IPR, the higher the eigenvector vector deviation. The inverse participation ratio $I^{k}$ is given by,
\begin{equation}
   I^{k}=\sum_{l=1}^{M}\left[u_{l}^{k}\right]^{4},
   \label{Eq:RMT:IPR}
\end{equation}
where $\left\{u_{l}^{k}: l, k=1, \cdots, M\right\}$ denote the $l$-th components of the $k$-th eigenvector $u^{k}$. If all the eigenvector elements in $u^{k}$ are equal and satisfy $u^k_l=1/\sqrt{M}$, then $I^{k}=1/M$. When one of the eigenvector elements of $u^{k}$ fulfills $u^k_l=1$ and all other eigenvector elements of $u^{k}$ are equal to 0, then $I^{k}=1$. As a result, the inverse participation ratio (IPR) can determine whether the number of eigenvector elements' inverse is noticeably non-zero.

In the financial markets, eigenportfolios are frequently built for each eigenvalue to reflect changes in the returns of a composite portfolio \cite{Dai-Huynh-Zheng-Zhou-2022-ResIntBusFinanc}. On this basis, we develop the composite index for assessing an economy's importance in global food trade networks. By constructing the corresponding eigenportfolio, we can obtain the composite importance index for an economy $i$ given the eigenvalue $lambda_k$,
\begin{equation}
G^{k}(i)=\sum_{j=1}^{M} u_{j}^{k} r_{j}(i) / \sum_{j=1}^{M} u_{j}^{k},
\end{equation}
where $\sum_{j=1}^{M} u_{j}^{k} r_{j}(i)$ is the projection of the economy's importance ranking series $r_{j}(t)$ on the $k$-th eigenvector $u_{j}^{k}$. The optimal eigenportfolio is found by scrutinizing the relationship between economies' importance rankings based on different eigenportfolios and average rankings $\left\langle r_{j}(i) \right\rangle$. Finally, we equate an economy's composite importance ranking with the results of the optimal eigenportfolio.

\section{Empirical results}
\label{S3:Results}

\subsection{Ranking of economic influence in the iCTN}
\label{S3-1:Ranking}

\begin{table}[!ht]
    \setlength\tabcolsep{1.97mm}
    \centering
    \caption{Top 10 economies identified based on different influence indicators in 2020. ID (in-degree centrality), OD (out-degree centrality), IS (in-strength), OS (out-strength), BC (betweenness centrality), EC (eigenvector centrality), KC (Katz centrality), CC (clustering centrality), IC (in-closeness centrality), OC (out-closeness centrality), PR (PageRank), HU (hubs), AU (authorities), CI (constraint index), IL (in-semi-local centrality), OL (out-semi-local centrality), MI (mutual information), IM (within-module), PT (participation coefficient), OM (outside-module).}
   \label{Table:iCTN:node:influence}
    \smallskip
    \tiny
    \begin{tabular}{cccccccccccccccccccccccc}
         \toprule
       Rank & ID & OD & IS & OS & BC & EC & KC & CC & IC & OC & PR & HU & AU & CI & IL & OL & MI & IM & PT & OM \\  
        \midrule
        \multicolumn{5}{l}{Panel A: Maize}\\
        1 & NLD & USA & MEX & USA & USA & NLD & HKG & HTI & USA & NLD & USA & GBR & USA & MEX & NLD & USA & USA & ARG & NPL & USA \\   
        2 & GBR & ARG & JPN & ARG & ARE & DEU & USA & SLE & ARG & USA & ARG & IRL & BRA & JPN & FRA & ARG & ARG & USA & MAR & ARG \\  
        3 & USA & BRA & VNM & BRA & ZAF & FRA & MYS & CUB & FRA & FRA & BRA & NLD & UKR & COL & DEU & FRA & BRA & ZAF & JPN & ZAF \\ 
        4 & FRA & FRA & CHN & UKR & FRA & GBR & KEN & ERI & BRA & ARE & FRA & DEU & ARG & CHN & GBR & BRA & FRA & BRA & MUS & ARE \\   
        5 & DEU & ZAF & KOR & FRA & NLD & ESP & AUS & VUT & DEU & GBR & ZAF & NER & ZAF & KOR & ARE & DEU & NLD & FRA & NGA & BRA \\  
        6 & ARE & UKR & EGY & HUN & TUR & DNK & ARE & LCA & ESP & DEU & UKR & BEN & RUS & VNM & USA & TUR & UKR & IND & HKG & FRA \\   
        7 & CAN & TUR & ESP & ZAF & CAN & BEL & THA & FJI & TUR & CAN & TUR & TUR & BGR & CAN & ESP & ESP & ZAF & UKR & AUS & NLD \\  
        8 & BEL & ESP & COL & BGR & GBR & ARE & PHL & LUX & UKR & JPN & IND & BEL & FRA & TWN & BEL & UKR & CAN & ESP & THA & IND \\  
        9 & ITA & IND & IRN & RUS & IND & ITA & IDN & TJK & MEX & KWT & ESP & ESP & PRY & IRN & DNK & MEX & DEU & NLD & GUY & TUR \\ 
        10 & ESP & HUN & ITA & PRY & DEU & HUN & UGA & TKM & ITA & SGP & DEU & FRA & HUN & GTM & CAN & ITA & IND & TUR & NZL & DEU \\  
        \multicolumn{5}{l}{Panel B: Rice}\\
        1 & NLD & IND & MEX & USA & ARE & NLD & NLD & AFG & ITA & GBR & IND & ESP & USA & MEX & NLD & ITA & IND & IND & JPN & IND \\   
        2 & GBR & ITA & NPL & BRA & ITA & DEU & DEU & ALB & NLD & NLD & ITA & NLD & BRA & COL & DEU & NLD & BRA & CHN & URY & CHN \\ 
        3 & DEU & BRA & VEN & IND & GBR & BEL & BEL & OMN & DEU & ITA & USA & DEU & PRY & VEN & BEL & DEU & USA & USA & GEO & NLD \\  
        4 & ARE & NLD & BRA & URY & NLD & GBR & GBR & SYC & IND & ESP & CHN & PRT & URY & BRA & GBR & BEL & NLD & ARE & PAN & USA \\    
        5 & BEL & USA & CRI & PRY & USA & POL & POL & DOM & BEL & ARE & BRA & GBR & GUY & HND & ITA & BRA & ITA & PAK & LBN & GBR \\   
        6 & ITA & DEU & COL & GUY & BRA & ITA & ITA & SVK & BRA & DEU  & PAK & BEL & RUS & NIC & POL & GBR & ARE & ZAF & LKA & ITA \\    
        7 & ESP & CHN & TUR & KHM & IND & DNK & DNK & LUX & GBR & IRL & ARE & IRL & BGR & GTM & ISL & IND & CHN & NLD & MRT & THA \\   
        8 & FRA & ARE & GTM & GRC & CHN & ISL & ISL & EST & ESP & FRA & GBR & TUR & GRC & CRI & IRL & ESP & ESP & ITA & GMB & ARE \\ 
        9 & ISL & TUR & HND & RUS & TUR & HUN & IRL & SVN & FRA & BEL & LKA & AUT & ITA & SLV & ESP & FRA & GBR & BRA & CYP & DEU \\ 
        10 & CHE & GBR & PAN & BGR & DEU & IRL & HUN & FIN & PRT & NOR  & BEL & ITA & ESP & TUR & FRA & PRT & PAK & ESP & NGA & JPN \\ 
        \multicolumn{5}{l}{Panel C: Soybean}\\
      1 & NLD & USA & CHN & BRA & USA & NLD & NLD & YEM & CAN & NLD & USA & ESP & BRA & CHN & NLD & CAN & USA & USA & SWE & USA \\   
        2 & DEU & CAN & ARG & USA & CAN & DEU & DEU & HND & USA & FRA  & CAN & PRT & USA & THA & DEU & USA & CAN & IND & JPN & CAN \\  
        3 & FRA & CHN & NLD & ARG & IND & FRA & AUT & JAM & CHN & DEU & CHN & DEU & ARG & NLD & FRA & CHN & NLD & CAN & JOR & NLD \\  
        4 & USA & BRA & EGY & PRY & CHN & ITA & HUN & PAN & BRA & ESP & BRA & FRA & URY & ESP & ITA & BRA & BRA & NLD & HRV & FRA \\ 
        5 & ESP & UKR & THA & CAN & NLD & ESP & POL & LBY & UKR & GBR & IND & CHN & RUS & TUR & ESP & UKR & DEU & UKR & CHN & DEU \\ 
        6 & ITA & FRA & MEX & URY & FRA & AUT & ITA & SUR & AUT & ITA & UKR & USA & CAN & MEX & GBR & AUT & CHN & DEU & MDA & CHN \\  
        7 & GBR & DEU & DEU & UKR & ZAF & GBR & FRA & BRB & ARG & CAN & AUT & BEL & UKR & TWN & AUT & NLD & IND & FRA & CYP & GBR \\ 
        8 & IND & AUT & ESP & NLD & ARE & BEL & DNK & TTO & FRA & USA & ARG & NLD & PRY & PAK & BEL & FRA & FRA & AUT & IRN & ITA \\ 
        9 & CAN & NLD & JPN & RUS & DEU & POL & SVK & OMN & NLD & BEL & NLD & IND & ETH & JPN & POL & ARG & UKR & BRA & CRI & AUT \\ 
        \multicolumn{5}{l}{Panel D: Wheat}\\  
        1 & GBR & RUS & IDN & RUS & FRA & GBR & BOL & VEN & FRA & GBR & CAN & ZAF & RUS & TUR & GBR & FRA & USA & USA & ISR & CAN \\  
        2 & NLD & CAN & TUR & CAN & USA & NLD & SYC & TJK & DEU & FRA & FRA & ITA & UKR & EGY & NLD & DEU & RUS & RUS & IRN & USA \\
        3 & DEU & FRA & EGY & USA & ARE & DEU & LKA & TKM & ITA & NLD & DEU & ESP & CAN & BGD & DEU & ITA & FRA & CAN & CYP & RUS \\ 
        4 & ITA & USA & CHN & FRA & CAN & ESP & AFG & SUR & POL & BEL & RUS & NLD & USA & IDN & ESP & POL & DEU & ITA & MLT & FRA \\ 
        5 & TUR & DEU & ITA & UKR & DEU & FRA & PHL & NPL & AUT & DEU & USA & TUR & FRA & CHN & FRA & AUT & CAN & UKR & ALB & DEU \\ 
        6 & BEL & UKR & DZA & ARG & GBR & ITA & HKG & LCA & CZE & MYS & UKR & DEU & ARG & PHL & ITA & CZE & UKR & DEU & EGY & ITA \\   
        7 & ESP & POL & BRA & AUS & TWN & CHE & MDV & TTO & NLD & ARE & ITA & BEL & AUS & NGA & BEL & NLD & AUS & AUS & CZE & UKR \\ 
        8 & FRA & IND & BGD & DEU & RUS & BEL & NZL & DOM & LTU & KEN & POL & ARE & LTU & MAR & CHE & LTU & NLD & POL & TUR & TUR \\
        9 & CHE & ITA & NGA & KAZ & ZAF & AUT & GTM & BRB & HUN & ESP & IND & GBR & DEU & JPN & AUT & HUN & POL & BGR & IRL & KAZ \\  
        10 & AUT & LTU & PHL & LTU & TUR & TUR & IND & FRO & RUS & THA & TUR & SGP & BGR & PAK & TUR & RUS & BEL & KAZ & MOZ & GBR \\ 
    \bottomrule
    \end{tabular}
    \label{Table:iCTN:node:influence:ranking}
\end{table}

Node importance metrics are widely used to identify critical economies in different trade networks \cite{Ausiello-Firmani-Laura-2013-IntWirelCommun,VidalHernandez-CantoLugo-CarmonaEscalante-HuertaQuintanilla-GarzaLagler-LopezRocha-2019-OceanCoastalManage,Wei-Xie-Zhou-2022-Energy}. Previous published studies are limited to one indicator or one type of indicator but rarely compare the various types of indicators. Meanwhile, fewer studies on food trade networks apply different indicators and comparatively analyze the trade influence of economies in different kinds of food networks. We calculate the influence of economies in different iCTNs from 1986 to 2020 based on the nodal importance indicators presented in Section~\ref{S2:DataMethod}. The trade impact of economies tends to change over time and varies across crops. We focus on economies' importance in maize, rice, soybean and wheat trade networks and show ranking results in Table~\ref{Table:iCTN:node:influence}.

Comparing the ranking results of different indicators, the four iCTNs have some interesting similarities. First, the indicators about degree scales based on the local structure of networks (in and out-degree centrality, in and out-semi-local centrality) only consider the number of export and import partners of the economies and do not take into account the weights and clustering distribution of trade links. As a result, the ranking results obtained based on the strength indicators are different from those obtained based on the degree indicators. However, the ranking results based on in-degree centrality and in-semi-local centrality are similar, as are those based on out-degree centrality and out-semi-local centrality. Secondly, we can see that the rankings obtained by in-degree centrality, in intensity, eigenvector centrality, in-closeness centrality, PageRank, authorities, and in semi-local centrality as metrics are relatively comparable. Due to the fact that they focus primarily on the economy's food imports, these indicators have a significant correlation. Similar features exist for the metrics that mainly consider the economy's export profile. The ranking results based on out-degree centrality, out-strength, Hub, out-closeness-centrality, and out-semi-local centrality are more similar. Likewise, rankings based on metrics that consider the distribution of other nodes connected to the nodes in the network have special properties, such as clustering coefficient, Katz centrality, within-module degree, participation coefficient, and out-module degree. The ranking results computed by these metrics vary widely. However, the ranking results of metrics based on within-module degree and outside-module degree are very similar.

\begin{figure}[h!]
     \subfigbottomskip=-2pt
     \subfigcapskip=-6pt
     \centering
     \subfigure[]{\label{level.sub.a}\includegraphics[width=0.233\linewidth]{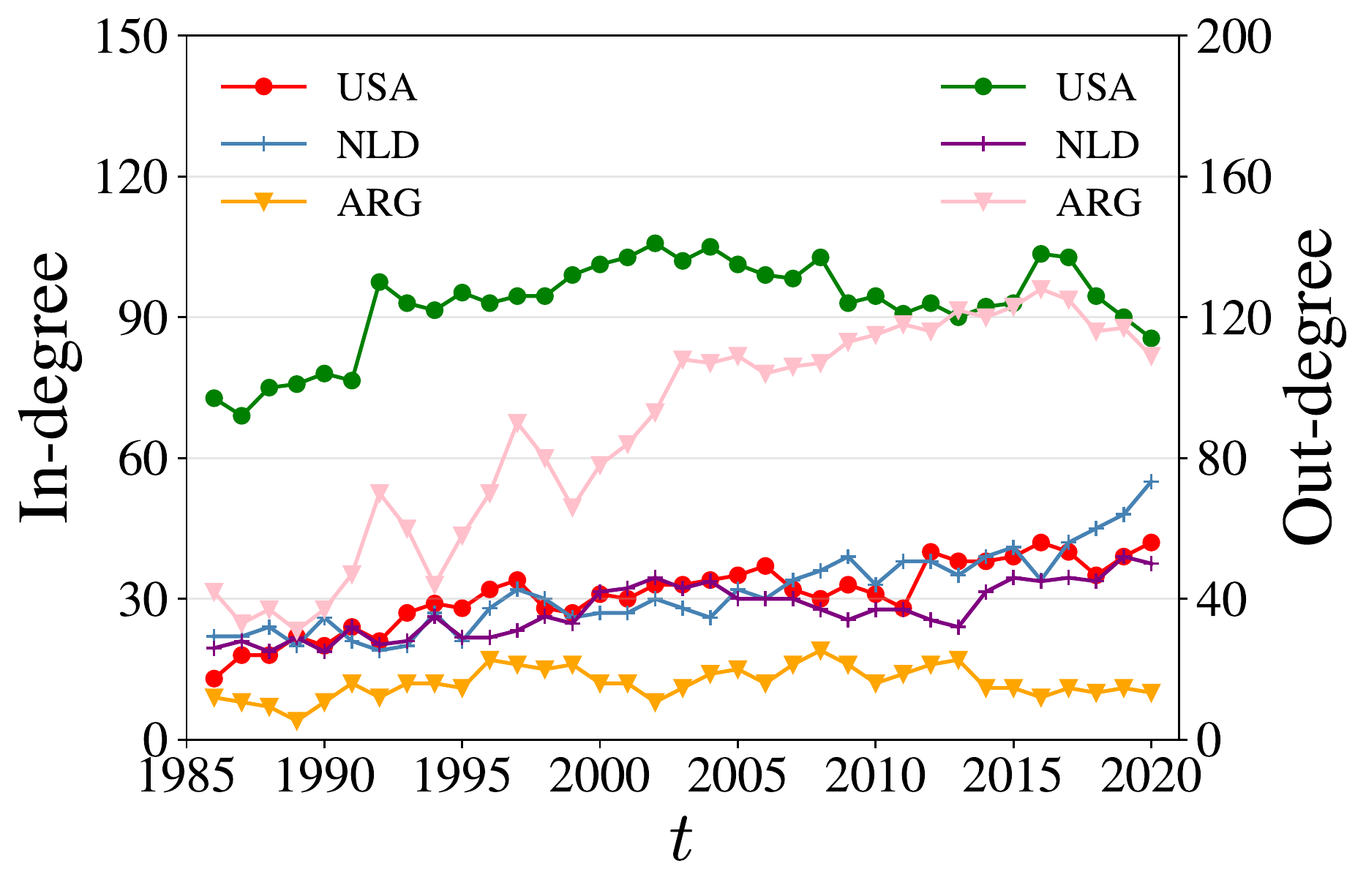}}
     \subfigure[]{\label{level.sub.b}\includegraphics[width=0.223\linewidth]{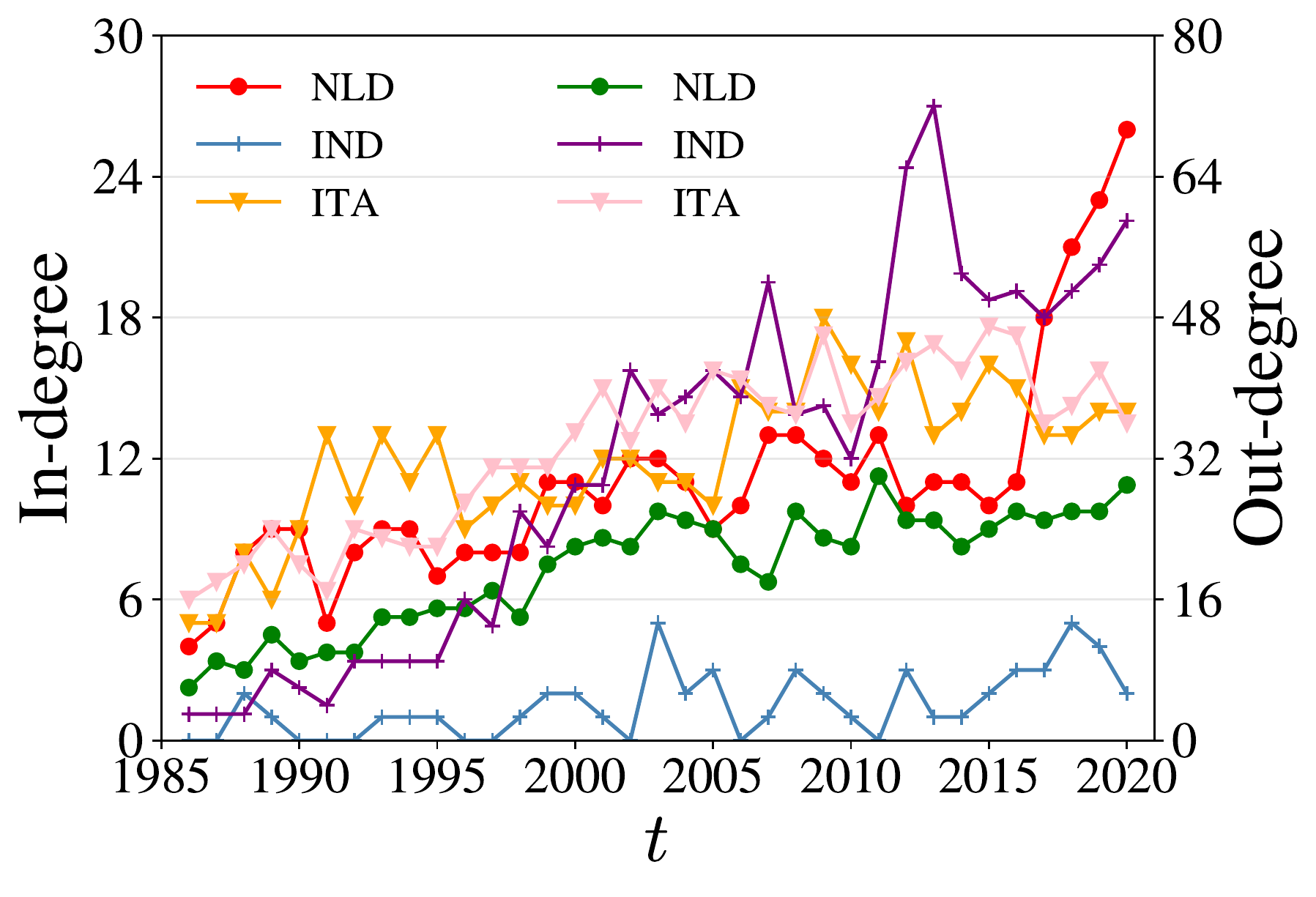}}
      \subfigure[]{\label{level.sub.c}\includegraphics[width=0.233\linewidth]{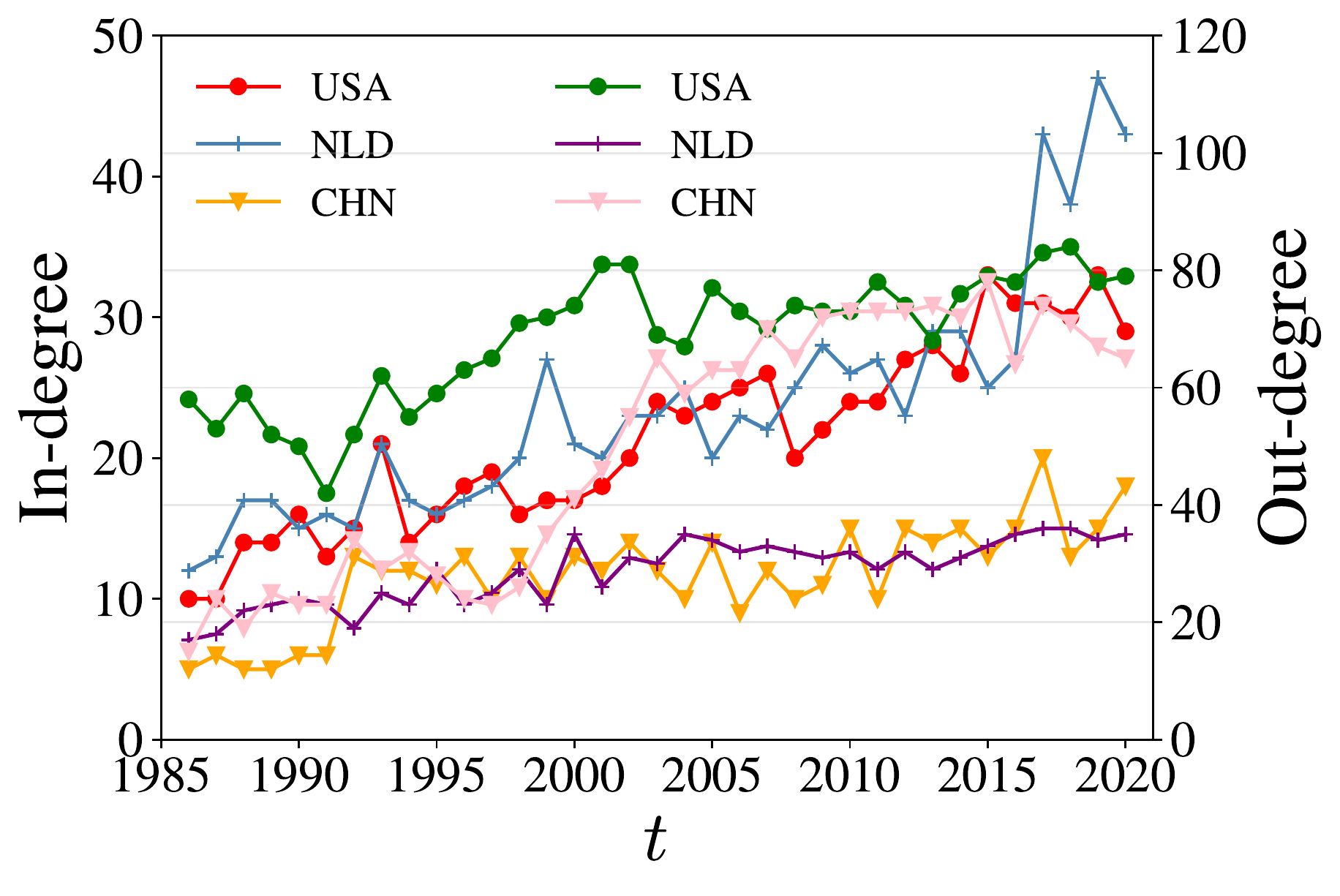}}
     \subfigure[]{\label{level.sub.d}\includegraphics[width=0.233\linewidth]{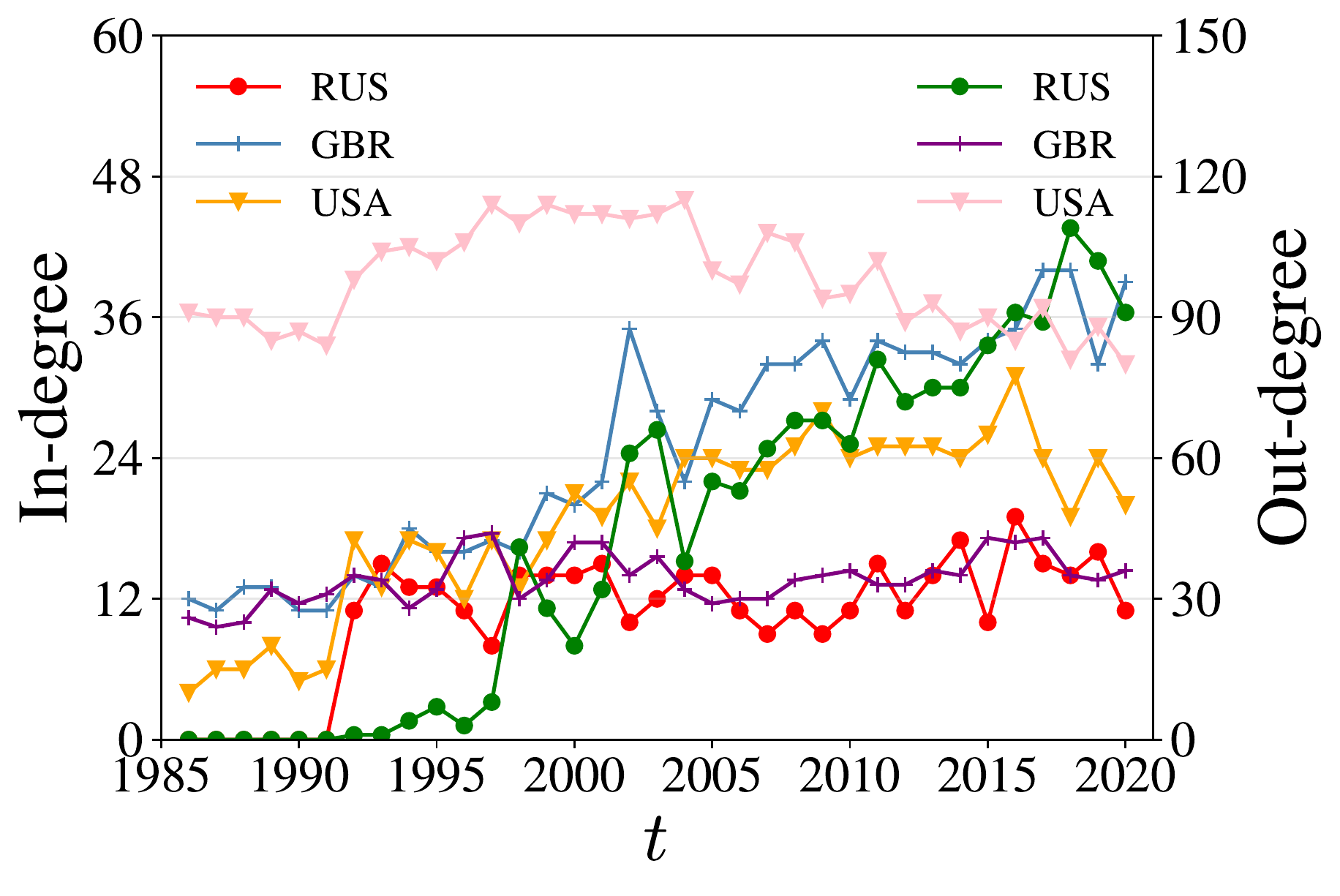}}
     \subfigure[]{\label{level.sub.e}\includegraphics[width=0.233\linewidth]{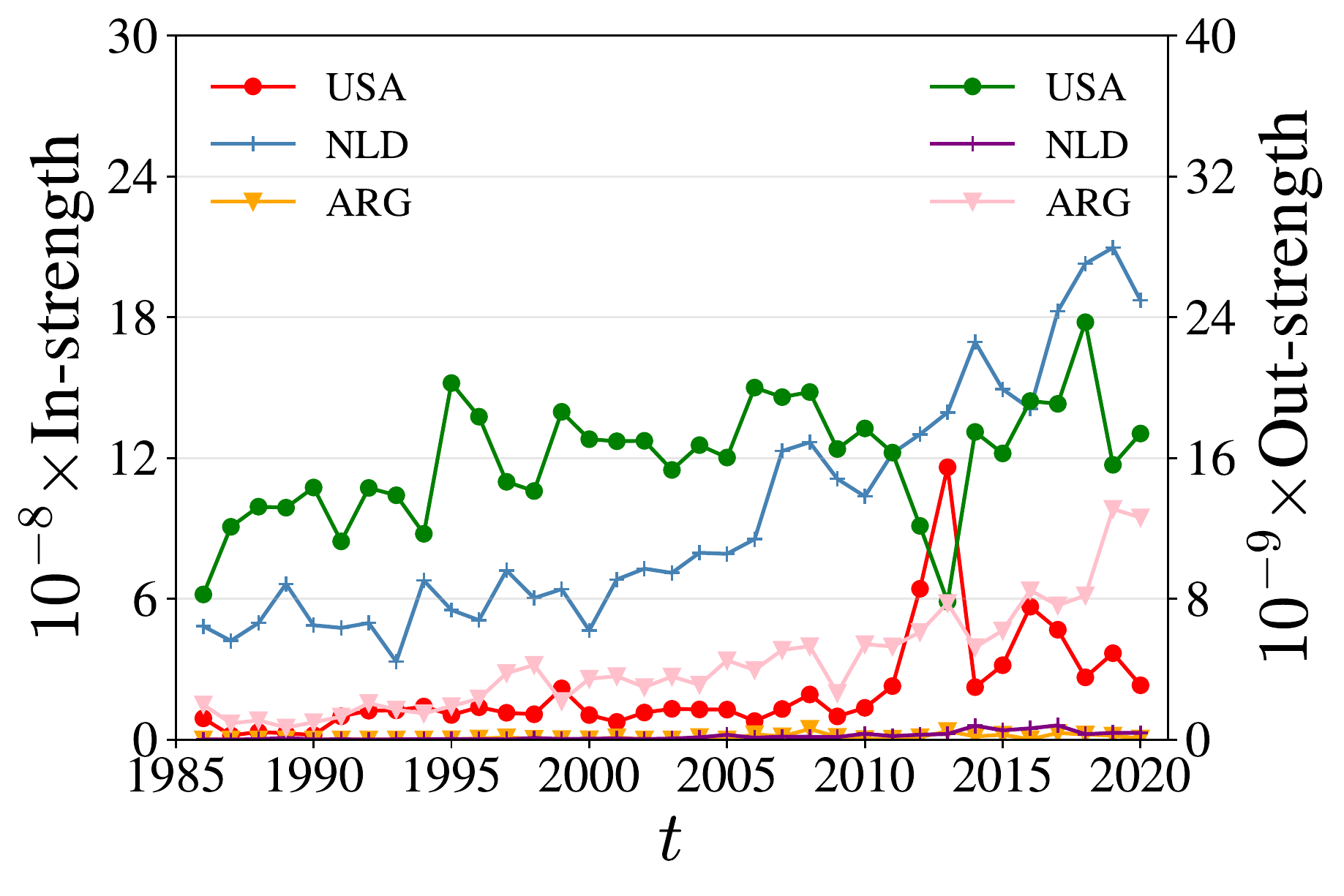}}
     \subfigure[]{\label{level.sub.f}\includegraphics[width=0.233\linewidth]{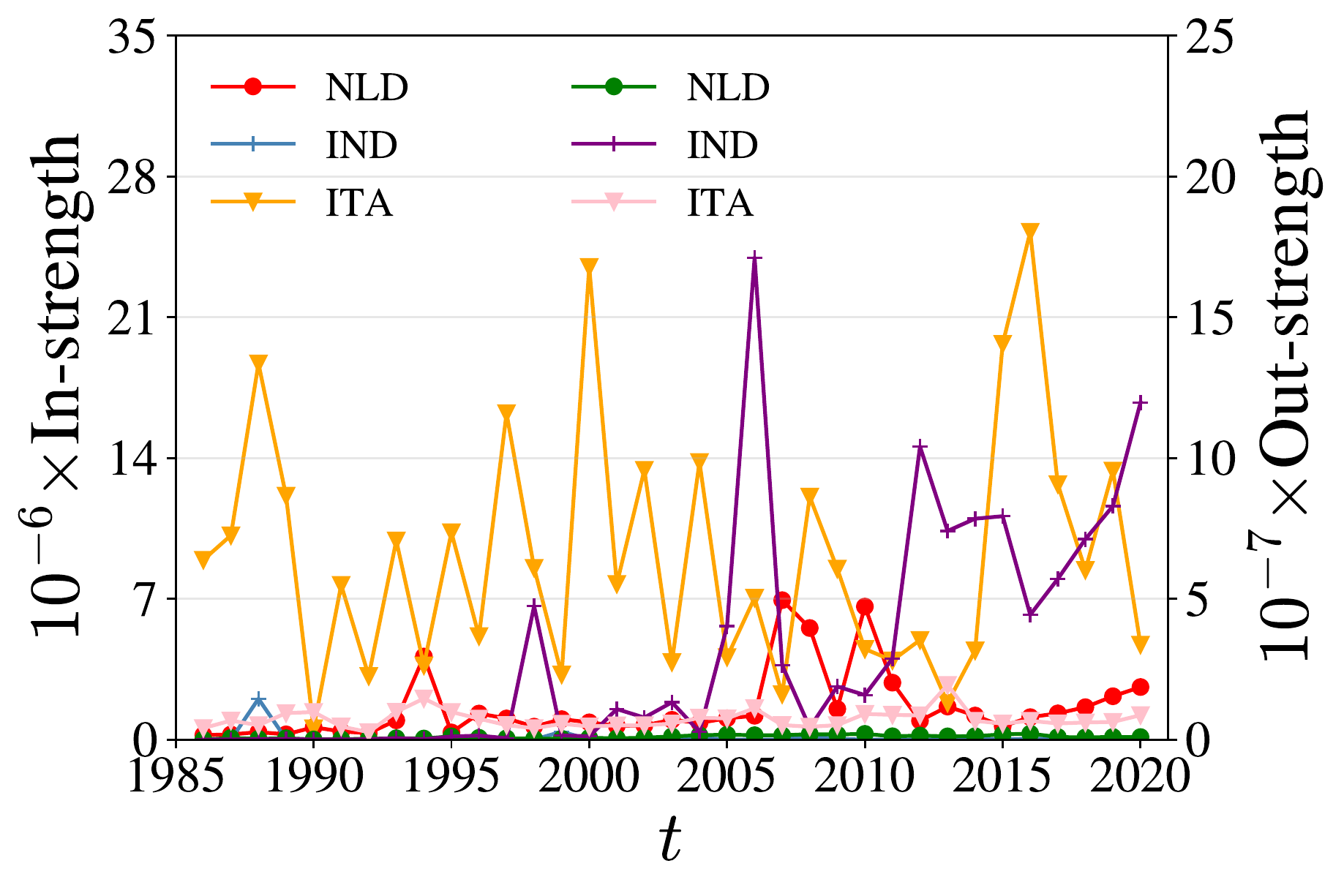}}
      \subfigure[]{\label{level.sub.g}\includegraphics[width=0.233\linewidth]{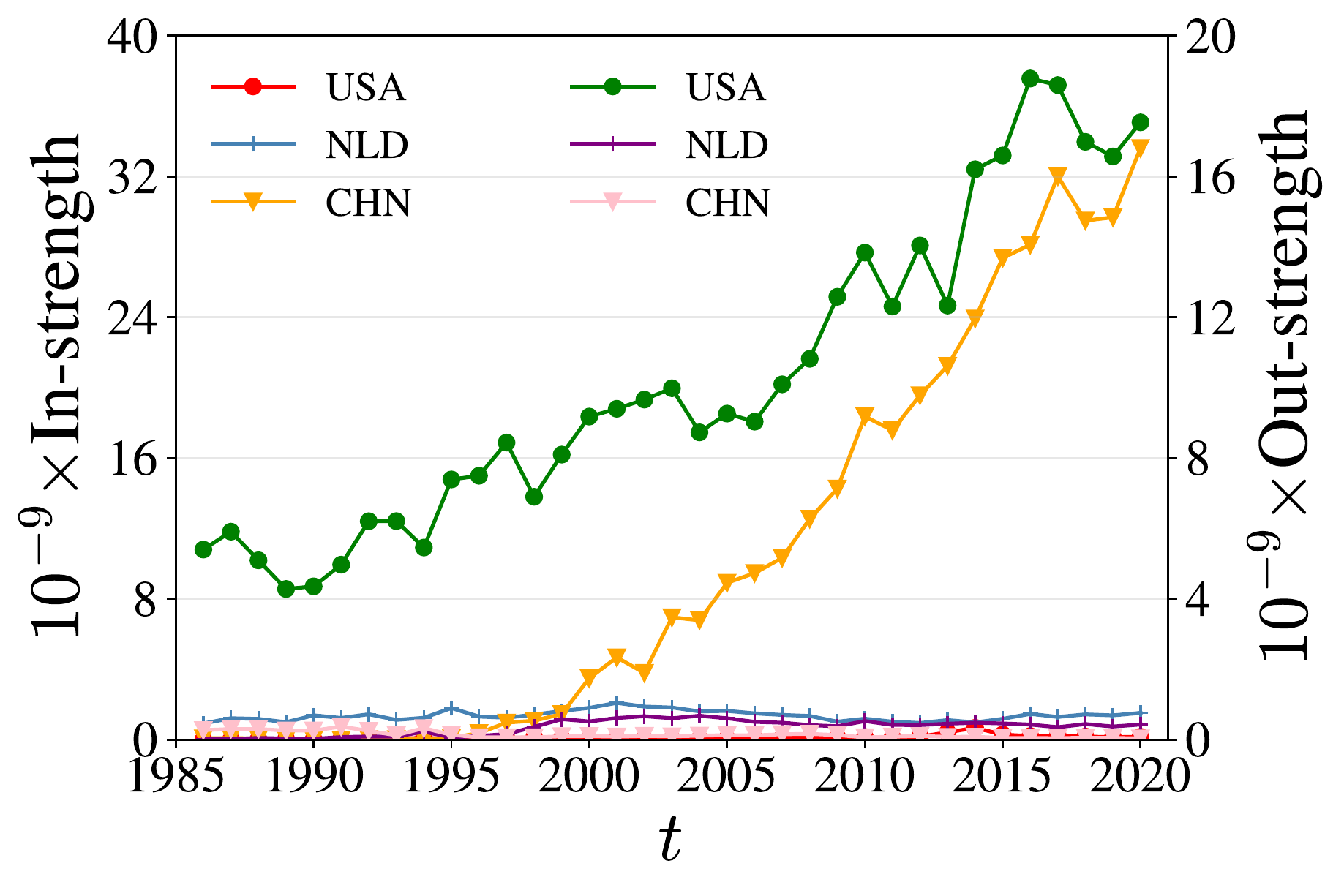}}
     \subfigure[]{\label{level.sub.h}\includegraphics[width=0.233\linewidth]{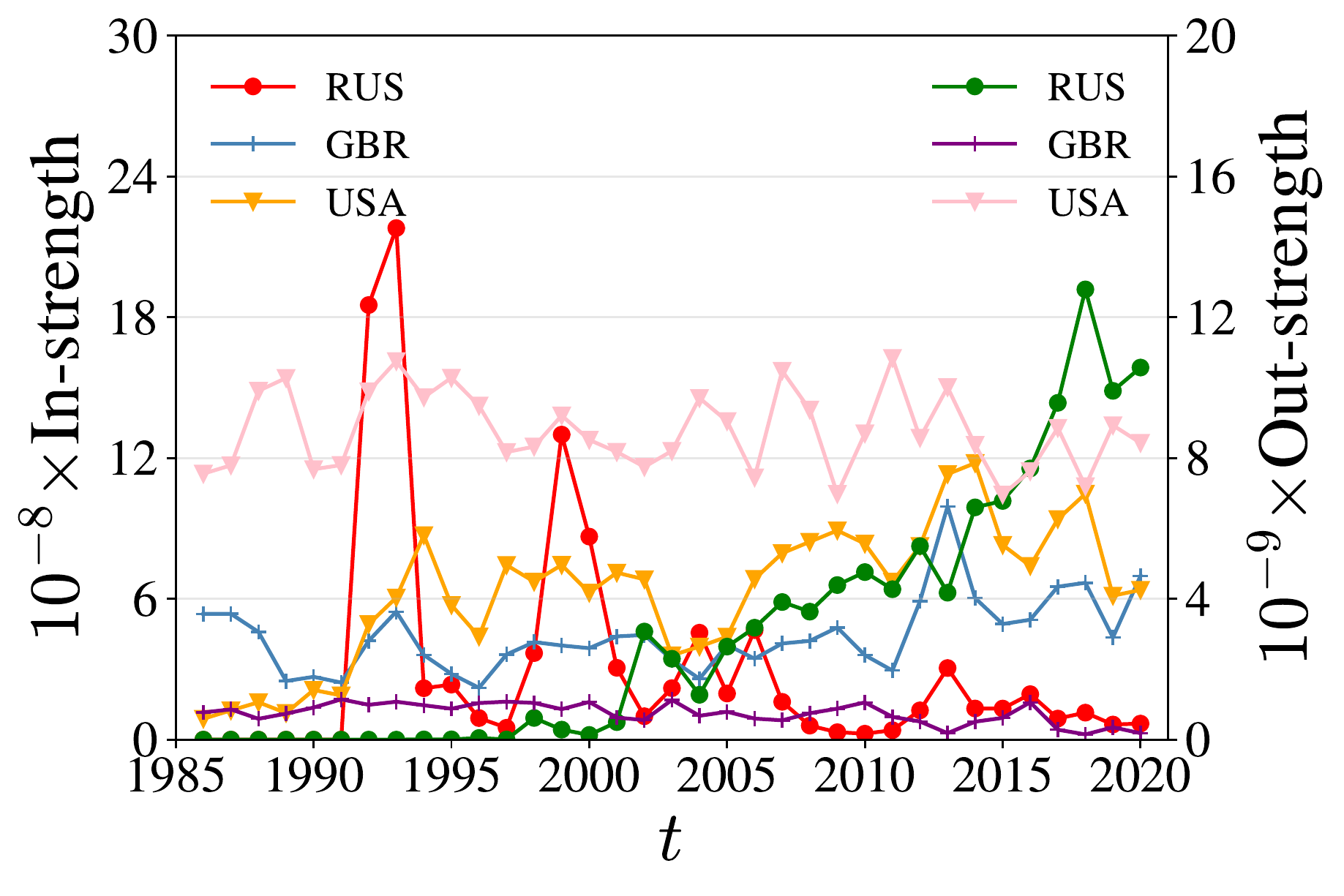}}
      \caption{Evolution of in-degree, out-degree, in-strength and out-strength of critical economies identified by node importance metrics in four iCTNs. Lines in different colors correspond to different economies.}
      \label{Fig:iCTN:Keynode:k:s:t}
\end{figure}

The trade influence of economies differs in different iCTNs. The trade influence of economies differs across different iCTNs. To get a clearer picture of food trade in key economies, we select three key economies in each iCTN in 2020 and portray the evolution of the in and out-degrees as well as the in and strengths of these economies from 1986-2020. Both the U.S.and the Netherlands have a significant impact on the maize and soybean trade, which can be seen in Panel A and Panel C of Table~\ref{Table:iCTN:node:influence:ranking}, present that the United States exports large amounts of maize and soybean to many economies. The Netherlands has a significant impact on the international maize trade network mainly because of its substantial maize imports from numerous economies. Interestingly, the Netherlands has a larger in-degree in the soybean trade network, i.e., it has many import trading partners, but its import volume is not large compared to China. For rice, Italy and India occupy a prominent position. The possible reason for this case is that India is the main rice exporter, and a large number of economies are highly dependent on rice imports from India. Italy imports and exports rice, with imports outgrowing exports (Fig.~\ref{Fig:iCTN:Keynode:k:s:t}(b)). China is one of the world's largest soybean markets, with many trade partners. Given the fact that China imports a lot of soybeans from the United States, China plays an important role in the global soybean import trade. British has a significant influence on the international wheat trade network due to its wheat imports. Concerning wheat exports, we found that the United States and Russia have an important impact on the wheat network. The possible explanation could be that these economies are principal wheat planting areas and suppliers. Based on different ranking comparisons, economies play divergent roles in different iCTNs. The trade influence of an economy is associated with its crop production. Trade could compensate for the gaps between domestic crop production and consumption of an economy. Thus an economy's demand for external food supplies determines its influence in the iCTNs. It is necessary to discuss the contribution of a particular economy to global food trade and food security across different crop types. Furthermore, the ranking results also show that some economies are indeed important, for instance, the United States and the Netherlands are identified as key economies in several food trade networks. The European economies have a relatively large influence on the global food trade. Risks to the global food supply must be tracked, and this requires keeping a watch on food production and trade in these economies.

\subsection{Correlation structure among node importance rank metrics}
\label{S3-2:Correlation}

\begin{figure}[h!]
     \subfigbottomskip=-1pt
     \subfigcapskip=-5pt
     \centering
     \subfigure[]{\label{level.sub.5}\includegraphics[width=0.243\linewidth]{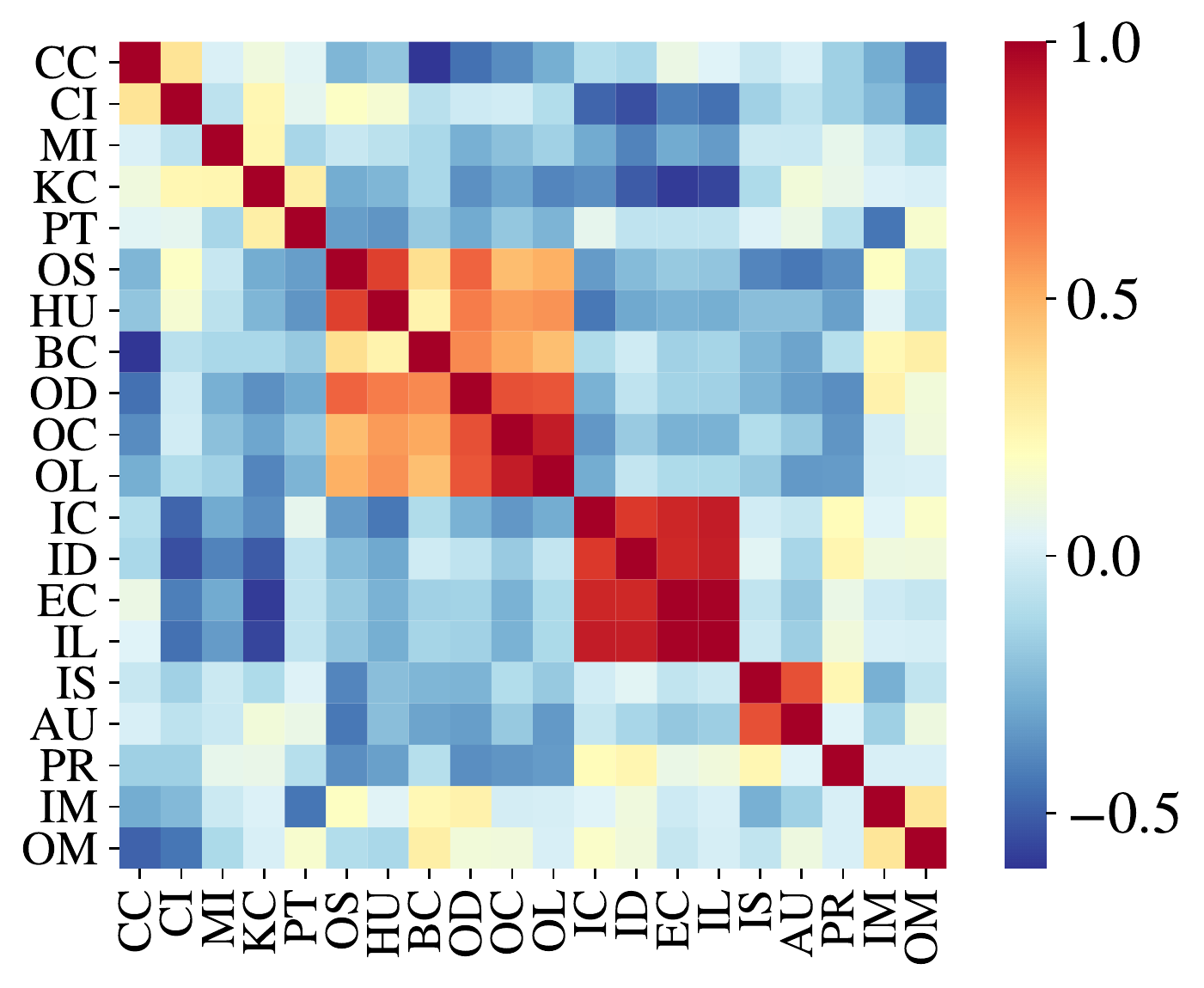}}
     \subfigure[]{\label{level.sub.6}\includegraphics[width=0.243\linewidth]{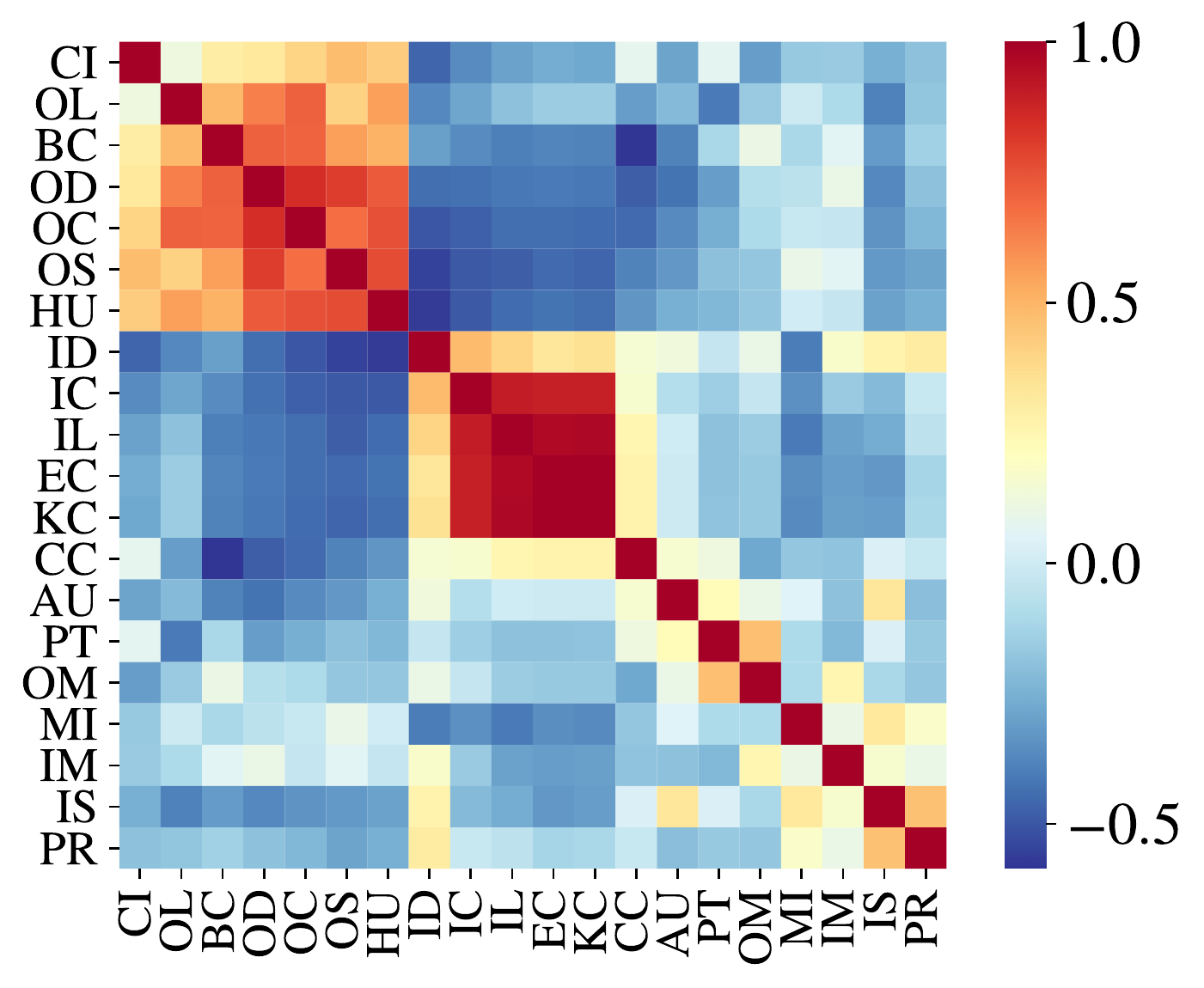}}
      \subfigure[]{\label{level.sub.7}\includegraphics[width=0.243\linewidth]{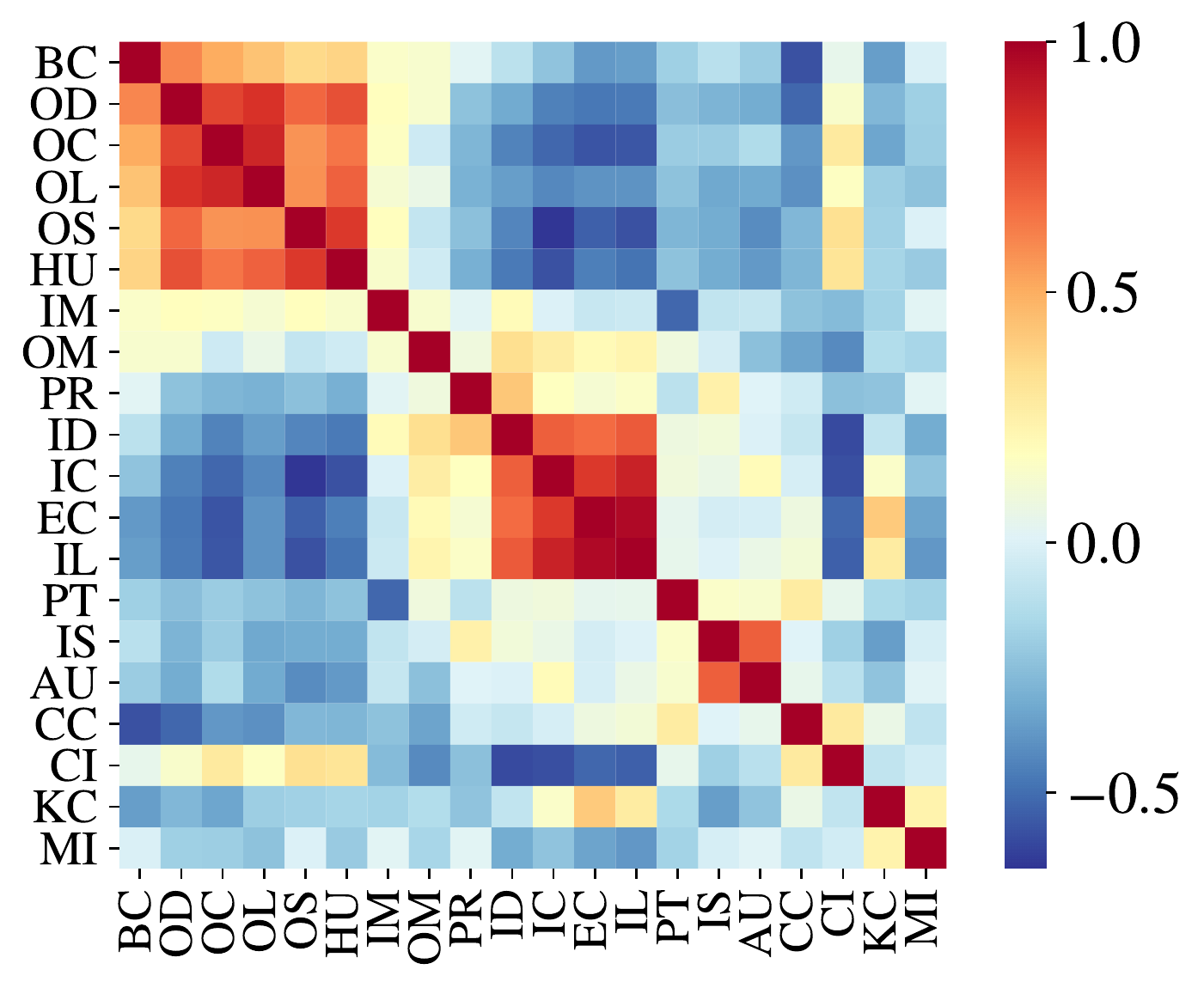}}
     \subfigure[]{\label{level.sub.8}\includegraphics[width=0.243\linewidth]{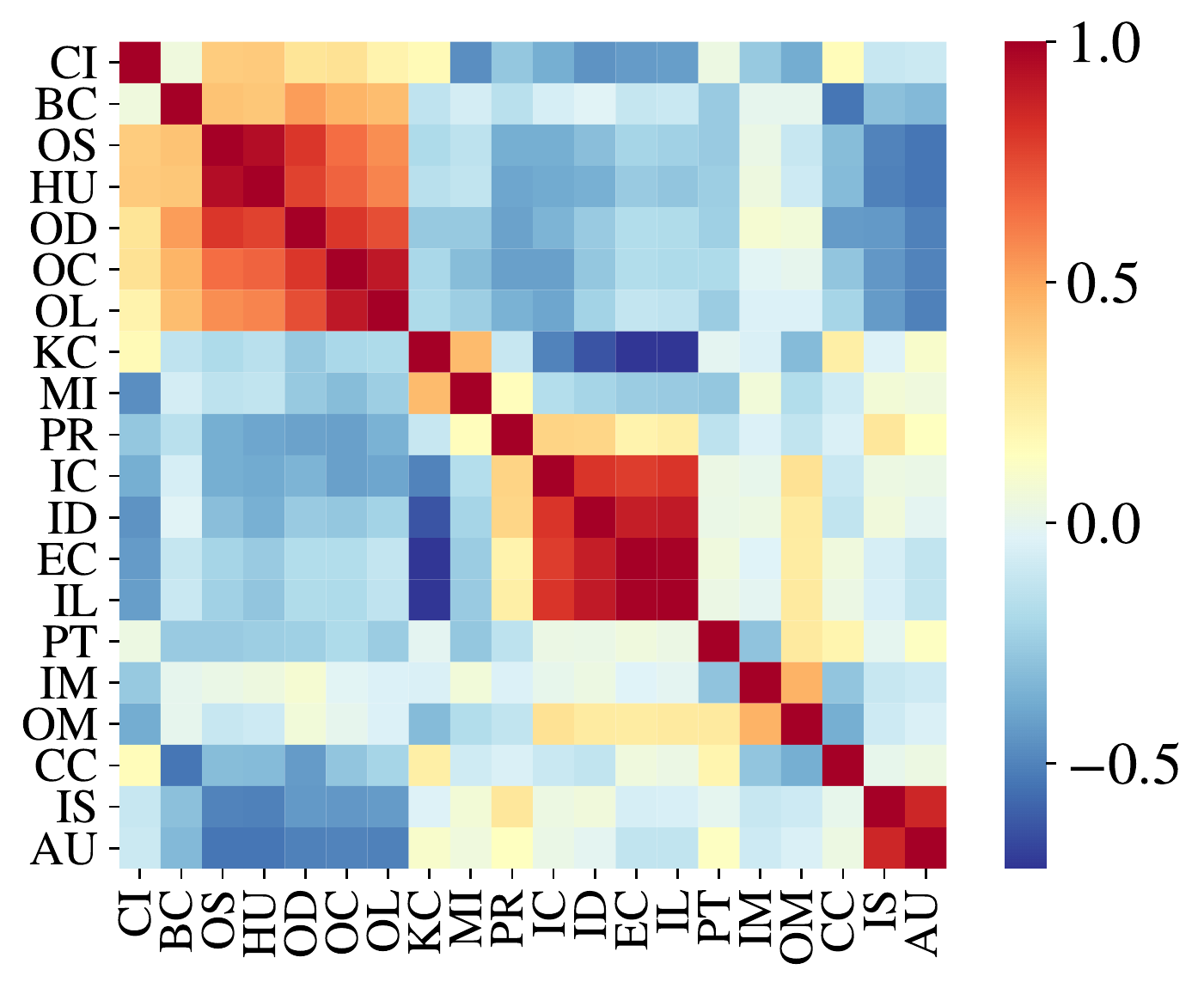}}
      \caption{Cross-correlation structure among 20 node importance rank metrics of the international trade network of maize (a), rice (b), soybean (c), and wheat (d) in 2020.}
      \label{Fig:iCTN:HeatMap:corrCoefficient:2020:20}
\end{figure}

Figure~\ref{Fig:iCTN:HeatMap:corrCoefficient:2020:20} shows the cross-correlation structure among 20 node importance rank metrics of the iCTNs. In each Heat map, blocks of related hues are easy to spot. We find that there are generally two significant clusters of indicators. {\color{red}{One cluster mainly contains node metrics related to import quantities, including in-closeness centrality, in-semi-local centrality, and eigenvector centrality. The other cluster mainly contains node metrics related to export quantities, including out-closeness centrality, out-semi-local centrality, betweenness centrality, and Hubs. }}The correlation between indicators in the same cluster is high. 

\begin{figure}[h!]
      \centering
      \includegraphics[width=0.233\linewidth]{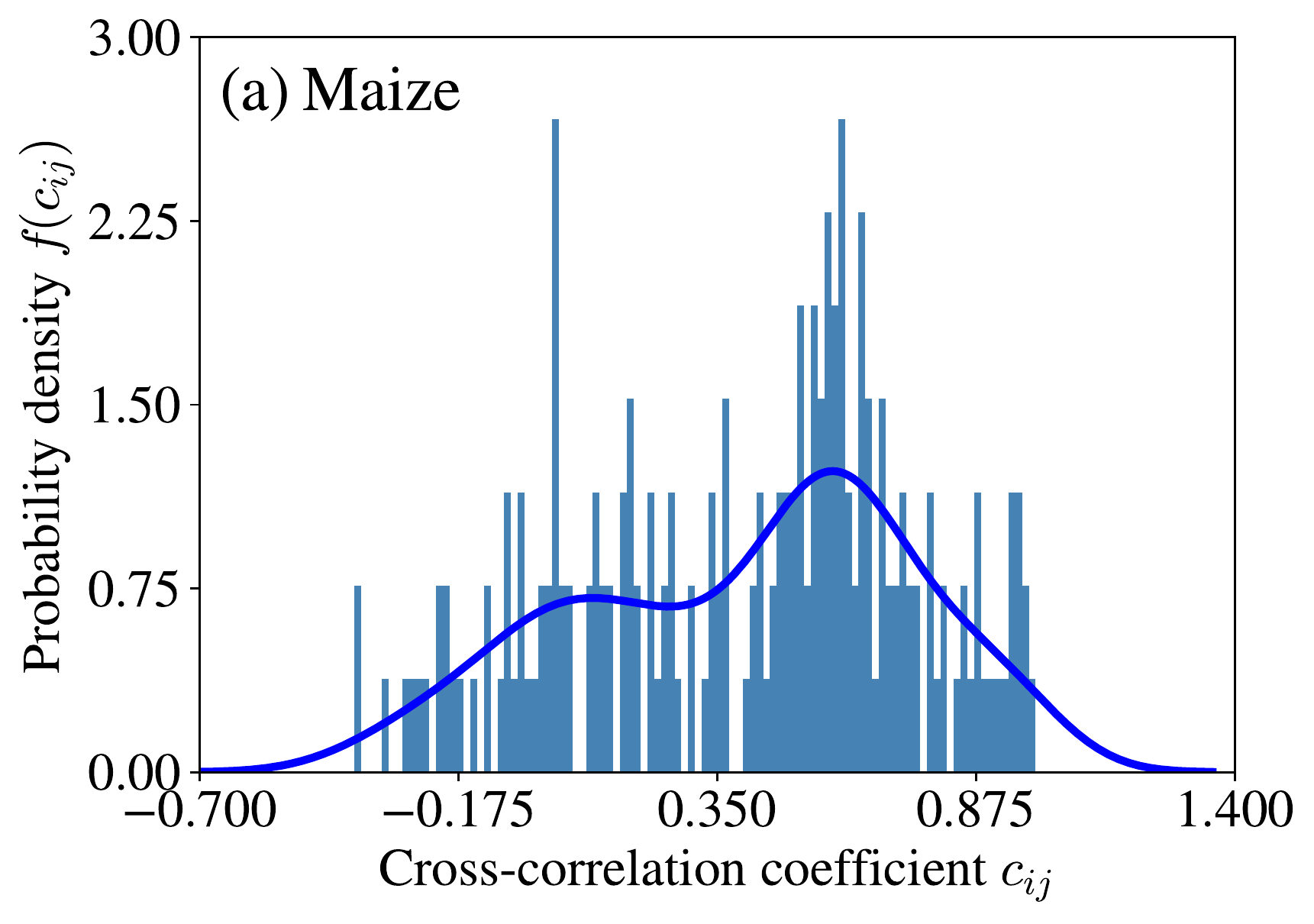}
      \includegraphics[width=0.233\linewidth]{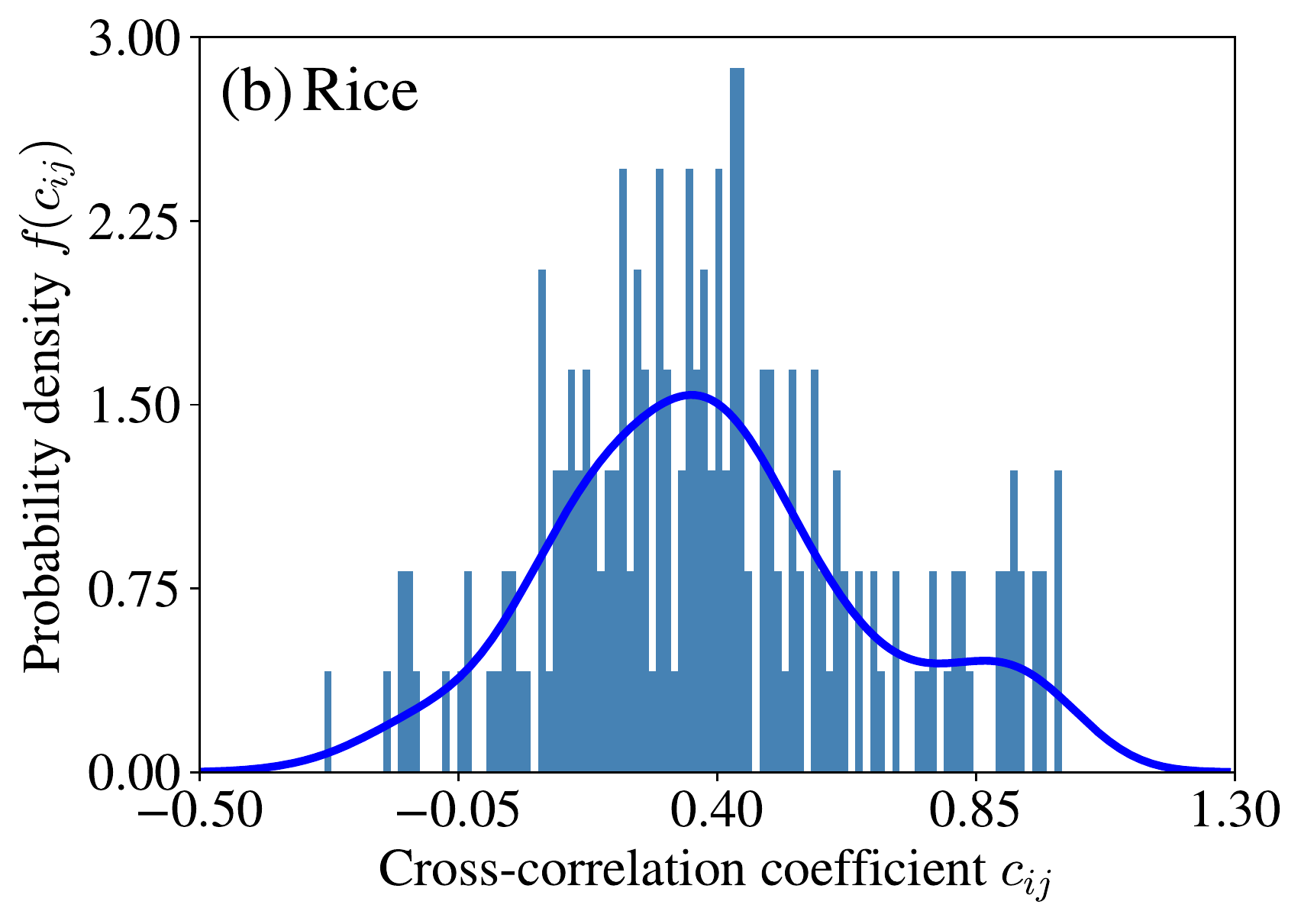}
      \includegraphics[width=0.233\linewidth]{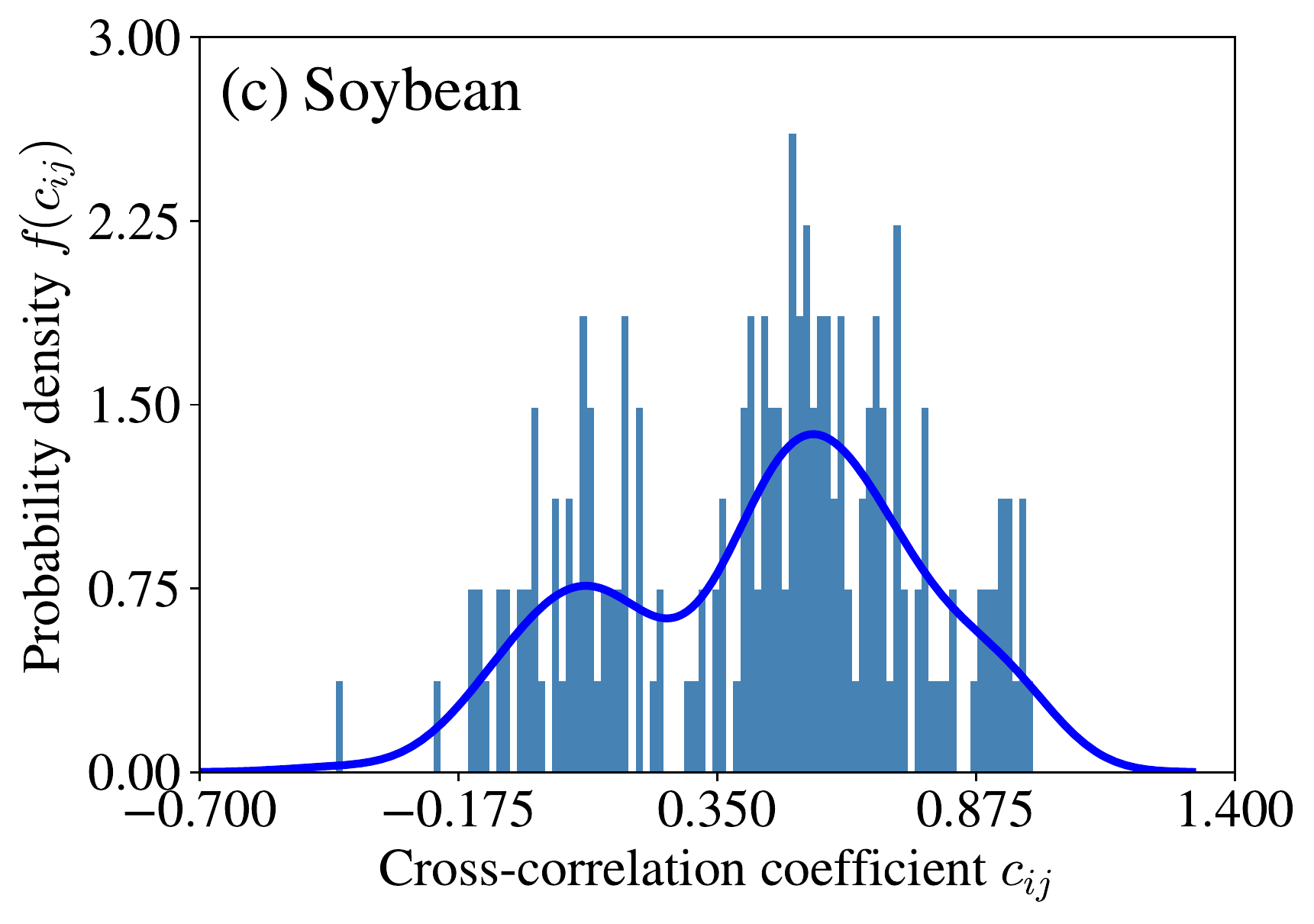}
      \includegraphics[width=0.233\linewidth]{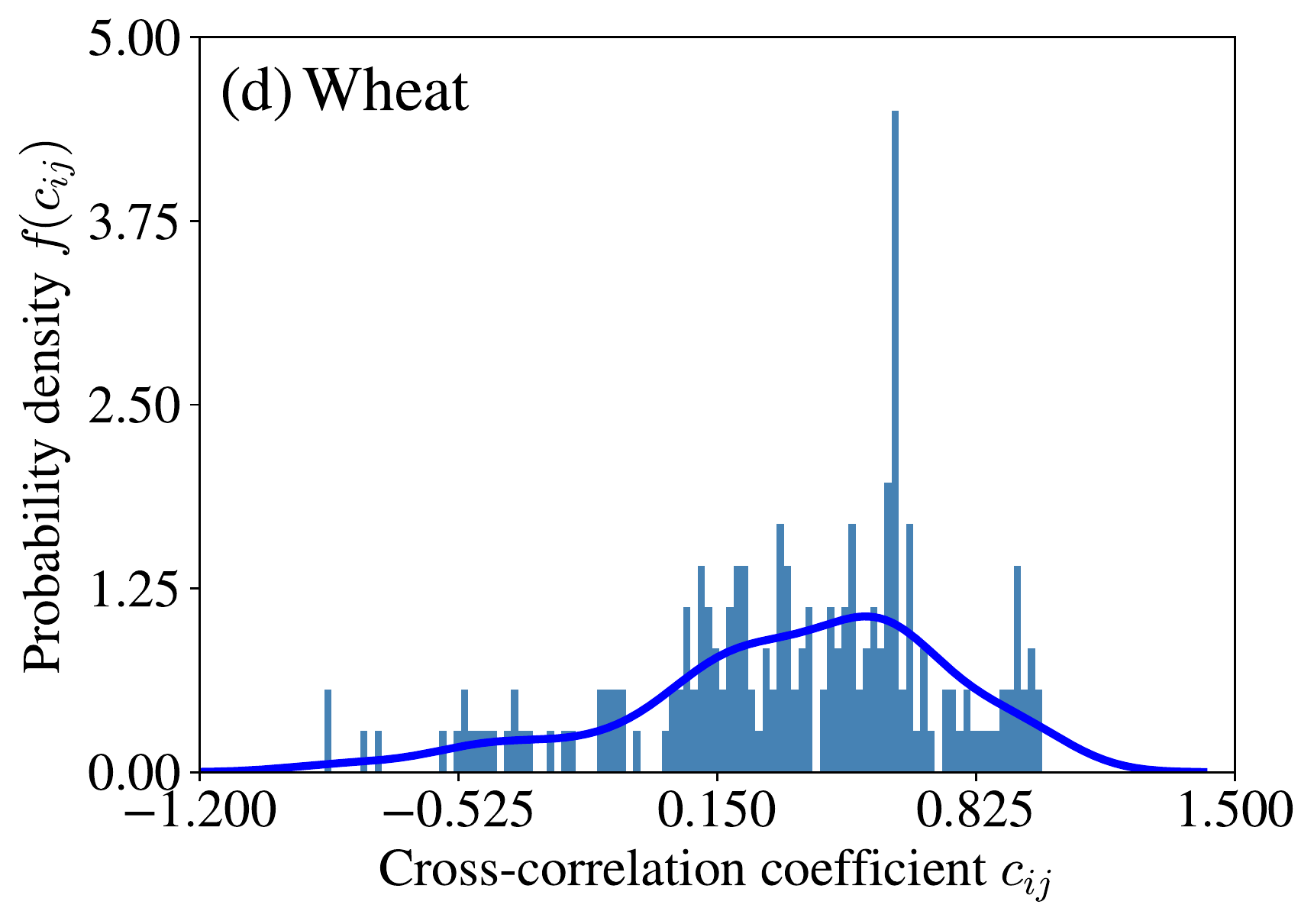}
      \caption{Distribution of the correlation coefficients $c_{ij}$ of the node importance rank series of four international crop trade networks in 2020. }
      \label{Fig:iCTN:PDF:corrCoefficient:2020}
\end{figure}

Figure~\ref{Fig:iCTN:PDF:corrCoefficient:2020} shows the distribution of the correlation coefficient $c_{ij}$. The histogram is the distribution of the probability density $f(c_{ij})$ of $c_{ij}$, and the curve is the corresponding kernel density distribution. As can be seen from Fig.~\ref{Fig:iCTN:PDF:corrCoefficient:2020}, positive correlations are rifer in the correlation structure of these indicators. Further, the kernel density distribution indicates that there are two peaks in the correlation structure for the maize, rice, and soybean trade networks. It is clear that some of the indicators do in fact significantly correlate with one another.

\begin{figure}[h!]
     \subfigbottomskip=-1pt
     \subfigcapskip=-5pt
     \centering
     \subfigure[]{\label{level.sub.9}\includegraphics[width=0.233\linewidth]{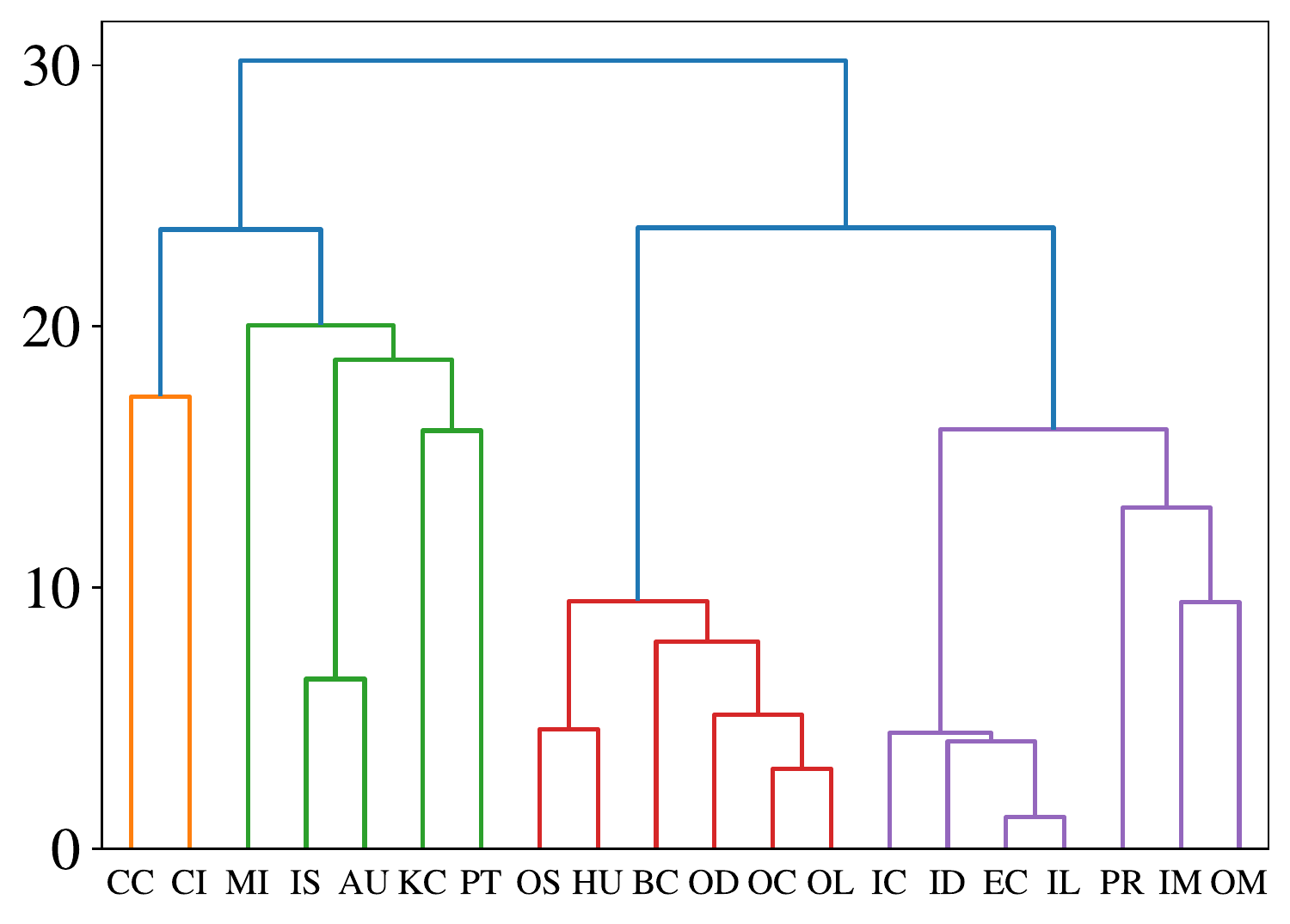}}
     \subfigure[]{\label{level.sub.10}\includegraphics[width=0.233\linewidth]{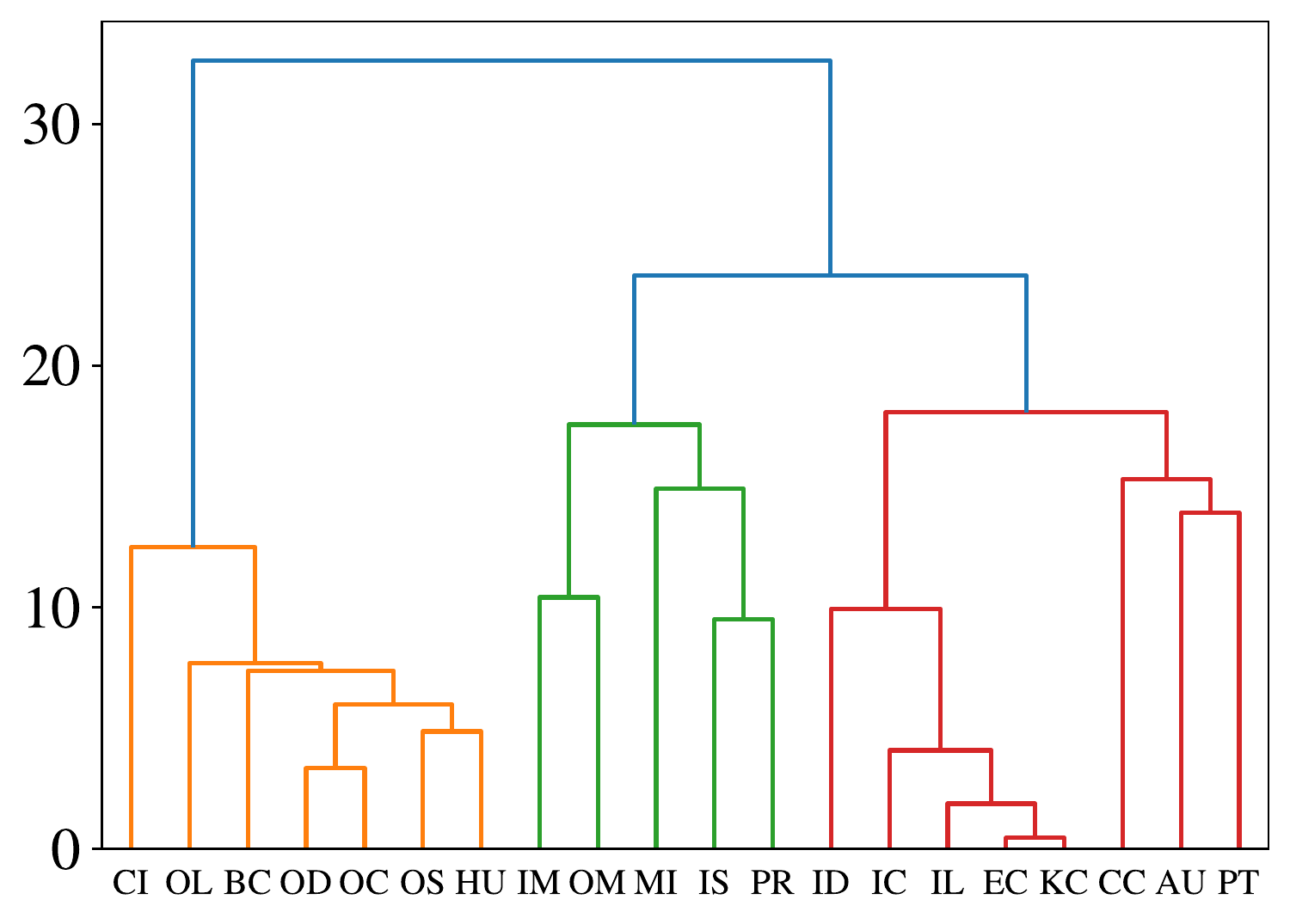}}
    \subfigure[]{\label{level.sub.11}\includegraphics[width=0.233\linewidth]{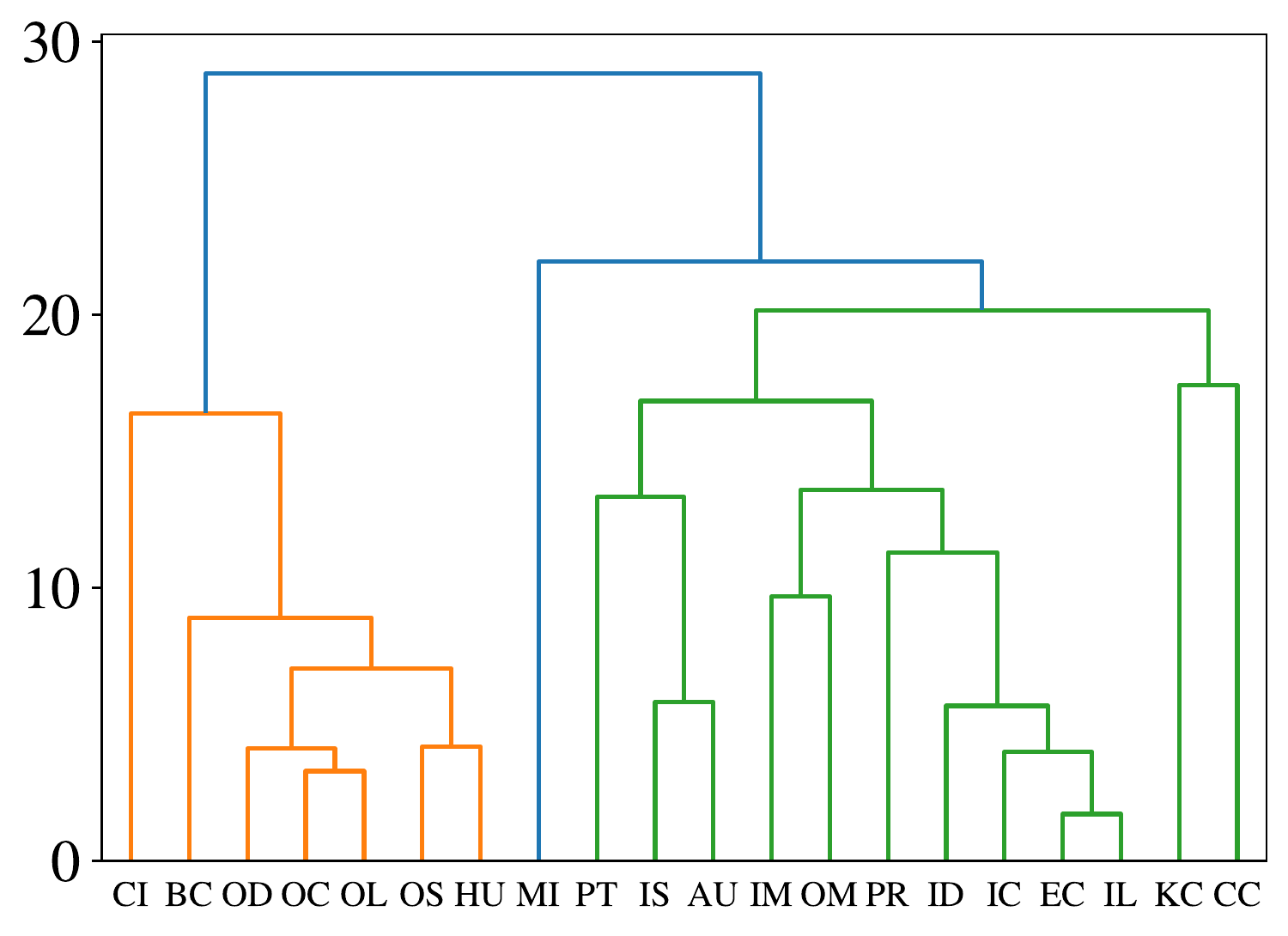}}
    \subfigure[]{\label{level.sub.12}\includegraphics[width=0.233\linewidth]{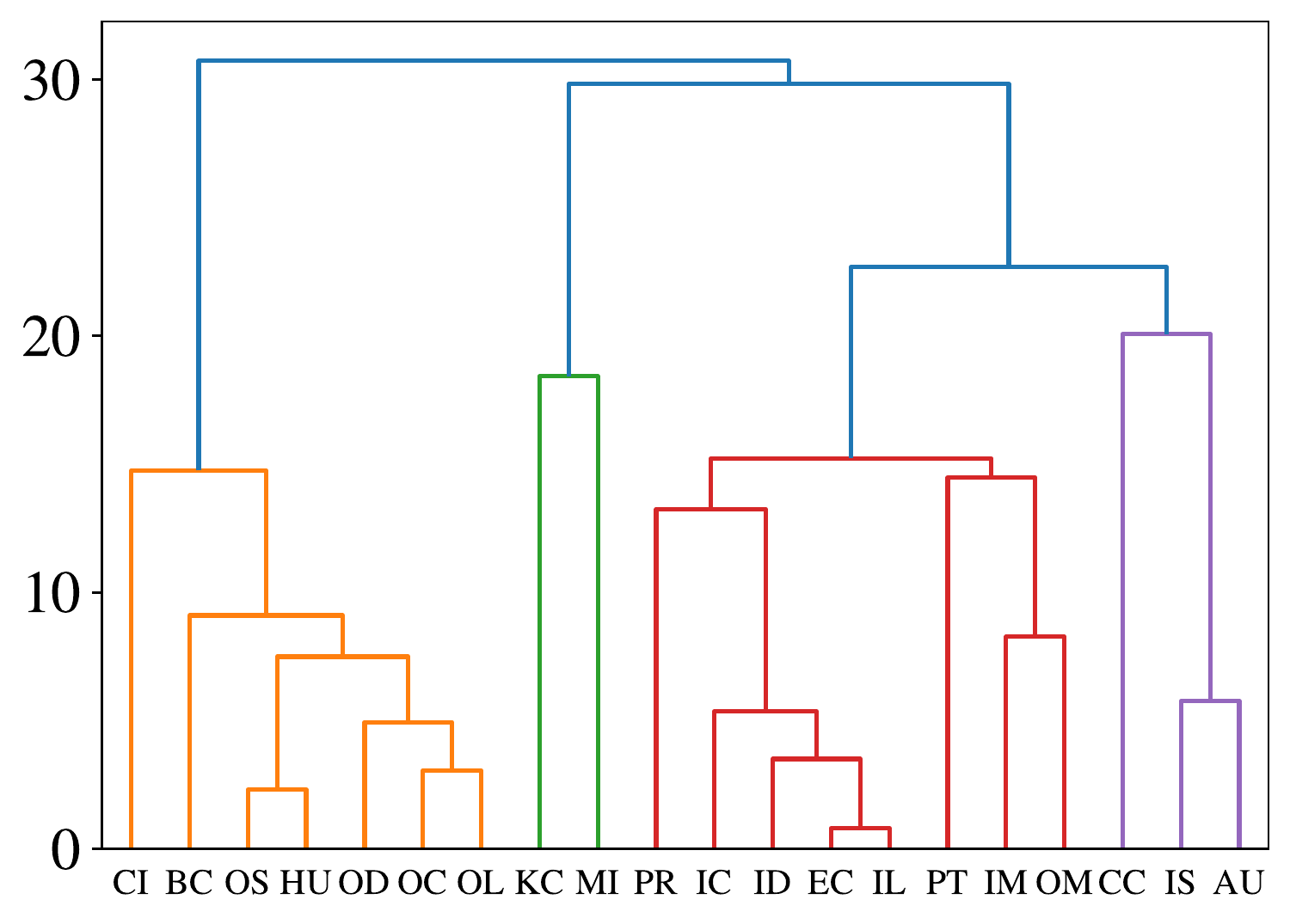}}
    \subfigure[]{\label{level.sub.13}\includegraphics[width=0.233\linewidth]{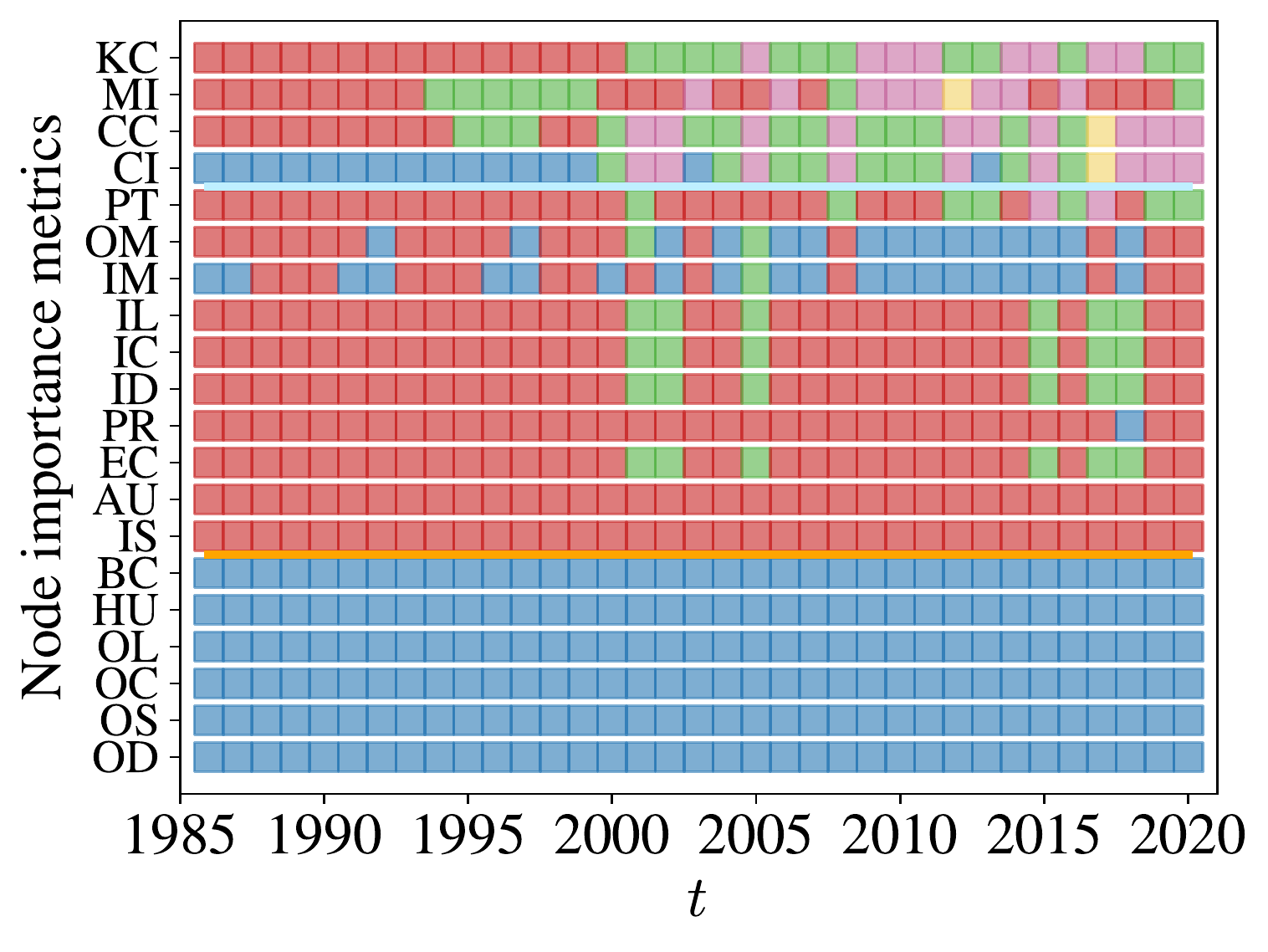}}
    \subfigure[]{\label{level.sub.14}\includegraphics[width=0.233\linewidth]{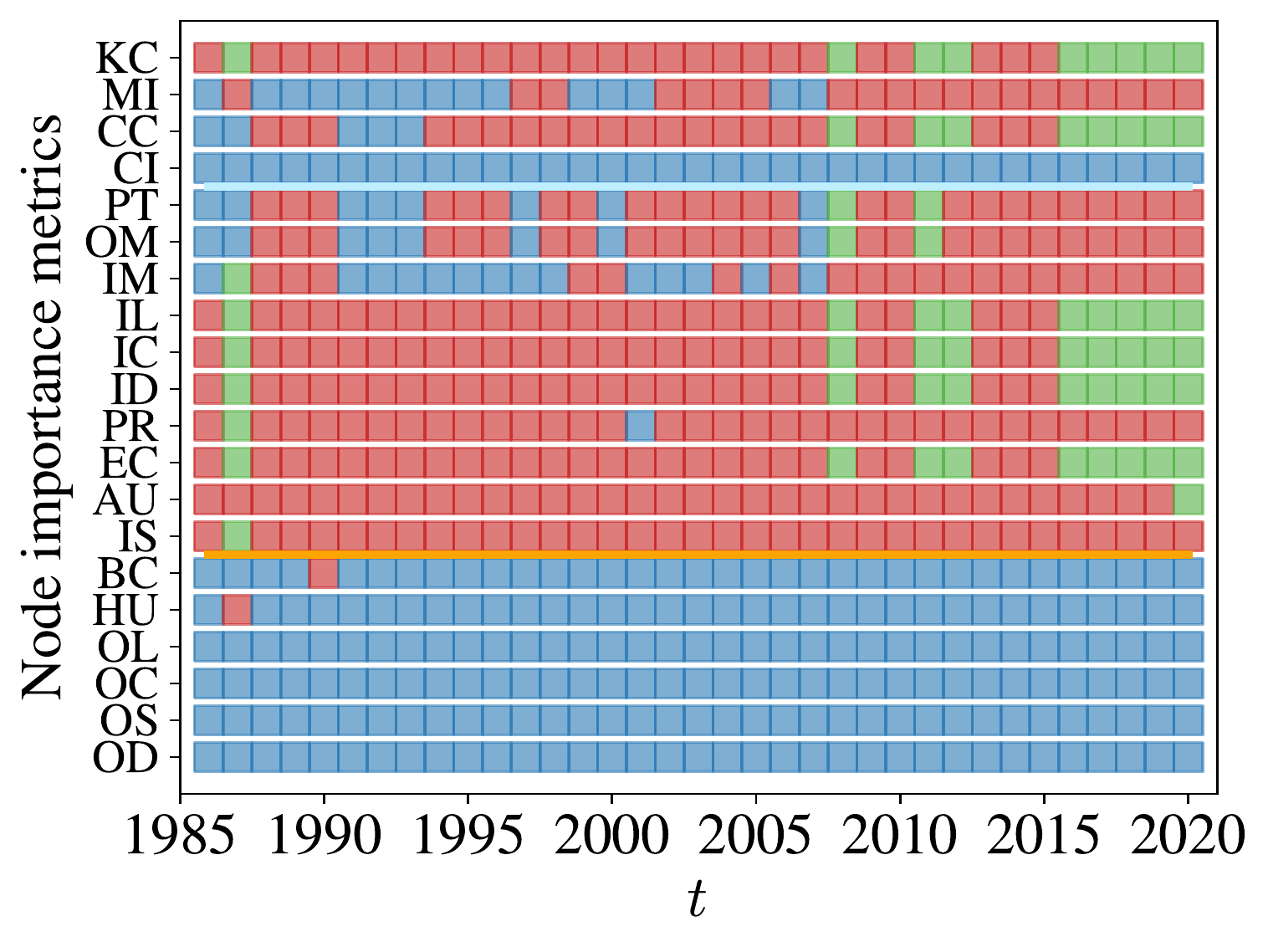}}
    \subfigure[]{\label{level.sub.15}\includegraphics[width=0.233\linewidth]{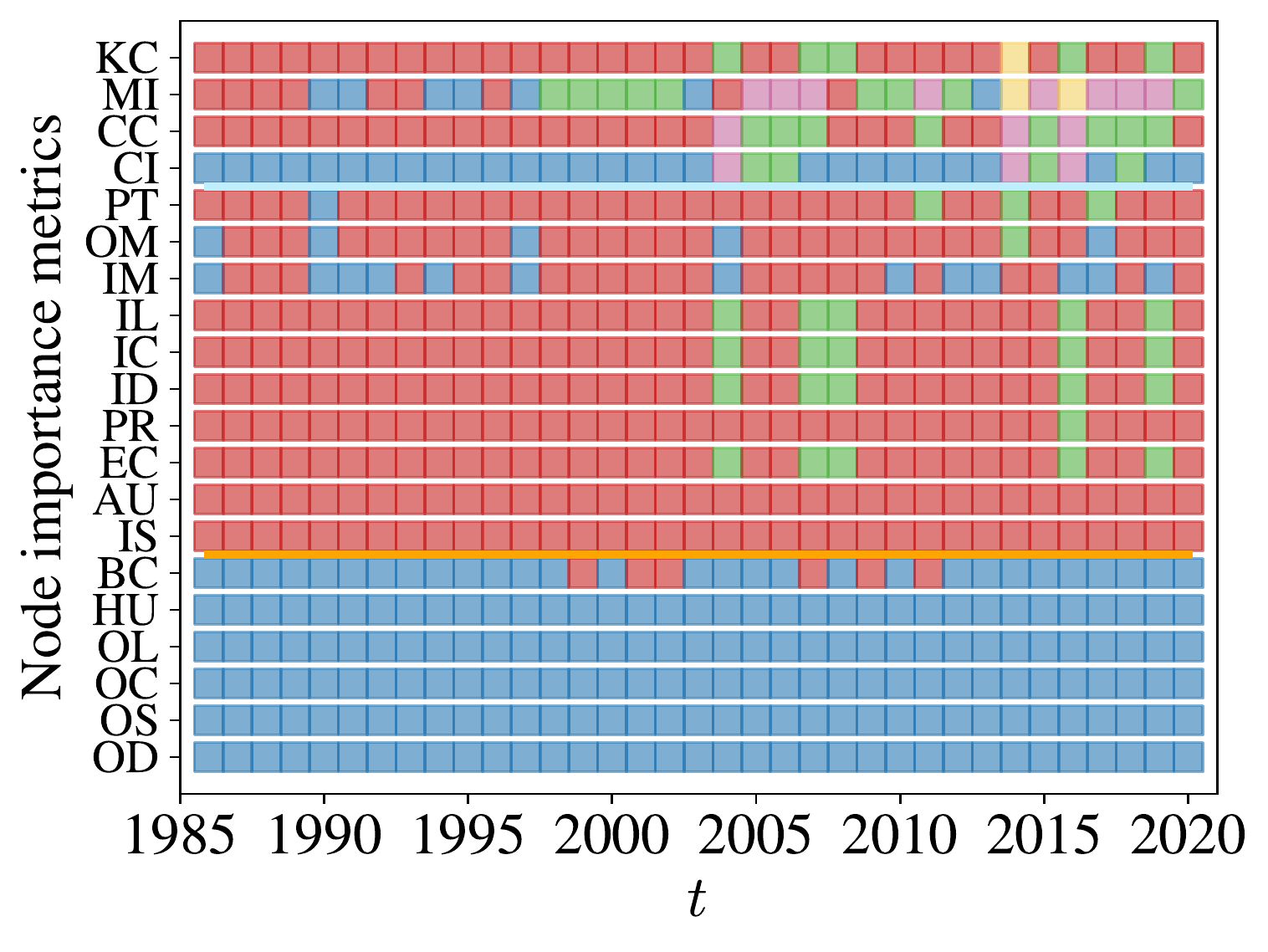}}
    \subfigure[]{\label{level.sub.16}\includegraphics[width=0.233\linewidth]{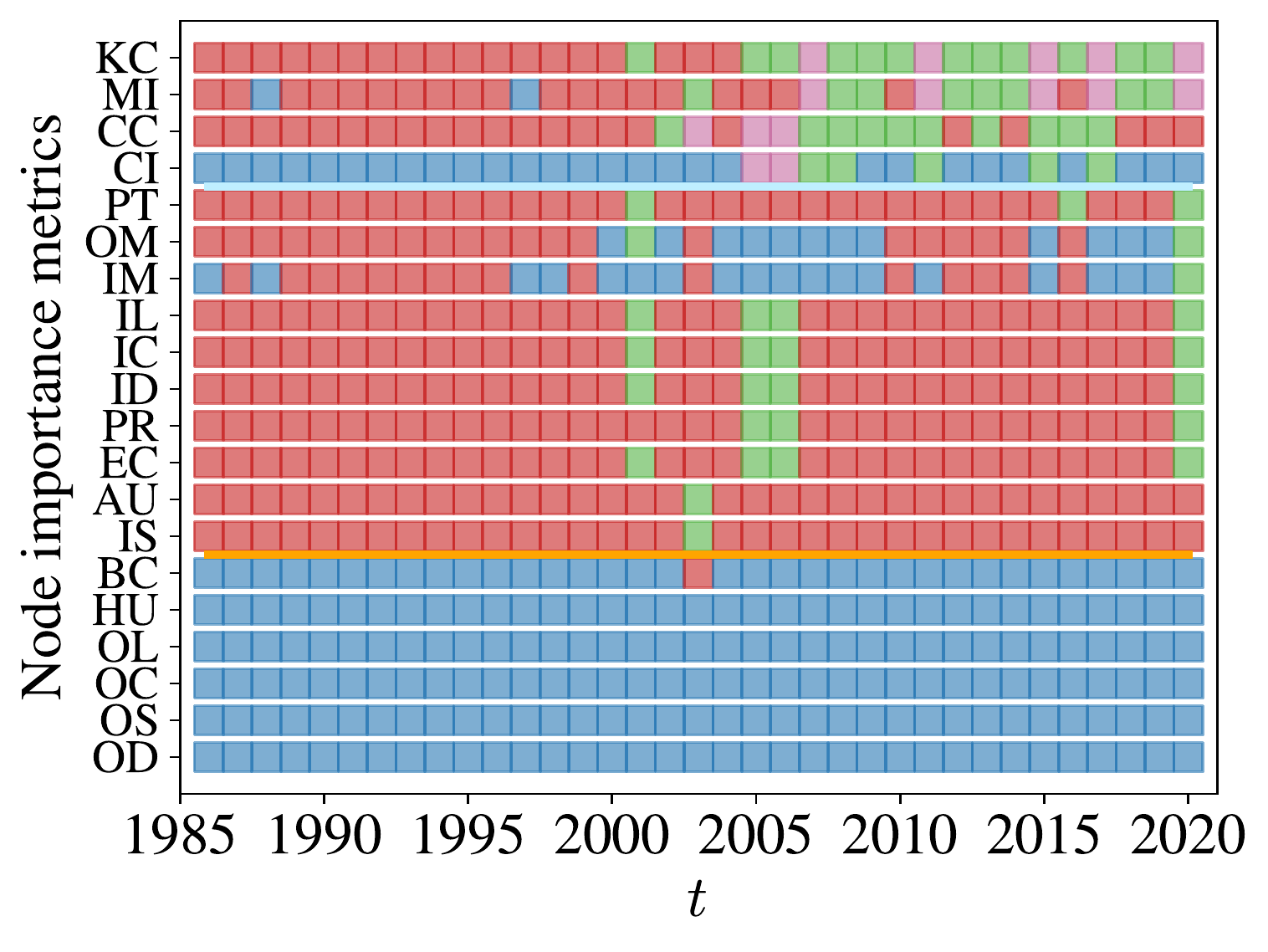}}
      \caption{Evolution of clusters identification of rankings of node importance using hierarchical clustering based on the leaf ordering algorithm. (a-d) show clusters identification of rankings of node importance in 2020. (e-h) show evolution of clusters identification from 1986-2020. The indicators corresponding to squares of the same color are sorted into the same cluster.}
      \label{Fig:iCTN:Cluster}
\end{figure}

To further characterize the correlation structure of node important metrics, we adopt hierarchical clustering based on the leaf ordering algorithm to search for clusters. Figure~\ref{Fig:iCTN:Cluster}(a-d) present cluster identification of rankings of node importance in 2020. We find that these metrics can be roughly divided into three groups. One is determined by the import scale, such as in-degree centrality, in-strength, in-closeness centrality, in-semi-local centrality, eigenvector centrality, Authorities, PageRank, within-module degree, outside-module degree, and participation coefficient. One is based on the export scale, namely, out-degree centrality, out-strength, out-closeness centrality, out-semi-local centrality, betweenness centrality, and Hubs. The remaining less relevant indicators are grouped into one group, i.e., those that present different characteristics in the iCTNs, such as constraint coefficient, clustering coefficient, Katz centrality, and mutual information. These clusters are not special cases in 2020. As shown in Fig.~\ref{Fig:iCTN:Cluster}(e-h), out-degree centrality, out-strength, out-closeness centrality, out-semi-local centrality, betweenness centrality, and Hubs (indicators marked under the orange line) are always in the same cluster, except for the occasional absence of betweenness centrality and Hubs. Indicators based on import scales are typically classified into the same cluster. Other indicators' behaviors are impenetrable. The results of clustering are also aligned with the results of our correlation analysis for these indicators.

\subsection{Distribution of eigenvalues}
\label{S3-3:Eigenvalues}

Different nodal importance indicators measure the influence of an economy in the iCTNs from different perspectives. It is comparatively biased to assess the importance of an economy using a single indicator. According to the correlation analysis above, it is worth mentioning that some indicators correlate with indicators based on the same flow directions. To probe into the correlation structure of these node importance metrics, we imitate the random matrix theory for constructing optimal eigenportfolios in the stock market to reduce the dimension and construct a composite node importance indicator.

 \begin{figure}[h!]
      \centering
      \includegraphics[width=0.233\linewidth]{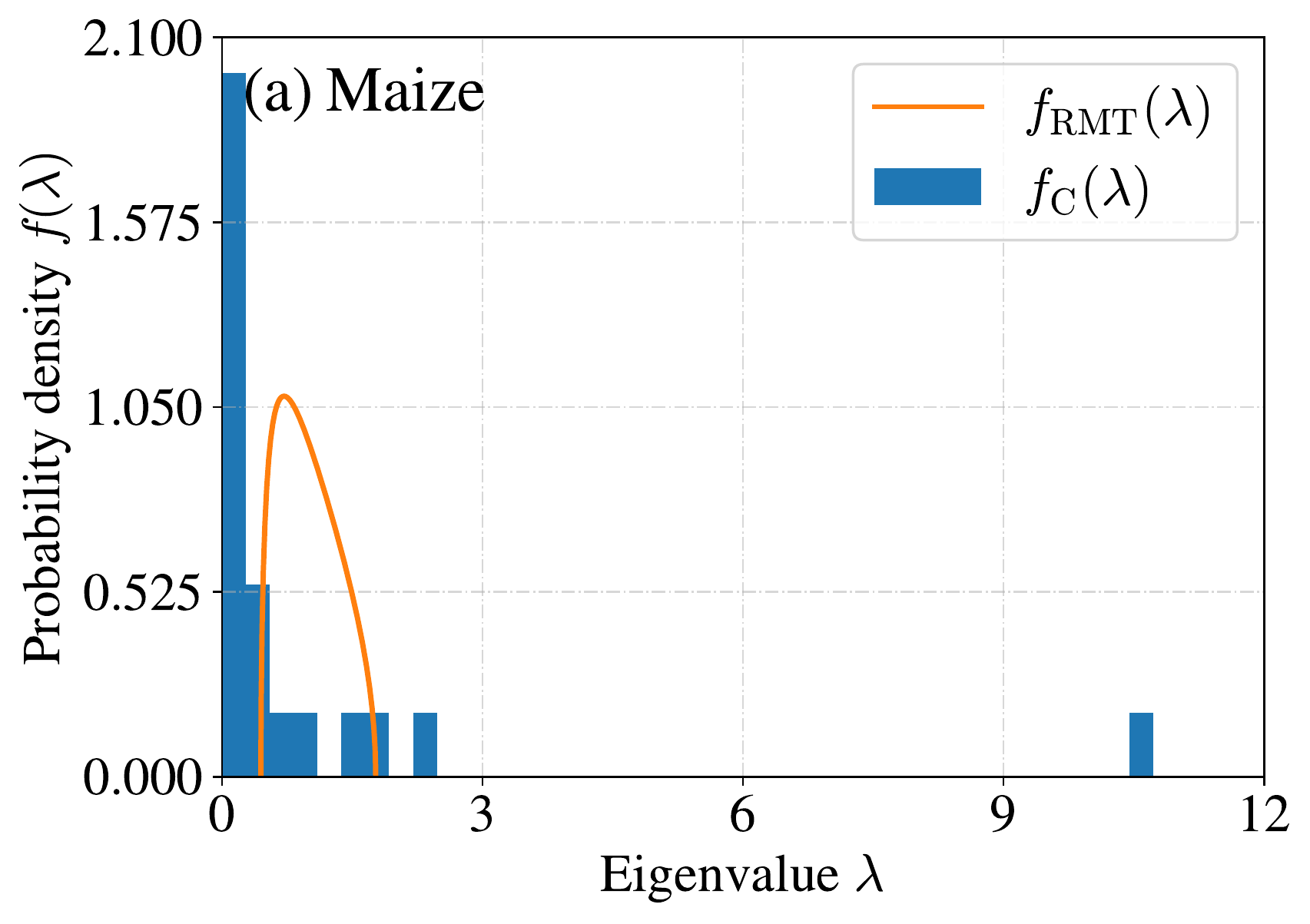}
      \includegraphics[width=0.233\linewidth]{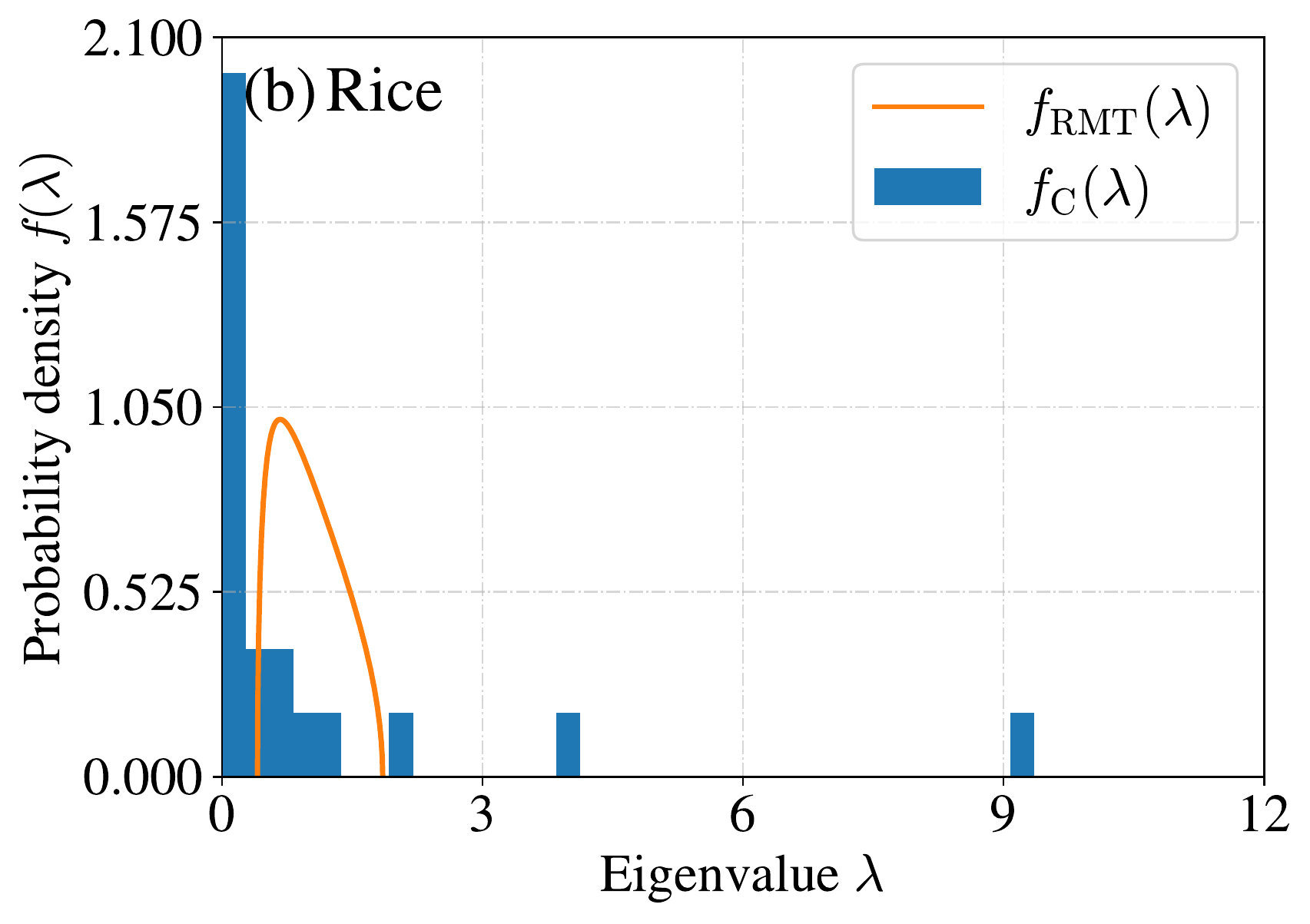}
      \includegraphics[width=0.233\linewidth]{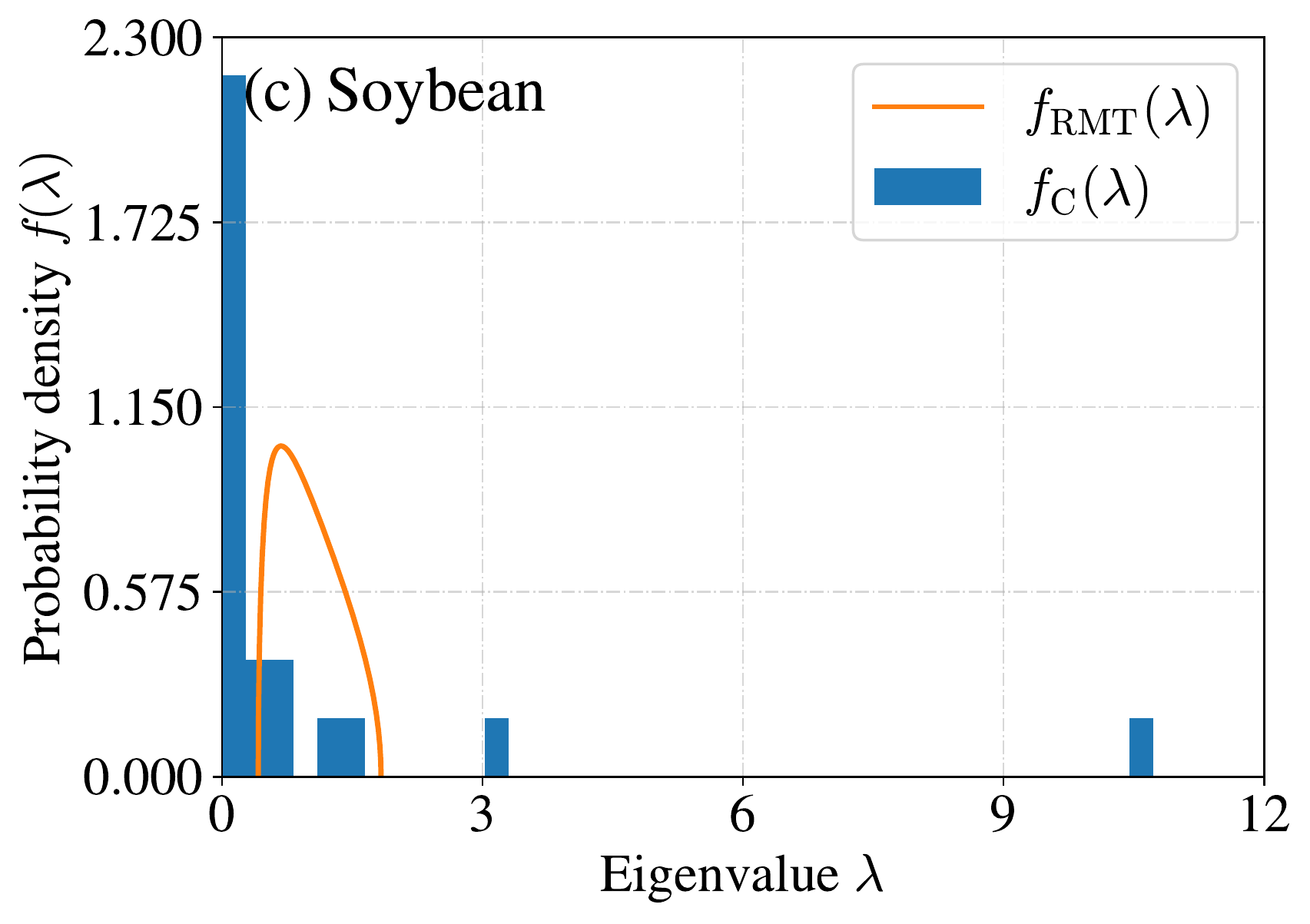}
      \includegraphics[width=0.233\linewidth]{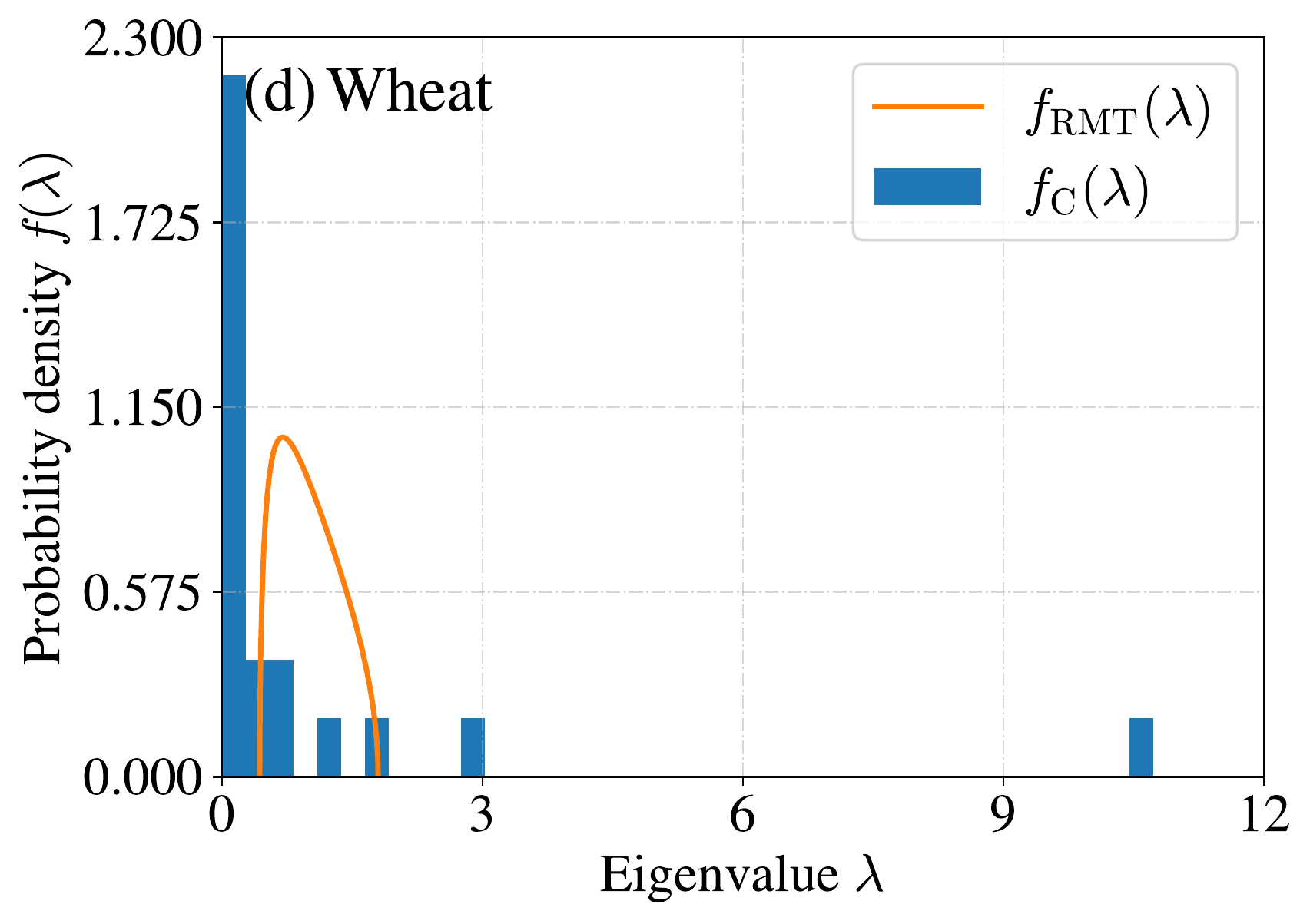}
      \caption{Empirical distribution of eigenvalues of the correlation matrix $C$. The solid curve is the theoretical distribution $f_{\mathrm{RMT}}(\lambda)$ given by Eq.~(\ref{Eq:RMT:PDF:eigenvalue}) of random matrix theory (RMT) in 2020.}
      \label{Fig:iCTN:PDF:eigenvalue:2020}
\end{figure}

We obtain eigenvalues and their corresponding eigenvectors of correlation matrices of economies' trade influence rankings in the four iCTNs. The largest eigenvalues $\lambda_{\max }^{\mathrm{RMT}}$ of four correlation matrices in 2020 are respectively 1.7705, 1.8506, 1.8321 and 1.7983. The smallest eigenvalues $\lambda_{\min }^{\mathrm{RMT}}$ are respectively 0.4481, 0.4091, 0.4179, 0.4343. We calculate 20 eigenvalues of the empirical correlation matrix $\mathrm{C}$ and sort $\lambda_{i}$ in descending order ($\lambda_i>\lambda_{i+1}$). For the international maize trade network, the smallest eigenvalue $\lambda_{20} = 0.0014$, which is about 1/320 of $\lambda_{\min }^{\mathrm{RMT}}=0.4481$, and the largest eigenvalue $\lambda_{20} = 10.6541$, which is about 6 times larger than $\lambda_{\max }^{\mathrm{RMT}}=1.7705$. Similar cases about the relationships between empirical correlation matrices and random matrices exist in the other three iCTNs. 
Figure~\ref{Fig:iCTN:PDF:eigenvalue:2020} compares the empirical distribution $P(\lambda)$ of these eigenvalues of $\mathrm{C}$ with the theoretical distribution $f_{\mathrm{RMT}}(\lambda)$ based on the random matrix $\mathrm{RMT}$. We find that most of the eigenvalues of $\mathrm{C}$ deviate from $f_{\mathrm{RMT}}(\lambda)$. For maize, rice, soybean and wheat, 3, 3, 3 and 3 minimum eigenvalues are respectively smaller than $\lambda_{\min }^{\mathrm{RMT}}$, 13, 12, 12 and 12 maximum eigenvalues are respectively larger than $\lambda_{\max }^{\mathrm{RMT}}$, and 4, 5, 6 and 5 eigenvalues are covered within $f_{\mathrm{RMT}}(\lambda)$.

 \begin{figure}[h!]
      \centering
      \includegraphics[width=0.475\linewidth]{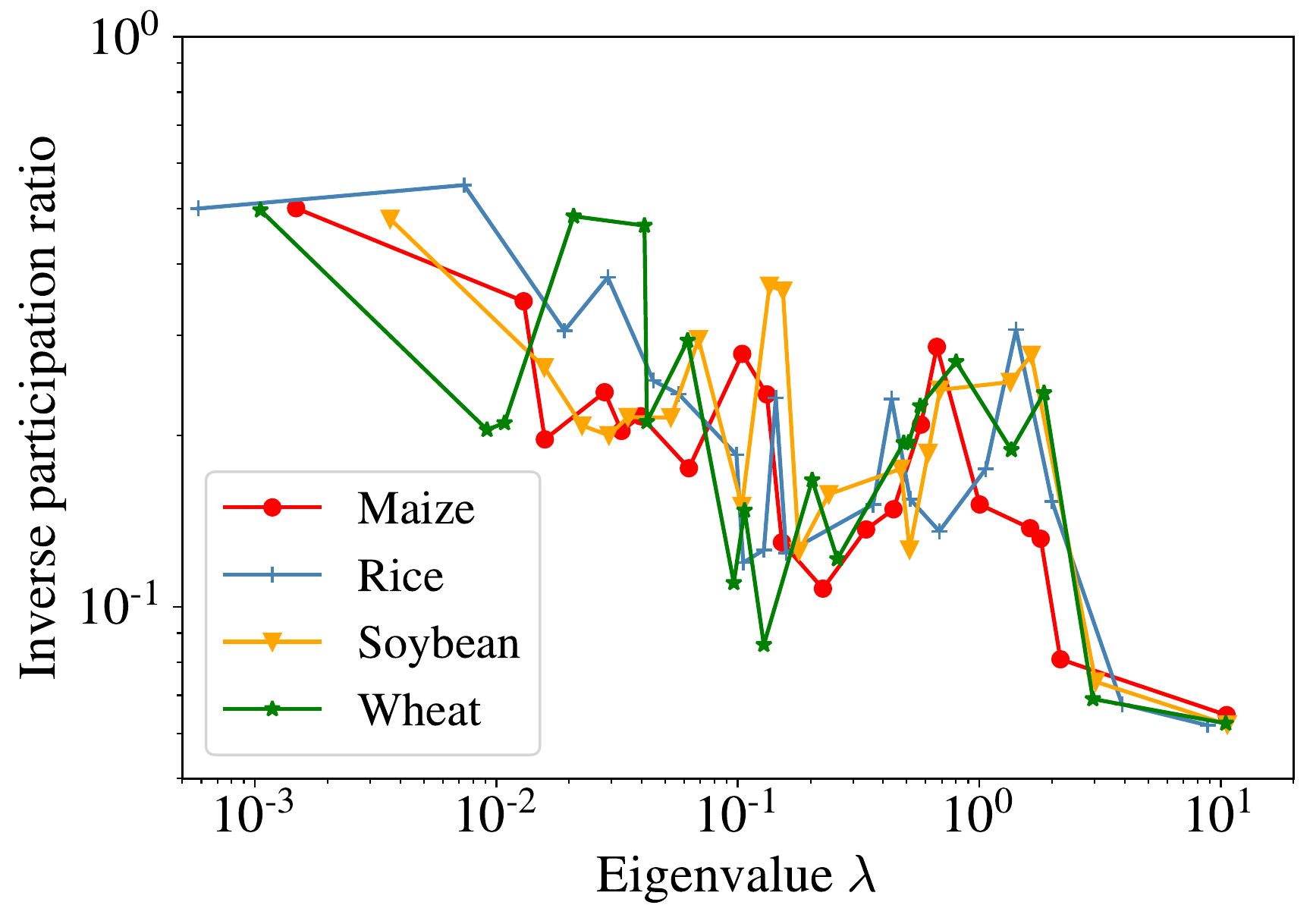}
      \caption{Inverse participation ratios of the eigenvalues for the correlation matrix $\mathrm{C}$ in 2020.}
      \label{Fig:iCTN:IPR:2020}
\end{figure}

To capture the deviation of the correlation matrix $\mathrm{C}$ from the results of the random matrix $\mathrm{RMT}$, we present the inverse participation ratio (IPR). Figure~\ref{Fig:iCTN:IPR:2020} shows the inverse participation ratios of all eigenvalues of the four correlation matrices. The average inverse participation ratios are 0.194, 0.222, 0.214, and 0.222, respectively, indicating that the average contribution is 5. The inverse participation ratios $I^1$ corresponding to the largest eigenvalue $\lambda_1$ are 0.065, 0.060, 0.062, and 0.062, respectively, and their inverse is about 16, indicating that about 16 indicators contribute to the eigenvector $u_1$. The inverse participation ratios corresponding to the second largest eigenvalue, $\lambda_2$, which falls in the upper $\lambda_{\max }^{\mathrm{RMT}}$, are 0.734, 0.066, 0.074, and 0.069, indicating that about 14 indicators contribute to the eigenvector $u_2$. The inverse participation ratios corresponding to the other eigenvalues vary and generally show that the smaller the eigenvalue, the larger the corresponding inverse participation ratio. This indicates that the major eigenvalues dedicate most information to relative eigenvectors.

 \begin{figure}[h!]
      \centering
      \includegraphics[width=0.233\linewidth]{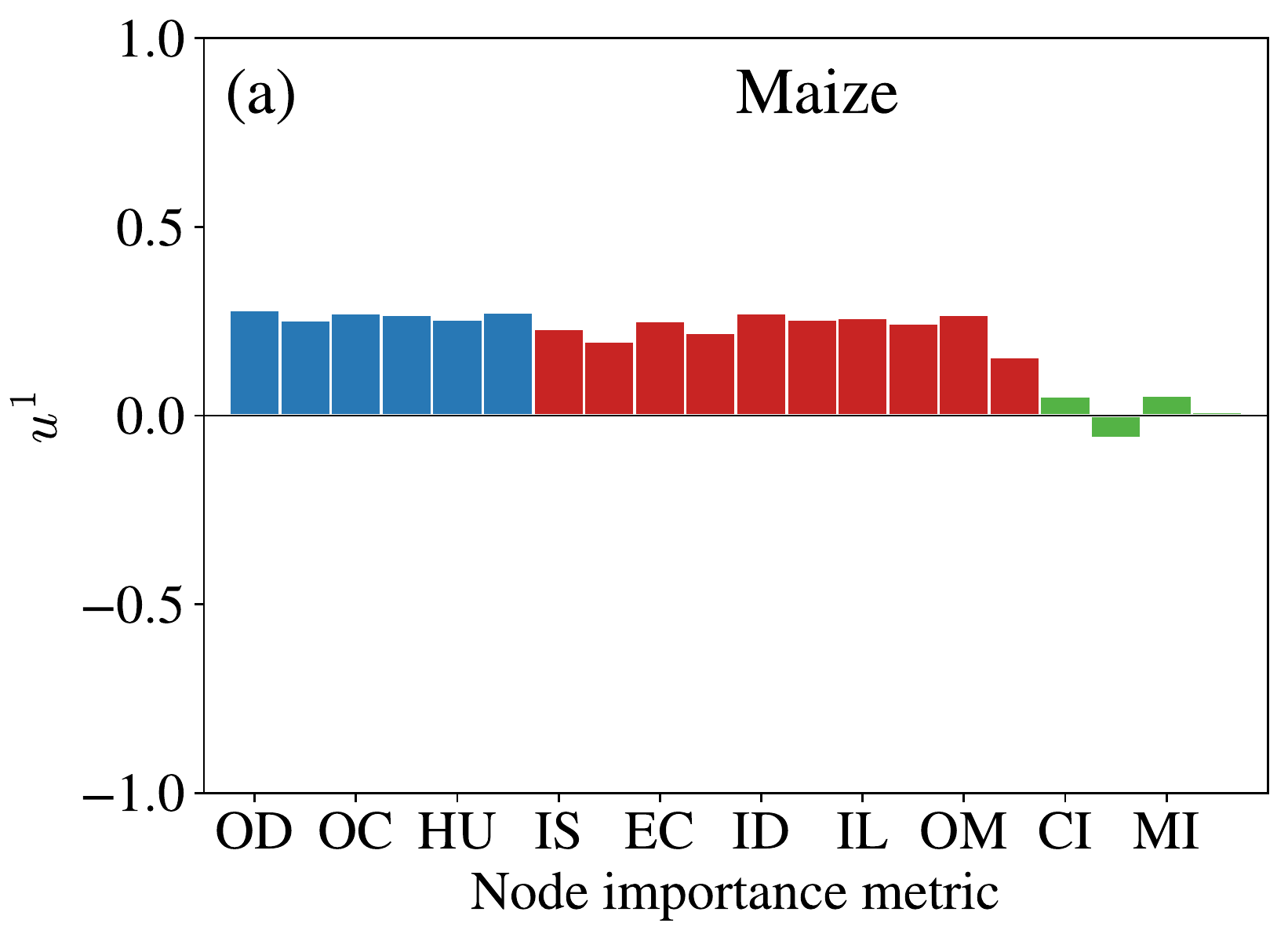}
      \includegraphics[width=0.233\linewidth]{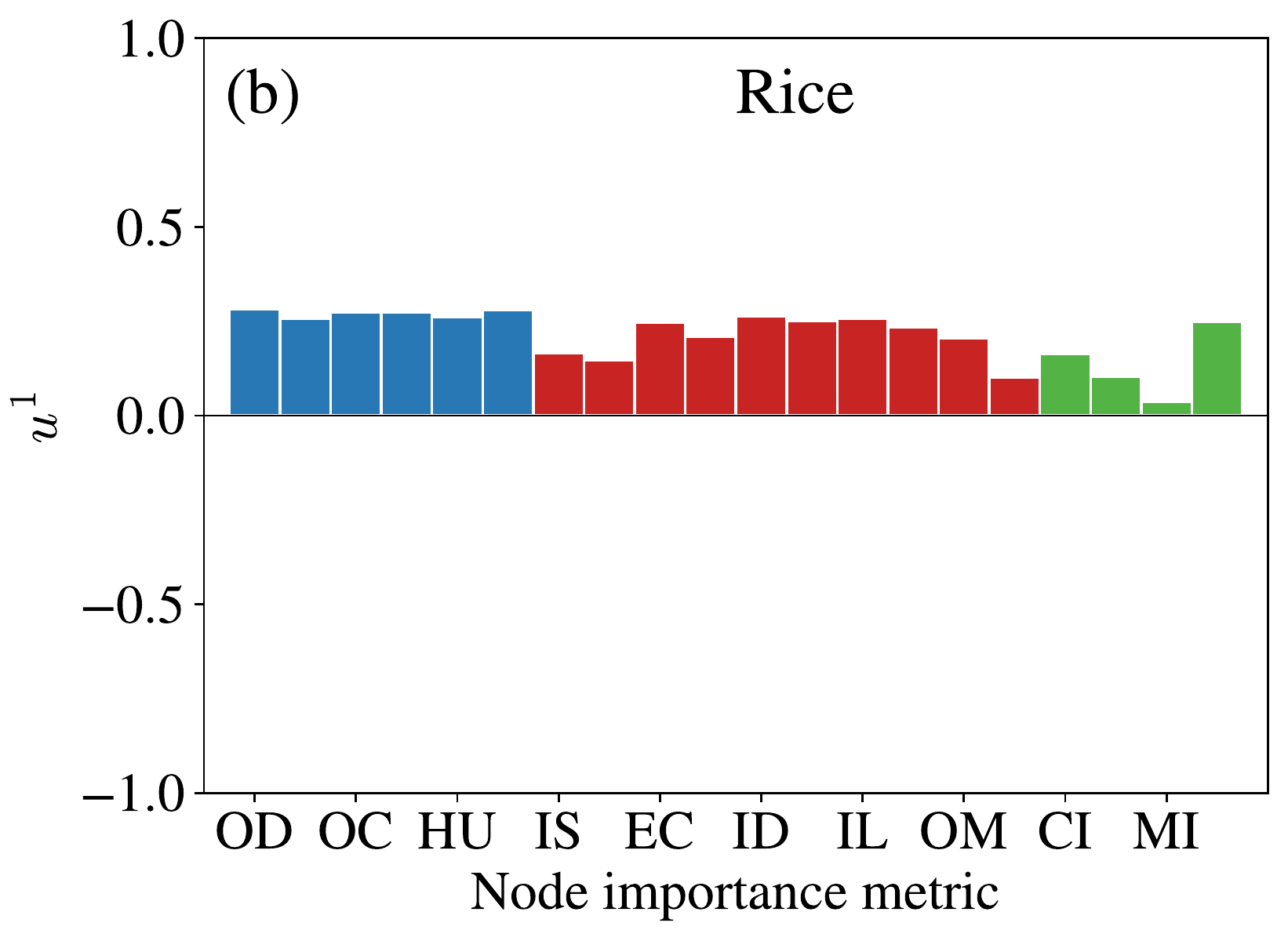}
      \includegraphics[width=0.233\linewidth]{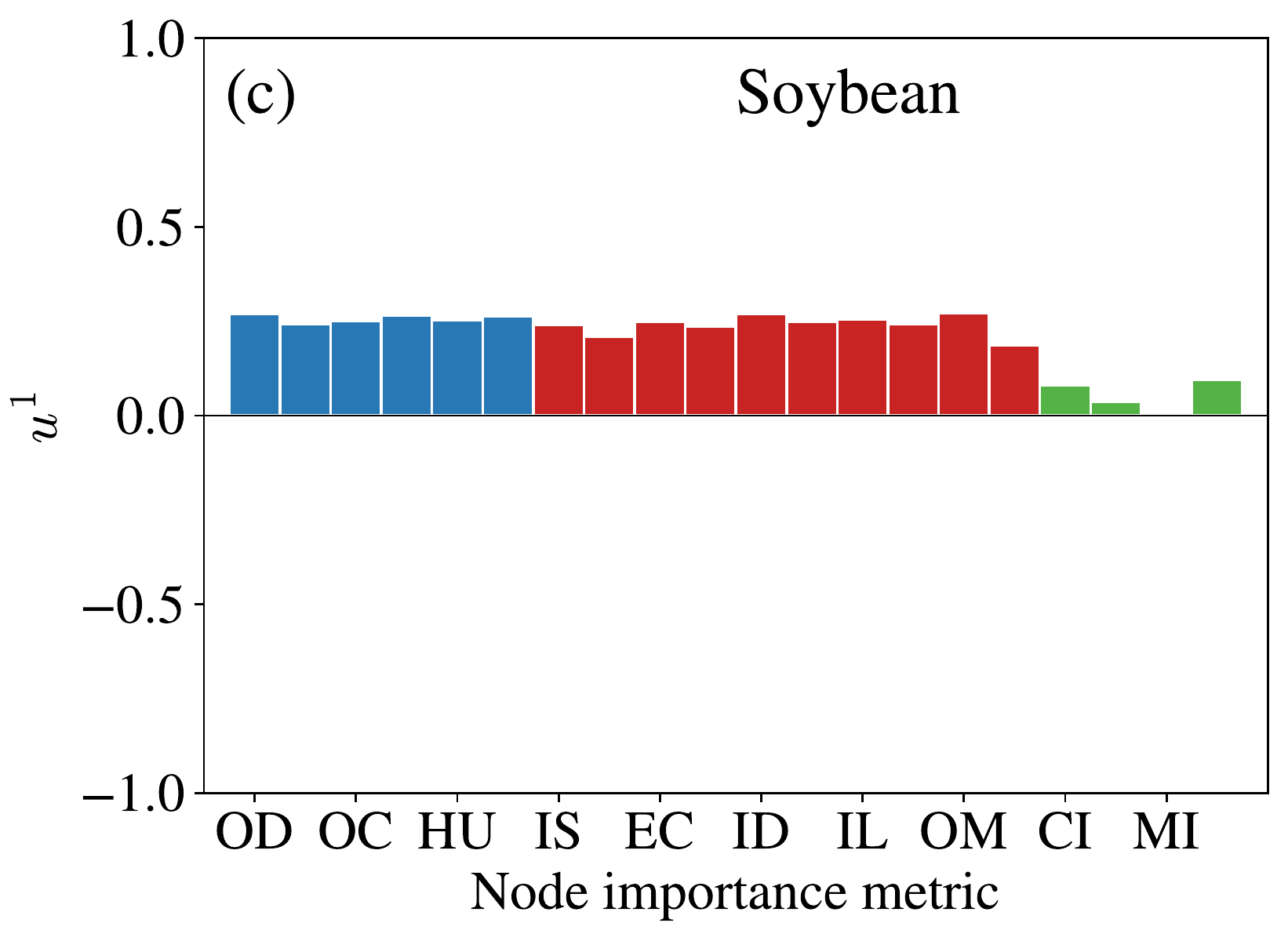}
      \includegraphics[width=0.233\linewidth]{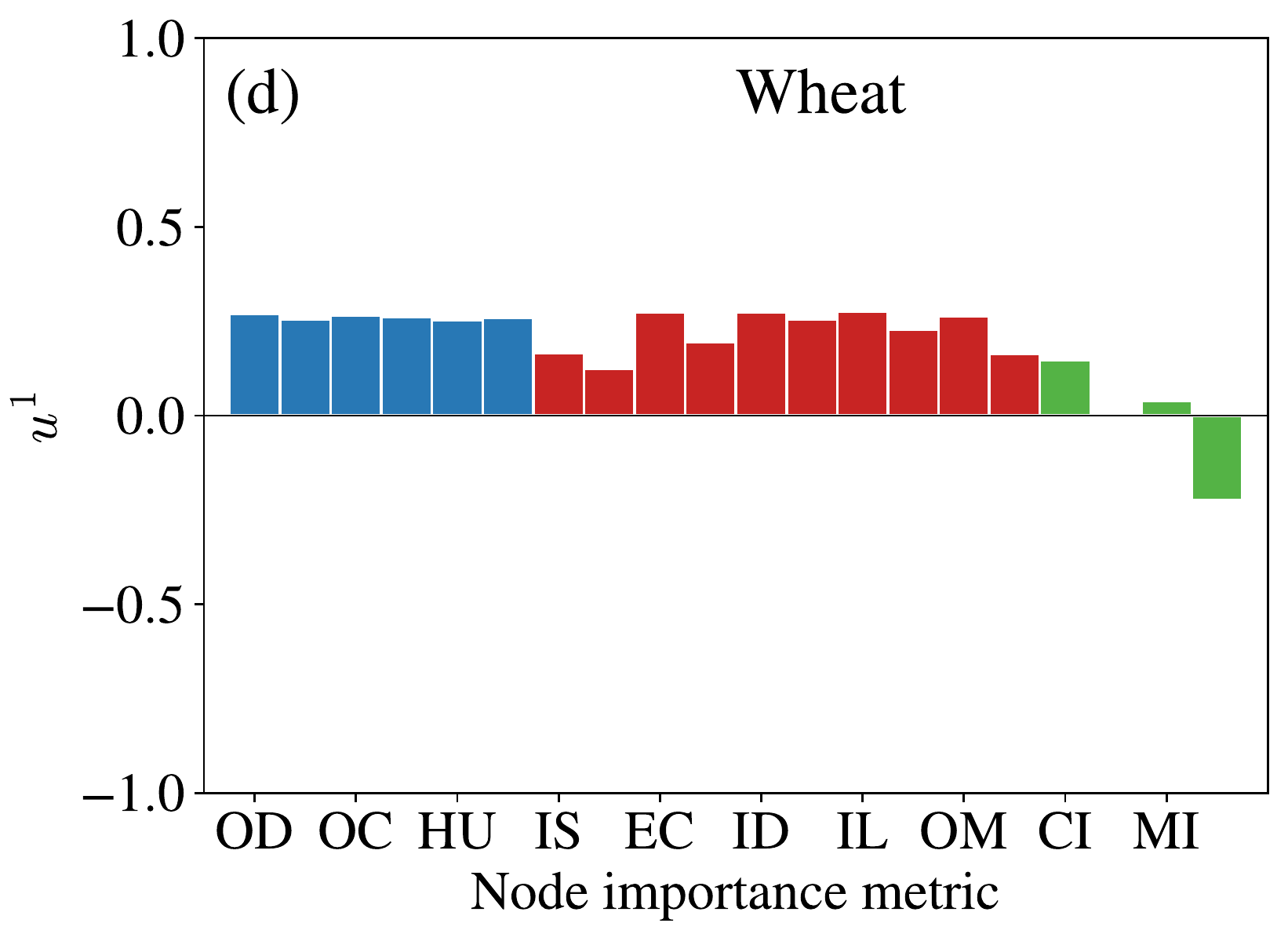}
       \\
      \includegraphics[width=0.233\linewidth]{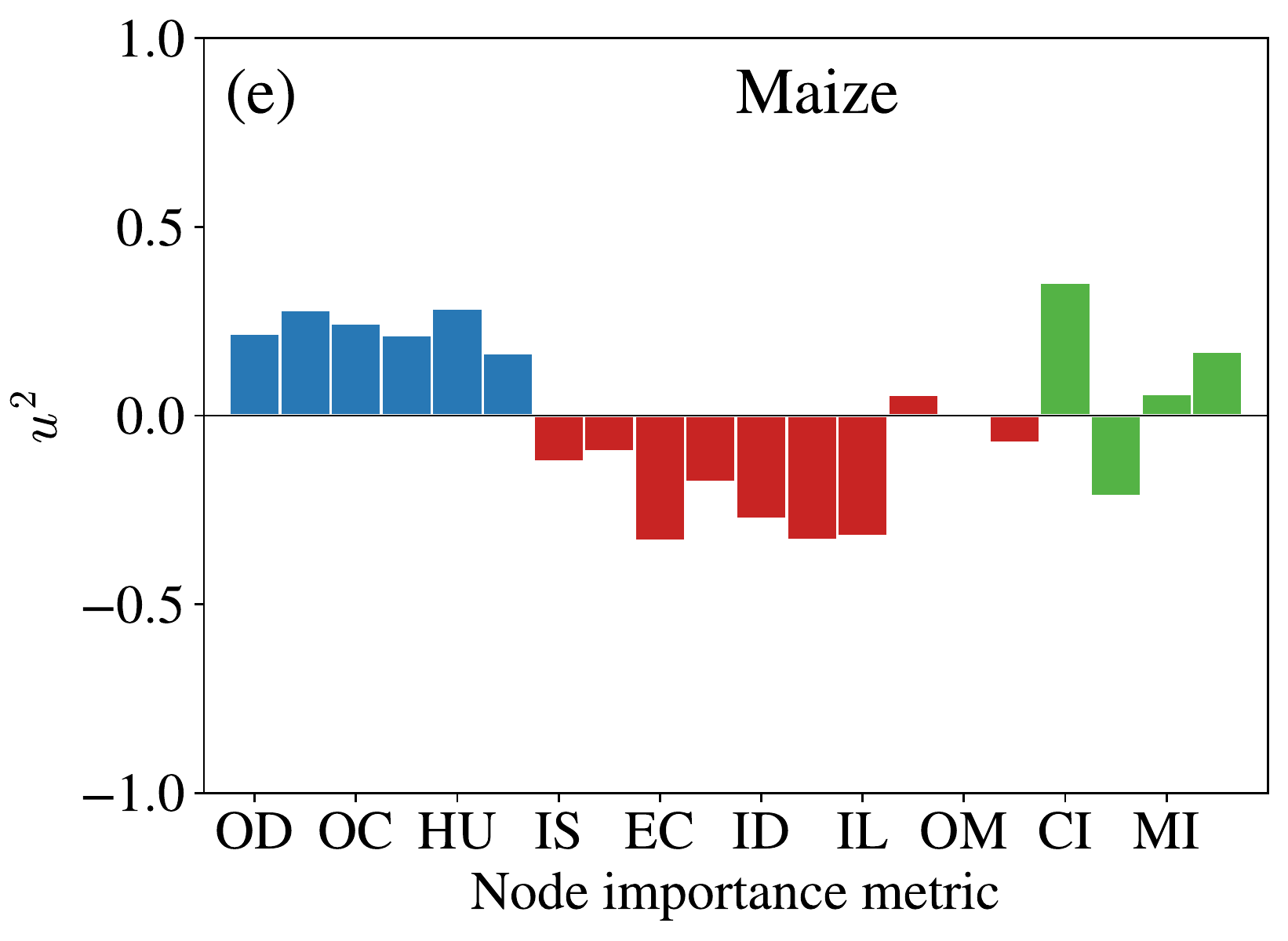}
      \includegraphics[width=0.233\linewidth]{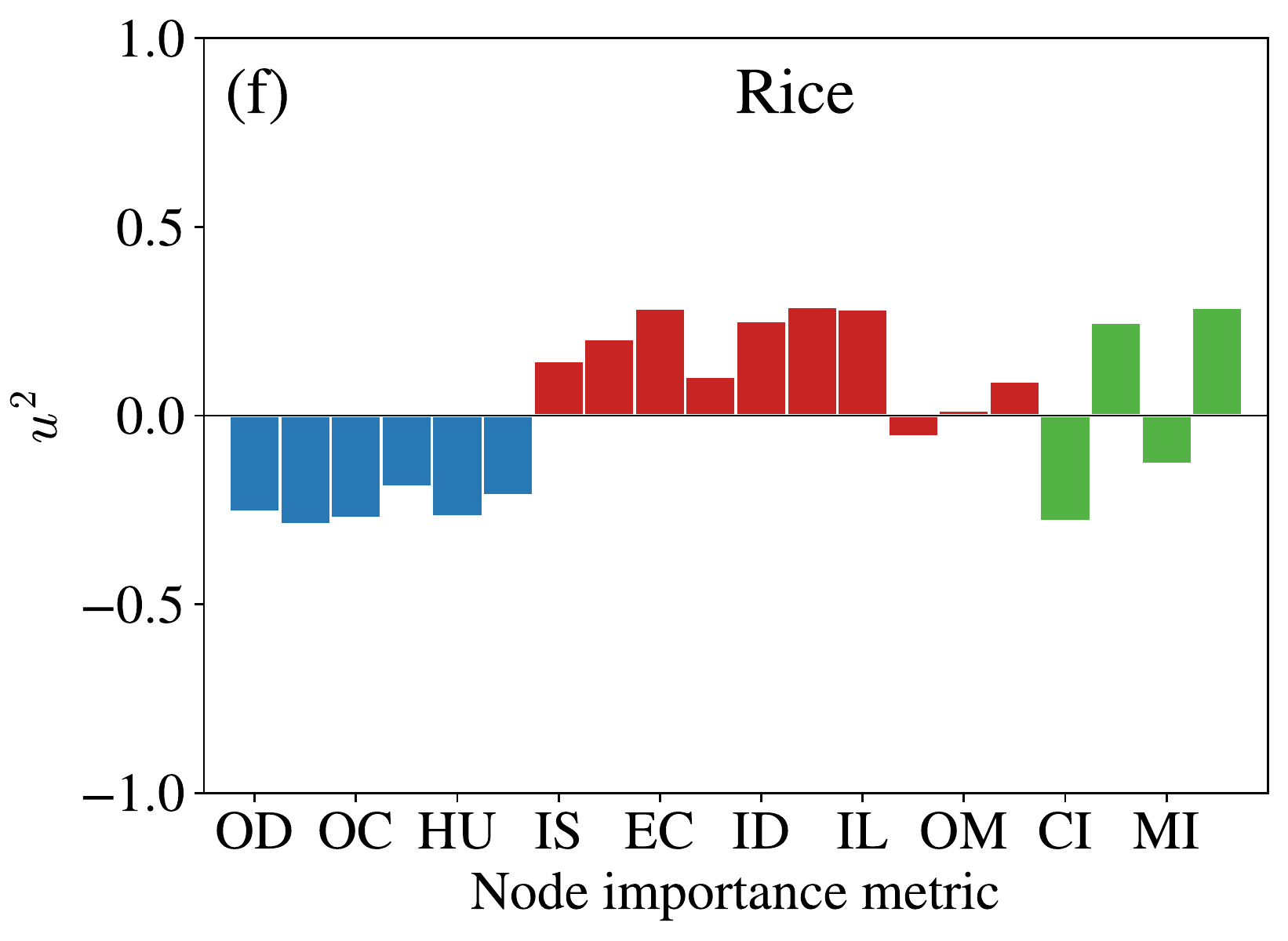}
      \includegraphics[width=0.233\linewidth]{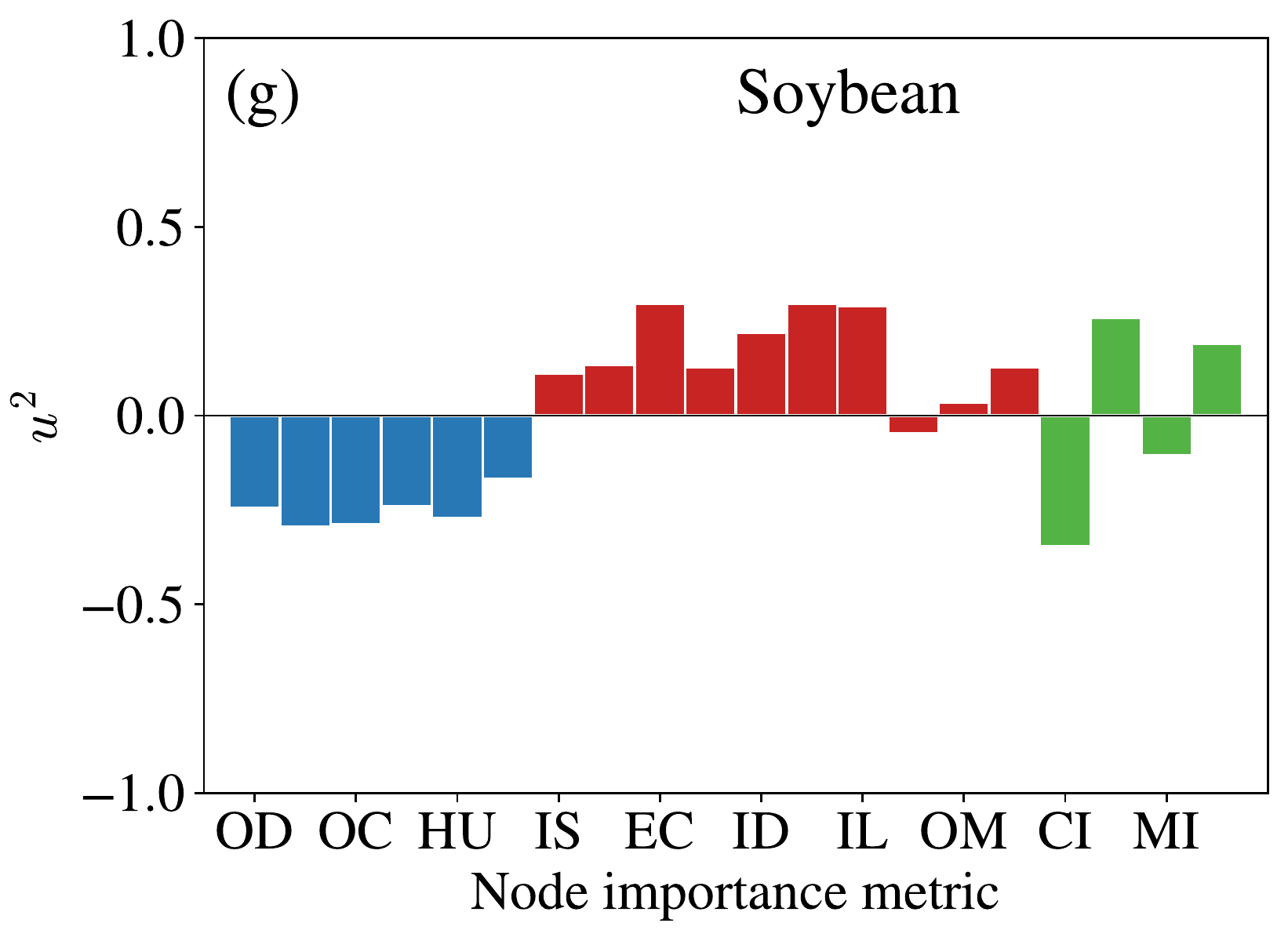}
      \includegraphics[width=0.233\linewidth]{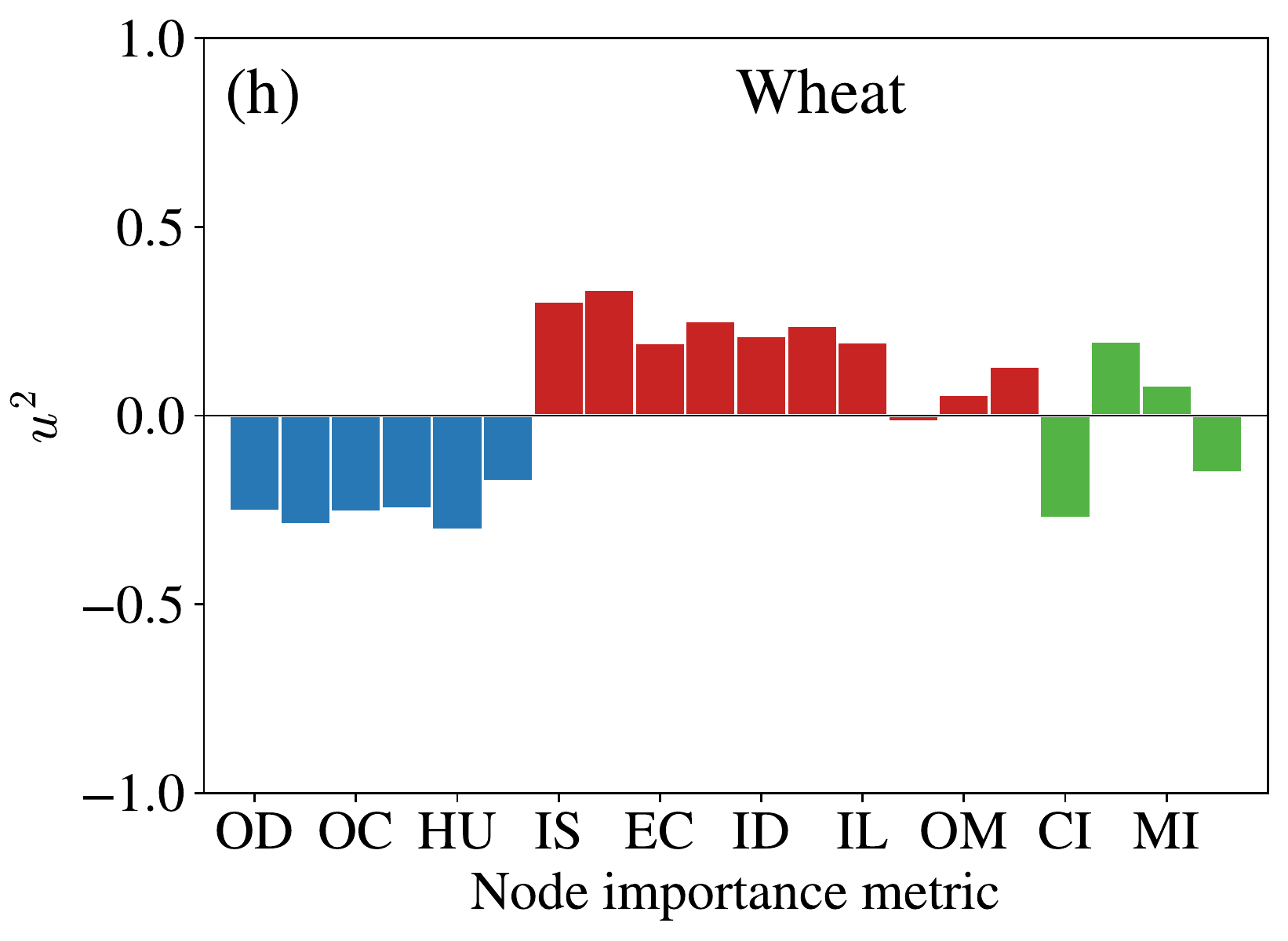}
       \\
      \includegraphics[width=0.233\linewidth]{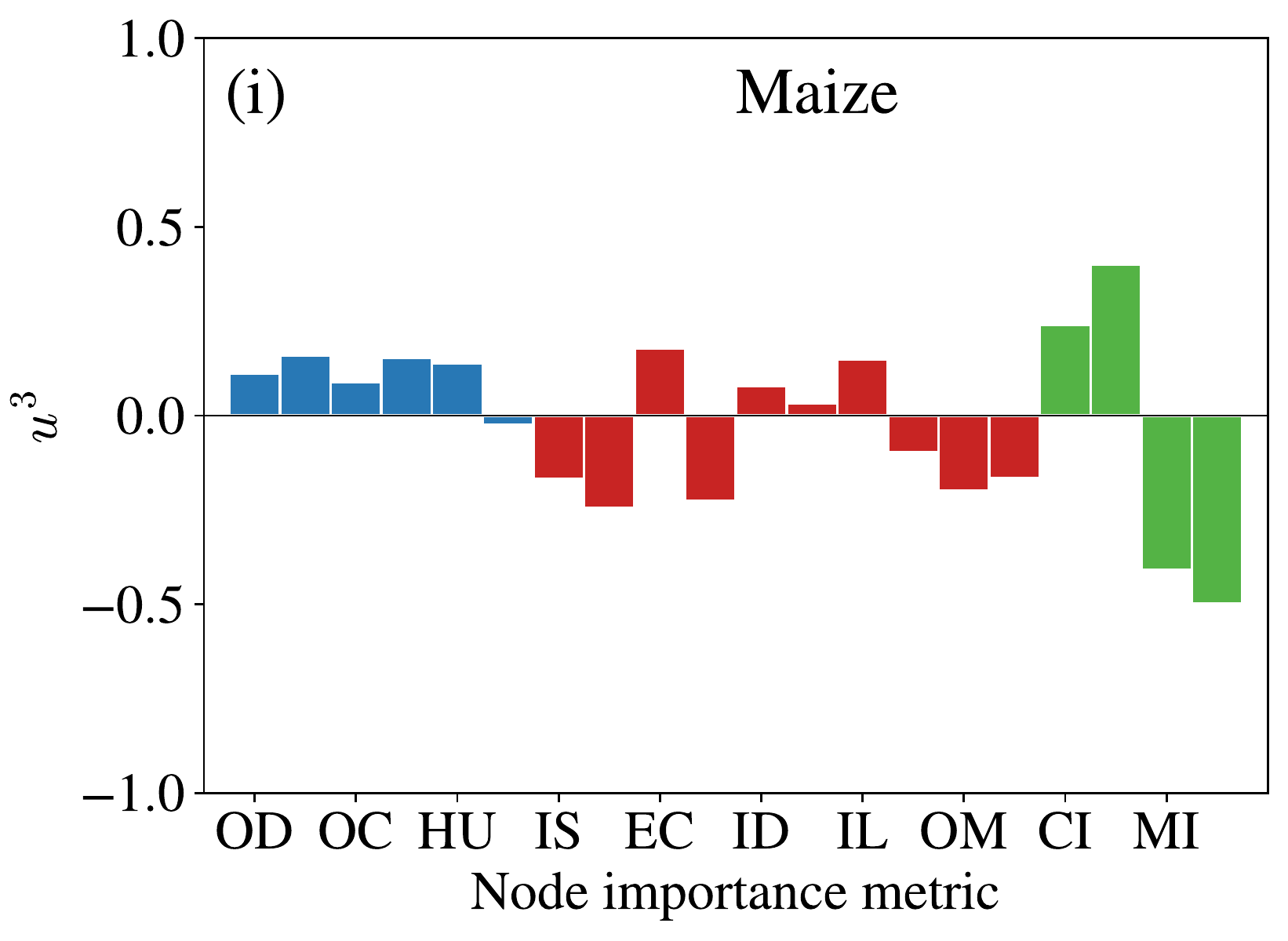}
      \includegraphics[width=0.233\linewidth]{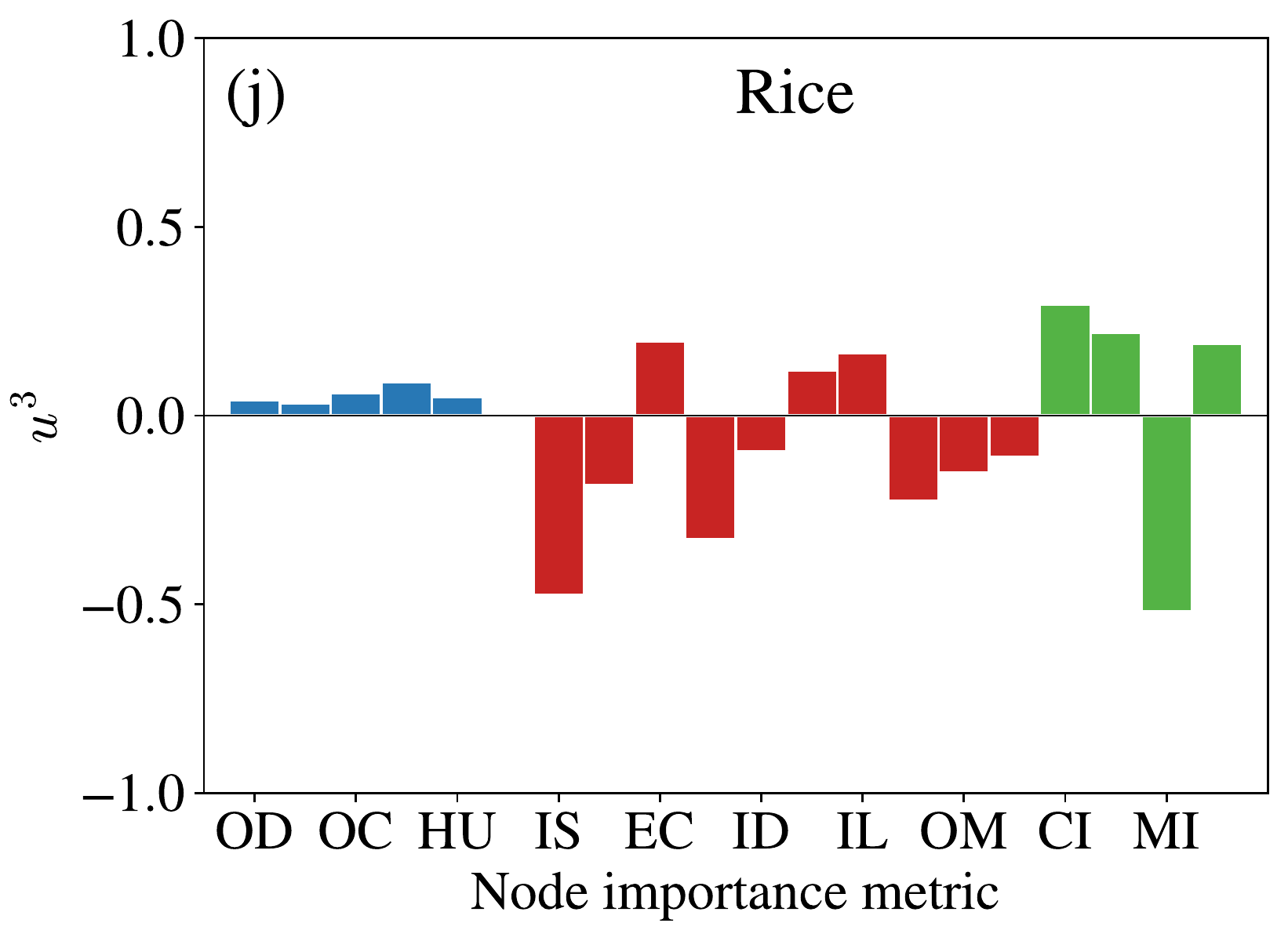}
      \includegraphics[width=0.233\linewidth]{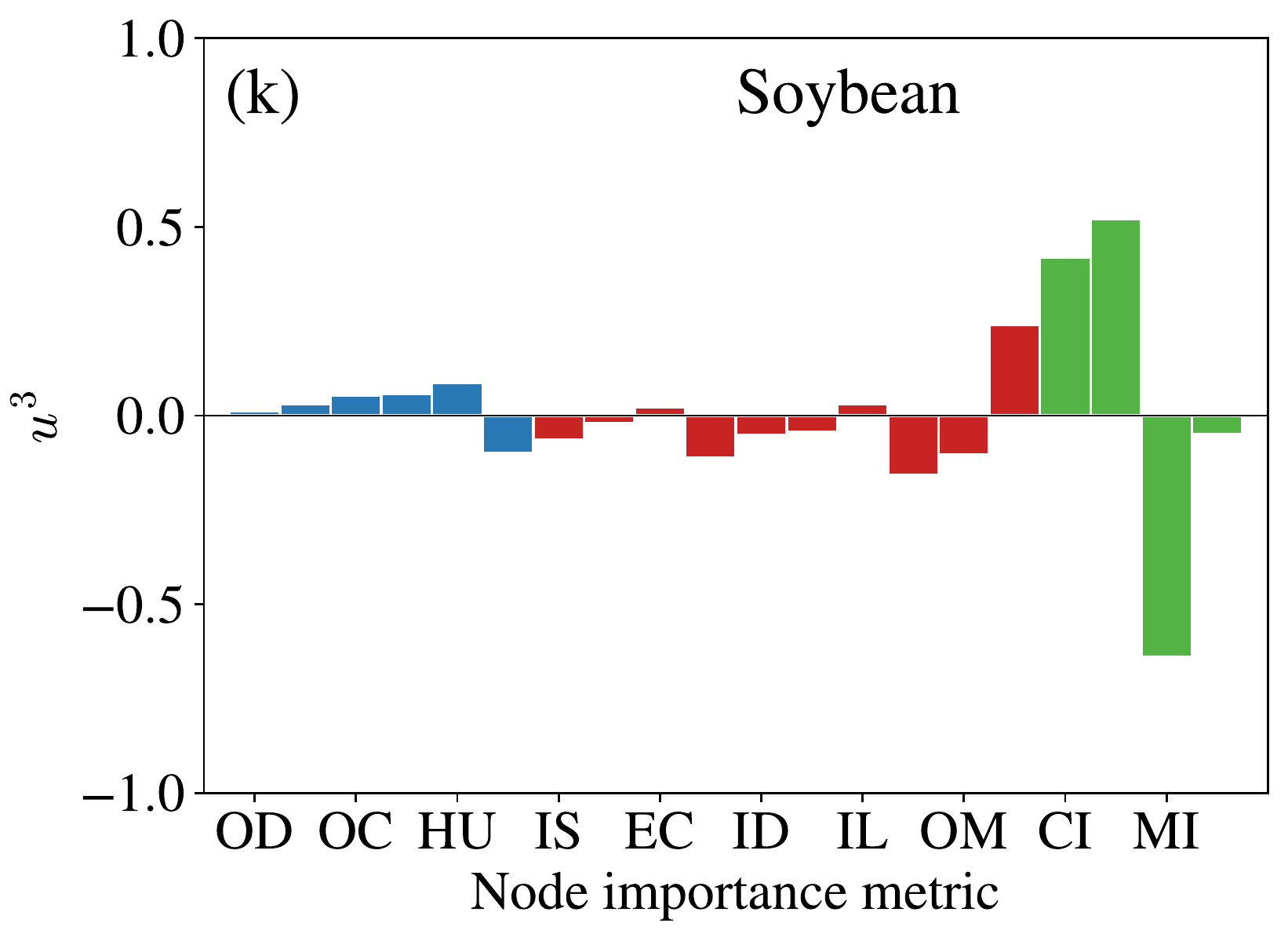}
      \includegraphics[width=0.233\linewidth]{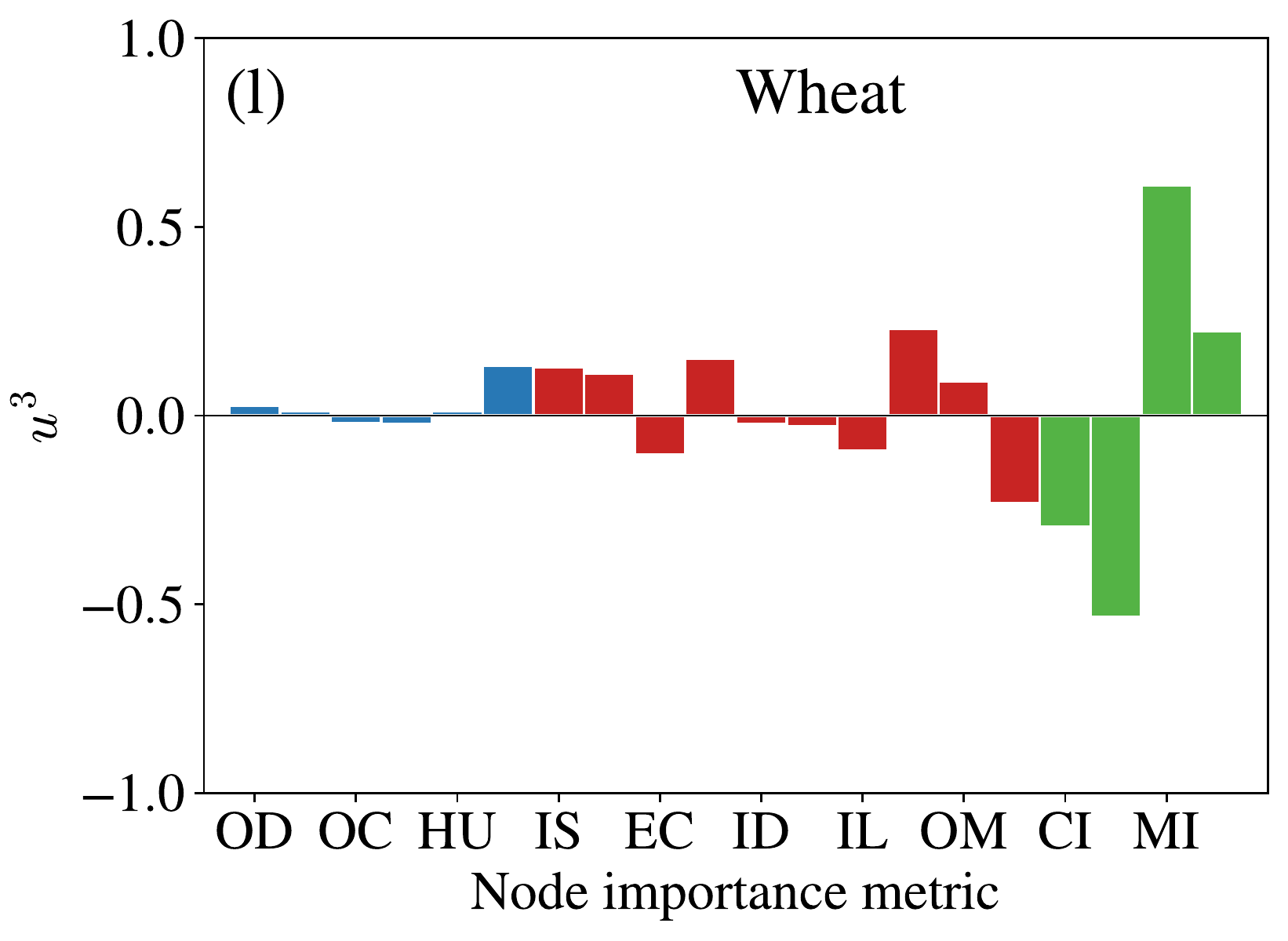}
       \\
      \includegraphics[width=0.233\linewidth]{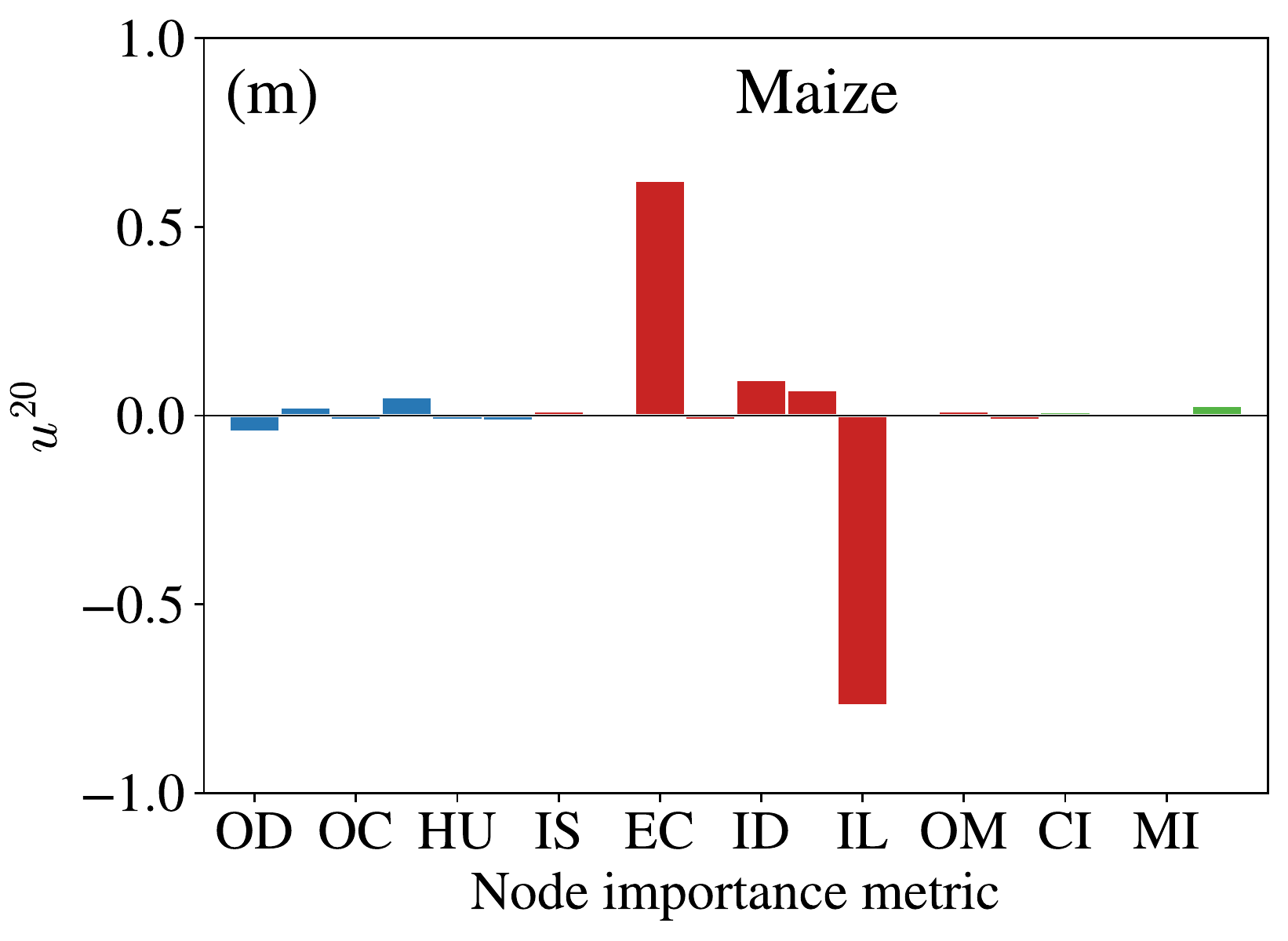}
      \includegraphics[width=0.233\linewidth]{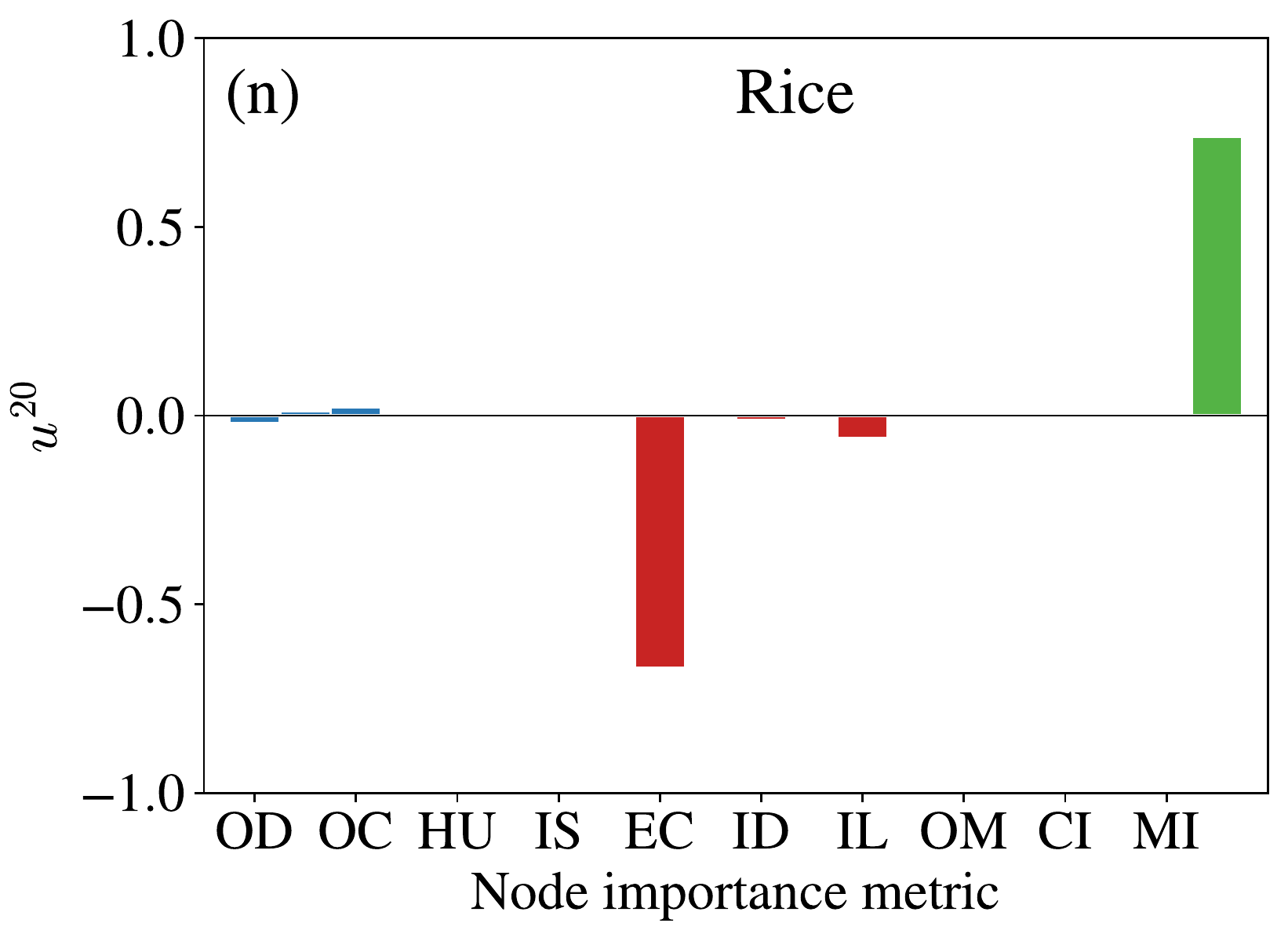}
      \includegraphics[width=0.233\linewidth]{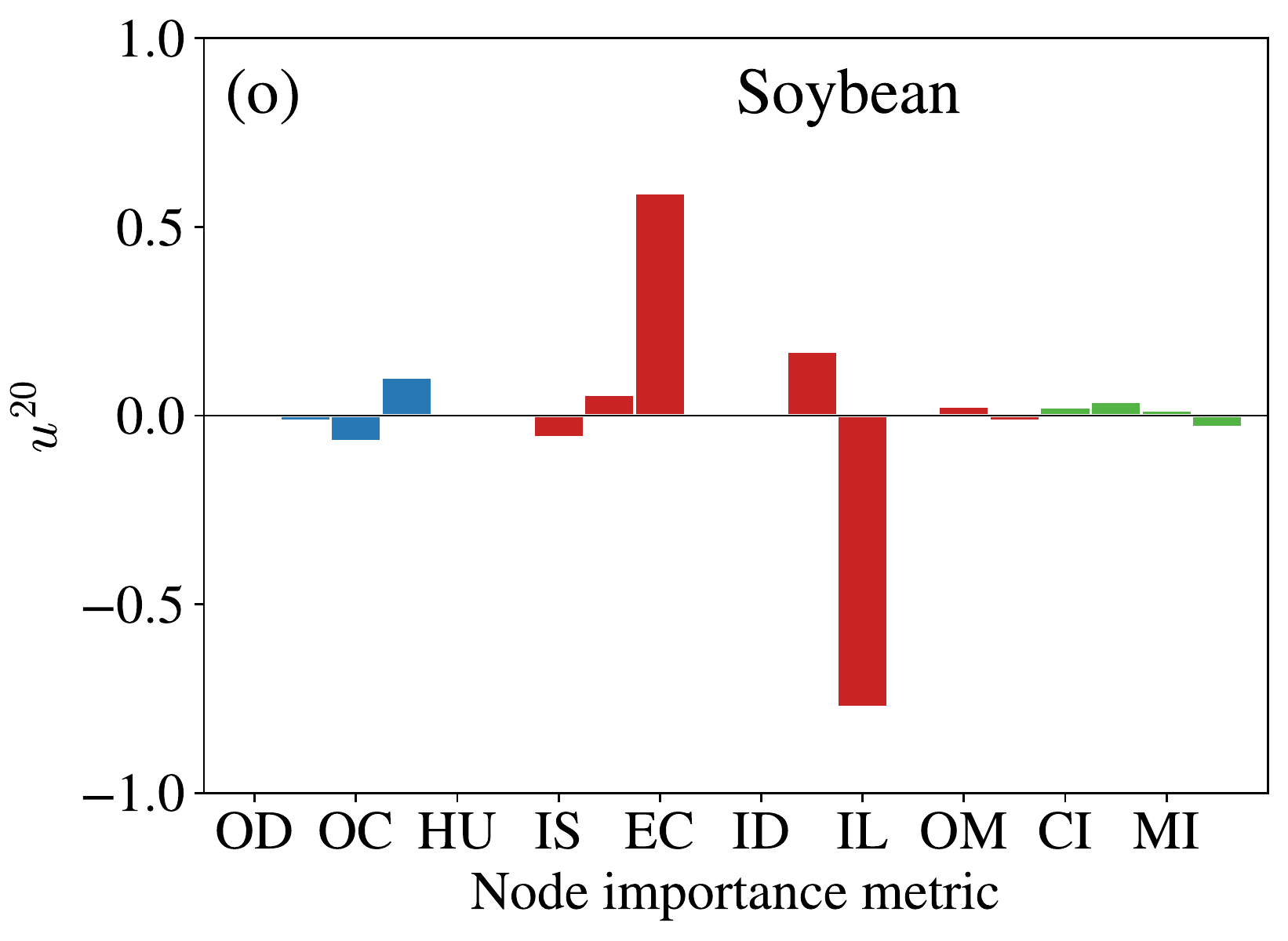}
      \includegraphics[width=0.233\linewidth]{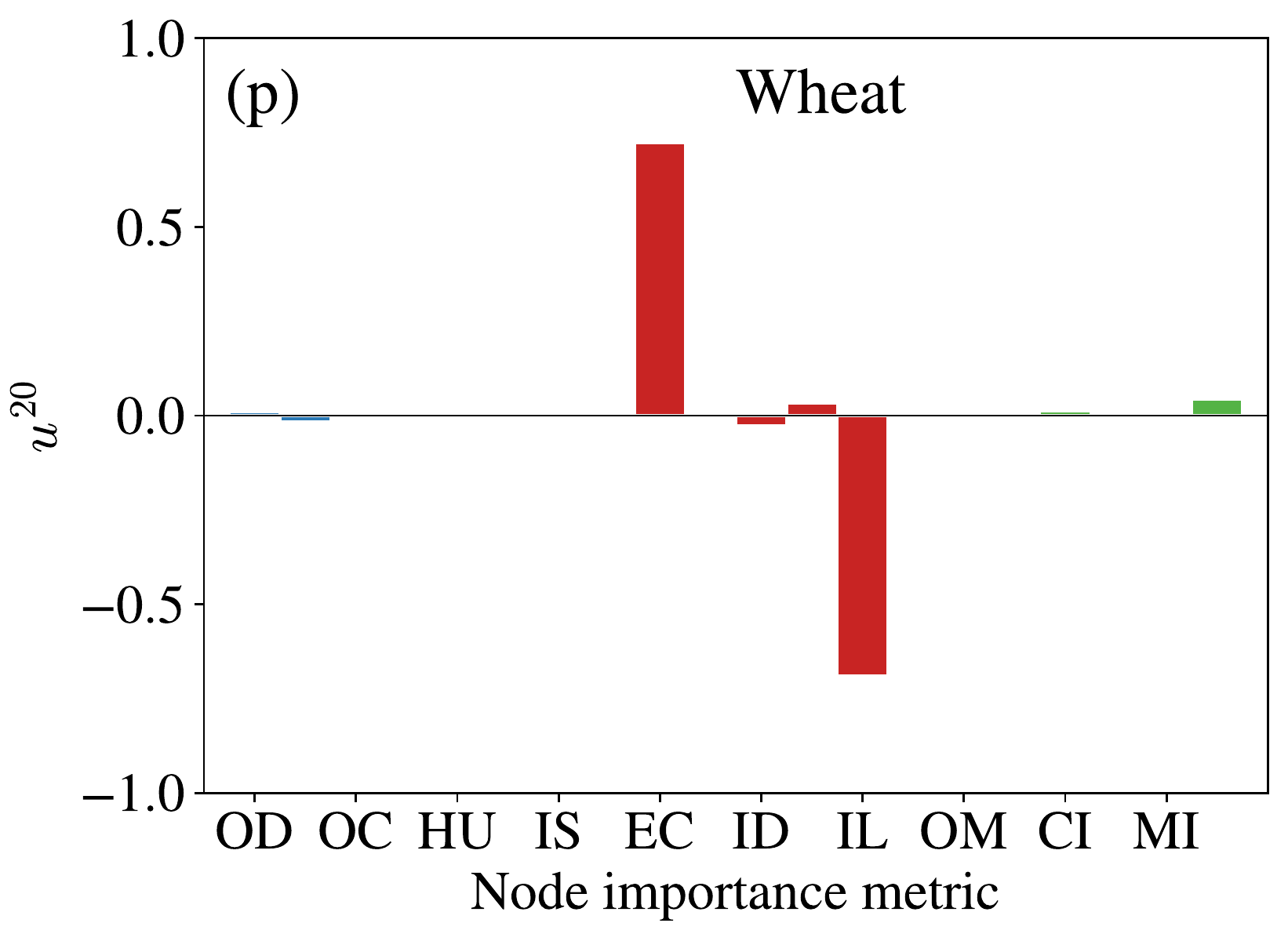}
      \\
      \caption{Components of the eigenvector $u_1$ of the largest eigenvalue $\lambda_1$, the eigenvector $u_2$ of the largest eigenvalue $\lambda_2$, the eigenvector $u_3$ of the largest eigenvalue $\lambda_3$ and the eigenvector $u_{20}$ of the smallest eigenvalue $\lambda_{20}$ given by Eq.~(\ref{Eq:RMT:PDF:eigenvalue}) of random matrix theory (RMT) in 2020.}
      \label{Fig:iCTN:PDF:eigenvector:component:2020}
\end{figure}

 \begin{figure}[h!]
      \centering
      \includegraphics[width=0.233\linewidth]{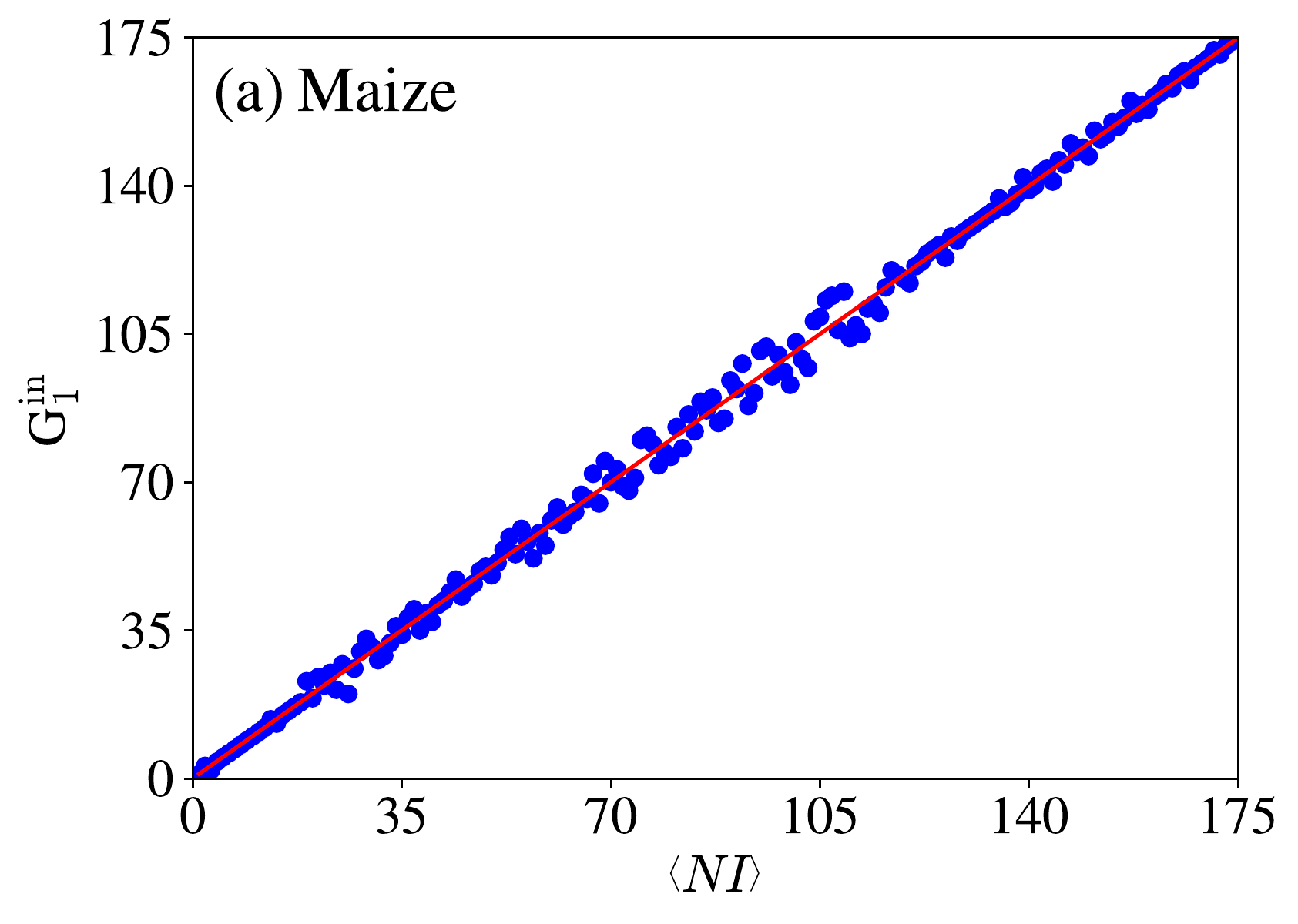}
      \includegraphics[width=0.233\linewidth]{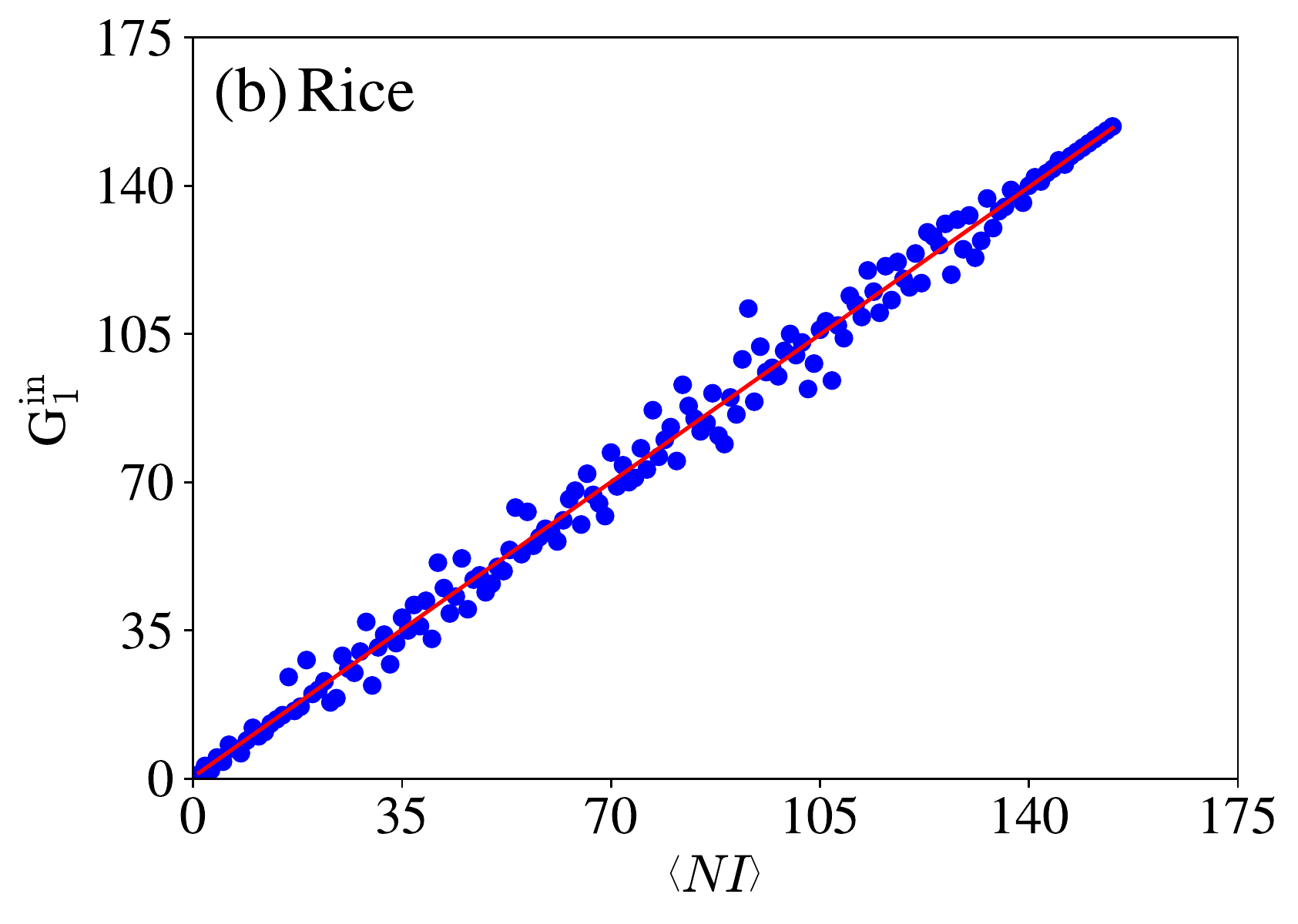}
      \includegraphics[width=0.233\linewidth]{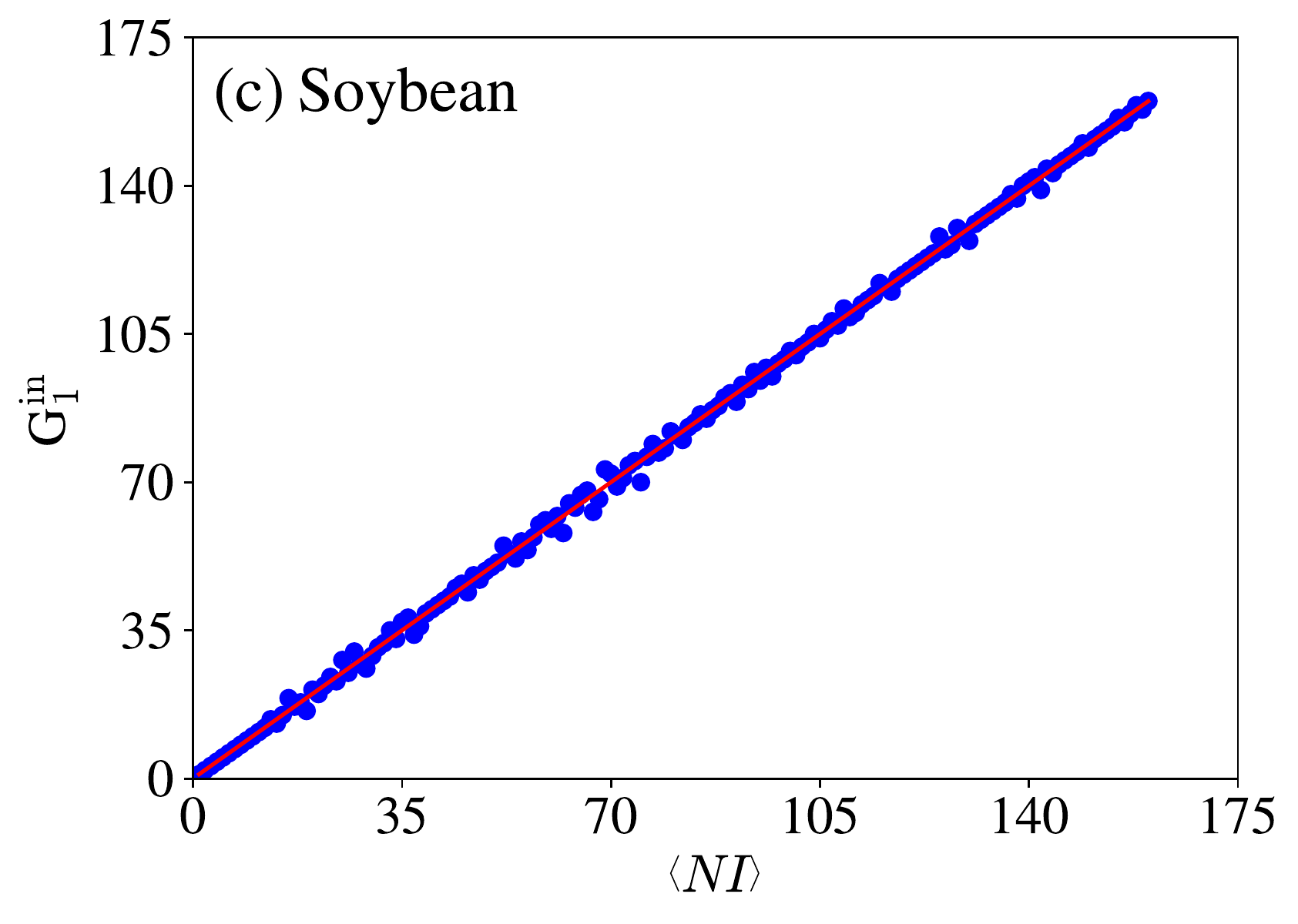}
      \includegraphics[width=0.233\linewidth]{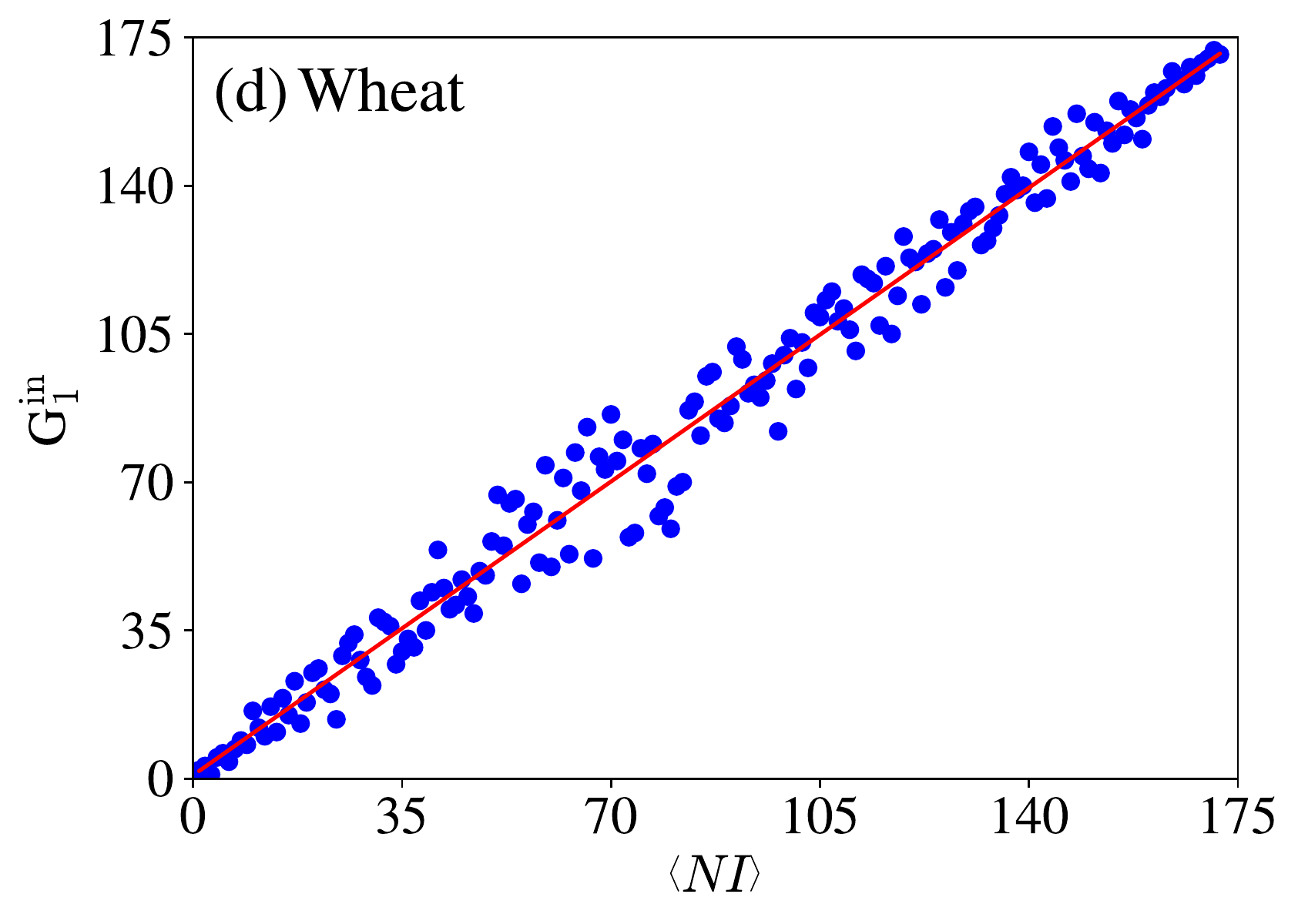}\\
      \includegraphics[width=0.233\linewidth]{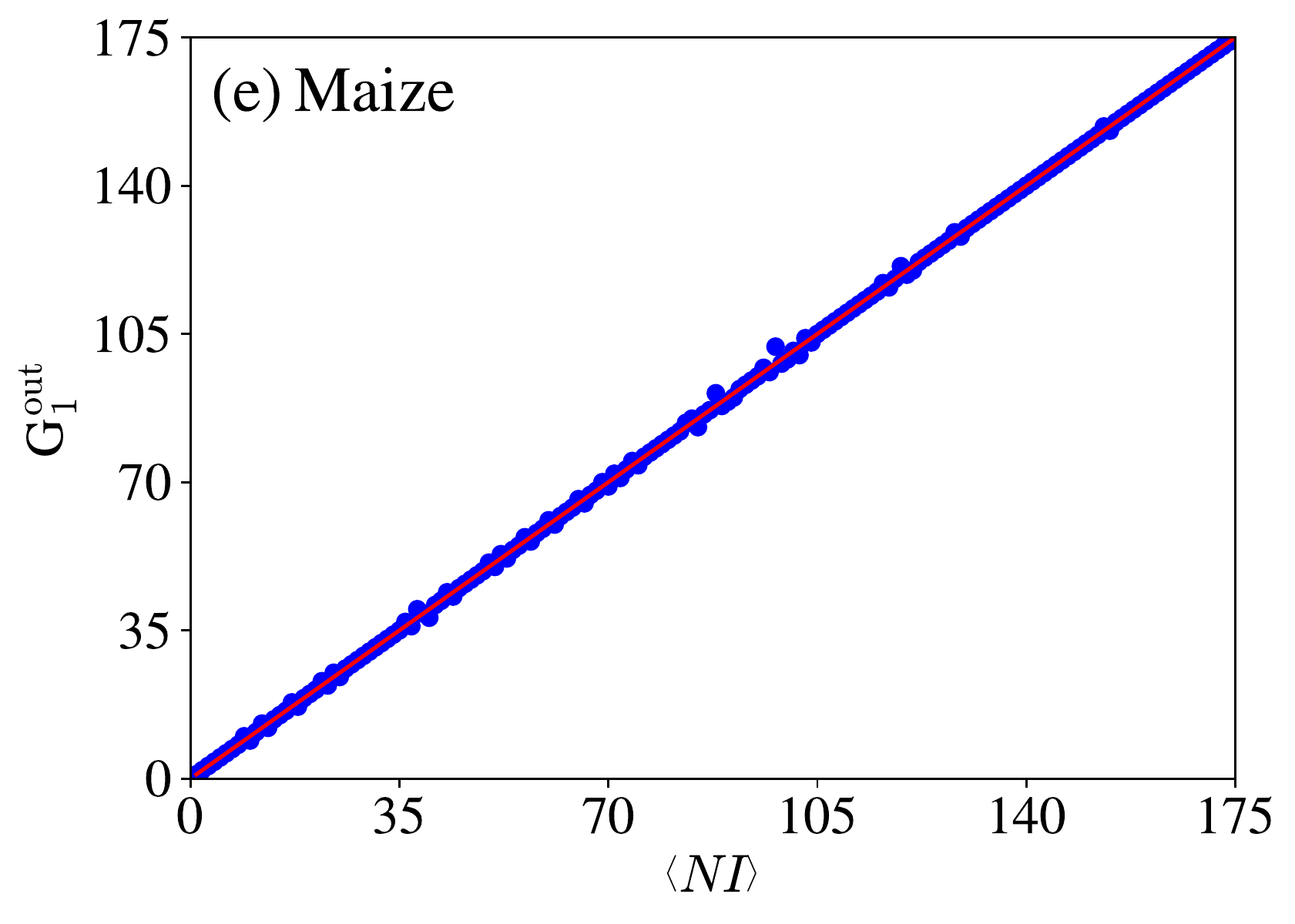}
      \includegraphics[width=0.233\linewidth]{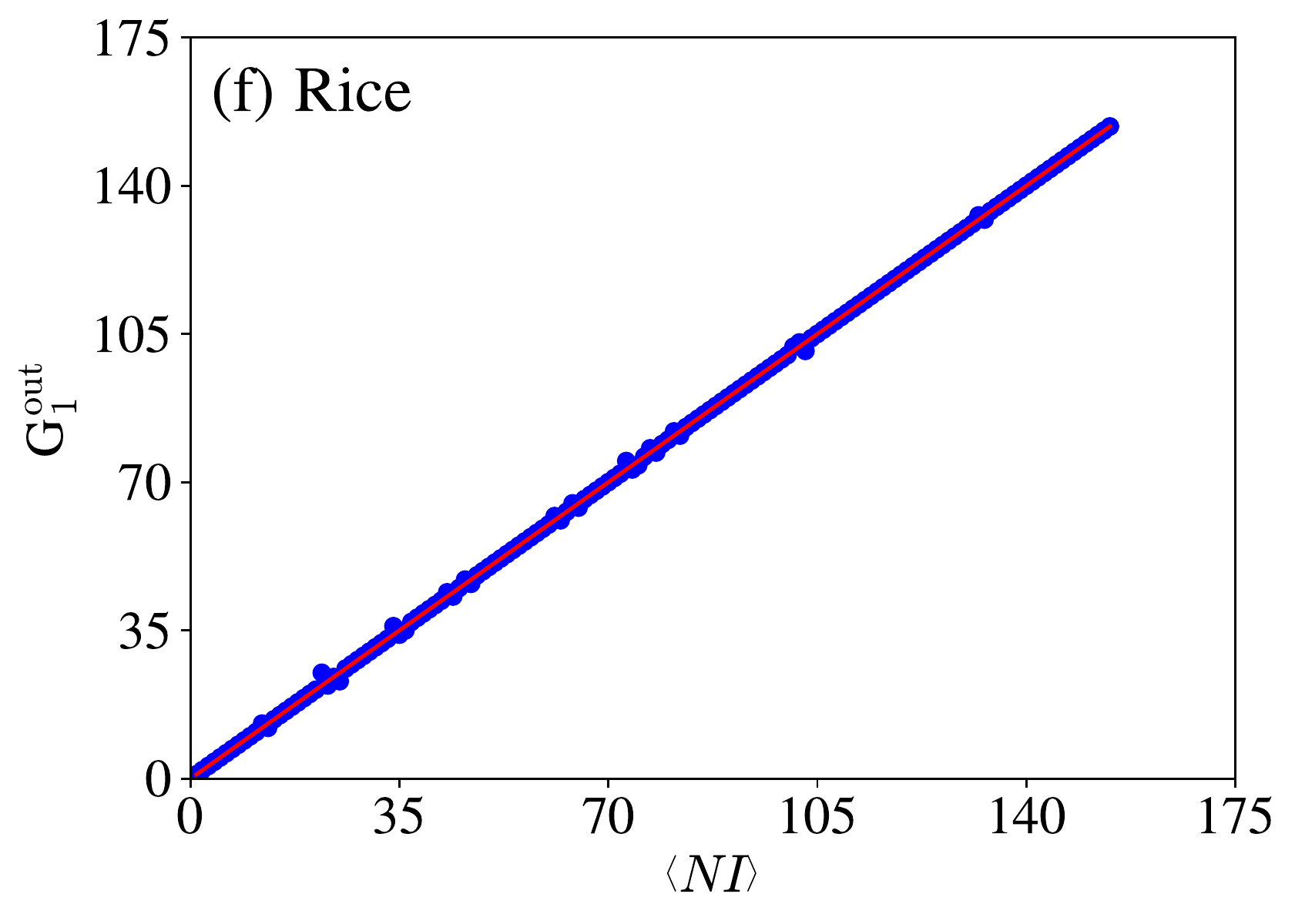}
      \includegraphics[width=0.233\linewidth]{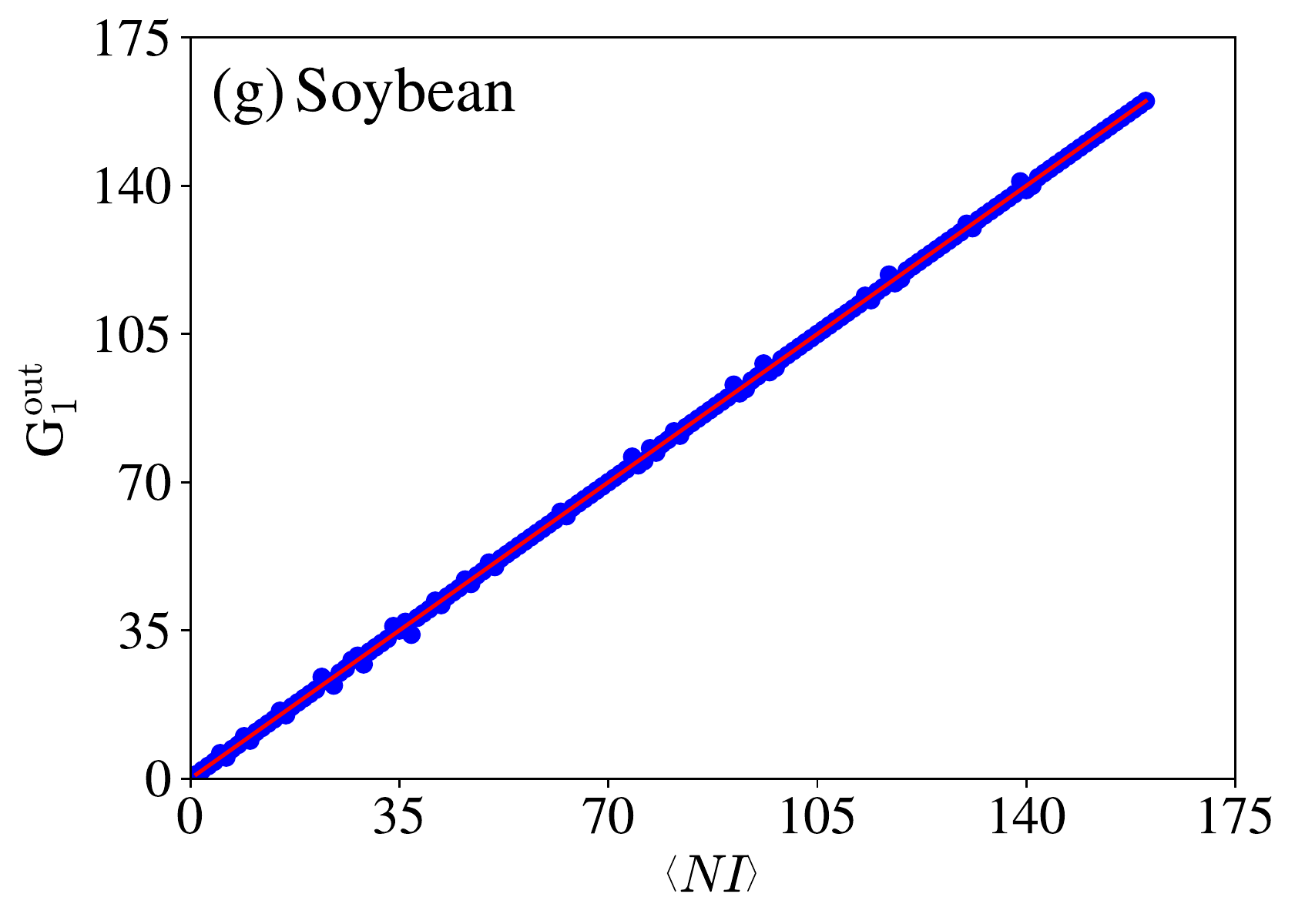}
      \includegraphics[width=0.233\linewidth]{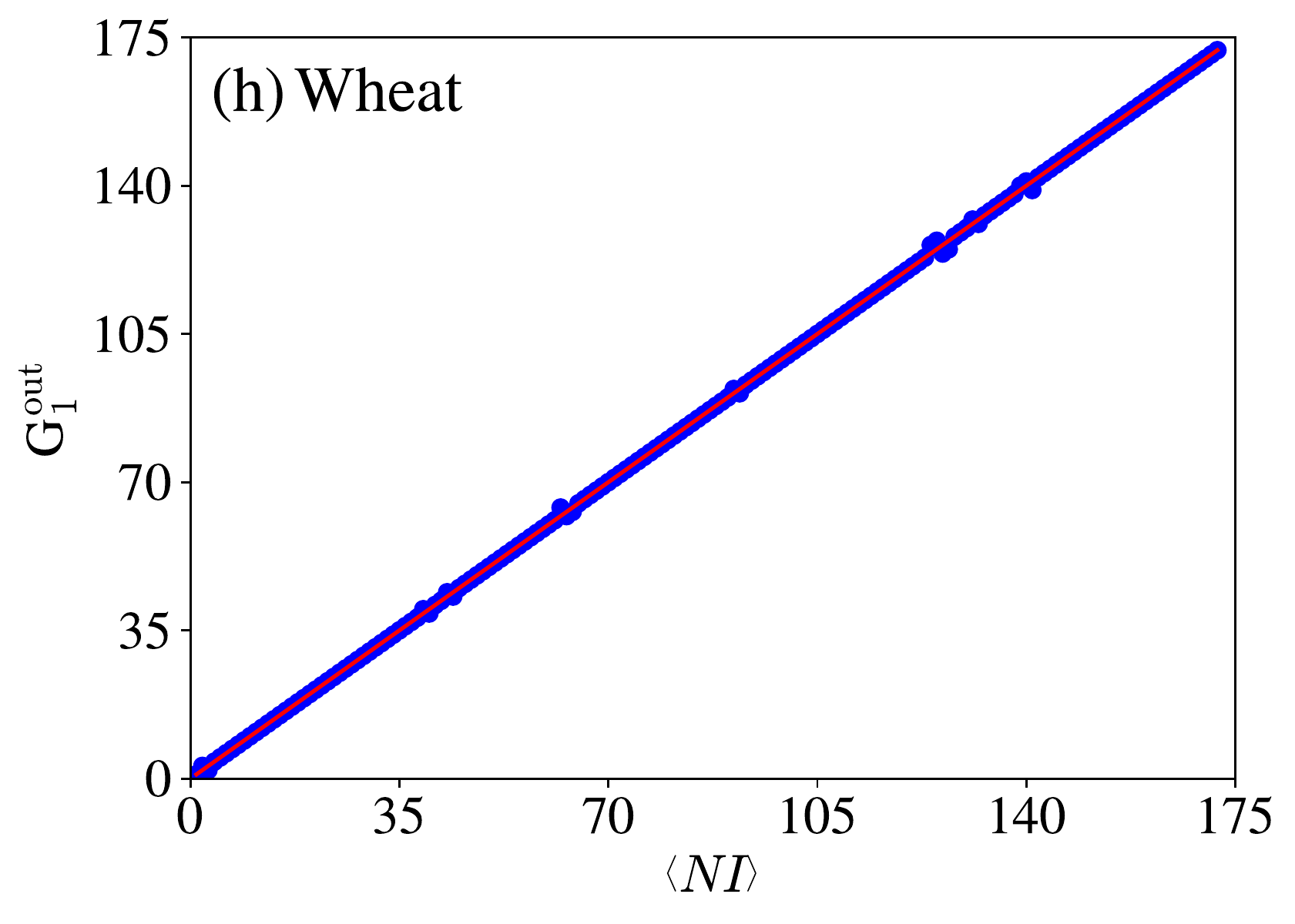}\\
      \includegraphics[width=0.233\linewidth]{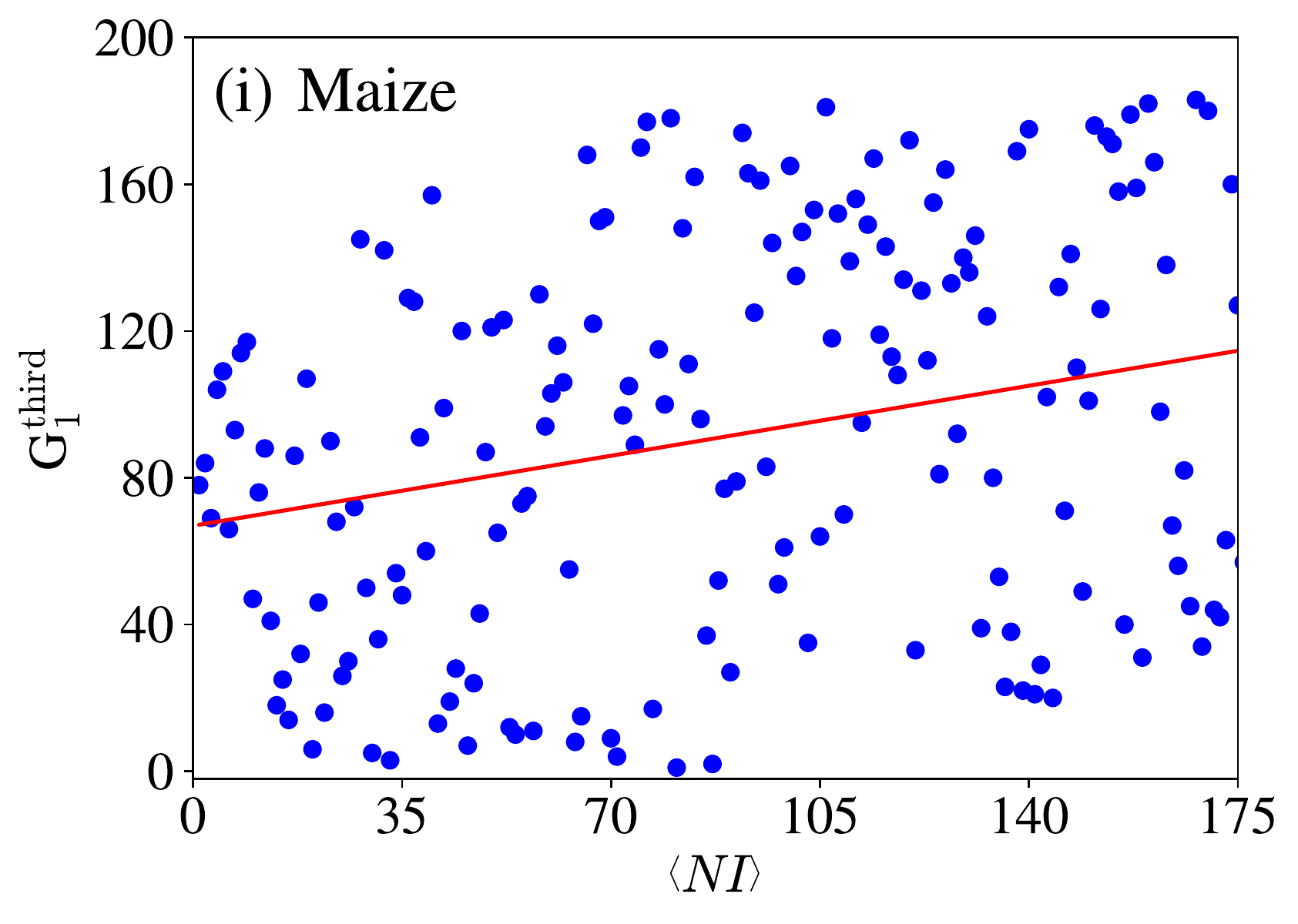}
      \includegraphics[width=0.233\linewidth]{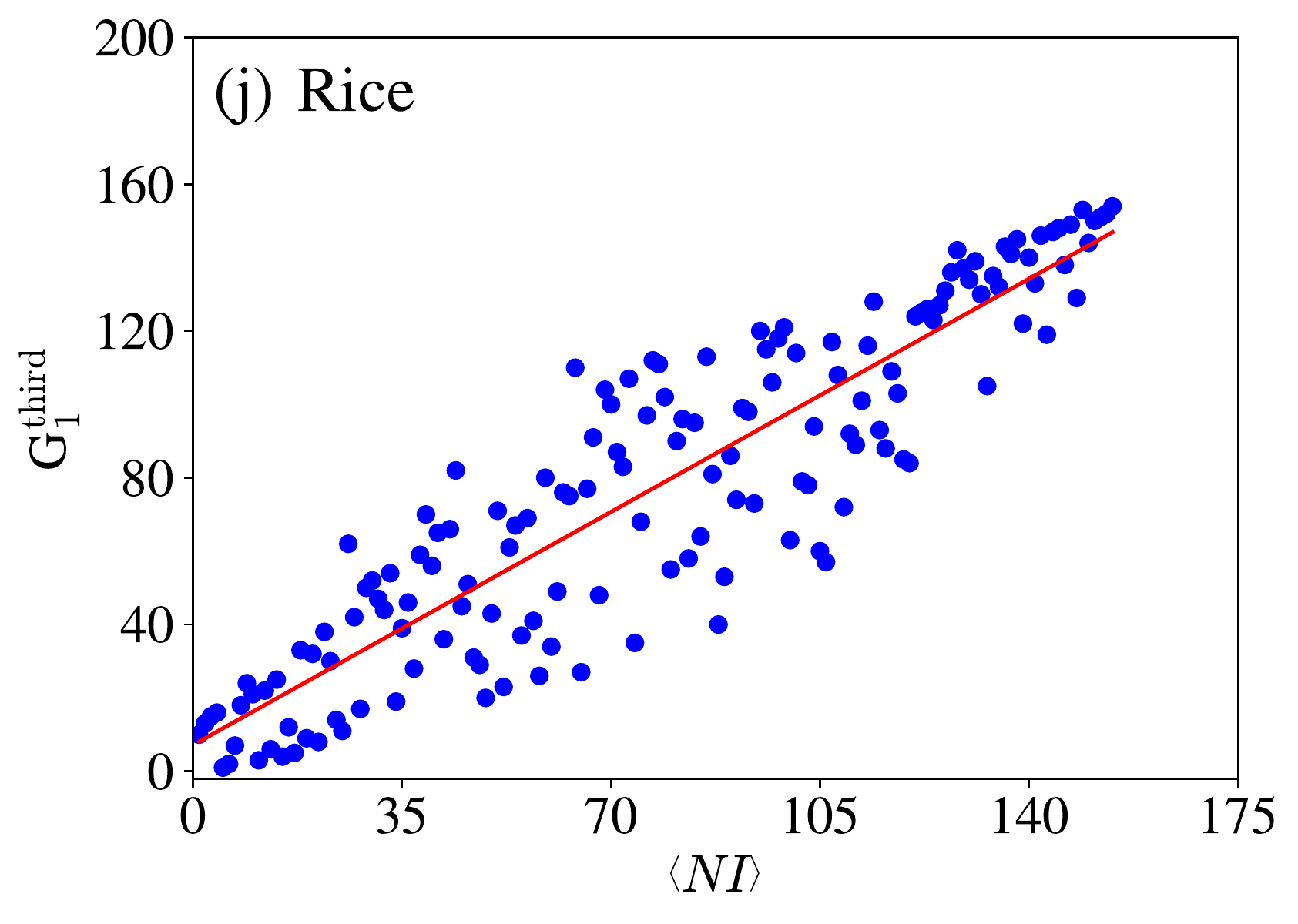}
      \includegraphics[width=0.233\linewidth]{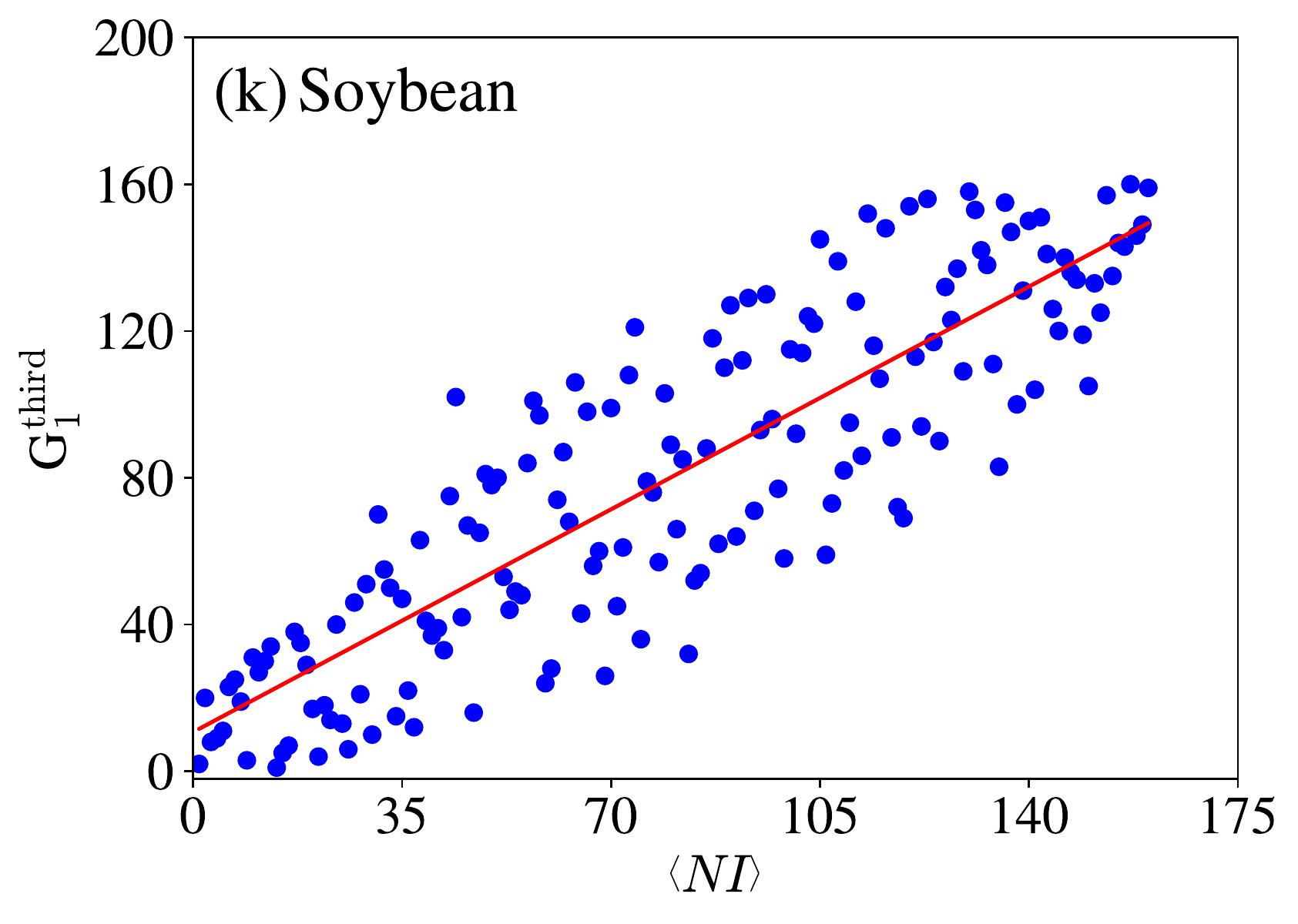}
      \includegraphics[width=0.233\linewidth]{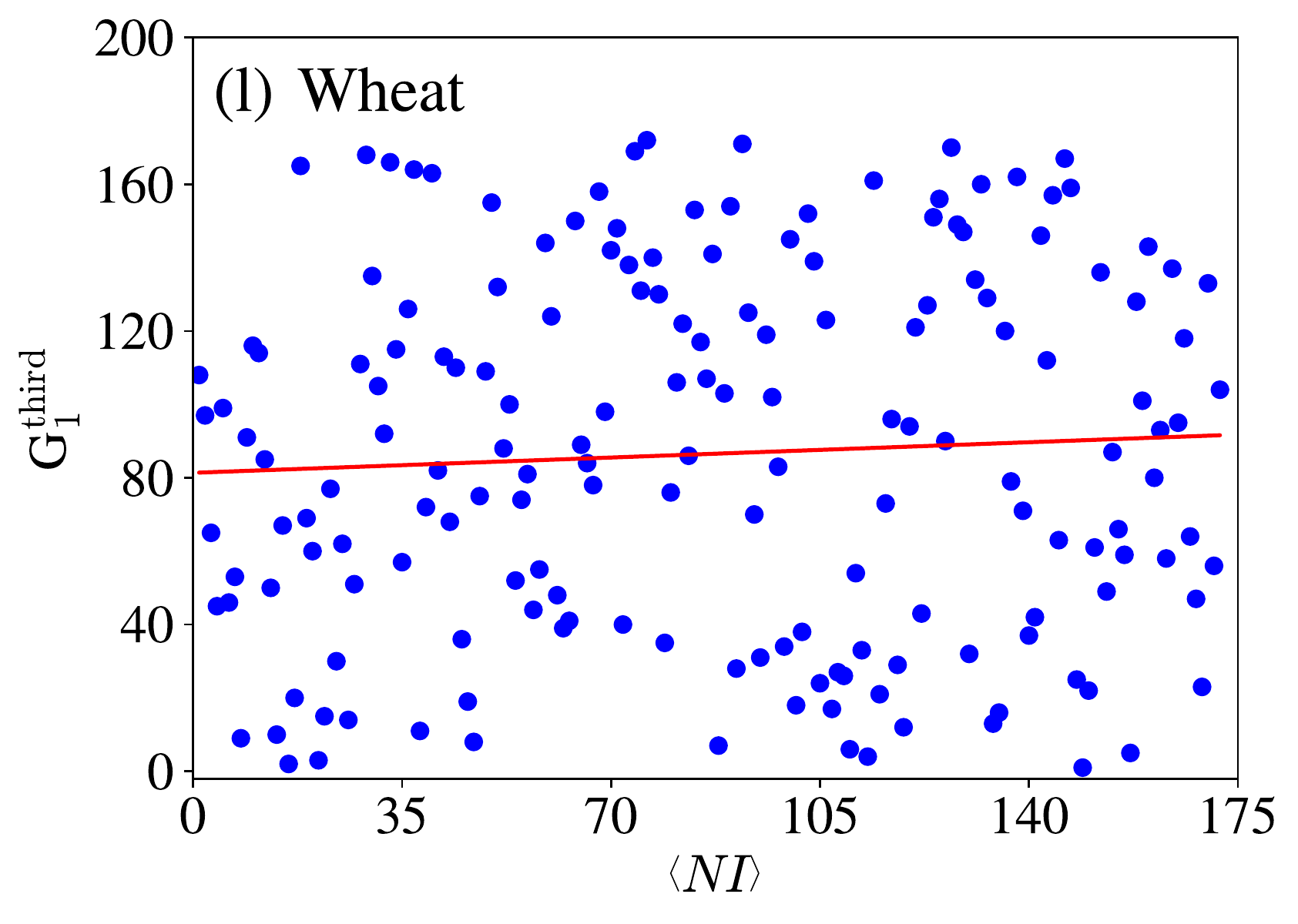}\\
      \caption{The relationship between the ranking of the maximum eigenportfolio and the average ranking of economic influence.}
      \label{Fig:G1:rank}
\end{figure}

 \begin{figure}[h!]
      \centering
      \includegraphics[width=0.233\linewidth]{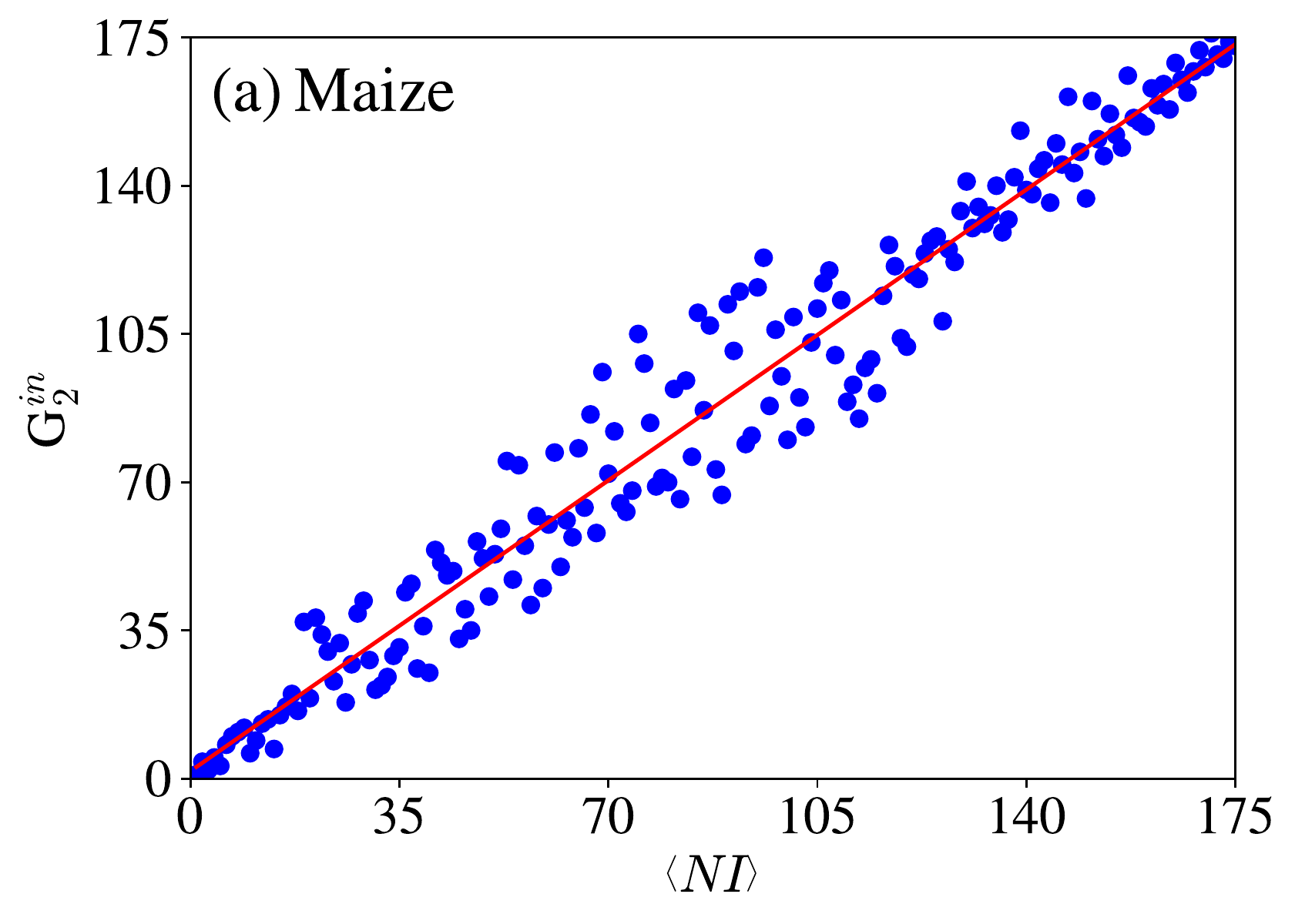}
      \includegraphics[width=0.233\linewidth]{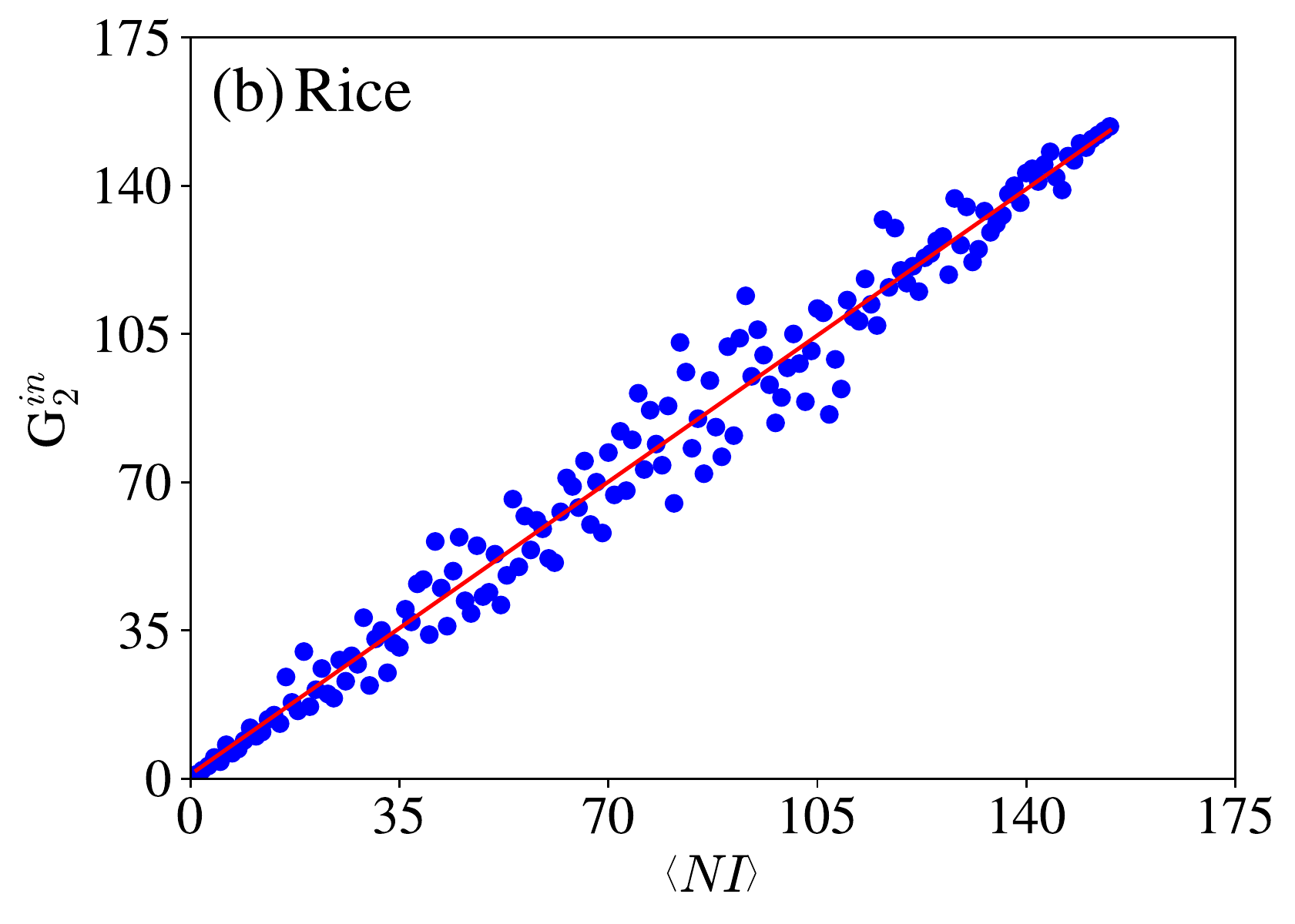}
      \includegraphics[width=0.233\linewidth]{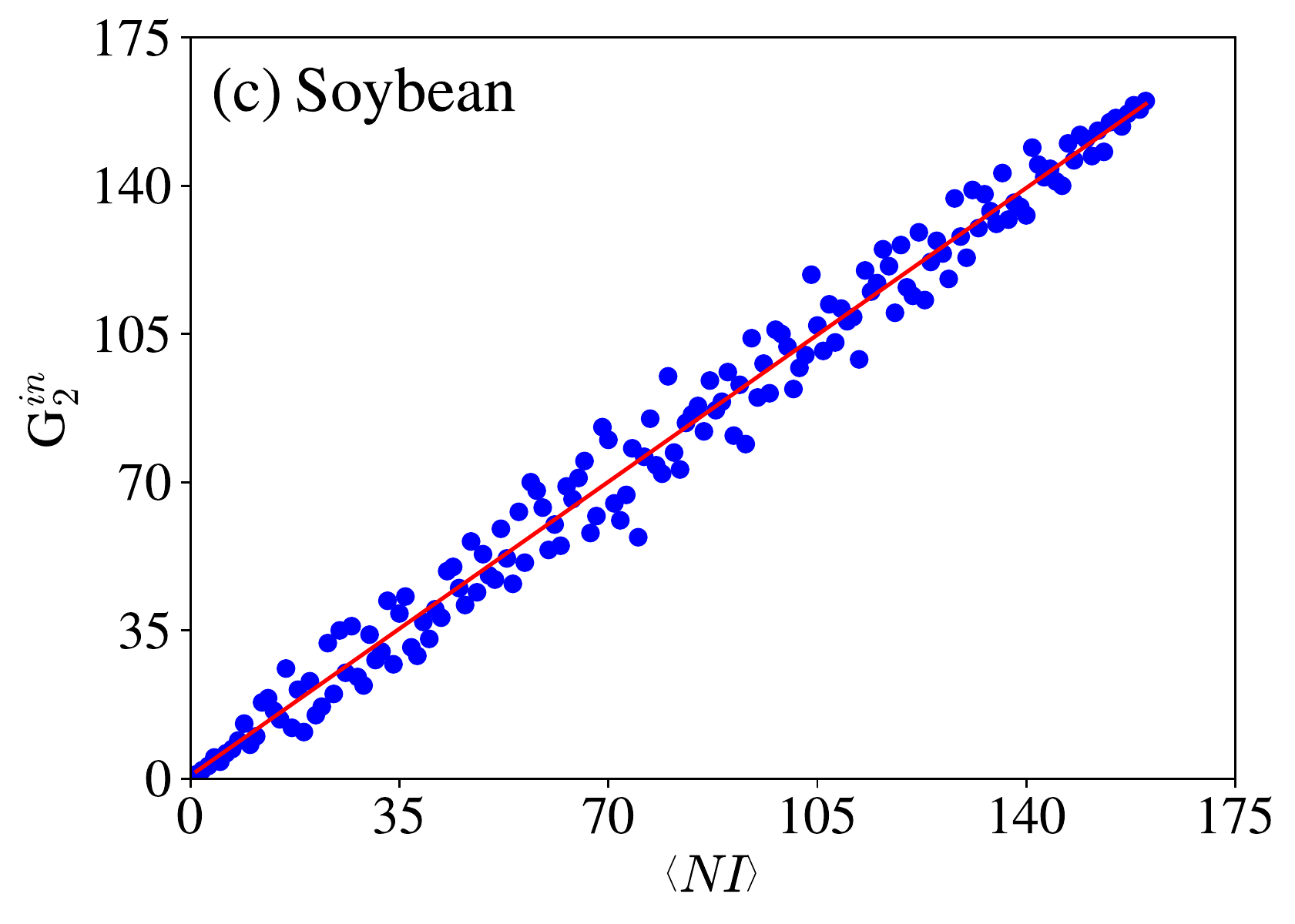}
      \includegraphics[width=0.233\linewidth]{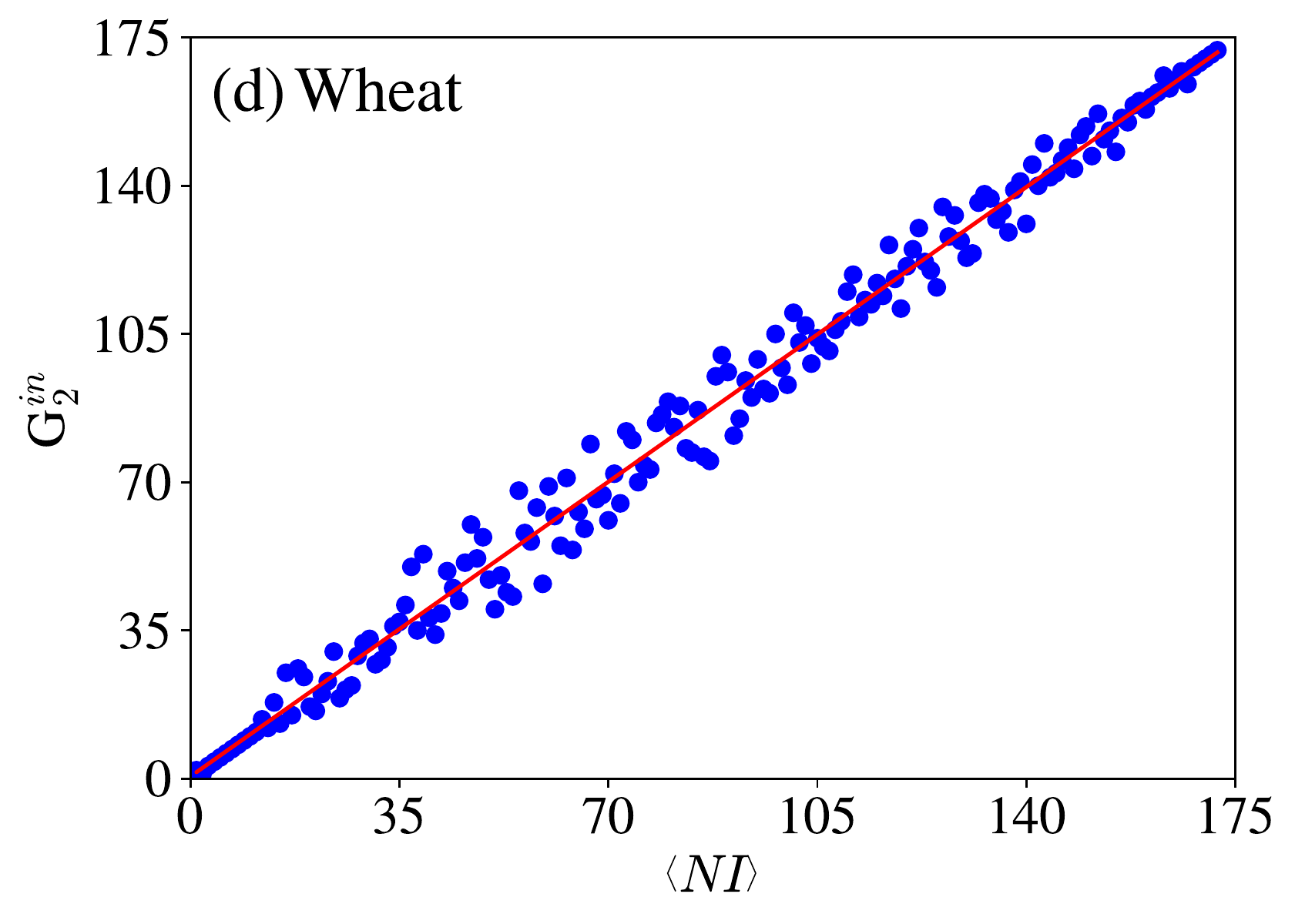}\\
      \includegraphics[width=0.233\linewidth]{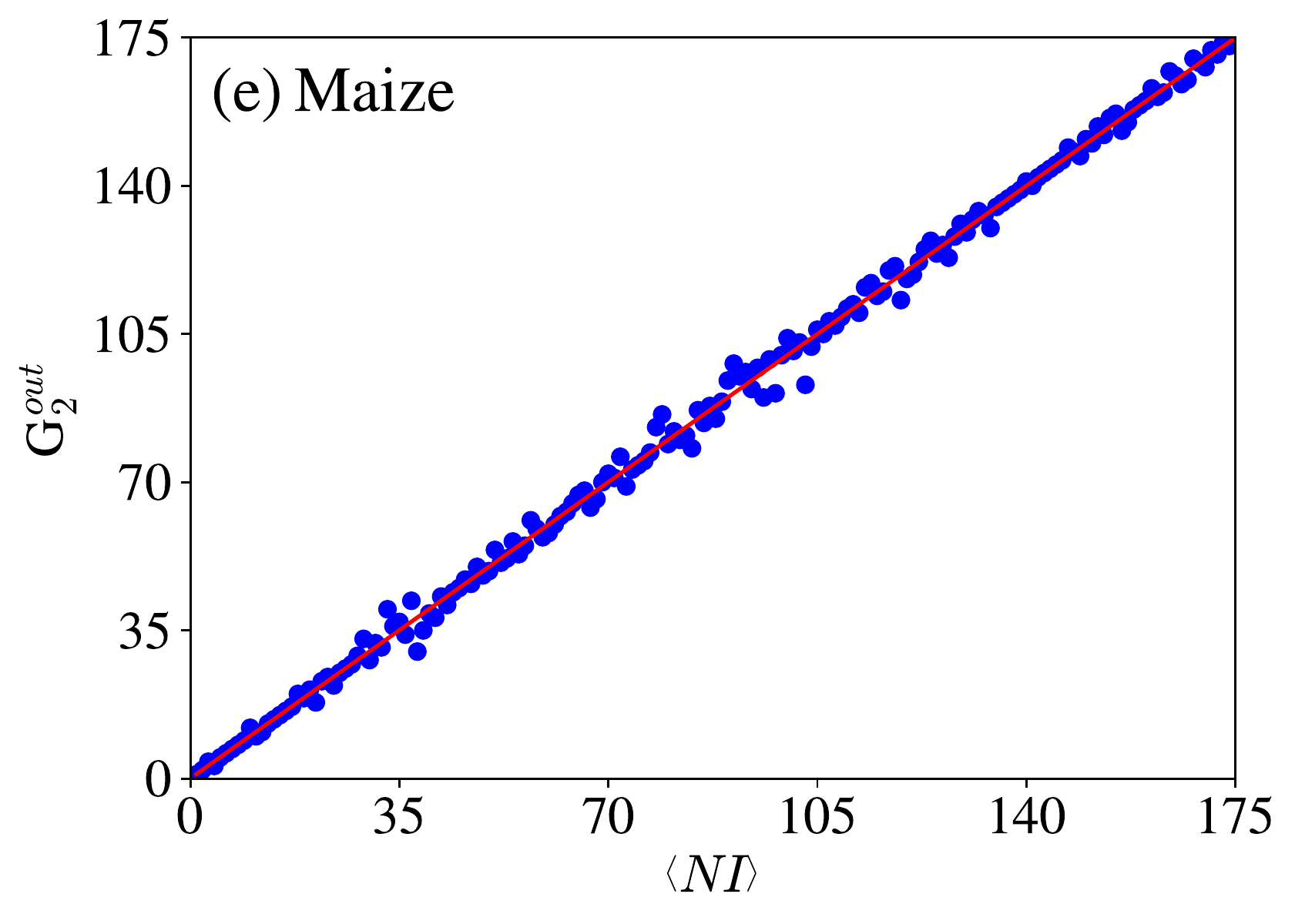}
      \includegraphics[width=0.233\linewidth]{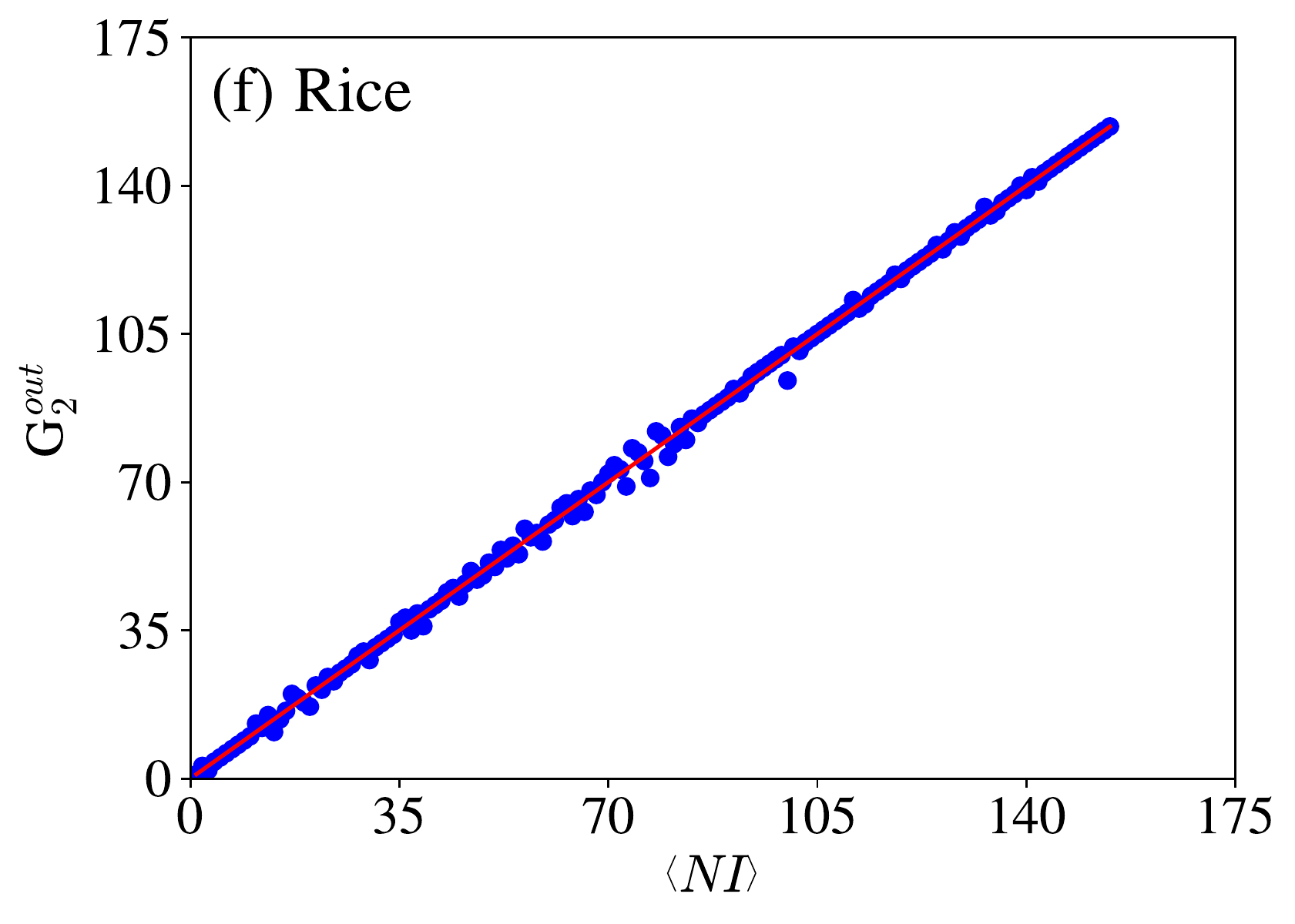}
      \includegraphics[width=0.233\linewidth]{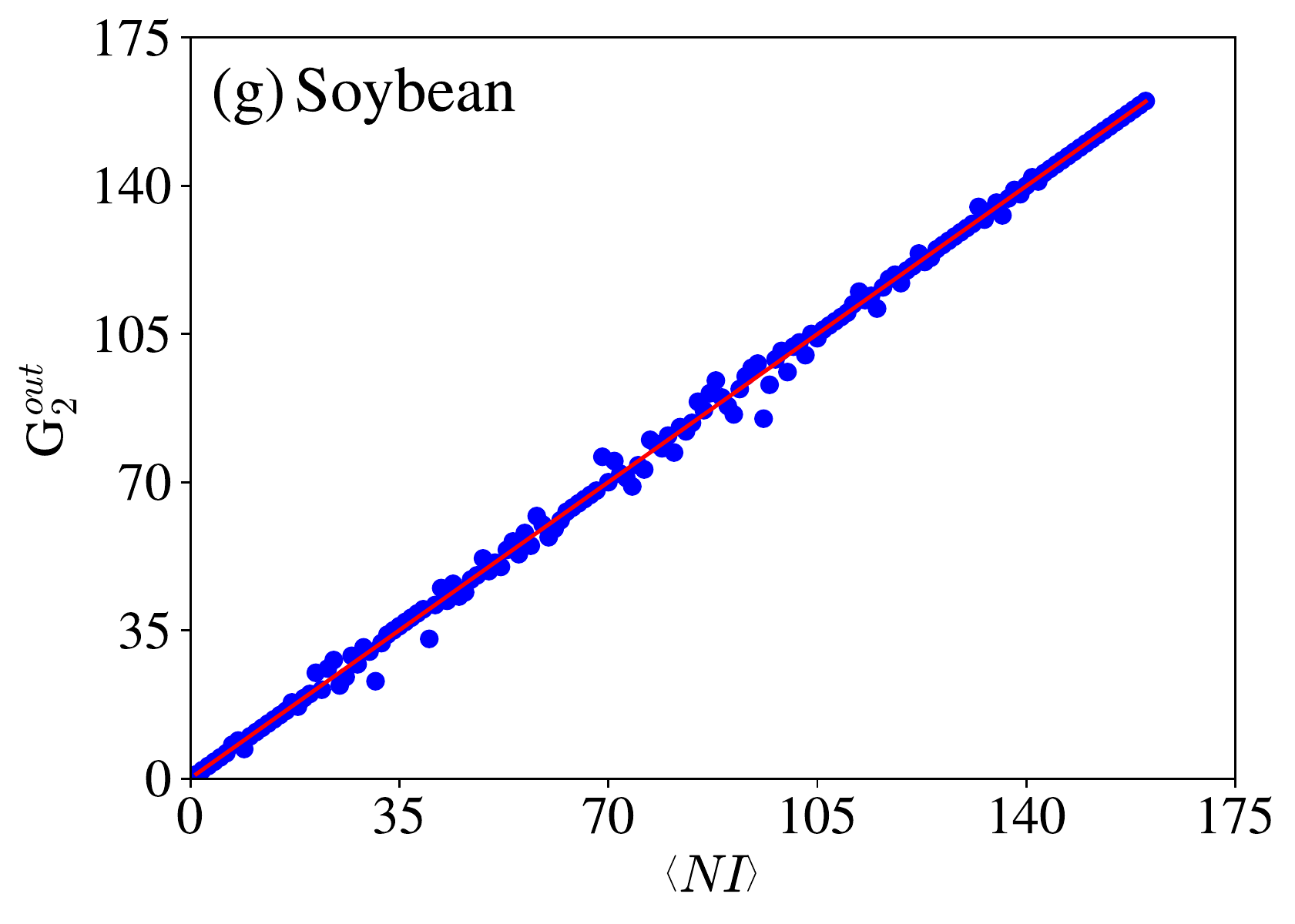}
      \includegraphics[width=0.233\linewidth]{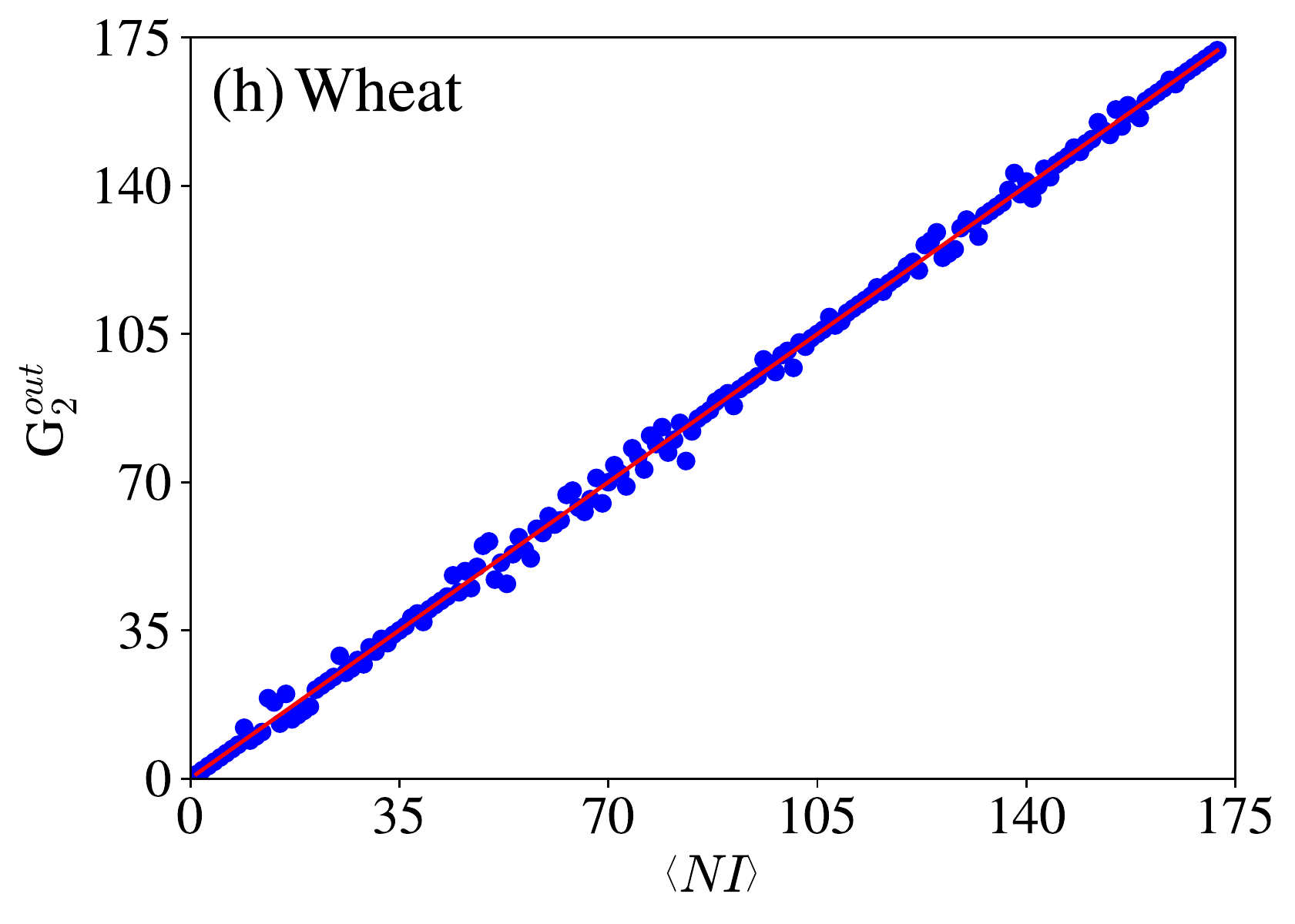}\\
      \includegraphics[width=0.233\linewidth]{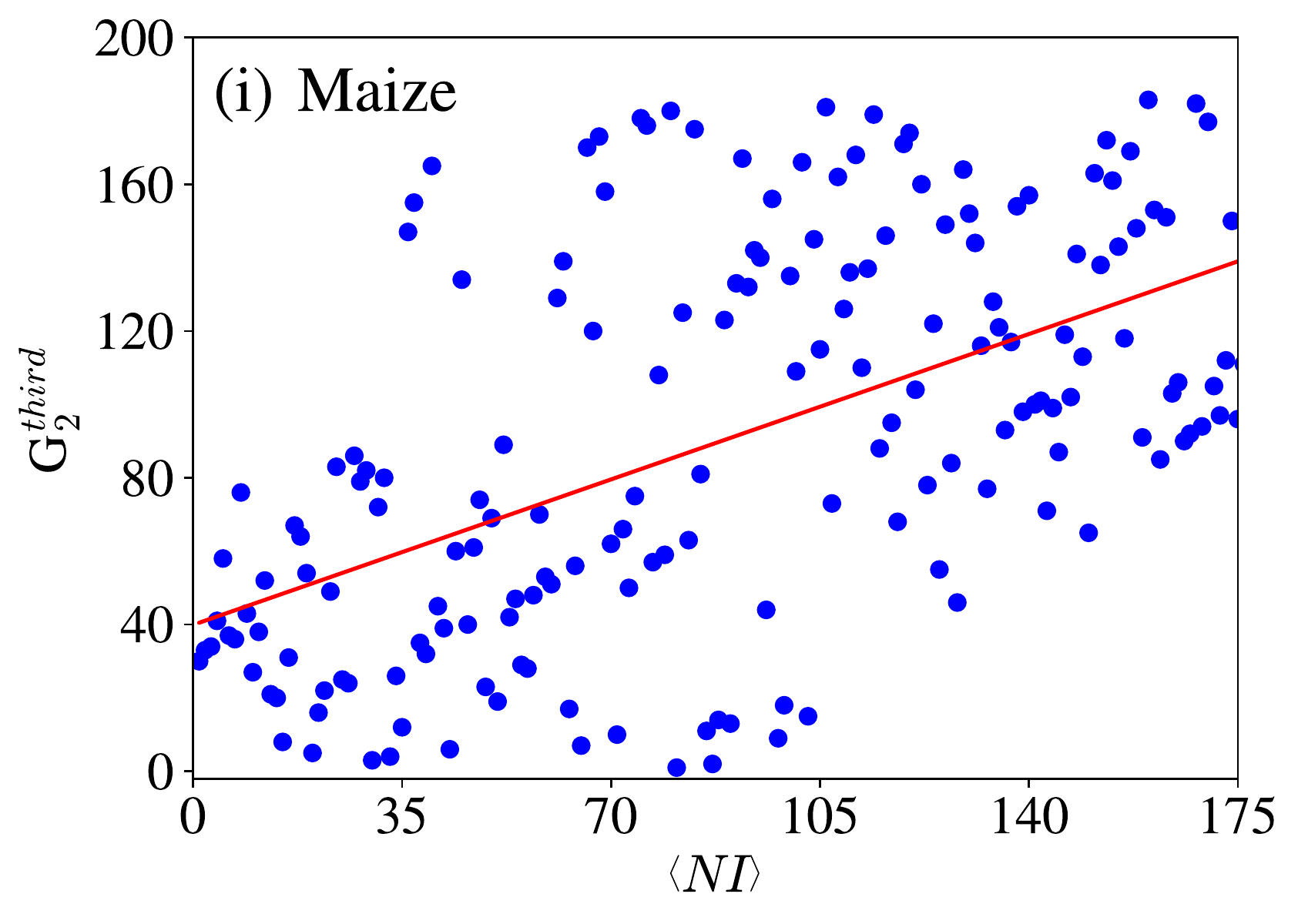}
      \includegraphics[width=0.233\linewidth]{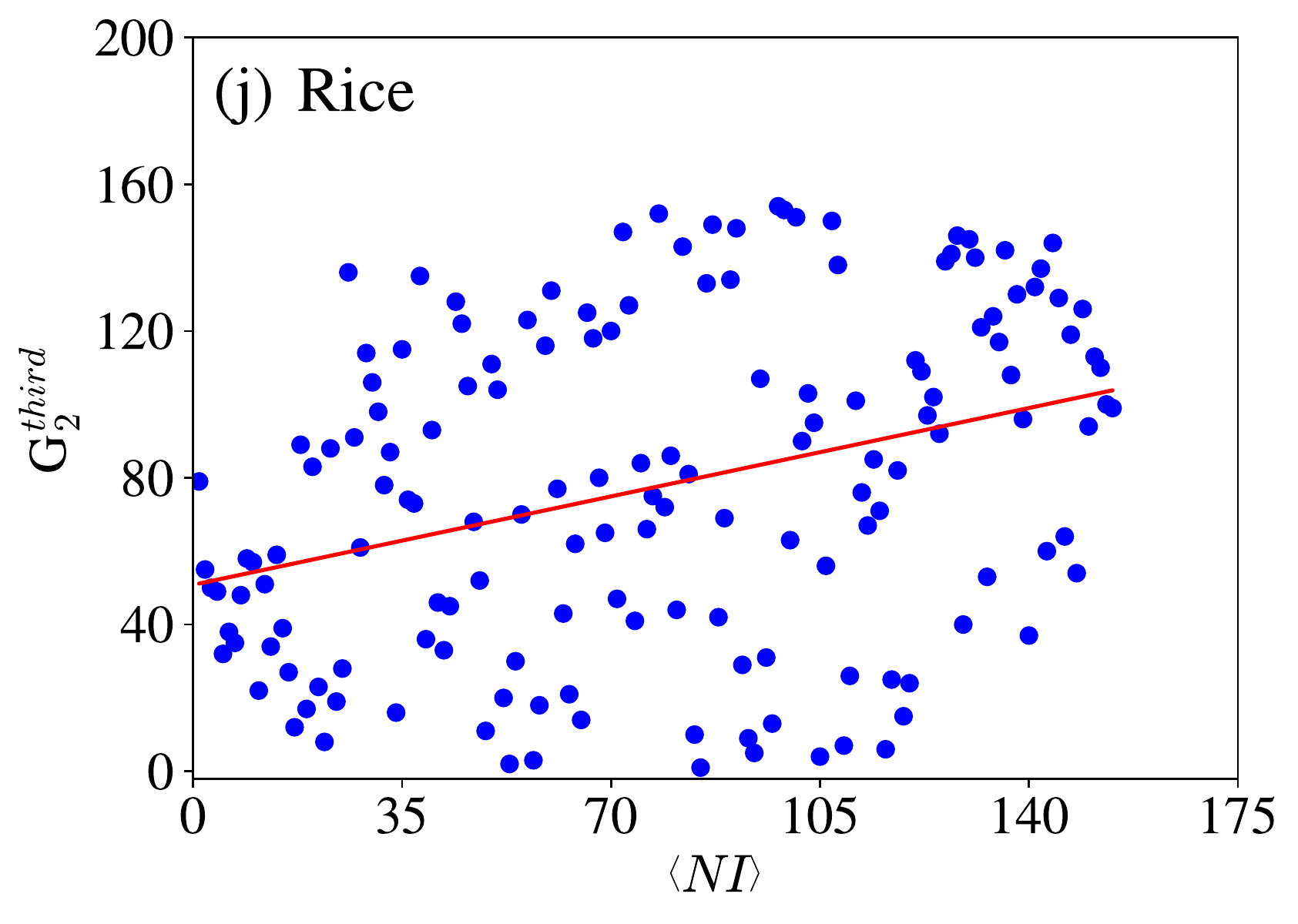}
      \includegraphics[width=0.233\linewidth]{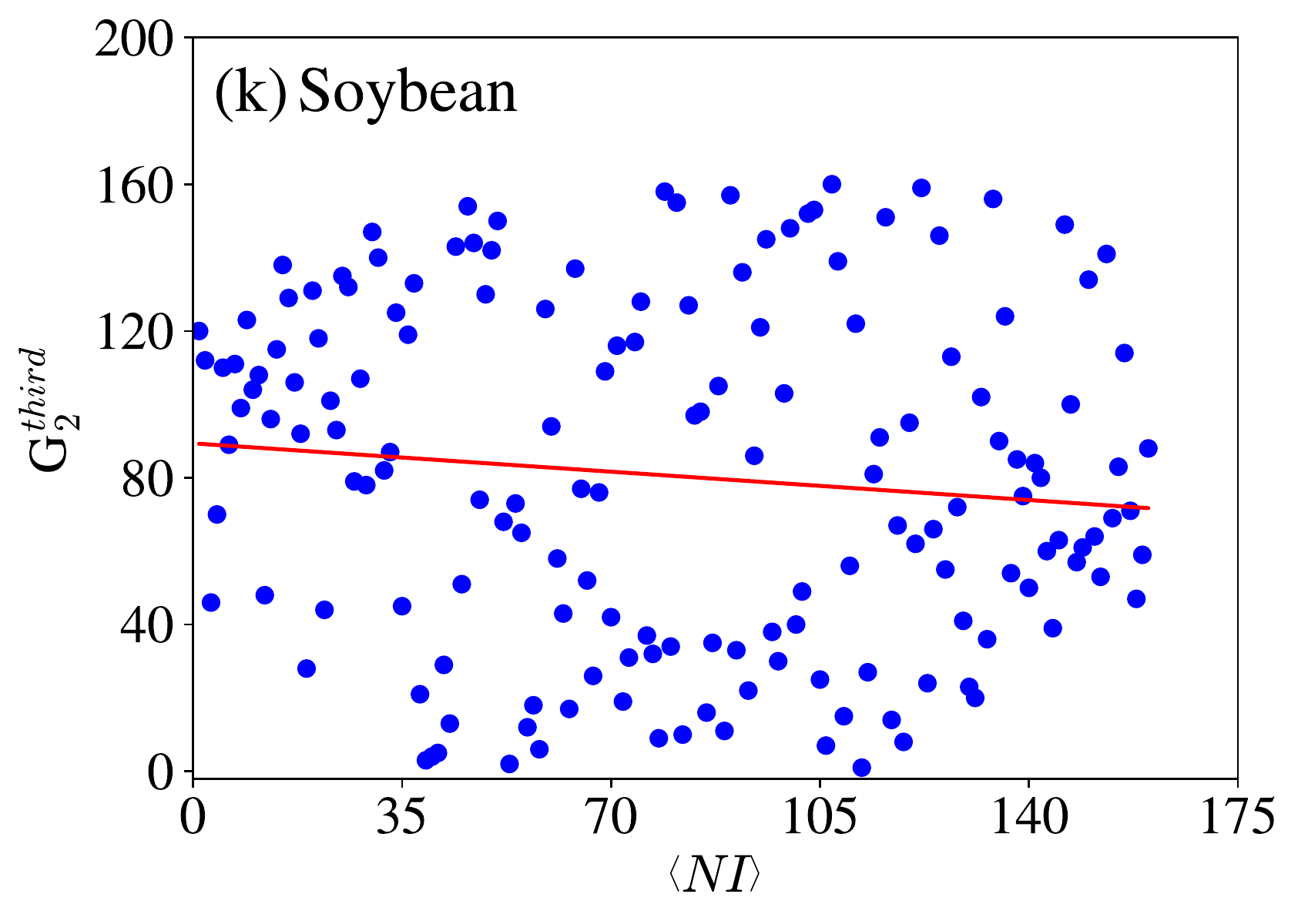}
      \includegraphics[width=0.233\linewidth]{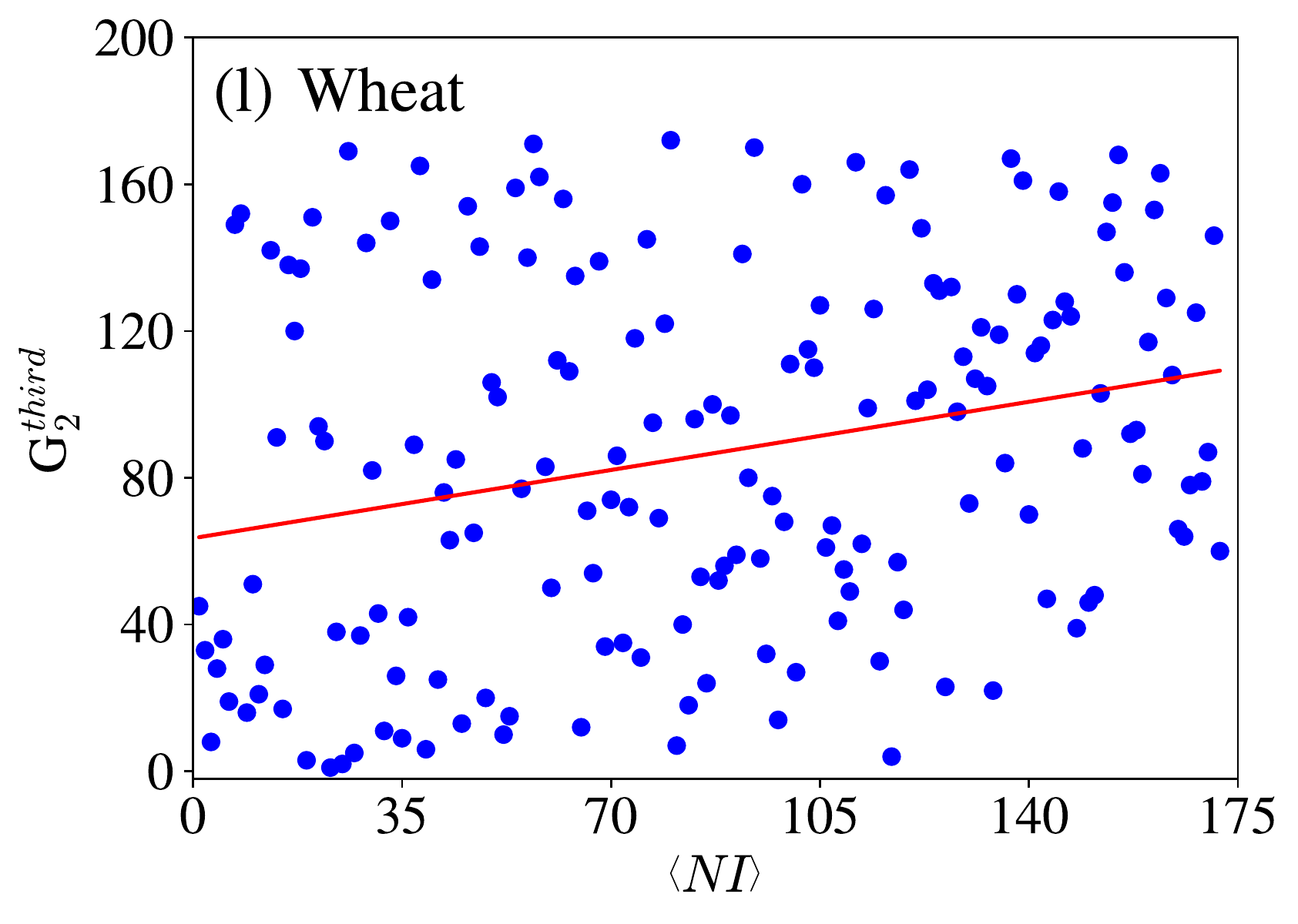}\\
      \caption{The relationship between the ranking of the second largest eigenportfolio and the average ranking of economic influence.}
      \label{Fig:G2:rank}
\end{figure}

{\color{red}{We present all the eigenvectors in \ref{sec5-2}. There are information contents in the components of eigenvector $u_1$ (associated with the largest eigenvalue $\lambda_1$), eigenvector $u_2$ (associated with the second largest eigenvalue $\lambda_2$), eigenvector $u_3$ (associated with the third largest eigenvalue $\lambda_3$), and eigenvector $u_{20}$ (associated with the smallest eigenvalue $\lambda_{20}$), as shown in Fig.~\ref{Fig:iCTN:PDF:eigenvector:component:2020}.}} Specifically, Fig.~\ref{Fig:iCTN:PDF:eigenvector:component:2020}(a-d) provide a detailed breakdown of the eigenvectors $u^1$ corresponding to eigenvalues for the maize, rice, soybean, and wheat networks. One notable finding is that the eigenvector of the largest eigenvalue contains the majority of the information, and most of the components of the eigenvectors have the same signs. Intriguingly, the components of $u_2$ (shown in Fig.~\ref{Fig:iCTN:PDF:eigenvector:component:2020}(e-h)) and $u_3$ (shown in Fig.~\ref{Fig:iCTN:PDF:eigenvector:component:2020}(i-l)) corresponding to the same group of indicators largely have the same signs. Additionally, the components of $u_2$ associated with indicators based on the import and export scale have opposite signs. Furthermore, the components corresponding to the indicators $KC$ and $CC$ (i.e., the sixth and seventh components) show the opposite sign with respect to $u_1$ for maize and wheat (as seen in Fig.~\ref{Fig:iCTN:PDF:eigenvector:component:2020}(a) and (d)). These observations suggest that the distribution of components corresponding to the eigenvectors of the correlation matrix $\mathrm{C}$ is likely related to the performance of the corresponding indicator that assesses the importance of economies. For example, Fig.~\ref{Fig:iCTN:PDF:eigenvector:component:2020}(e,g,h) clearly demonstrate that the absolute values of the eigenvector components corresponding to $EC$ and $IL$ are large, while others are close to zero, indicating a significant correlation. Moreover, the analysis reveals that $u_{20}$ covers less information, and components corresponding to highly relevant metrics tend to exhibit similar features. Overall, these findings shed new light on the relationship between eigenvector components and economic indicators. They provide a compelling illustration of the extent to which the eigenvector components can reveal important insights into the behavior and performance of different node importance metrics, thereby enhancing our understanding of the economies' impact on the iCTNs.

\subsection{Composite trade influence of economies}
\label{S3-3:Compositenodeimportance}

The results obtained from analyzing the correlation coefficient matrix $\mathrm{C}$ and $\mathrm{RMT}$ demonstrate that the information provided by the indicators is predominantly contained in the largest and second largest eigenvalues, and this finding holds not only for the 2020 data but also for data from 1986-2020. Specifically, our analysis reveals that the first and second largest eigenvalues of the correlation matrix consistently exceed the maximum eigenvalue of the random matrix, while the third to twentieth eigenvalues are consistently smaller, except for some cases where the third largest eigenvalue slightly exceeds the maximum eigenvalue (Fig.~\ref{Fig:iCTN:eigenvalue:t}). Therefore, we employ the method in line with constructing eigenportfolios based on the random matrix theory in financial markets. We calculate the largest eigenvalue combination $\mathrm{G}^{1}$, and the second largest eigenportfolio $\mathrm{G}^{2}$ and compare them with the average node importance rankings $\left\langle{NI}\right \rangle$ to obtain the optimal eigenportfolio as the composite node importance metric $\mathrm{CNI}$. In recognition of the existence of indicator clusters, we divide the eigenvalues and corresponding eigenvectors into three categories, leading to the construction of three types of eigenportfolios $\mathrm{G}^{\mathrm{in}}_k$, $\mathrm{G}^{\mathrm{out}}_k$, and $\mathrm{G}^{\mathrm{third}}_k$, which correspond to the three categories of node importance metrics identified in Section~\ref{S3-2:Correlation}.

Figure~\ref{Fig:G1:rank} illustrates the relationship between the largest eigenvalue portfolios and the average node importance rankings, while Fig.~\ref{Fig:G2:rank} displays the relationship between the second-largest eigenvalue portfolios and the average node importance rankings. A notable observation arising from Fig.~\ref{Fig:G1:rank} is that both $\mathrm{G}^{\mathrm{in}}_1$ and $\mathrm{G}^{\mathrm{out}}_1$ exhibit a linear relationship with $\left \langle {NI}\right \rangle$, with the exception of $\mathrm{G}^{\mathrm{third}}_1$, particularly for maize (Fig.~\ref{Fig:G1:rank}(a)) and wheat (Fig.~\ref{Fig:G1:rank}(d)). The results presented in Fig.~\ref{Fig:G2:rank} are comparable to the above descriptions, except that $\mathrm{G}^{\mathrm{in}}_1$, $\mathrm{G}^{\mathrm{out}}_1$, and $\mathrm{G}^{\mathrm{third}}_1$ are more proximate to $\left\langle { NI}\right \rangle$. These findings suggest that the largest eigenportfolio embodies a collective effect of multidimensional node importance metrics. Taken as a whole, these results encourage us to convert the maximum eigenportfolios into composite node importance metrics, which enable us to rank economies' composite trade influence based on three types of composite node importance metrics.

\begin{table}[!ht]
    \centering
    \renewcommand\tabcolsep{1mm}
    \caption{Top 10 economies identified based on composite node importance index (CNI).}
    \smallskip
    \begin{tabular}{cccccccccccccccccccccccc}
         \toprule
        \multirow{2}{*}{Rank}~&~&Maize & ~ && ~ & Rice & ~ && ~ &Soybean & ~ && ~ & Wheat~&~& \\  
        \cline{2-4} \cline{6-8}\cline{10-12}\cline{14-16}
         & $\mathrm{CNI^{\mathrm{in}}}$ & $\mathrm{CNI^{\mathrm{out}}}$ & $\mathrm{CNI^{third}}$ &&  $\mathrm{CNI^{\mathrm{in}}}$ & $\mathrm{CNI^{\mathrm{out}}}$ & $\mathrm{CNI^{third}}$ & & $\mathrm{CNI^{\mathrm{in}}}$ & $\mathrm{CNI^{\mathrm{out}}}$ & $\mathrm{CNI^{third}}$  & & $\mathrm{CNI^{\mathrm{in}}}$ & $\mathrm{CNI^{\mathrm{out}}}$ & $\mathrm{CNI^{third}}$ \\  
                 \midrule
        1 & NLD & USA & LSO &~& GBR & BRA & CZE &~& NLD & USA & MLT &~& NLD & FRA & AUT~& \\  
        2 & GBR & ARG & TON &~& NLD & ITA & SVK  &~& DEU & CAN & HUN &~& ITA & CAN & ESP~& \\  
        3 & JPN & FRA & NER &~& ITA & USA & HUN &~& ESP & BRA & SVN &~& TUR & RUS & NLD~& \\   
        4 & ESP & BRA & BRB &~& DEU & IND &EST
        &~& ITA & UKR & LVA &~& GBR & DEU & GRC~& \\
        5 & DEU & ZAF & MOZ &~& ESP & GBR & LUX &~& FRA & CHN & SVK &~& ESP & USA & MAR~& \\    
        6 & ITA & UKR & NAM &~& BEL & NLD & FIN &~& GBR & NLD & EST &~& BEL & UKR & DEU~& \\ 
        7 & CAN & IND & USA &~& FRA & CHN & POL  &~& BEL & AUT & IRL &~& DEU & POL & CZE~& \\  
        8 & BEL & NLD & ARG &~& PRT & ESP & SVN &~& CHN & DEU & BLR &~& ARE & LTU & DZA~& \\  
        9 & TUR & TUR & BRA &~& IRL & DEU & AUT &~& PRT & FRA & BIH &~& MAR & GBR & BIH~& \\  
        10 & ARE & DEU & ZAF &~& ARE & PAK & BGR &~& CAN & ARG & CZE &~& ISR & IND & HUN~& \\  
         \bottomrule
    \end{tabular}
    \label{Table:iCTN:CNI:rank}
\end{table}

Table~\ref{Table:iCTN:CNI:rank} presents the top ten most important economies in the four iCTNs for 2020, using composite node importance metrics. It is evident that the ranking results based on $\mathrm{CNI}^{\mathrm{third}}$ are perplexing and divergent. Hence, we focus on the ranking results based on $\mathrm{CNI}^{\mathrm{in}}$ and $\mathrm{CNI}^{\mathrm{out}}_1$. Notably, European economies have a significant influence on the iCTNs, with the Netherlands, Britain, and Germany being particularly influential.

 \begin{figure}[h!]
      \centering
      \includegraphics[width=0.233\linewidth]{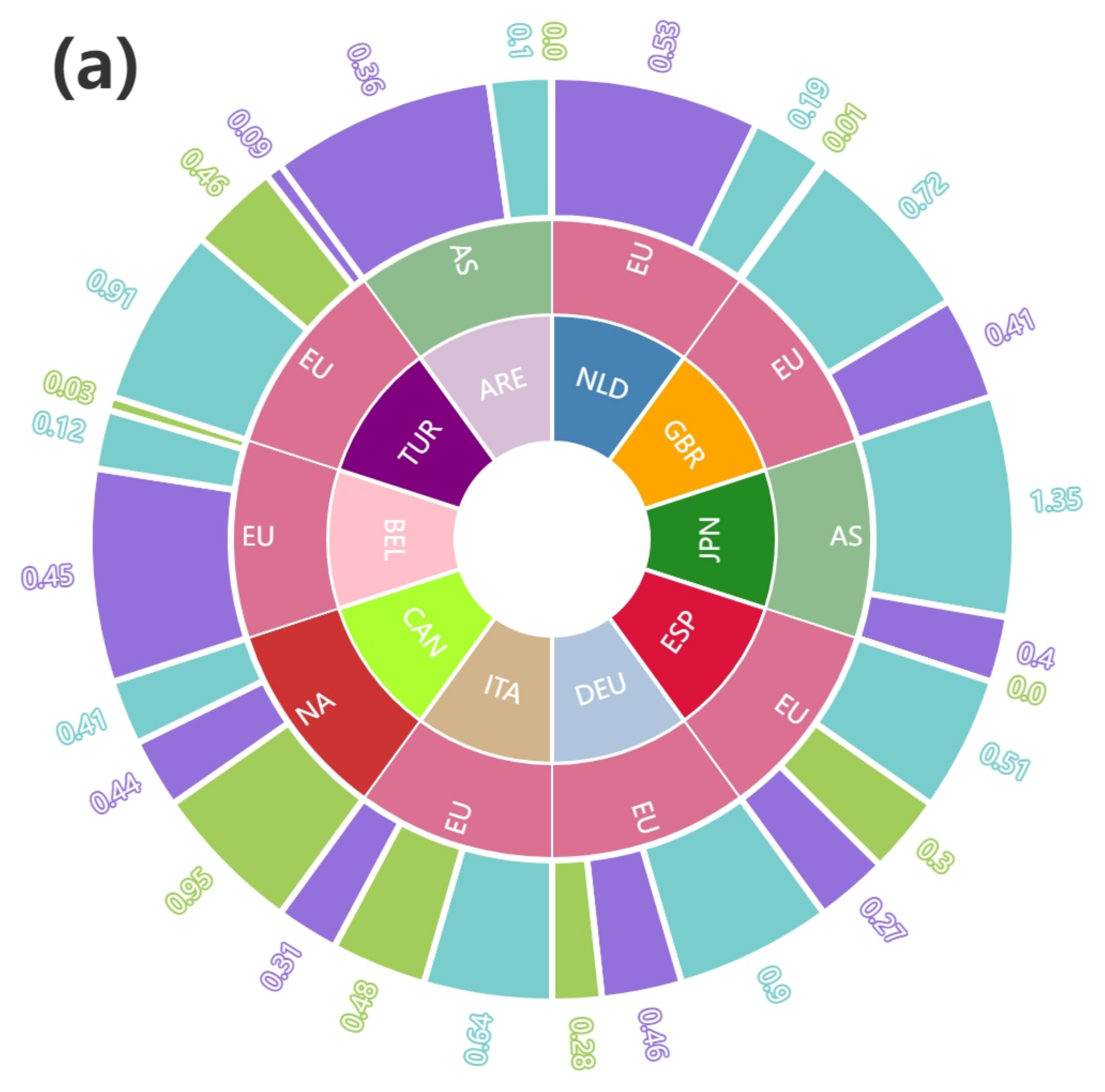}
      \includegraphics[width=0.233\linewidth]{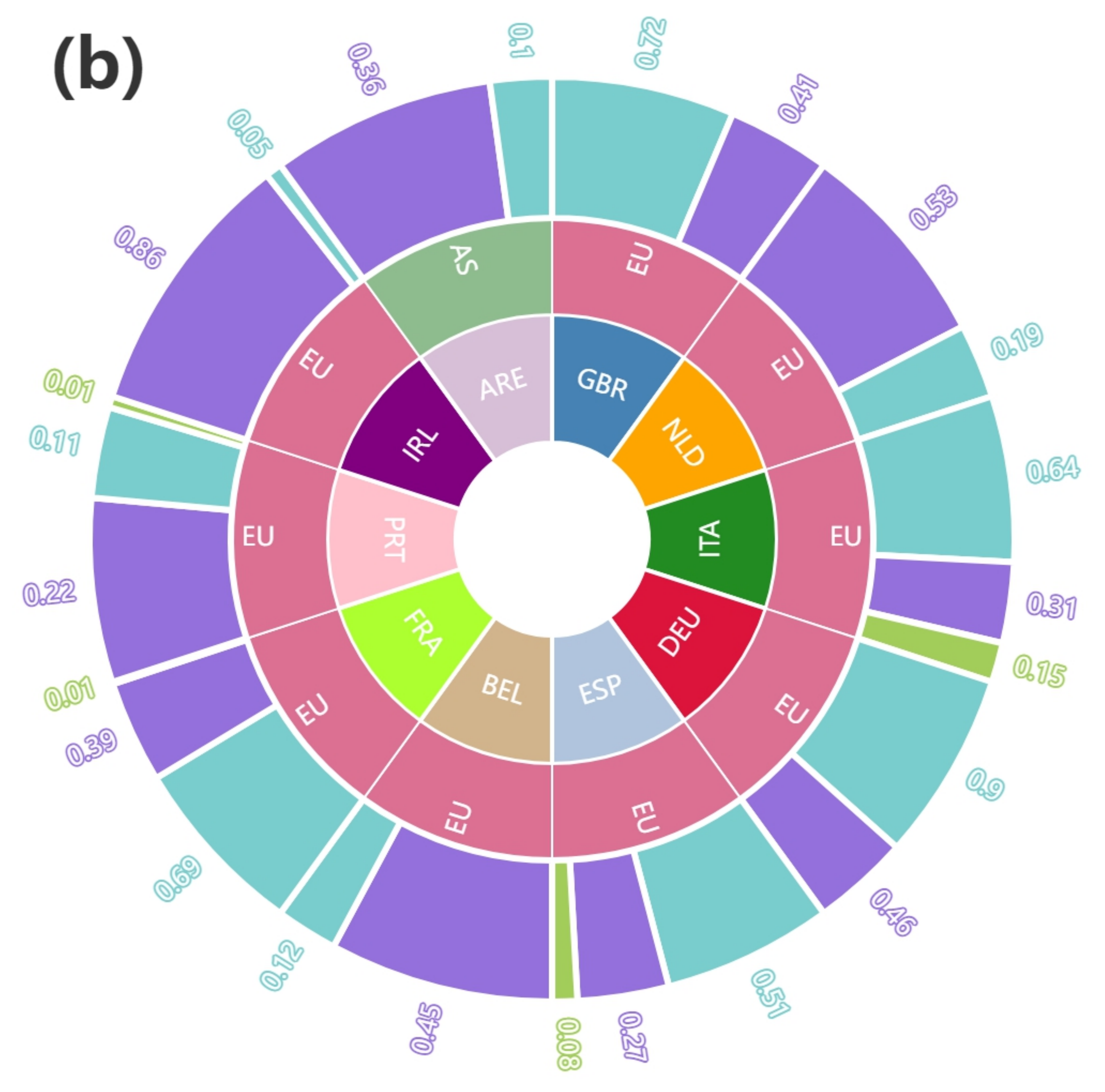}
      \includegraphics[width=0.233\linewidth]{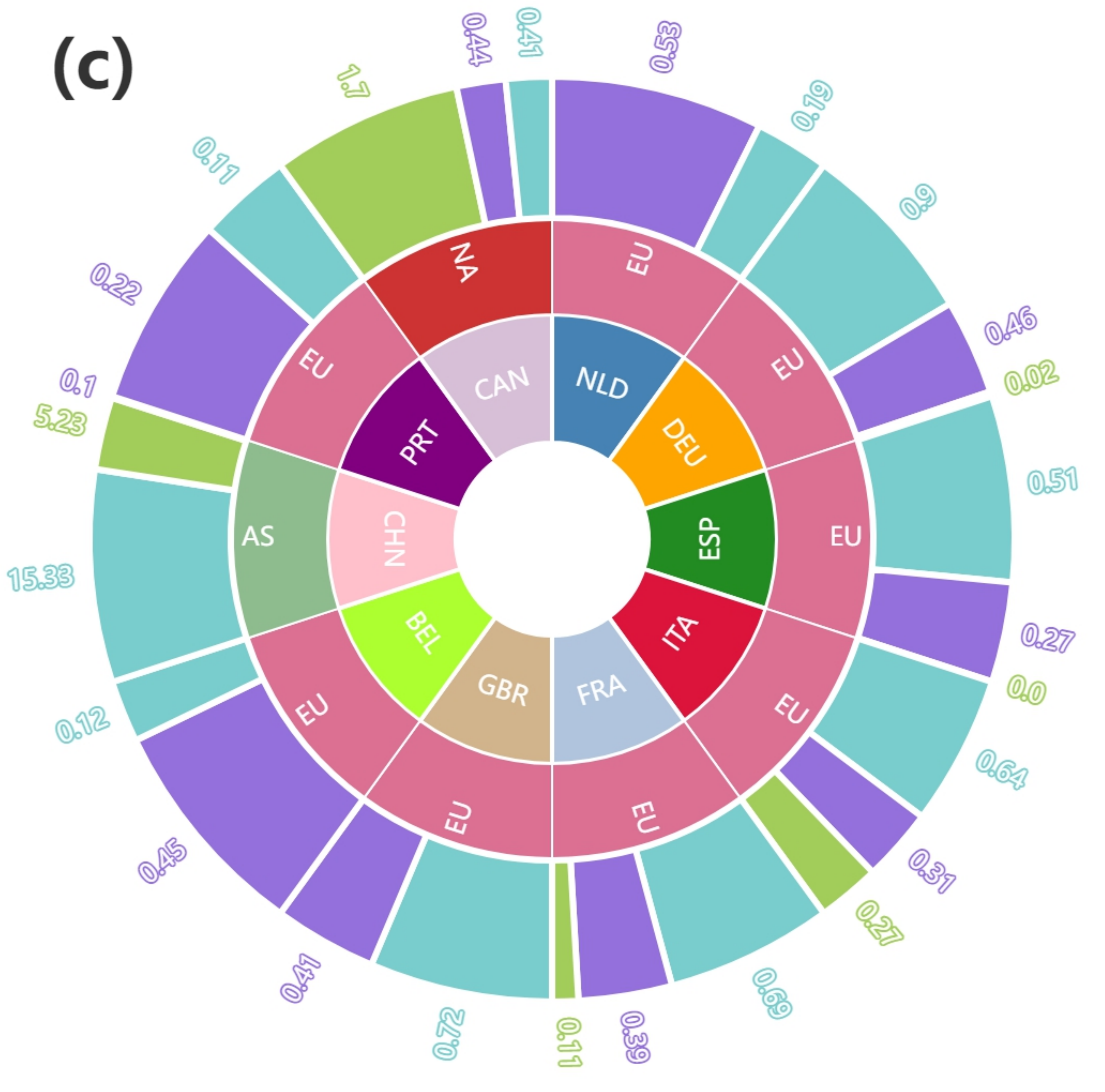}
      \includegraphics[width=0.233\linewidth]{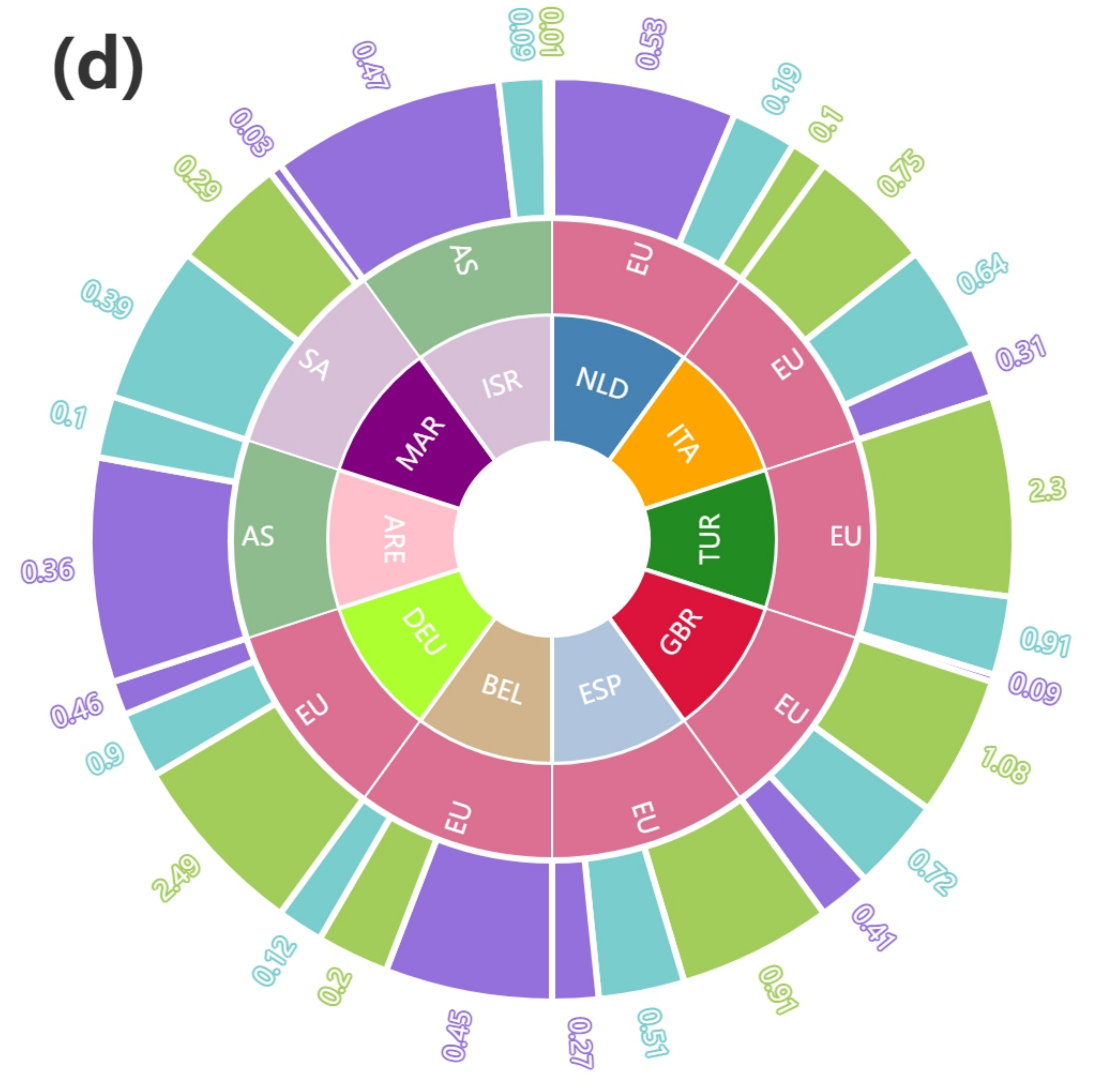}\\
      \includegraphics[width=0.233\linewidth]{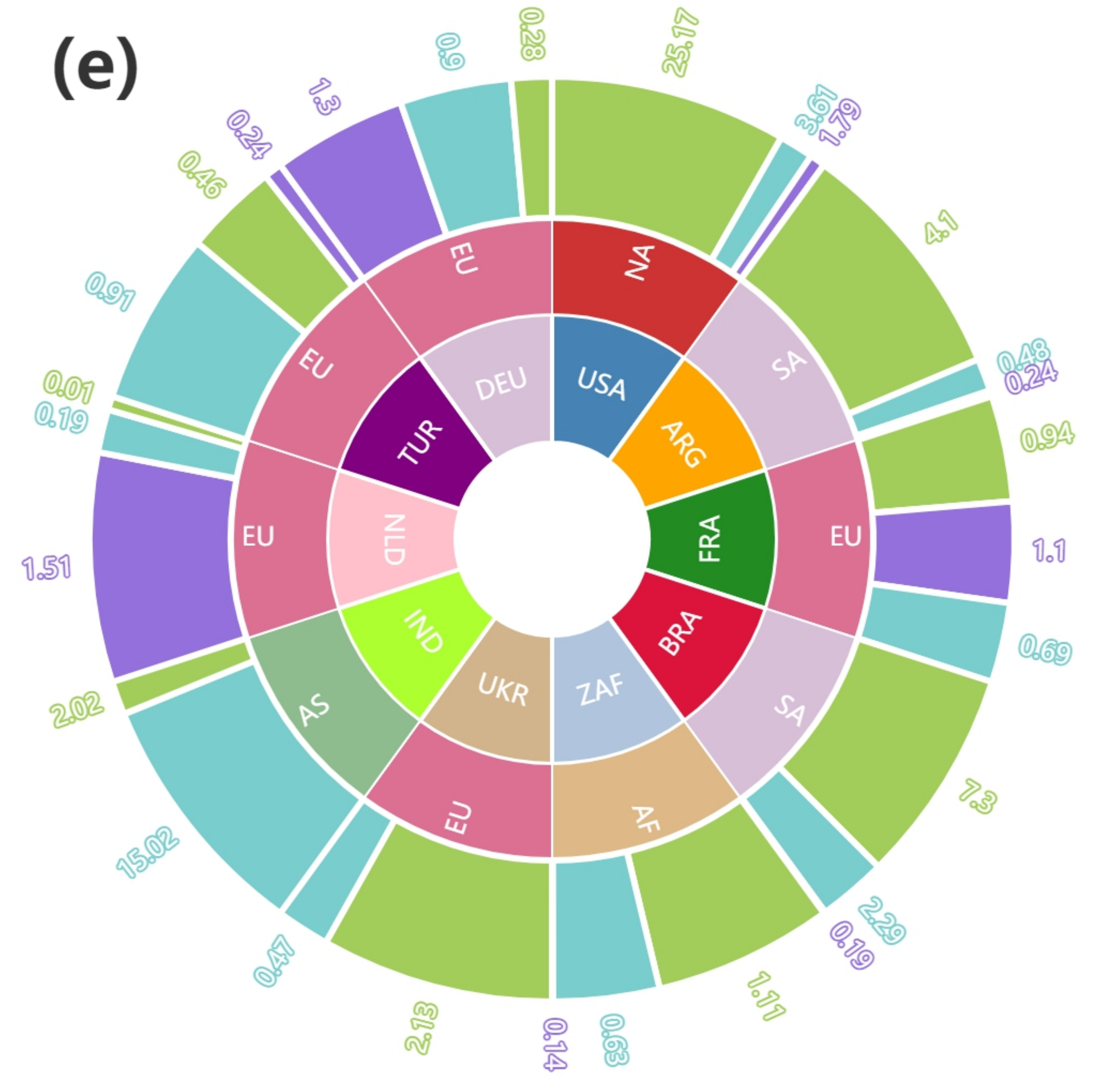}
      \includegraphics[width=0.233\linewidth]{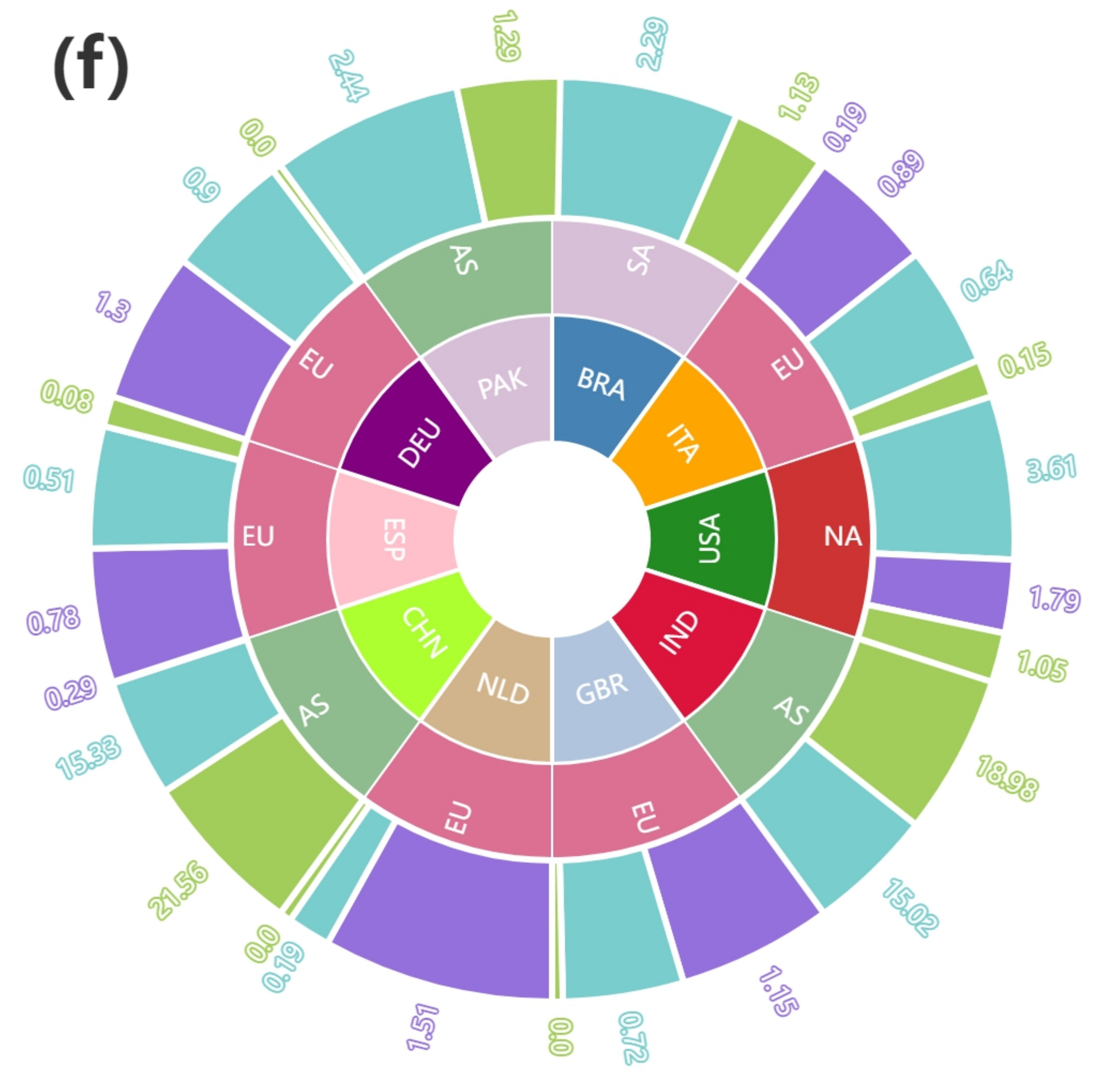}
      \includegraphics[width=0.233\linewidth]{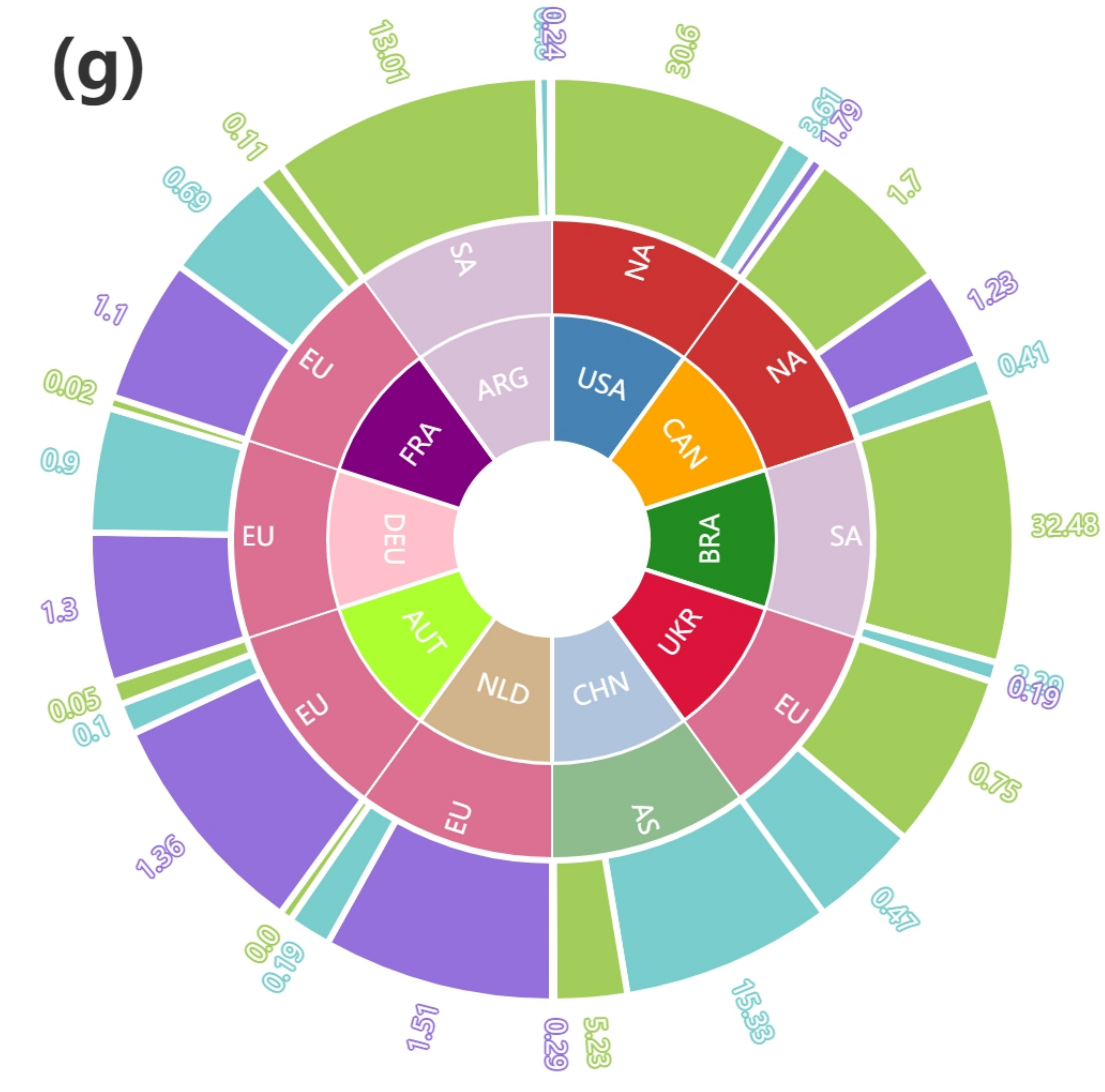}
      \includegraphics[width=0.233\linewidth]{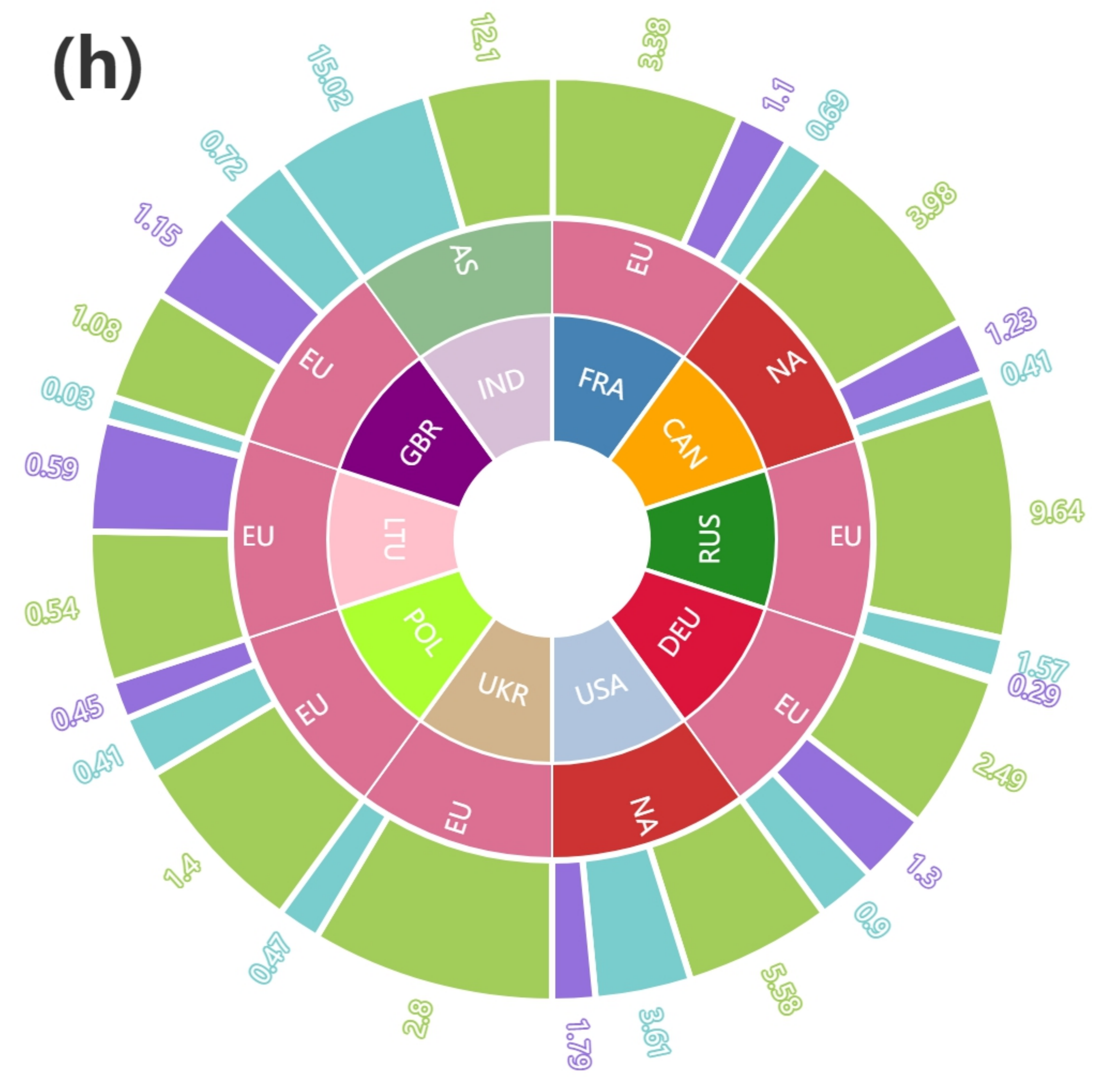}\\
      \caption{Top 10 important economies identified by $\mathrm{CNI^{\mathrm{in}}}$ (a-d) and $\mathrm{CNI^{\mathrm{out}}}$ (e-h). The innermost circle shows the top 10 important economies. The second circle shows regions where economies are located, with different colors corresponding to different regions. The third circle shows basic descriptions of economies. The green panels correspond to the ratio of the economy's production to the world's total production. The blue panels correspond to the ratio of the economy's population to the world's total population. The purple panel corresponds to the GDP per capital of the economy, in \$100,000 units. }
      \label{Fig:keycountry:CNI}
\end{figure}

For a more detailed analysis of the factors contributing to the influence of economies, we have plotted the regional distribution, crop production ratio (the ratio between the crop production of a given economy and that of all economies), population ratio (the ratio between the population of a given economy and that of all economies), and GDP per capita of the key economies in Fig.~\ref{Fig:keycountry:CNI}. According to the ranking results based on $\mathrm{CNI}^{\mathrm{in}}$, economies with low crop production exert a greater impact on import trade, as exemplified by the Netherlands, the UK, and Spain importing maize (Fig.~\ref{Fig:keycountry:CNI}(a)), and Germany, Netherlands, and Italy importing rice (Fig.~\ref{Fig:keycountry:CNI}(b)). It should be noted that all of these are developed economies with high GDP per capita. At the same time, Asian economies also wield significant influence on food imports. This could be attributed to their low food production, such as Japan importing maize, the United Arab Emirates importing maize, rice, and wheat, and Israel importing wheat. Alternatively, it could be due to the large population size leading to gaps between local production and domestic food demand, as is the case with China's importation of soybeans (Fig.~\ref{Fig:keycountry:CNI}(c)).

 According to ranking results based on $\mathrm{CNI}^{\mathrm{out}}$, we can discern that economies with significant export influences are typically those with higher food production or smaller populations. For instance, the United States and Brazil, with their high maize (Fig.~\ref{Fig:keycountry:CNI}(e)) and soybean production (Fig.~\ref{Fig:keycountry:CNI}(g)), India, with its high rice (Fig.~\ref{Fig:keycountry:CNI}(f)) and wheat production (Fig.~\ref{Fig:keycountry:CNI}(h)), and Russia, with its high wheat production (Fig.~\ref{Fig:keycountry:CNI}(h)), have significant influence on the export trade. Some European economies with low crop production may export more crops, owing to their small population and food surpluses. Notably, France and the Netherlands, without much maize production (Fig.~\ref{Fig:keycountry:CNI}(e)), Italy and Spain, without much rice production (Fig.~\ref{Fig:keycountry:CNI}(f)), Austria and Germany, without much soybean production (Fig.~\ref{Fig:keycountry:CNI}(g)), and Lithuania, without much wheat production (Fig.~\ref{Fig:keycountry:CNI}(h)), play important roles in the trade. The result is somewhat counterintuitive that some economies without crop production are still identified as key economies based on $\mathrm{CNI}^{\mathrm{out}}$. For example, the United Kingdom without rice and soybean production in 2020 (Fig.~\ref{Fig:keycountry:CNI}(f, g)), and the Netherlands without soybean production in 2020 (Fig.~\ref{Fig:keycountry:CNI}(g)), remain vital to the trade network. These two economies have higher values of $OC$, which indicates their significant impact on the structure and function of the iCTNs. European economies, in general, occupy prominent trade positions in the iCTNs, and Asian and North American economies also hold important roles. Furthermore, economies with high food production wield significant influence over export trade, whereas those with insufficient food production have considerable sway over import trade.

\section{Discussion}
\label{S4:Discussion}

Prior studies have noted that food security is a grave concern in the wake of extreme weather, conflict, and epidemics. External food supply through international trade is of fundamental importance for economies with insufficient domestic production. Trade restrictions imposed by economies with a significant impact on food trade may exacerbate the volatility in global food markets. It has rekindled political and scientific interest in measuring economies' trade influence. Complex network analysis has been proposed to investigate food trade patterns. In this paper, we assemble international crop trade networks using maize, rice, soybean, and wheat trade datasets from 1986 to 2020. Our initial objective is to identify key economies that have a substantial impact on global food trade. We evaluate the importance of economies in the iCTNs and identify significant economies using multidimensional node importance metrics. Through analyzing the correlation structure of different node importance metrics based on random matrix theory, we incorporate 20 metrics into a single composite metric. Our analysis yields some intriguing results.

Firstly, economies' impact on the iCTNs depends on the focus of node important metrics. The findings unequivocally indicate that how important economies are in the iCTNs is a multidimensional problem. It motivates our focus on the correlation structure of those indicators. Despite indicators in our work gauging economies' importance from different aspects, some of them have significant correlations. Based on correlation analysis and cluster identification, we classify indicators into three groups. One is based on the import scale such as in-degree centrality, in-strength, in-closeness centrality, in-semi-local centrality, eigenvector centrality, Authorities, PageRank,within-module degree, outside-module degree, and participation coefficient. Another is based on the export scale, namely, out-degree centrality, out-strength, out-closeness centrality, out-semi-local centrality, betweenness centrality, and Hubs. The remaining less relevant indicators are grouped into one category, such as constraint coefficient, clustering coefficient, Katz centrality, and mutual information. These findings facilitate our understanding of node importance metrics and 
enlighten us to construct composite metrics.

In terms of composite node importance metrics following optimal eigenportfolios in the financial market, we present the ranking characteristics of influential economies in the iCTNs. It should be pointed out that European economies have a significant influence in the iCTNs, especially the Netherlands, Britain and Germany. In addition, economies with low crop production have a greater influence on the import trade. At the same time, Asian economies also have an important influence on food imports. A possible explanation for this might be the low food production. Another possible explanation is that the large population results in gaps between local production and domestic food demands. Another important finding is that influential economies on food exporting are generally those with higher food production or smaller populations. What is counterintuitive is that some economies without crop production are still identified as key economies since their trading partners
are important in the network. 

Finally, in view of the unique trade patterns of different crops, economies' trade impact differs across crops. These findings might be explained by the fact that economies' food trade is tied to crop diversity and nutrition demand. Therefore, it is necessary to talk about the contribution of an economy to global food trade and food security across different crop types.

Since the criteria for importance are diverse, identifying key nodes is not a trivial job. It is impossible to find a perfect index to quantify a node's importance. Our motivation for this paper is to make extensive empirical comparisons with multi-dimensional metrics. Thus, we can go a step closer to a comprehensive grasp of the similarities and differences across various approaches. Despite we just qualitatively evaluate node importance metrics instead of designing a convincing index, our study nevertheless helps identify vital nodes in every situation. Further investigation about objective functions considering metrics’ performances should, however, be the subject of future work.

This paper reaffirms the existing research measuring node importance based on different methods. Overall, vital node identification is challenging and meaningful. We find some approaches have many similarities and dissimilarities using the network theory and random matrix theory. We also find that European economies have a significant impact on the iCTNs. Additionally, economies with poor crop production play a major role in the import trade, whereas economies with higher food production or smaller populations are crucial to the export trade.



\section*{CRediT authorship contribution statement}

Funding acquisition, Wei-Xing Zhou; Investigation, Yin-Ting Zhang; Methodology, Yin-Ting Zhang and Wei-Xing Zhou; Supervision, Wei-Xing Zhou; Writing – original draft, Yin-Ting Zhang and Wei-Xing Zhou; Writing – review \& editing, Yin-Ting Zhang and Wei-Xing Zhou.

\section*{Declaration of competing interest}

The authors declare that they have no known competing financial interests or personal relationships that could have appeared to
influence the work reported in this paper.

\section*{Data Availability}
Publicly available datasets were analyzed in this study. This data can be found here: \href{https://www.fao.org}{ https://www.fao.org}.

\section*{Acknowledgements}
\label{S6:Acknowledgements}

This work was partly supported by the National Natural Science Foundation of China (72171083), the Shanghai Outstanding Academic Leaders Plan, and the Fundamental Research Funds for the Central Universities.


\clearpage

\appendix

\renewcommand\thetable{\Alph{section}\arabic{table}}    
\section{Correlation of the influence indicators}
\label{sec5-1}
\setcounter{table}{0}

\begin{table}[!ht]
    \centering
    \renewcommand\tabcolsep{2mm}
    \caption{Correlation of the influence indicators in 2020.}
    \tiny
    \smallskip
    \begin{tabular}{cccccccccccccccccccccccc}
         \toprule
            & OD & ID & IS & OS & BC & EC & KC & CC & IC & OC & PR & HU & AU & CI & IL & OL & MI & IM & PT & OM \\ 
         \midrule
        OD & 1.000  & 0.698  & 0.571  & 0.916  & 0.898  & 0.626  & -0.015  & -0.241  & 0.602  & 0.942  & 0.488  & 0.906  & 0.465  & 0.276  & 0.644  & 0.937  & 0.101  & 0.740  & 0.372  & 0.758 \\  
        ID & 0.698  & 1.000  & 0.654  & 0.588  & 0.707  & 0.957  & -0.110  & -0.053  & 0.945  & 0.637  & 0.714  & 0.573  & 0.518  & -0.041  & 0.971  & 0.661  & 0.036  & 0.673  & 0.449  & 0.746  \\  
        IS & 0.571  & 0.654  & 1.000  & 0.443  & 0.558  & 0.577  & 0.120  & 0.027  & 0.603  & 0.602  & 0.670  & 0.515  & 0.884  & 0.135  & 0.602  & 0.565  & 0.240  & 0.471  & 0.458  & 0.618  \\  
        OS & 0.916  & 0.588  & 0.443  & 1.000  & 0.800  & 0.548  & -0.040  & -0.196  & 0.508  & 0.850  & 0.408  & 0.943  & 0.343  & 0.361  & 0.563  & 0.857  & 0.172  & 0.677  & 0.294  & 0.639  \\  
        BC & 0.898  & 0.707  & 0.558  & 0.800  & 1.000  & 0.614  & 0.094  & -0.372  & 0.642  & 0.867 & 0.590  & 0.779  & 0.454  & 0.218  & 0.638  & 0.841  & 0.155  & 0.730  & 0.392  & 0.792  \\  
        EC & 0.626  & 0.957  & 0.577  & 0.548  & 0.614  & 1.000  & -0.268  & 0.037  & 0.956  & 0.554  & 0.617  & 0.521  & 0.442  & -0.056  & 0.996  & 0.592  & 0.015  & 0.580  & 0.408  & 0.654  \\  
        KC & -0.015  & -0.110  & 0.120  & -0.040  & 0.094  & -0.268  & 1.000  & -0.198  & -0.084  & 0.019  & 0.190  & 0.000  & 0.259  & 0.121  & -0.212  & -0.080  & 0.003  & 0.120  & 0.299  & 0.192  \\   
        CC & -0.241  & -0.053  & 0.027  & -0.196  & -0.372  & 0.037  & -0.198  & 1.000  & -0.075  & -0.211  & -0.182  & -0.137  & 0.024  & 0.181  & 0.019  & -0.167  & -0.386  & -0.276  & -0.056  & -0.307  \\  
        IC & 0.602  & 0.945  & 0.603  & 0.508  & 0.642  & 0.956  & -0.084  & -0.075  & 1.000  & 0.542  & 0.687  & 0.482  & 0.515  & -0.072  & 0.967  & 0.548  & 0.051  & 0.613  & 0.489  & 0.734  \\ 
        OC & 0.942  & 0.637  & 0.602  & 0.850  & 0.867  & 0.554  & 0.019  & -0.211  & 0.542  & 1.000  & 0.475  & 0.885  & 0.502  & 0.302  & 0.577  & 0.974  & 0.110  & 0.648  & 0.390  & 0.738  \\  
        PR & 0.488  & 0.714  & 0.670  & 0.408  & 0.590  & 0.617  & 0.190  & -0.182  & 0.687  & 0.475  & 1.000  & 0.428  & 0.527  & 0.044  & 0.645  & 0.463  & 0.261  & 0.547  & 0.345  & 0.631  \\ 
        HU & 0.906  & 0.573  & 0.515  & 0.943  & 0.779  & 0.521  & 0.000  & -0.137  & 0.482  & 0.885  & 0.428  & 1.000  & 0.442  & 0.360  & 0.538  & 0.885  & 0.166  & 0.636  & 0.295  & 0.643  \\   
        AU & 0.465  & 0.518  & 0.884  & 0.343  & 0.454  & 0.442  & 0.259  & 0.024  & 0.515  & 0.502  & 0.527  & 0.442  & 1.000  & 0.132  & 0.472  & 0.429  & 0.178  & 0.459  & 0.431  & 0.610  \\   
        CI & 0.276  & -0.041  & 0.135  & 0.361  & 0.218  & -0.056  & 0.121  & 0.181  & -0.072  & 0.302  & 0.044  & 0.360  & 0.132  & 1.000  & -0.045  & 0.240  & -0.256  & 0.034  & 0.158  & 0.017  \\  
        IL & 0.644  & 0.971  & 0.602  & 0.563  & 0.638  & 0.996  & -0.212  & 0.019  & 0.967  & 0.577  & 0.645  & 0.538  & 0.472  & -0.045  & 1.000  & 0.610  & 0.022  & 0.610  & 0.427  & 0.686 \\
        OL & 0.937  & 0.661  & 0.565  & 0.857  & 0.841  & 0.592  & -0.080  & -0.167  & 0.548  & 0.974 & 0.463  & 0.885  & 0.429  & 0.240  & 0.610  & 1.000  & 0.105  & 0.637  & 0.335  & 0.694  \\  
        MI & 0.101  & 0.036  & 0.240  & 0.172  & 0.155  & 0.015  & 0.003  & -0.386  & 0.051  & 0.110 & 0.261  & 0.166  & 0.178  & -0.256  & 0.022  & 0.105  & 1.000  & 0.208  & -0.052  & 0.190  \\  
        IM & 0.740  & 0.673  & 0.471  & 0.677  & 0.730  & 0.580  & 0.120  & -0.276  & 0.613  & 0.648 & 0.547  & 0.636  & 0.459  & 0.034  & 0.610  & 0.637  & 0.208  & 1.000  & 0.165  & 0.756  \\ 
        PT & 0.372  & 0.449  & 0.458  & 0.294  & 0.392  & 0.408  & 0.299  & -0.056  & 0.489  & 0.390  & 0.345  & 0.295  & 0.431  & 0.158  & 0.427  & 0.335  & -0.052  & 0.165  & 1.000  & 0.548  \\  
        OM & 0.758  & 0.746  & 0.618  & 0.639  & 0.792  & 0.654  & 0.192  & -0.307  & 0.734  & 0.738  & 0.631  & 0.643  & 0.610  & 0.017  & 0.686  & 0.694  & 0.190  & 0.756  & 0.548  & 1.000 \\   
    \bottomrule
    \end{tabular}
    \label{Table:iCTN:node:influence:correlation:maize}
\end{table}

\begin{table}[!ht]
    \centering
    \renewcommand\tabcolsep{2mm}
    \caption{Correlation of the influence indicators in 2020.}
    \tiny
    \smallskip
    \begin{tabular}{cccccccccccccccccccccccc}
         \toprule
            & OD & ID & IS & OS & BC & EC & KC & CC & IC & OC & PR & HU & AU & CI & IL & OL & MI & IM & PT & OM \\  
         \midrule
        OD & 1.000  & 0.402  & 0.221  & 0.932  & 0.904  & 0.347  & 0.350  & 0.043  & 0.348  & 0.964 & 0.378  & 0.912  & 0.157  & 0.666  & 0.371  & 0.888  & 0.171  & 0.615  & 0.238  & 0.574  \\ 
        ID & 0.402  & 1.000  & 0.611  & 0.302  & 0.455  & 0.763  & 0.773  & 0.448  & 0.831  & 0.355  & 0.673  & 0.318  & 0.544  & 0.067  & 0.799  & 0.417  & -0.038  & 0.603  & 0.496  & 0.651  \\  
        IS & 0.221  & 0.611  & 1.000  & 0.206  & 0.252  & 0.300  & 0.311  & 0.218  & 0.380  & 0.194  & 0.707  & 0.236  & 0.595  & -0.044  & 0.355  & 0.189  & 0.429  & 0.480  & 0.237  & 0.307  \\ 
        OS & 0.932  & 0.302  & 0.206  & 1.000  & 0.839  & 0.260  & 0.260  & 0.028  & 0.258  & 0.898  & 0.305  & 0.924  & 0.140  & 0.746  & 0.279  & 0.780  & 0.253  & 0.565  & 0.151  & 0.440  \\  
        BC & 0.904  & 0.455  & 0.252  & 0.839  & 1.000  & 0.374  & 0.377  & -0.039  & 0.395  & 0.891  & 0.426  & 0.810  & 0.178  & 0.588  & 0.398  & 0.818  & 0.149  & 0.581  & 0.284  & 0.614  \\  
        EC & 0.347  & 0.763  & 0.300  & 0.260  & 0.374  & 1.000  & 0.999  & 0.482  & 0.958  & 0.308 & 0.447  & 0.275  & 0.396  & 0.132  & 0.990  & 0.430  & -0.150  & 0.357  & 0.416  & 0.496  \\  
        KC & 0.350  & 0.773  & 0.311  & 0.260  & 0.377  & 0.999  & 1.000  & 0.482  & 0.961  & 0.310  & 0.455  & 0.276  & 0.402  & 0.129  & 0.992  & 0.434  & -0.148  & 0.364  & 0.422  & 0.502  \\ 
        CC & 0.043  & 0.448  & 0.218  & 0.028  & -0.039  & 0.482  & 0.482  & 1.000  & 0.425  & 0.031 & 0.182  & 0.090  & 0.367  & 0.116  & 0.485  & 0.099  & -0.283  & 0.127  & 0.179  & 0.062  \\   
        IC & 0.348  & 0.831  & 0.380  & 0.258  & 0.395  & 0.958  & 0.961  & 0.425  & 1.000  & 0.306  & 0.495  & 0.264  & 0.399  & 0.096  & 0.968  & 0.394  & -0.116  & 0.442  & 0.447  & 0.573  \\  
        OC & 0.964  & 0.355  & 0.194  & 0.898  & 0.891  & 0.308  & 0.310  & 0.031  & 0.306  & 1.000 & 0.335  & 0.913  & 0.144  & 0.689  & 0.335  & 0.910  & 0.159  & 0.569  & 0.225  & 0.538  \\  
        PR & 0.378  & 0.673  & 0.707  & 0.305  & 0.426  & 0.447  & 0.455  & 0.182  & 0.495  & 0.335  & 1.000  & 0.322  & 0.267  & 0.090  & 0.495  & 0.369  & 0.341  & 0.510  & 0.237  & 0.405  \\   
        HU & 0.912  & 0.318  & 0.236  & 0.924  & 0.810  & 0.275  & 0.276  & 0.090  & 0.264  & 0.913  & 0.322  & 1.000  & 0.209  & 0.713  & 0.299  & 0.827  & 0.170  & 0.548  & 0.157  & 0.427  \\   
        AU & 0.157  & 0.544  & 0.595  & 0.140  & 0.178  & 0.396  & 0.402  & 0.367  & 0.399  & 0.144 & 0.267  & 0.209  & 1.000  & -0.075  & 0.426  & 0.210  & 0.050  & 0.201  & 0.372  & 0.280  \\    
        CI & 0.666  & 0.067  & -0.044  & 0.746  & 0.588  & 0.132  & 0.129  & 0.116  & 0.096  & 0.689 & 0.090  & 0.713  & -0.075  & 1.000  & 0.140  & 0.530  & -0.130  & 0.231  & 0.012  & 0.097  \\   
        IL & 0.371  & 0.799  & 0.355  & 0.279  & 0.398  & 0.990  & 0.992  & 0.485  & 0.968  & 0.335  & 0.495  & 0.299  & 0.426  & 0.140  & 1.000  & 0.446  & -0.133  & 0.391  & 0.444  & 0.525  \\
        OL & 0.888  & 0.417  & 0.189  & 0.780  & 0.818  & 0.430  & 0.434  & 0.099  & 0.394  & 0.910  & 0.369  & 0.827  & 0.210  & 0.530  & 0.446  & 1.000  & 0.124  & 0.537  & 0.176  & 0.517  \\ 
        MI & 0.171  & -0.038  & 0.429  & 0.253  & 0.149  & -0.150  & -0.148  & -0.283  & -0.116  & 0.159 & 0.341  & 0.170  & 0.050  & -0.130  & -0.133  & 0.124  & 1.000  & 0.235  & -0.175  & 0.019  \\  
        IM & 0.615  & 0.603  & 0.480  & 0.565  & 0.581  & 0.357  & 0.364  & 0.127  & 0.442  & 0.569 & 0.510  & 0.548  & 0.201  & 0.231  & 0.391  & 0.537  & 0.235  & 1.000  & 0.104  & 0.648  \\  
        PT & 0.238  & 0.496  & 0.237  & 0.151  & 0.284  & 0.416  & 0.422  & 0.179  & 0.447  & 0.225 & 0.237  & 0.157  & 0.372  & 0.012  & 0.444  & 0.176  & -0.175  & 0.104  & 1.000  & 0.620  \\ 
        OM & 0.574  & 0.651  & 0.307  & 0.440  & 0.614  & 0.496  & 0.502  & 0.062  & 0.573  & 0.538 & 0.405  & 0.427  & 0.280  & 0.097  & 0.525  & 0.517  & 0.019  & 0.648  & 0.620  & 1.000 \\  
    \bottomrule
    \end{tabular}
    \label{Table:iCTN:node:influence:correlation:rice}
\end{table}

\begin{table}[!ht]
    \centering
    \renewcommand\tabcolsep{2mm}
    \caption{Correlation of the influence indicators in 2020.}
    \tiny
    \smallskip

    \begin{tabular}{cccccccccccccccccccccccc}
         \toprule
            & OD & ID & IS & OS & BC & EC & KC & CC & IC & OC & PR & HU & AU & CI & IL & OL & MI & IM & PT & OM \\  
         \midrule
        OD & 1.000  & 0.609  & 0.575  & 0.913  & 0.880  & 0.506  & 0.162  & -0.086  & 0.504  & 0.941 & 0.556  & 0.939  & 0.467  & 0.449  & 0.527  & 0.963  & 0.056  & 0.693  & 0.445  & 0.753  \\  
        ID & 0.609  & 1.000  & 0.719  & 0.501  & 0.669  & 0.914  & 0.359  & 0.225  & 0.927  & 0.511  & 0.805  & 0.523  & 0.618  & -0.019  & 0.928  & 0.579  & -0.017  & 0.683  & 0.586  & 0.807  \\  
        IS & 0.575  & 0.719  & 1.000  & 0.469  & 0.631  & 0.637  & 0.076  & 0.188  & 0.678  & 0.548  & 0.708  & 0.499  & 0.894  & 0.092  & 0.663  & 0.550  & 0.148  & 0.566  & 0.585  & 0.657  \\ 
        OS & 0.913  & 0.501  & 0.469  & 1.000  & 0.775  & 0.402  & 0.168  & -0.034  & 0.365  & 0.874 & 0.460  & 0.945  & 0.344  & 0.530  & 0.413  & 0.887  & 0.144  & 0.658  & 0.364  & 0.641  \\  
        BC & 0.880  & 0.669  & 0.631  & 0.775  & 1.000  & 0.537  & 0.097  & -0.146  & 0.568  & 0.829  & 0.650  & 0.791  & 0.507  & 0.343  & 0.560  & 0.831  & 0.139  & 0.664  & 0.438  & 0.746  \\  
        EC & 0.506  & 0.914  & 0.637  & 0.402  & 0.537  & 1.000  & 0.581  & 0.304  & 0.951  & 0.404 & 0.688  & 0.456  & 0.563  & -0.052  & 0.991  & 0.506  & -0.128  & 0.553  & 0.550  & 0.739  \\ 
        KC & 0.162  & 0.359  & 0.076  & 0.168  & 0.097  & 0.581  & 1.000  & 0.051  & 0.415  & 0.073 & 0.159  & 0.191  & 0.022  & -0.077  & 0.515  & 0.188  & -0.022  & 0.146  & 0.128  & 0.295  \\  
        CC & -0.086  & 0.225  & 0.188  & -0.034  & -0.146  & 0.304  & 0.051  & 1.000  & 0.233  & -0.096 & 0.158  & -0.014  & 0.168  & 0.083  & 0.324  & -0.046  & -0.423  & 0.000  & 0.308  & 0.043  \\  
        IC & 0.504  & 0.927  & 0.678  & 0.365  & 0.568  & 0.951  & 0.415  & 0.233  & 1.000  & 0.408  & 0.718  & 0.415  & 0.640  & -0.113  & 0.970  & 0.487  & -0.076  & 0.580  & 0.554  & 0.766  \\  
        OC & 0.941  & 0.511  & 0.548  & 0.874  & 0.829  & 0.404  & 0.073  & -0.096  & 0.408  & 1.000 & 0.485  & 0.903  & 0.477  & 0.503  & 0.427  & 0.966  & 0.019  & 0.660  & 0.403  & 0.665  \\    
        PR & 0.556  & 0.805  & 0.708  & 0.460  & 0.650  & 0.688  & 0.159  & 0.158  & 0.718  & 0.485  & 1.000  & 0.467  & 0.571  & 0.082  & 0.712  & 0.518  & 0.124  & 0.563  & 0.427  & 0.660  \\  
        HU & 0.939  & 0.523  & 0.499  & 0.945  & 0.791  & 0.456  & 0.191  & -0.014  & 0.415  & 0.903 & 0.467  & 1.000  & 0.390  & 0.513  & 0.468  & 0.928  & 0.020  & 0.665  & 0.418  & 0.679  \\ 
        AU & 0.467  & 0.618  & 0.894  & 0.344  & 0.507  & 0.563  & 0.022  & 0.168  & 0.640  & 0.477 & 0.571  & 0.390  & 1.000  & 0.059  & 0.604  & 0.453  & 0.044  & 0.481  & 0.529  & 0.532  \\  
        CI & 0.449  & -0.019  & 0.092  & 0.530  & 0.343  & -0.052  & -0.077  & 0.083  & -0.113  & 0.503 & 0.082  & 0.513  & 0.059  & 1.000  & -0.038  & 0.453  & -0.219  & 0.068  & 0.151  & 0.083  \\   
        IL & 0.527  & 0.928  & 0.663  & 0.413  & 0.560  & 0.991  & 0.515  & 0.324  & 0.970  & 0.427  & 0.712  & 0.468  & 0.604  & -0.038  & 1.000  & 0.525  & -0.129  & 0.574  & 0.565  & 0.762  \\  
        OL & 0.963  & 0.579  & 0.550  & 0.887  & 0.831  & 0.506  & 0.188  & -0.046  & 0.487  & 0.966  & 0.518  & 0.928  & 0.453  & 0.453  & 0.525  & 1.000  & -0.002  & 0.673  & 0.432  & 0.723  \\   
        MI & 0.056  & -0.017  & 0.148  & 0.144  & 0.139  & -0.128  & -0.022  & -0.423  & -0.076  & 0.019  & 0.124  & 0.020  & 0.044  & -0.219  & -0.129  & -0.002  & 1.000  & 0.167  & -0.165  & 0.095  \\ 
        IM & 0.693  & 0.683  & 0.566  & 0.658  & 0.664  & 0.553  & 0.146  & 0.000  & 0.580  & 0.660  & 0.563  & 0.665  & 0.481  & 0.068  & 0.574  & 0.673  & 0.167  & 1.000  & 0.195  & 0.707  \\  
        PT & 0.445  & 0.586  & 0.585  & 0.364  & 0.438  & 0.550  & 0.128  & 0.308  & 0.554  & 0.403 & 0.427  & 0.418  & 0.529  & 0.151  & 0.565  & 0.432  & -0.165  & 0.195  & 1.000  & 0.593  \\  
        OM & 0.753  & 0.807  & 0.657  & 0.641  & 0.746  & 0.739  & 0.295  & 0.043  & 0.766  & 0.665 & 0.660  & 0.679  & 0.532  & 0.083  & 0.762  & 0.723  & 0.095  & 0.707  & 0.593  & 1.000 \\  
    \bottomrule
    \end{tabular}
    \label{Table:iCTN:node:influence:correlation:soybean}
\end{table}

\begin{table}[!ht]
    \centering
    \renewcommand\tabcolsep{2mm}
    \caption{Correlation of the influence indicators in 2020.}
    \tiny
    \smallskip
    \begin{tabular}{cccccccccccccccccccccccc}
         \toprule
            & OD & ID & IS & OS & BC & EC & KC & CC & IC & OC & PR & HU & AU & CI & IL & OL & MI & IM & PT & OM \\  
         \midrule
         OD & 1.000  & 0.598  & 0.292  & 0.949  & 0.859  & 0.613  & -0.508  & -0.142  & 0.517  & 0.951 & 0.363  & 0.937  & 0.151  & 0.606  & 0.614  & 0.929  & 0.078  & 0.692  & 0.394  & 0.710  \\  
        ID & 0.598  & 1.000  & 0.550  & 0.545  & 0.647  & 0.971  & -0.763  & 0.111  & 0.941  & 0.584  & 0.697  & 0.517  & 0.452  & 0.207  & 0.975  & 0.587  & 0.133  & 0.653  & 0.582  & 0.814  \\  
        IS & 0.292  & 0.550  & 1.000  & 0.219  & 0.319  & 0.489  & -0.331  & 0.119  & 0.501  & 0.277  & 0.613  & 0.198  & 0.903  & 0.201  & 0.493  & 0.269  & 0.232  & 0.350  & 0.321  & 0.417  \\  
        OS & 0.949  & 0.545  & 0.219  & 1.000  & 0.809  & 0.563  & -0.483  & -0.119  & 0.464  & 0.906  & 0.328  & 0.985  & 0.074  & 0.644  & 0.563  & 0.879  & 0.104  & 0.639  & 0.344  & 0.629  \\  
        BC & 0.859  & 0.647  & 0.319  & 0.809  & 1.000  & 0.618  & -0.449  & -0.247  & 0.589  & 0.840   & 0.436  & 0.805  & 0.205  & 0.449  & 0.624  & 0.827  & 0.178  & 0.630  & 0.366  & 0.673  \\  
        EC & 0.613  & 0.971  & 0.489  & 0.563  & 0.618  & 1.000  & -0.874  & 0.197  & 0.932  & 0.605  & 0.631  & 0.540  & 0.381  & 0.216  & 0.998  & 0.613  & 0.077  & 0.637  & 0.593  & 0.815  \\  
        KC & -0.508  & -0.763  & -0.331  & -0.483  & -0.449  & -0.874  & 1.000  & -0.287  & -0.734  & -0.510  & -0.405  & -0.465  & -0.222  & -0.159  & -0.857  & -0.528  & 0.013  & -0.430  & -0.543  & -0.645  \\   
        CC & -0.142  & 0.111  & 0.119  & -0.119  & -0.247  & 0.197  & -0.287  & 1.000  & 0.100  & -0.091  & 0.056  & -0.132  & 0.075  & 0.145  & 0.191  & -0.067  & -0.370  & -0.138  & 0.179  & -0.074  \\  
        IC & 0.517  & 0.941  & 0.501  & 0.464  & 0.589  & 0.932  & -0.734  & 0.100  & 1.000  & 0.494 & 0.693  & 0.448  & 0.427  & 0.183  & 0.941  & 0.484  & 0.110  & 0.608  & 0.539  & 0.784  \\   
        OC & 0.951  & 0.584  & 0.277  & 0.906  & 0.840  & 0.605  & -0.510  & -0.091  & 0.494  & 1.000  & 0.339  & 0.913  & 0.139  & 0.608  & 0.606  & 0.973  & 0.028  & 0.641  & 0.418  & 0.692  \\ 
        PR & 0.363  & 0.697  & 0.613  & 0.328  & 0.436  & 0.631  & -0.405  & 0.056  & 0.693  & 0.339 & 1.000  & 0.312  & 0.476  & 0.163  & 0.643  & 0.346  & 0.311  & 0.438  & 0.328  & 0.496  \\  
        HU & 0.937  & 0.517  & 0.198  & 0.985  & 0.805  & 0.540  & -0.465  & -0.132  & 0.448  & 0.913 & 0.312  & 1.000  & 0.056  & 0.651  & 0.539  & 0.886  & 0.095  & 0.630  & 0.347  & 0.627  \\  
        AU & 0.151  & 0.452  & 0.903  & 0.074  & 0.205  & 0.381  & -0.222  & 0.075  & 0.427  & 0.139 & 0.476  & 0.056  & 1.000  & 0.118  & 0.384  & 0.115  & 0.123  & 0.326  & 0.338  & 0.388  \\  
        CI & 0.606  & 0.207  & 0.201  & 0.644  & 0.449  & 0.216  & -0.159  & 0.145  & 0.183  & 0.608  & 0.163  & 0.651  & 0.118  & 1.000  & 0.221  & 0.569  & -0.373  & 0.234  & 0.227  & 0.210  \\  
        IL & 0.614  & 0.975  & 0.493  & 0.563  & 0.624  & 0.998  & -0.857  & 0.191  & 0.941  & 0.606 & 0.643  & 0.539  & 0.384  & 0.221  & 1.000  & 0.612  & 0.083  & 0.643  & 0.589  & 0.820  \\ 
        OL & 0.929  & 0.587  & 0.269  & 0.879  & 0.827  & 0.613  & -0.528  & -0.067  & 0.484  & 0.973  & 0.346  & 0.886  & 0.115  & 0.569  & 0.612  & 1.000  & 0.041  & 0.624  & 0.384  & 0.670  \\  
        MI & 0.078  & 0.133  & 0.232  & 0.104  & 0.178  & 0.077  & 0.013  & -0.370  & 0.110  & 0.028 & 0.311  & 0.095  & 0.123  & -0.373  & 0.083  & 0.041  & 1.000  & 0.227  & -0.100  & 0.146  \\    
        IM & 0.692  & 0.653  & 0.350  & 0.639  & 0.630  & 0.637  & -0.430  & -0.138  & 0.608  & 0.641  & 0.438  & 0.630  & 0.326  & 0.234  & 0.643  & 0.624  & 0.227  & 1.000  & 0.323  & 0.801  \\
        PT & 0.394  & 0.582  & 0.321  & 0.344  & 0.366  & 0.593  & -0.543  & 0.179  & 0.539  & 0.418 & 0.328  & 0.347  & 0.338  & 0.227  & 0.589  & 0.384  & -0.100  & 0.323  & 1.000  & 0.663  \\ 
        OM & 0.710  & 0.814  & 0.417  & 0.629  & 0.673  & 0.815  & -0.645  & -0.074  & 0.784  & 0.692 & 0.496  & 0.627  & 0.388  & 0.210  & 0.820  & 0.670  & 0.146  & 0.801  & 0.663  & 1.000 \\  
    \bottomrule
    \end{tabular}
    \label{Table:iCTN:node:influence:correlation:wheat}
\end{table}

\clearpage

\renewcommand\thefigure{\Alph{section}\arabic{figure}}    
\section{Eigenvalues and components of the corresponding eigenvector}
\label{sec5-2}
\setcounter{figure}{0}    

 \begin{figure}[h]
      \centering
     \subfigbottomskip=-1pt
     \subfigcapskip=-5pt
      \subfigure[]{\includegraphics[width=0.233\linewidth]{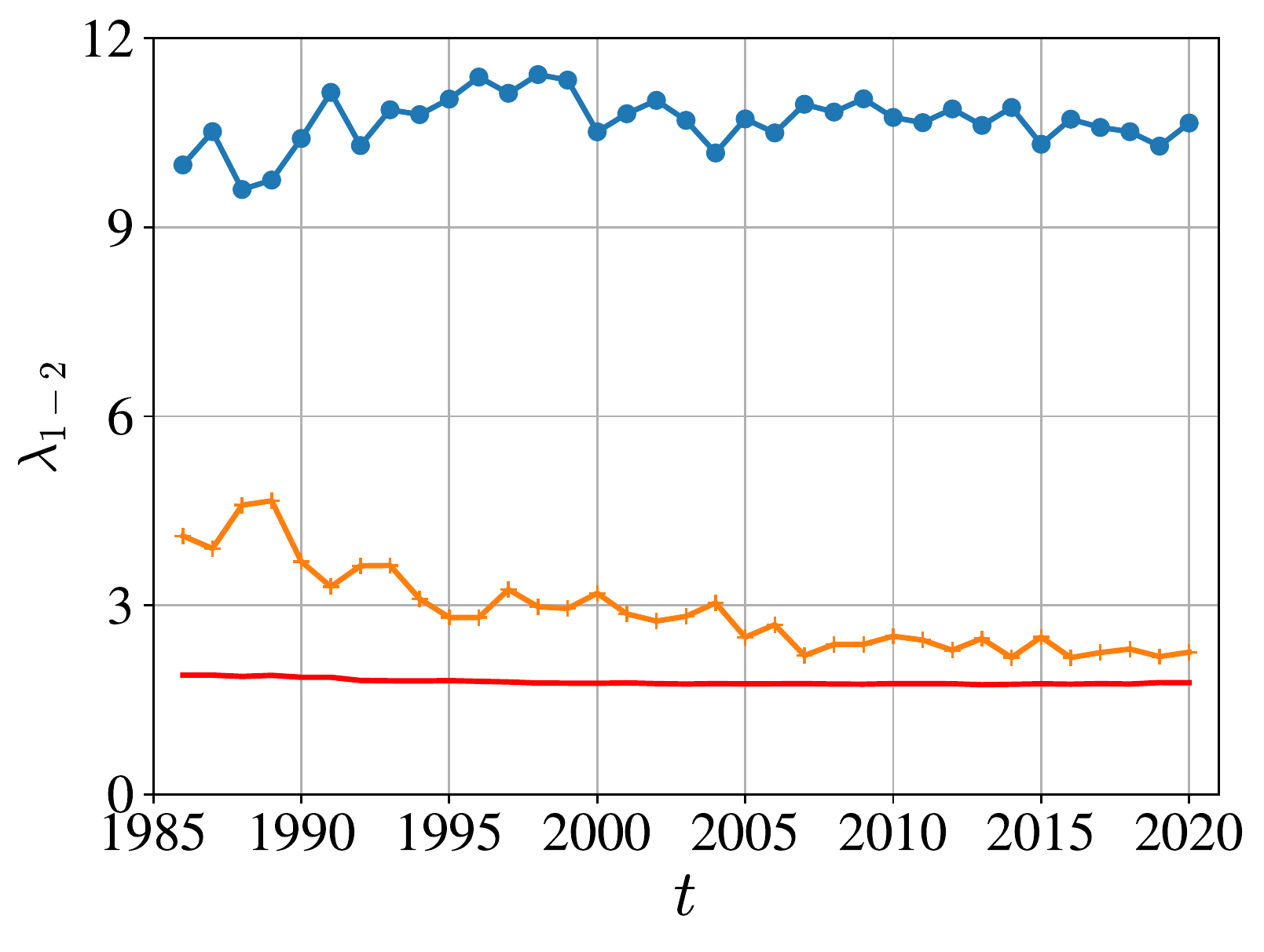}}
      \subfigure[]{\includegraphics[width=0.233\linewidth]{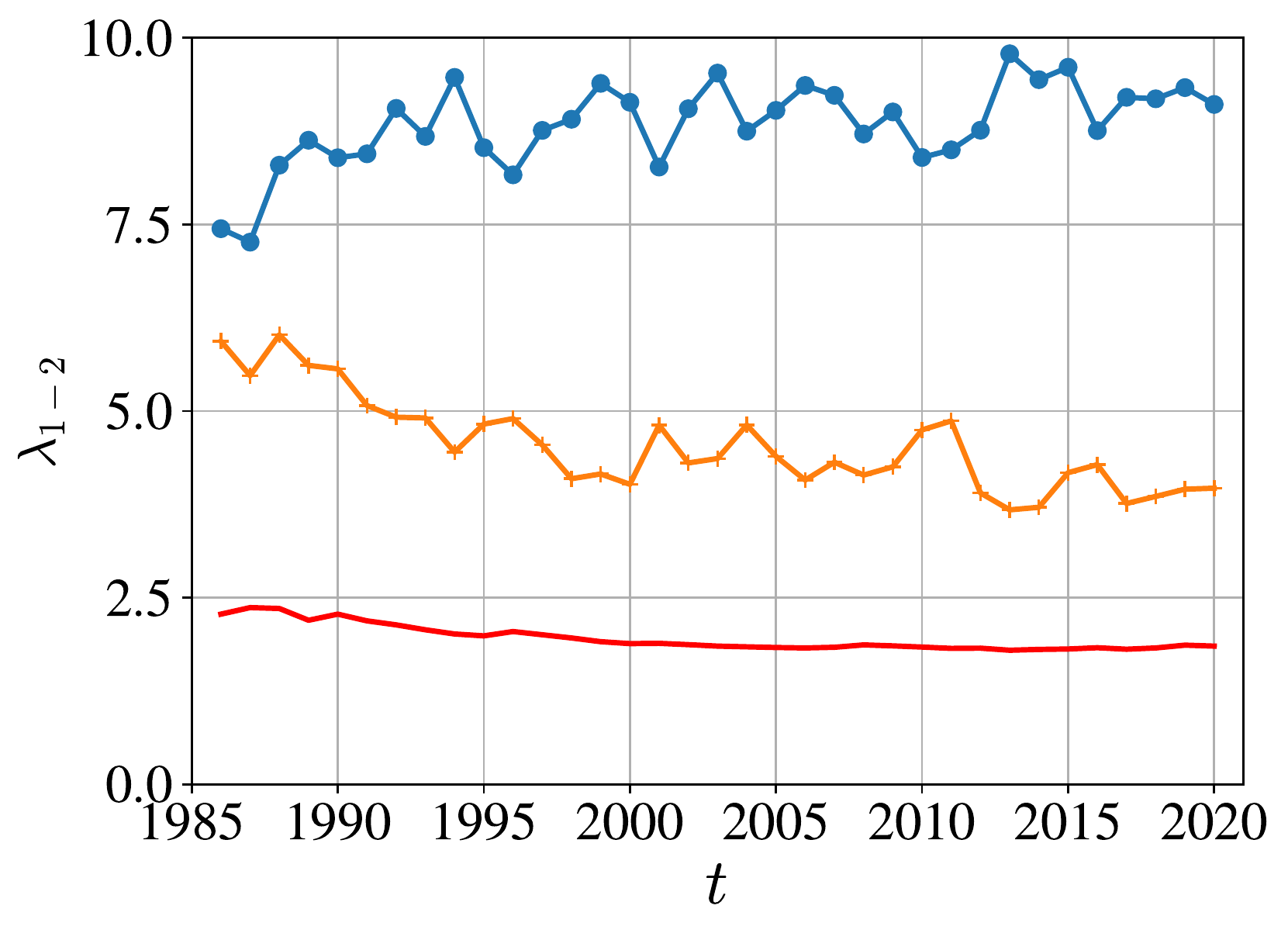}}
      \subfigure[]{\includegraphics[width=0.233\linewidth]{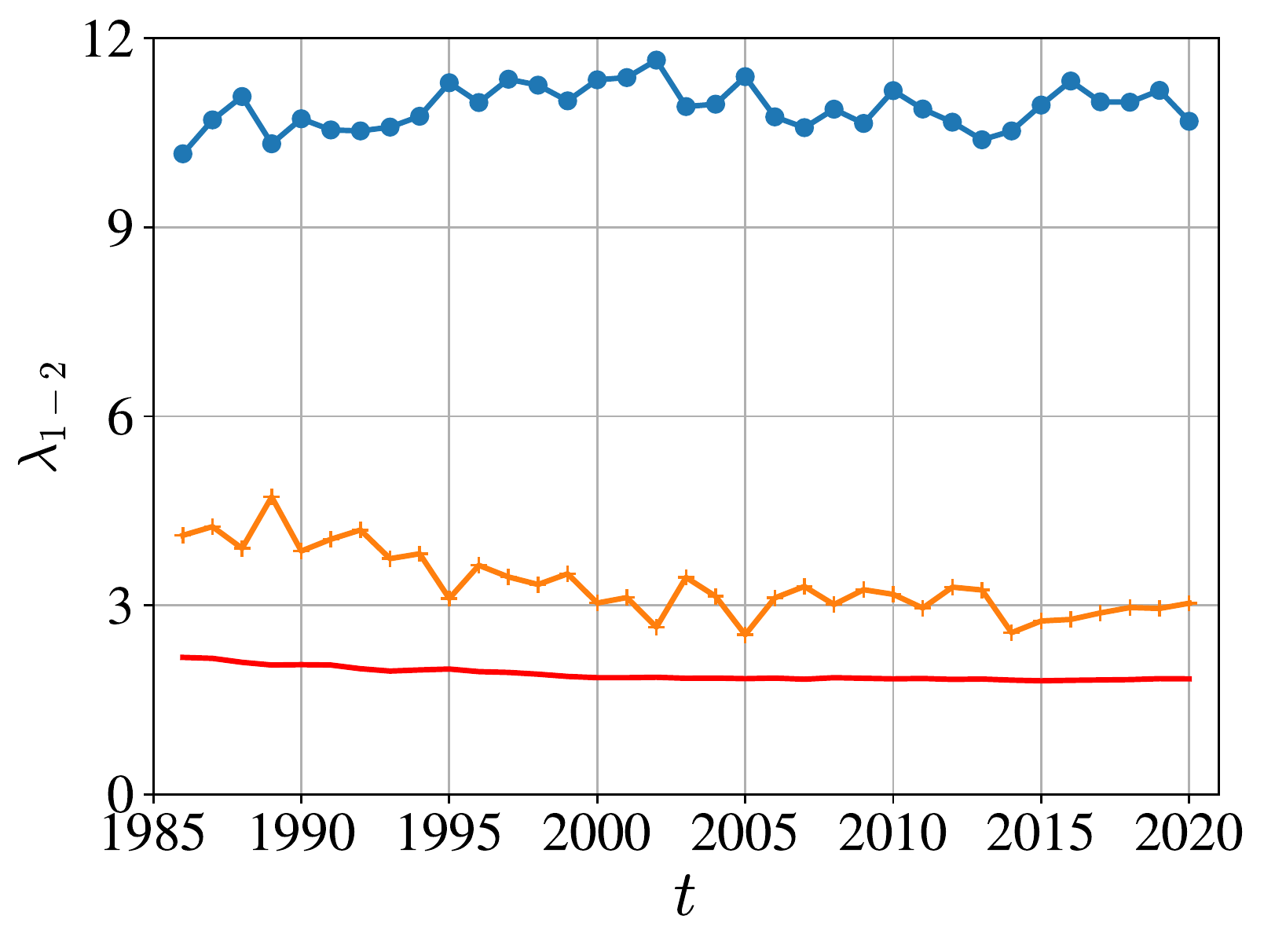}}
      \subfigure[]{\includegraphics[width=0.233\linewidth]{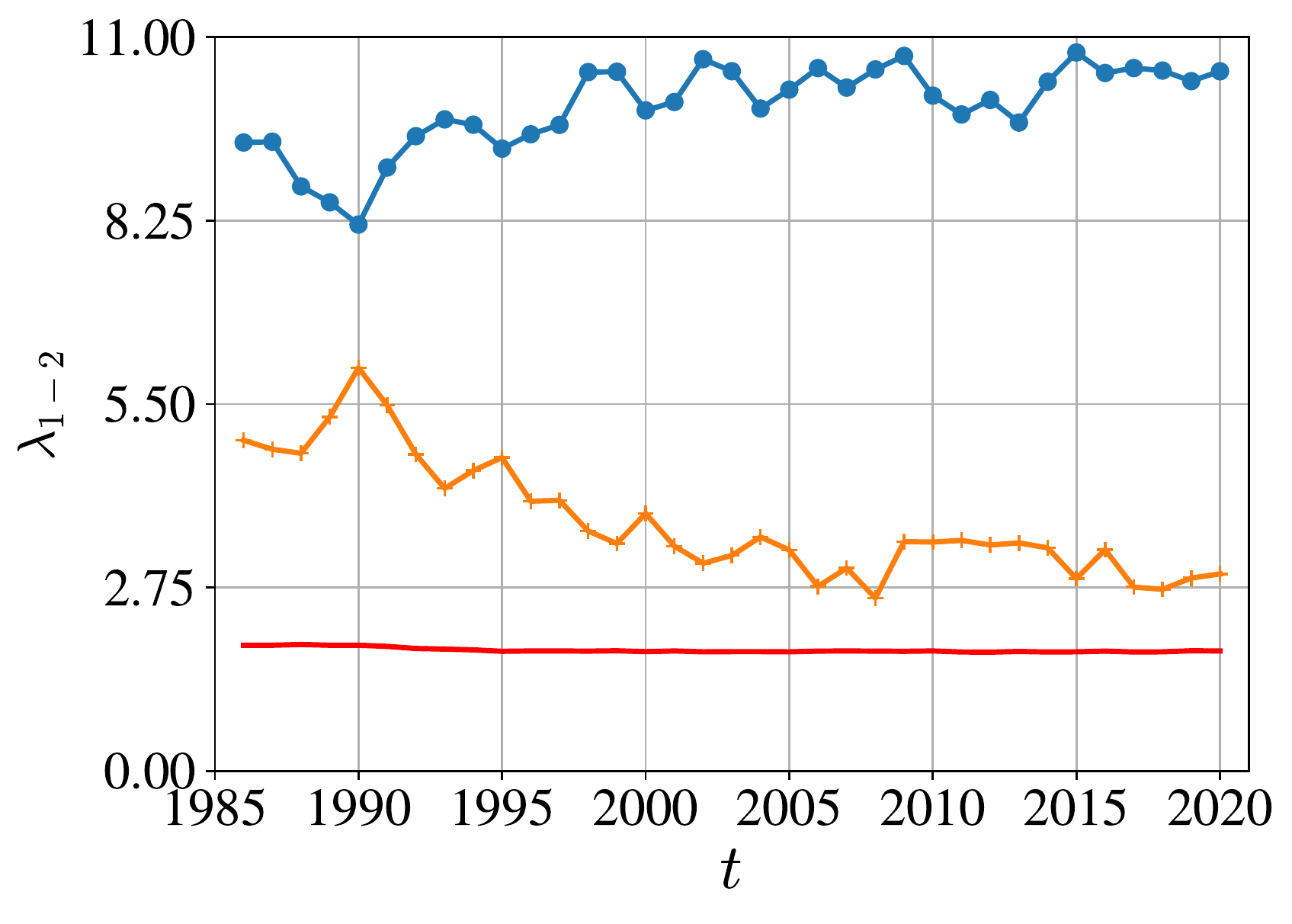}}\\
      \subfigure[]{\includegraphics[width=0.233\linewidth]{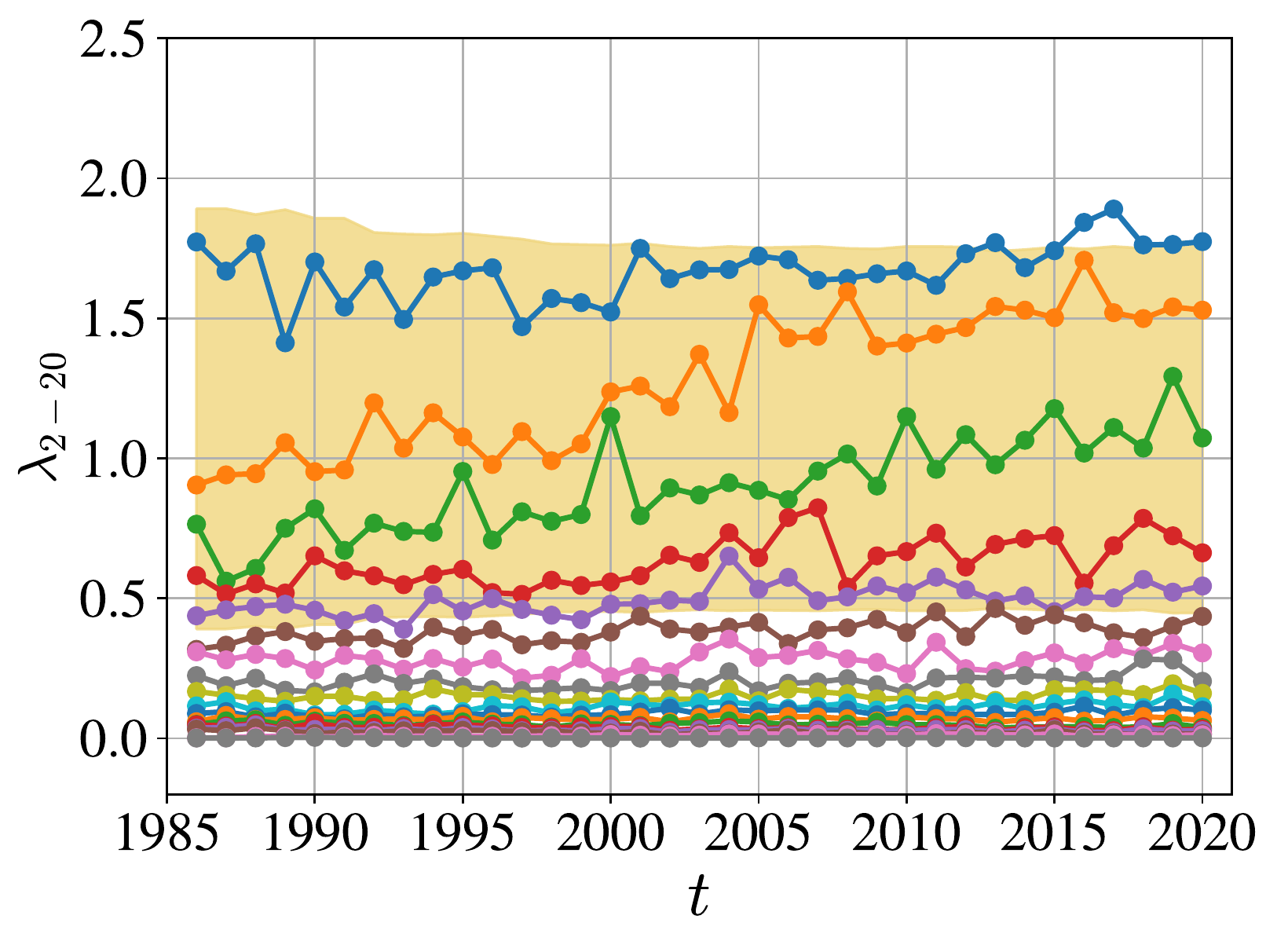}}
      \subfigure[]{\includegraphics[width=0.233\linewidth]{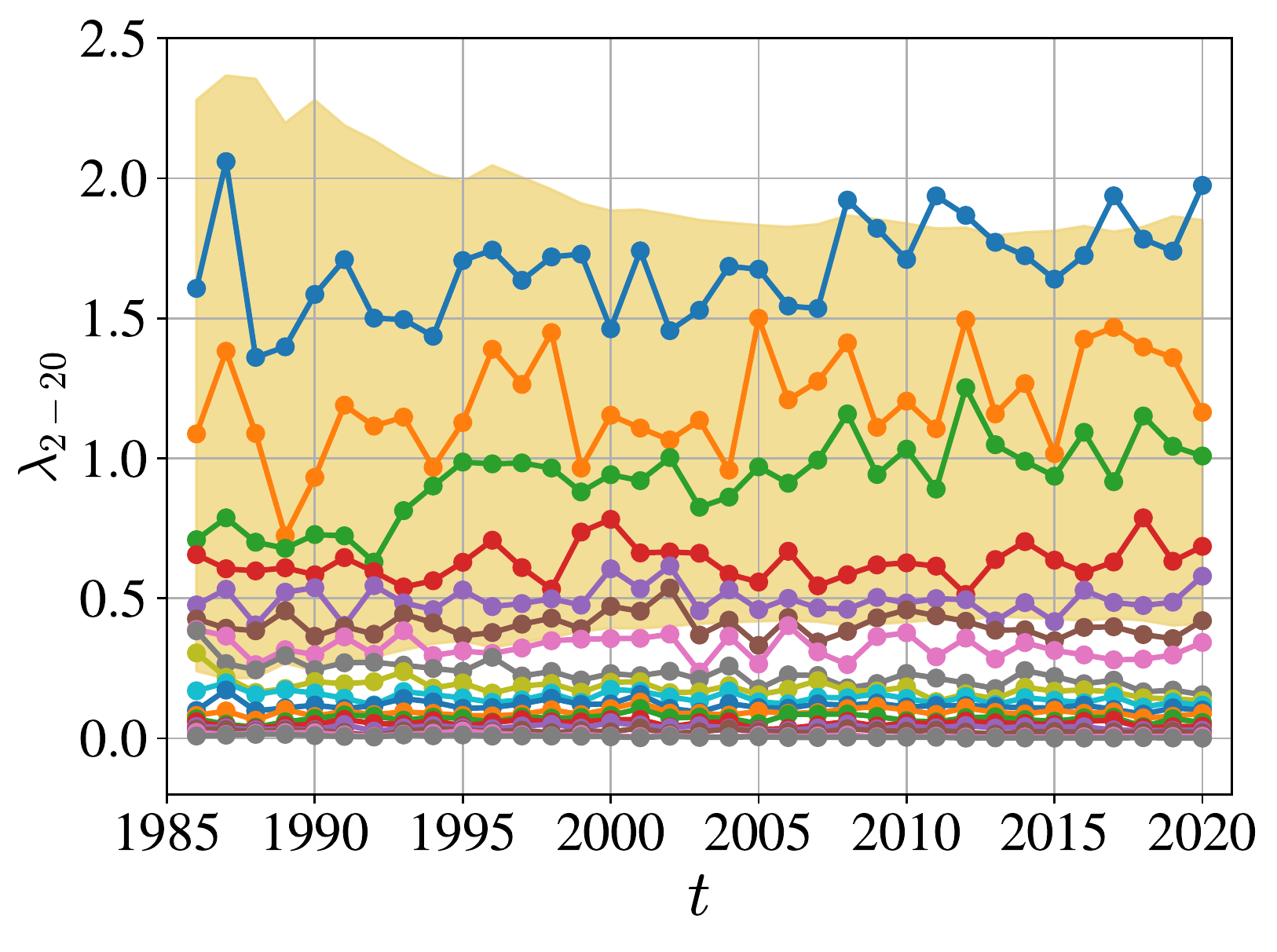}}
      \subfigure[]{\includegraphics[width=0.233\linewidth]{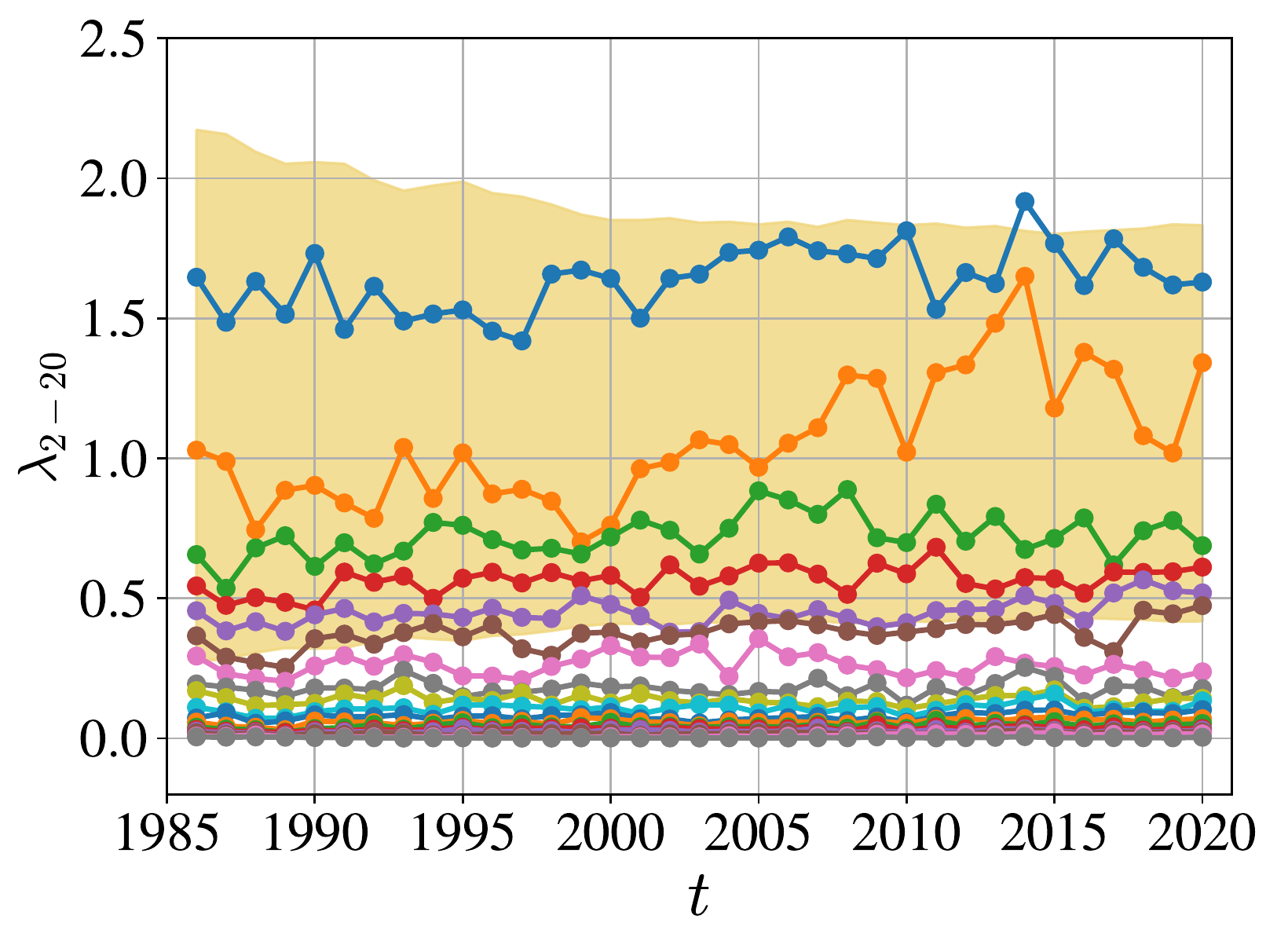}}
      \subfigure[]{\includegraphics[width=0.233\linewidth]{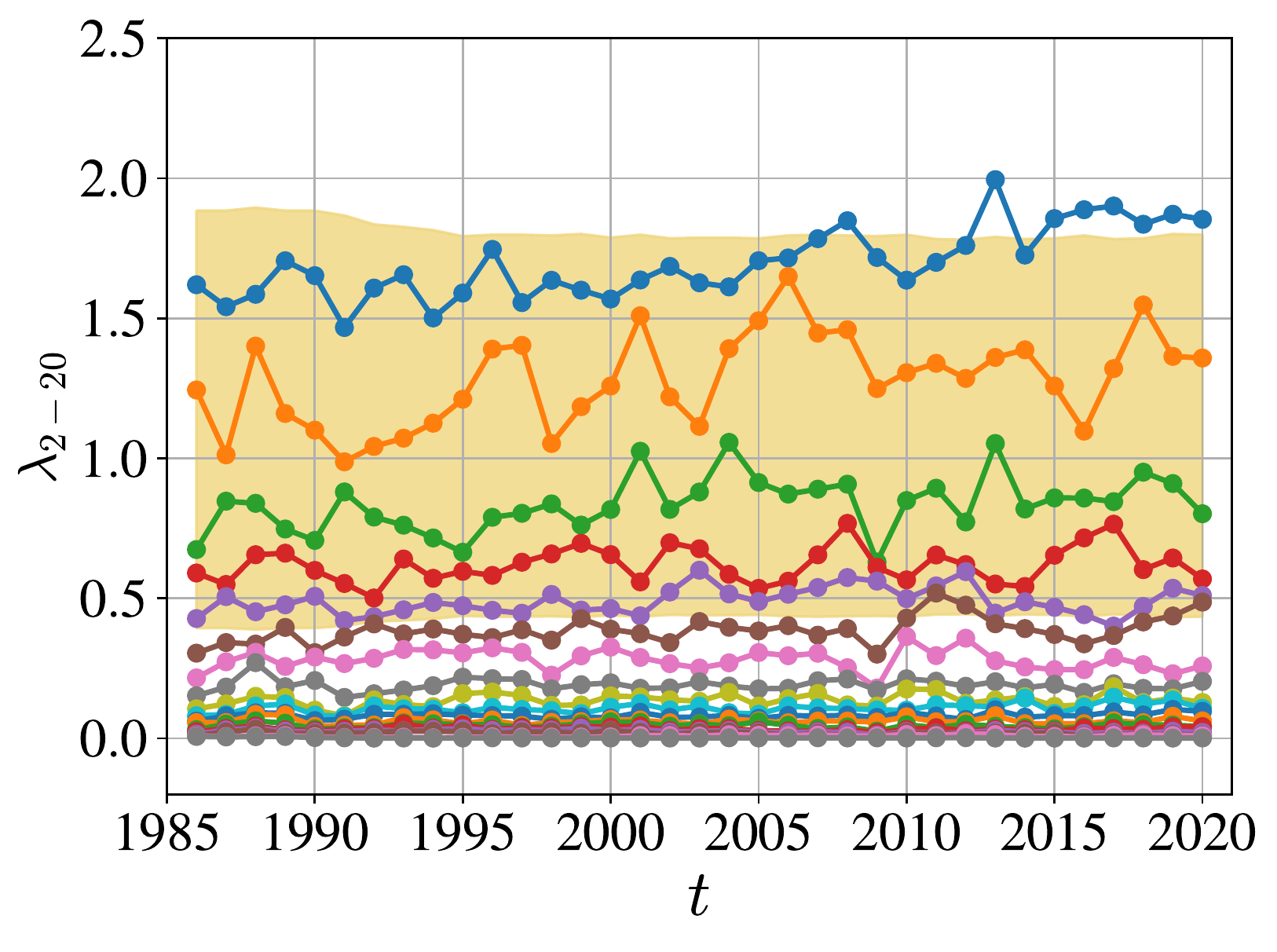}}
      \caption{Evolution of 20 eigenvalues corresponding to different node importance indexes the correlation matrix $\mathrm{C}$ from 1986 to 2020. Different colors correspond to different indicators. (a, c, e, g) are the evolution of the first and second eigenvalues, and the red line describes the largest eigenvalue $\lambda_{\max }^{\mathrm{RMT}}$ the random correlation matrix $R$. (b, d, f, h) are the evolution of the other eigenvalues. The light gold band describes the largest eigenvalue $\lambda_{\max }^{\mathrm{RMT}}$ and the smallest eigenvalue $\lambda_{\min }^{\mathrm{RMT}}$ of the random correlation matrix $R$.}
      \label{Fig:iCTN:eigenvalue:t}
\end{figure}

 \begin{figure}[h!]
      \centering
      \includegraphics[width=0.233\linewidth]{Fig_iCTN_u1component_B_Maize_2020.pdf}
      \includegraphics[width=0.233\linewidth]{Fig_iCTN_u1component_B_Rice_2020.pdf}
      \includegraphics[width=0.233\linewidth]{Fig_iCTN_u1component_B_Soybean_2020.pdf}
      \includegraphics[width=0.233\linewidth]{Fig_iCTN_u1component_B_Wheat_2020.pdf}
       \\
      \includegraphics[width=0.233\linewidth]{Fig_iCTN_u2component_B_Maize_2020.pdf}
      \includegraphics[width=0.233\linewidth]{Fig_iCTN_u2component_B_Rice_2020.pdf}
      \includegraphics[width=0.233\linewidth]{Fig_iCTN_u2component_B_Soybean_2020.pdf}
      \includegraphics[width=0.233\linewidth]{Fig_iCTN_u2component_B_Wheat_2020.pdf}
       \\
      \includegraphics[width=0.233\linewidth]{Fig_iCTN_u3component_B_Maize_2020.pdf}
      \includegraphics[width=0.233\linewidth]{Fig_iCTN_u3component_B_Rice_2020.pdf}
      \includegraphics[width=0.233\linewidth]{Fig_iCTN_u3component_B_Soybean_2020.pdf}
      \includegraphics[width=0.233\linewidth]{Fig_iCTN_u3component_B_Wheat_2020.pdf}
       \\
      \includegraphics[width=0.233\linewidth]{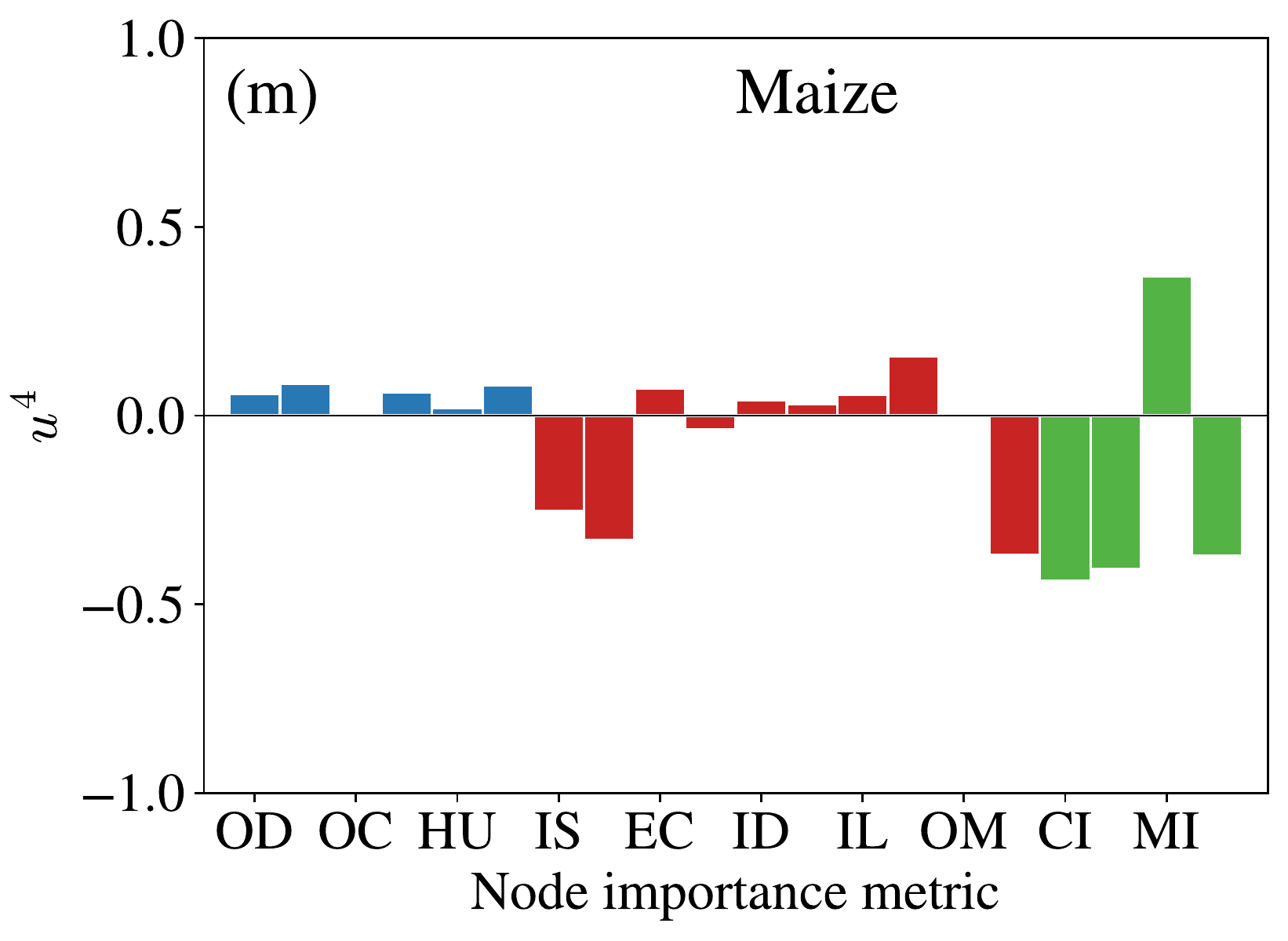}
      \includegraphics[width=0.233\linewidth]{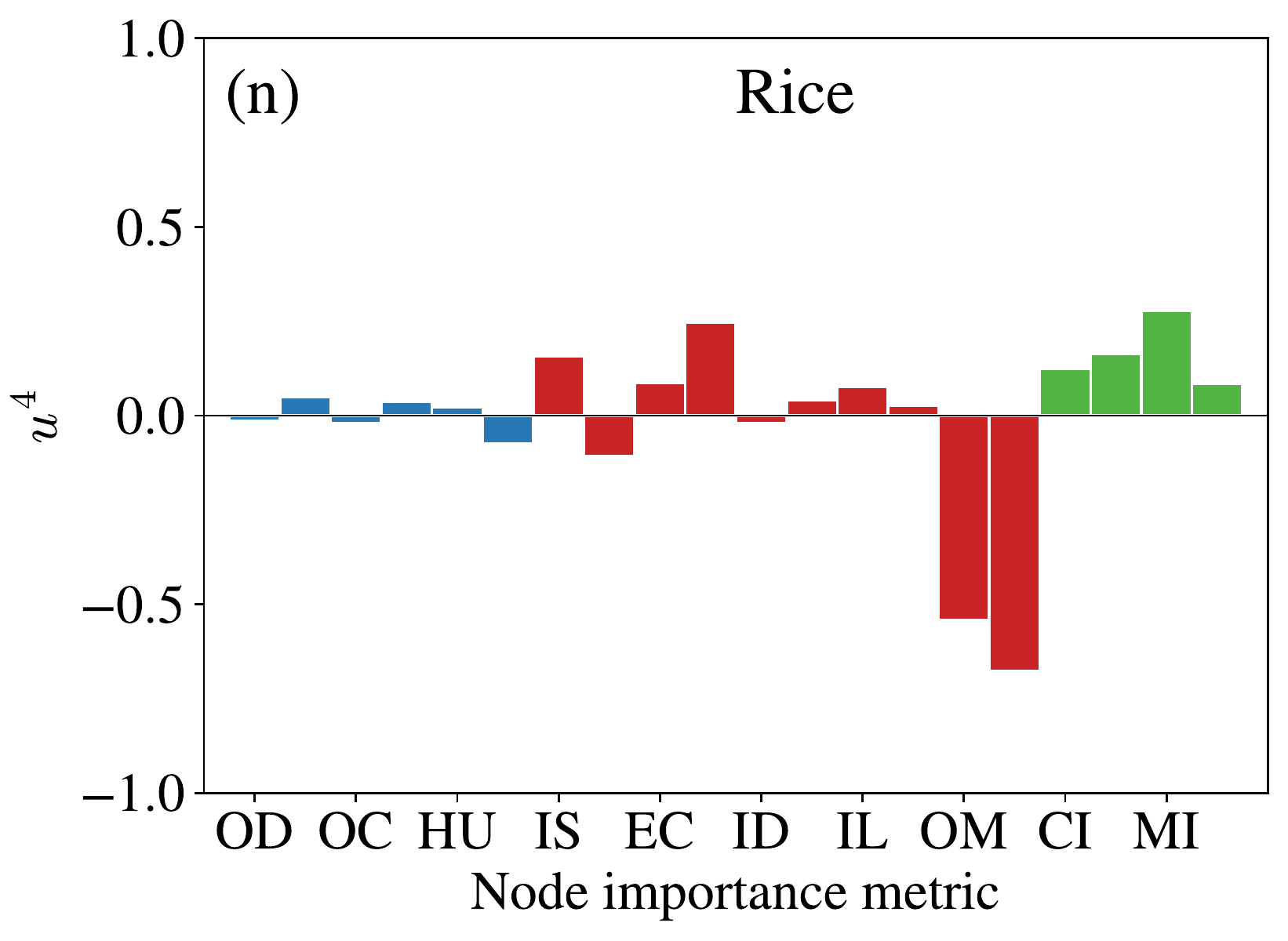}
      \includegraphics[width=0.233\linewidth]{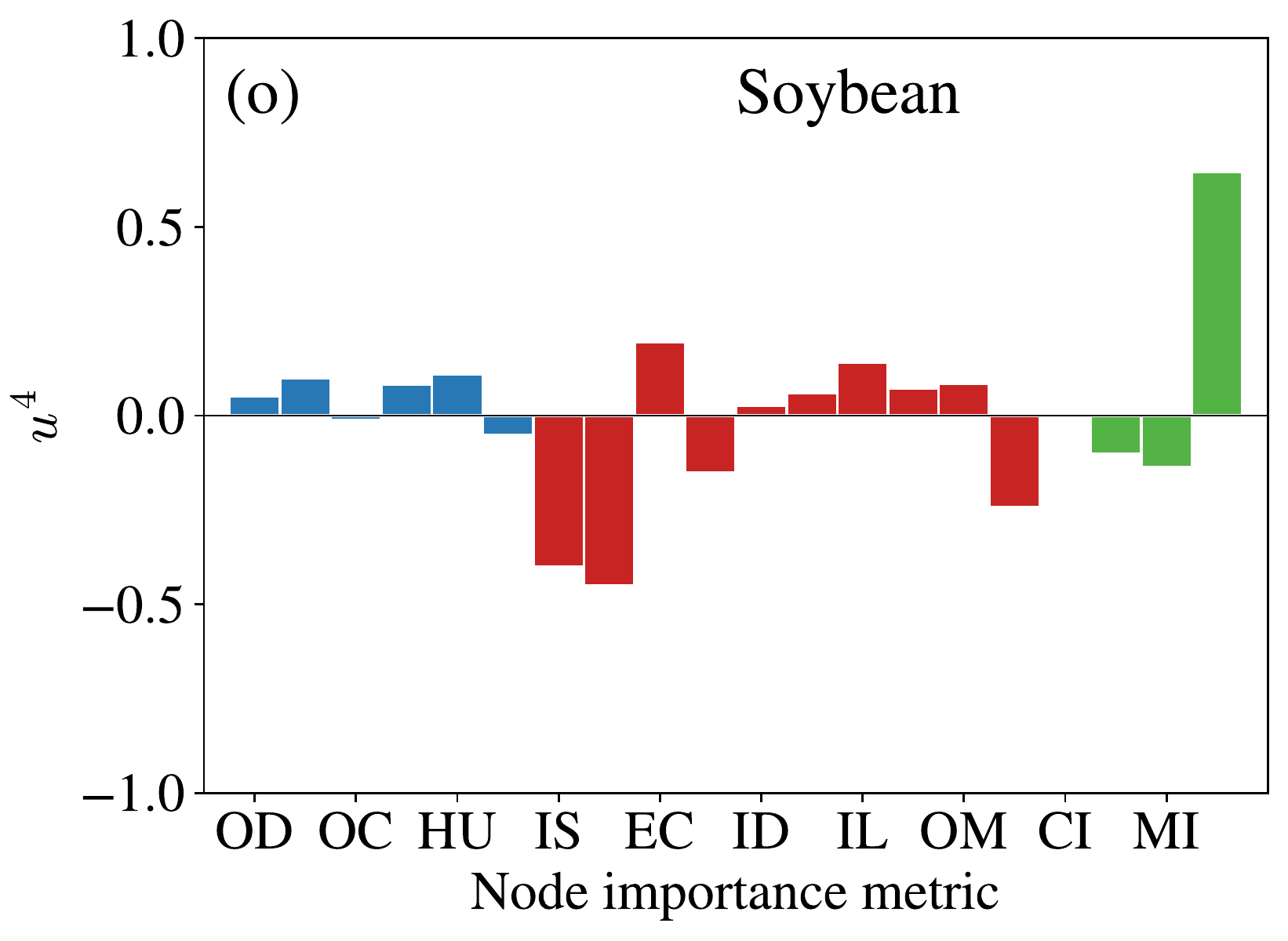}
      \includegraphics[width=0.233\linewidth]{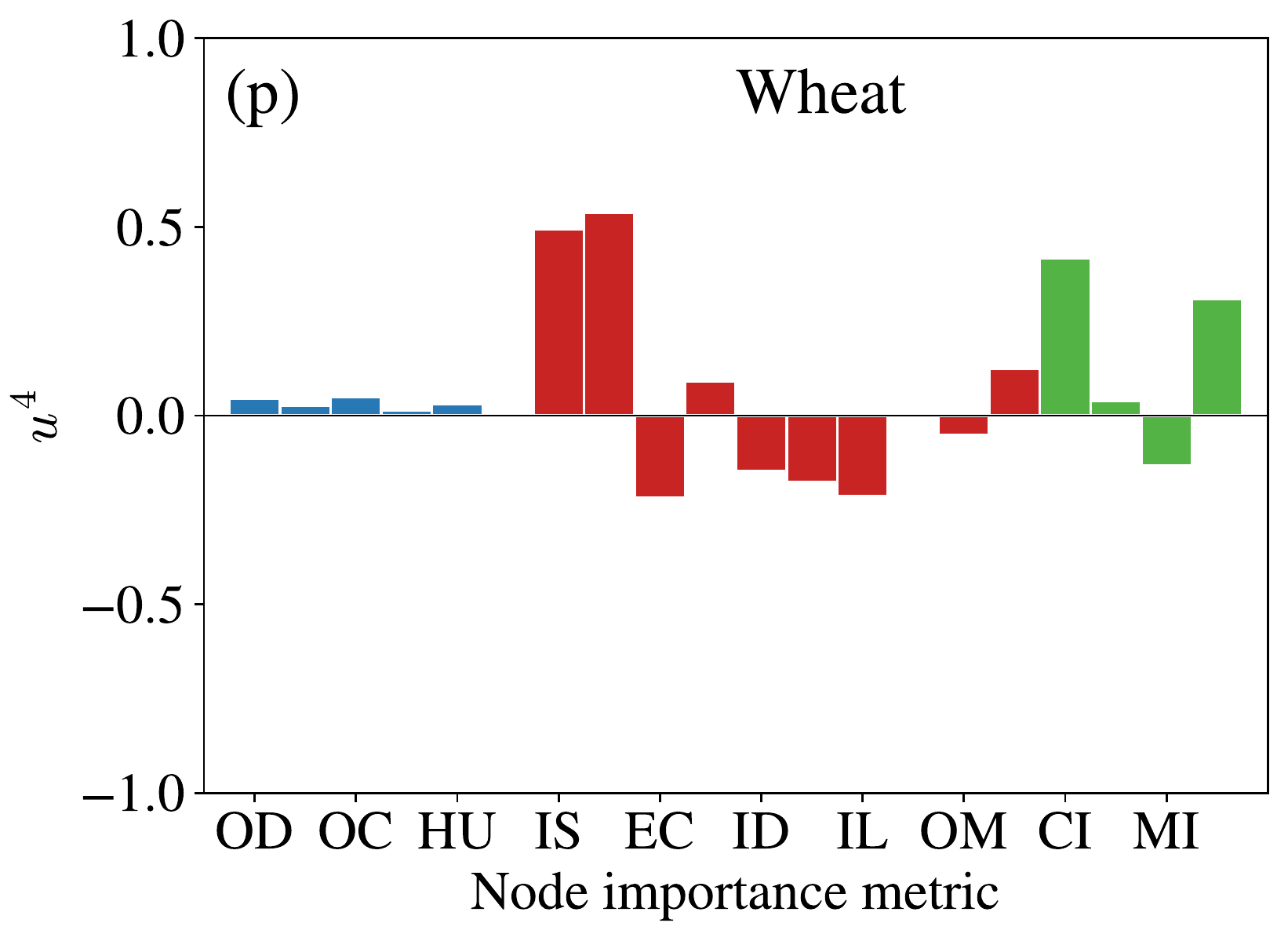}
       \\
      \includegraphics[width=0.233\linewidth]{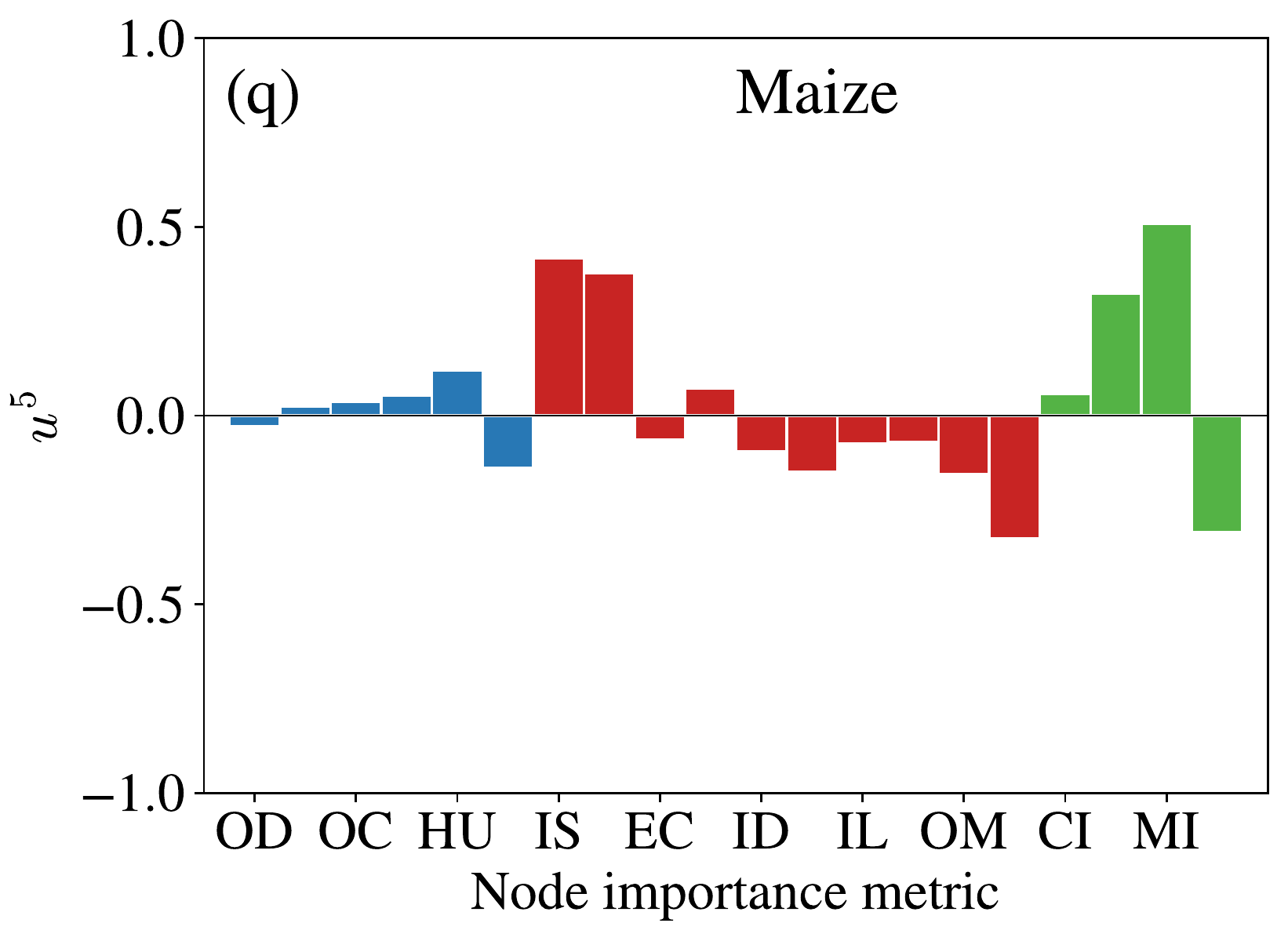}
      \includegraphics[width=0.233\linewidth]{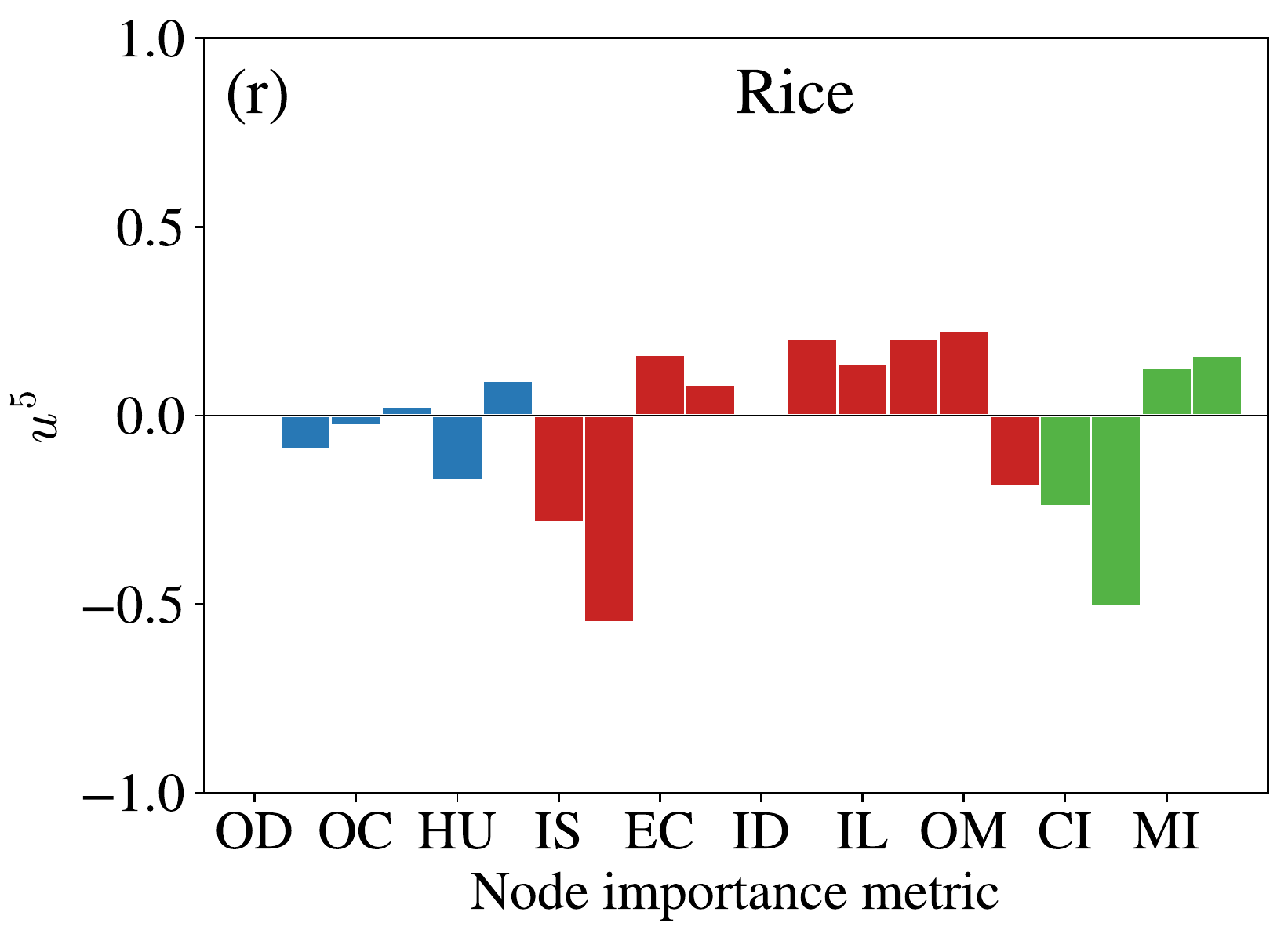}
      \includegraphics[width=0.233\linewidth]{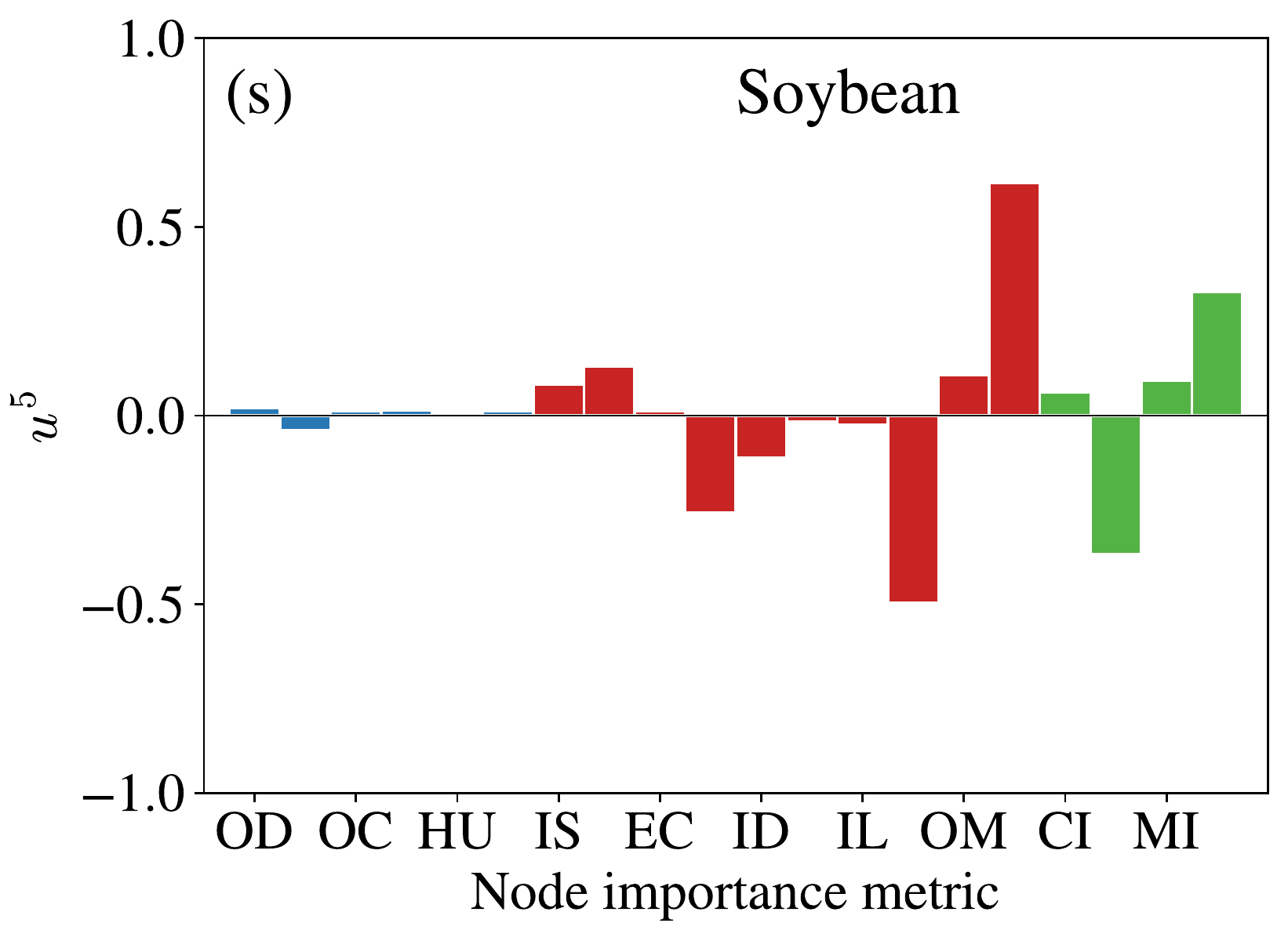}
      \includegraphics[width=0.233\linewidth]{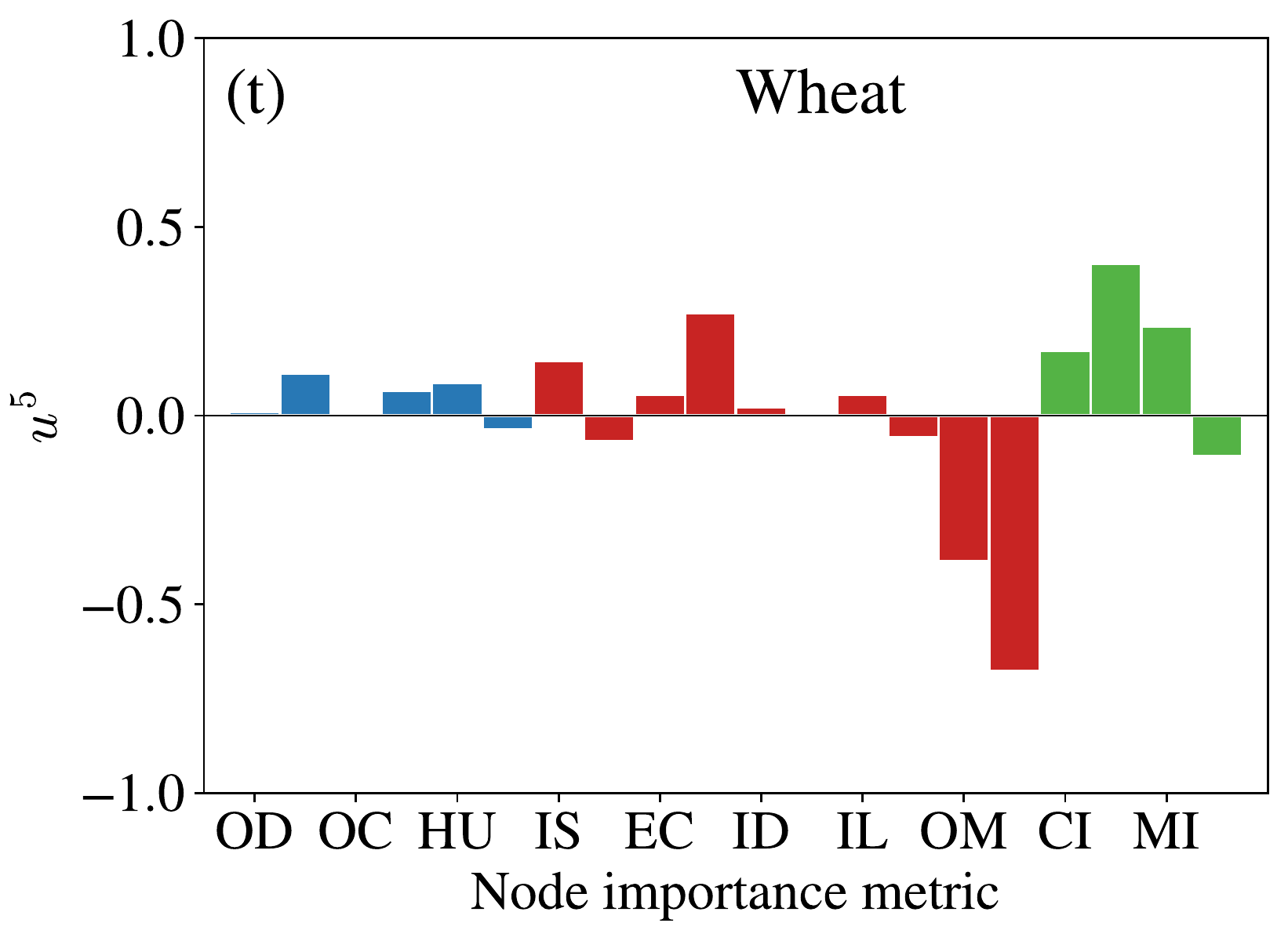}
      \caption{Components of the eigenvector $u_1-u_5$ the top five largest  eigenvalues $\lambda_1-\lambda_5$ given by Eq.~(\ref{Eq:RMT:PDF:eigenvalue}) of random matrix theory (RMT) in 2020.}
      \label{Fig:iCTN:PDF:eigenvalue:1-5:2020}
\end{figure}

 \begin{figure}[h!]
      \centering
      \includegraphics[width=0.233\linewidth]{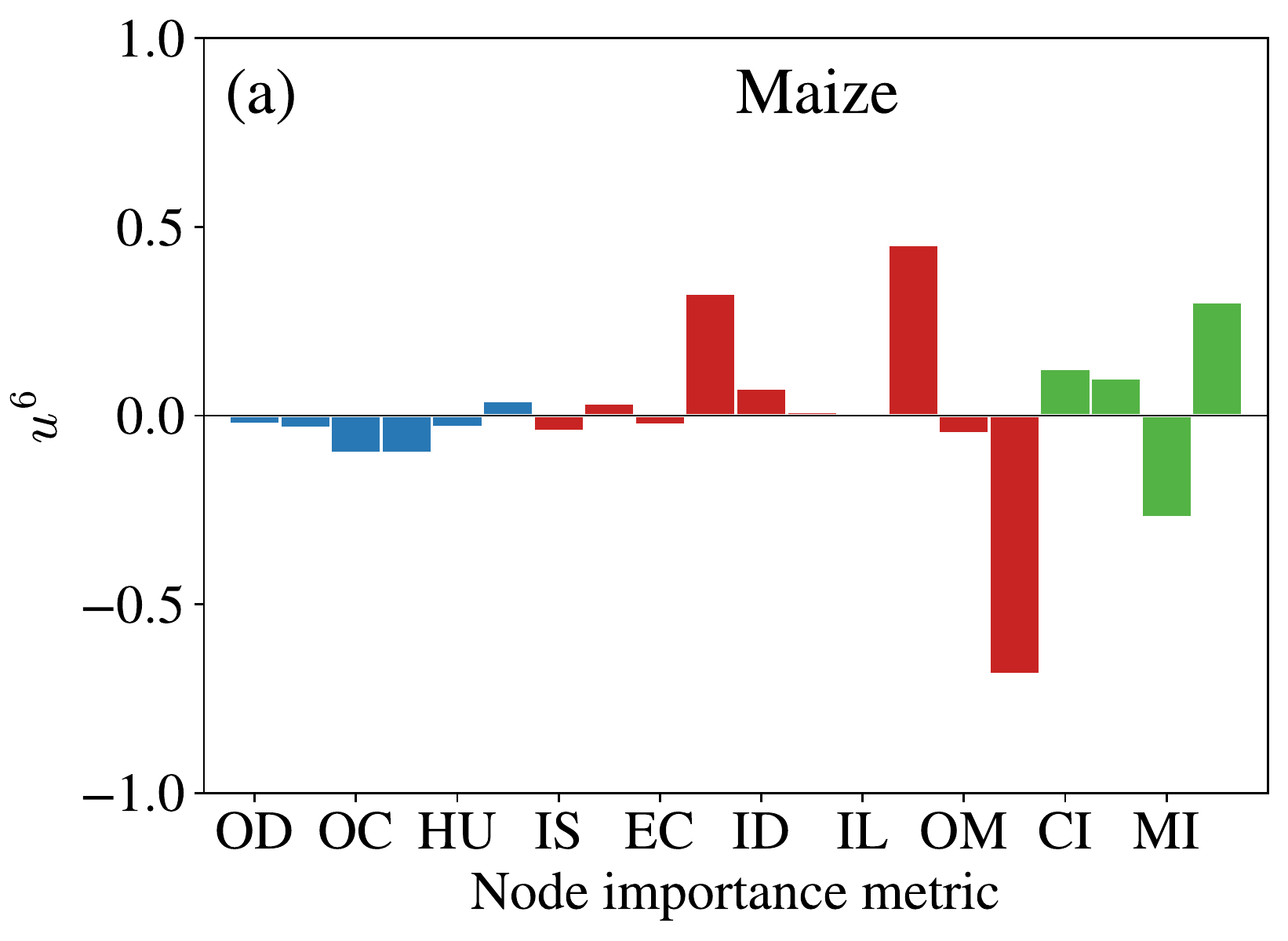}
      \includegraphics[width=0.233\linewidth]{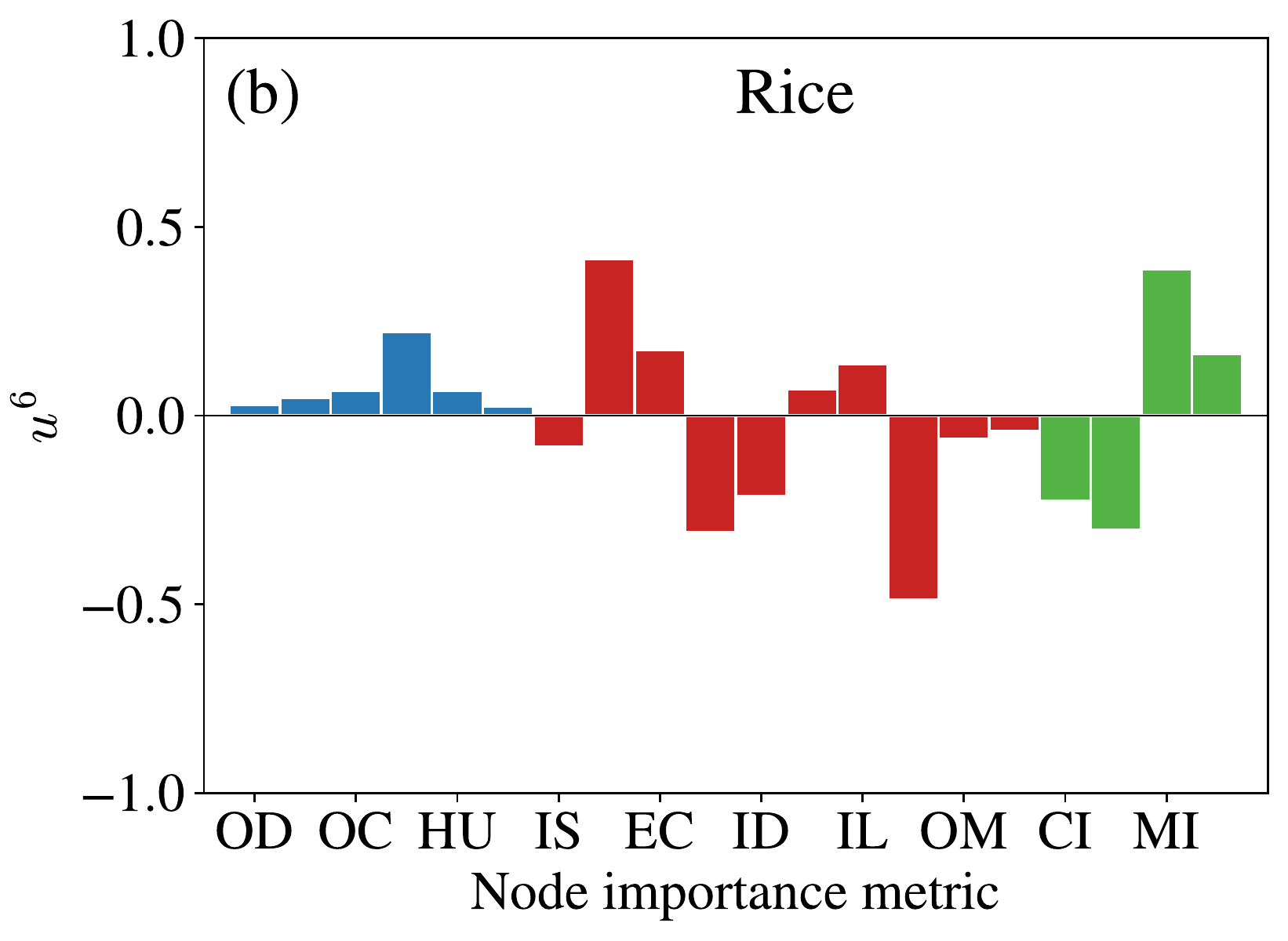}
      \includegraphics[width=0.233\linewidth]{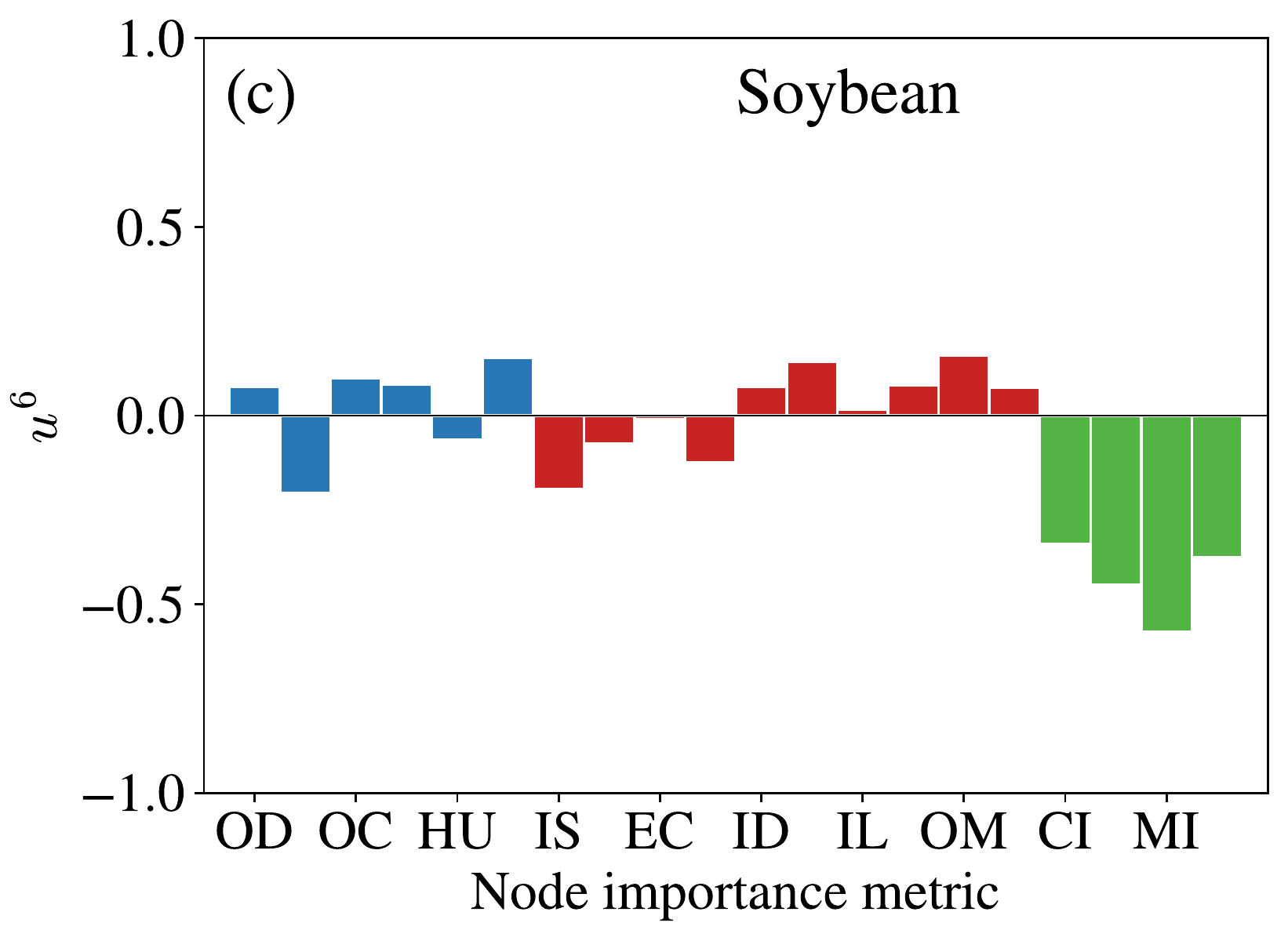}
      \includegraphics[width=0.233\linewidth]{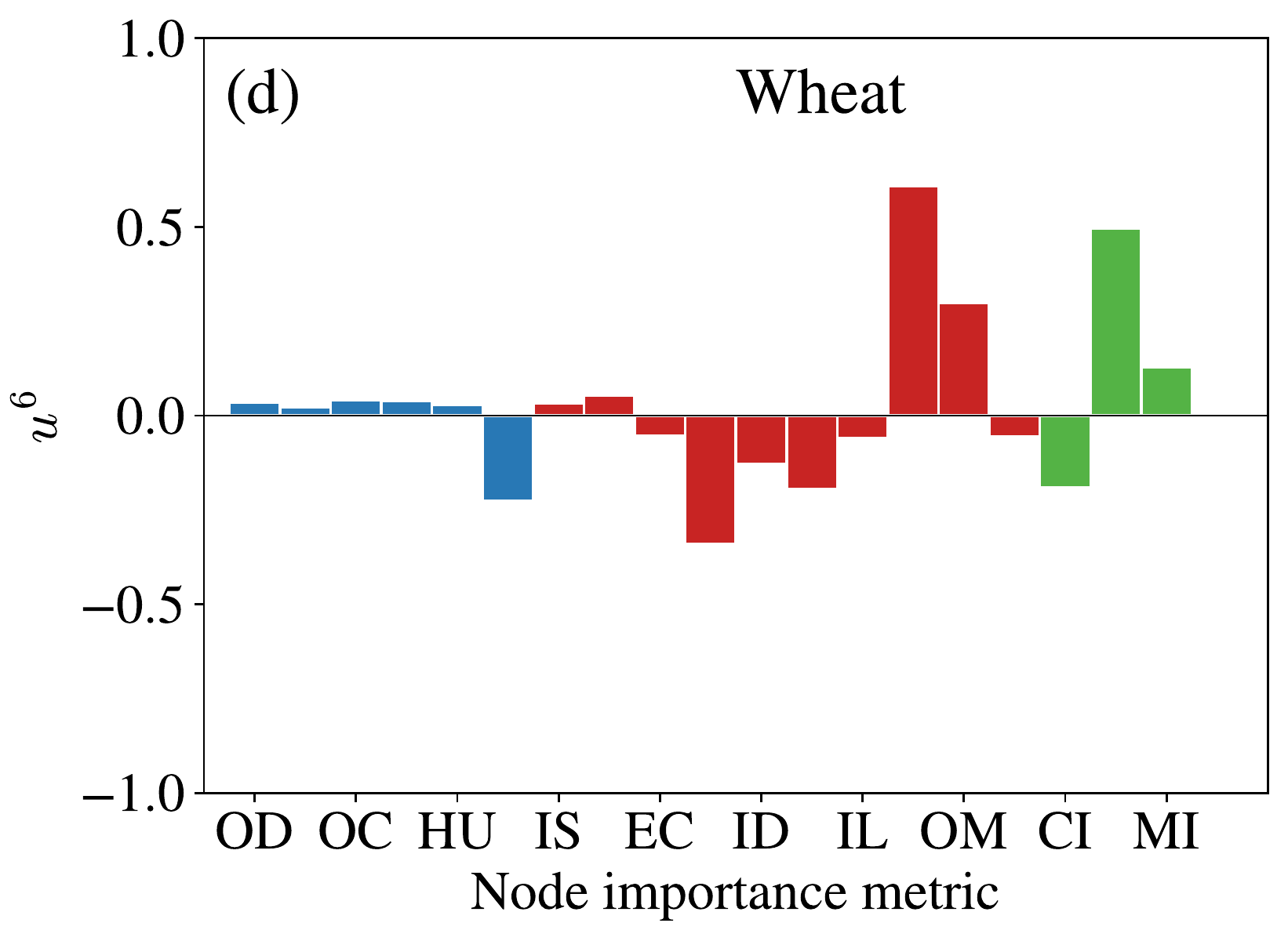}
      \includegraphics[width=0.233\linewidth]{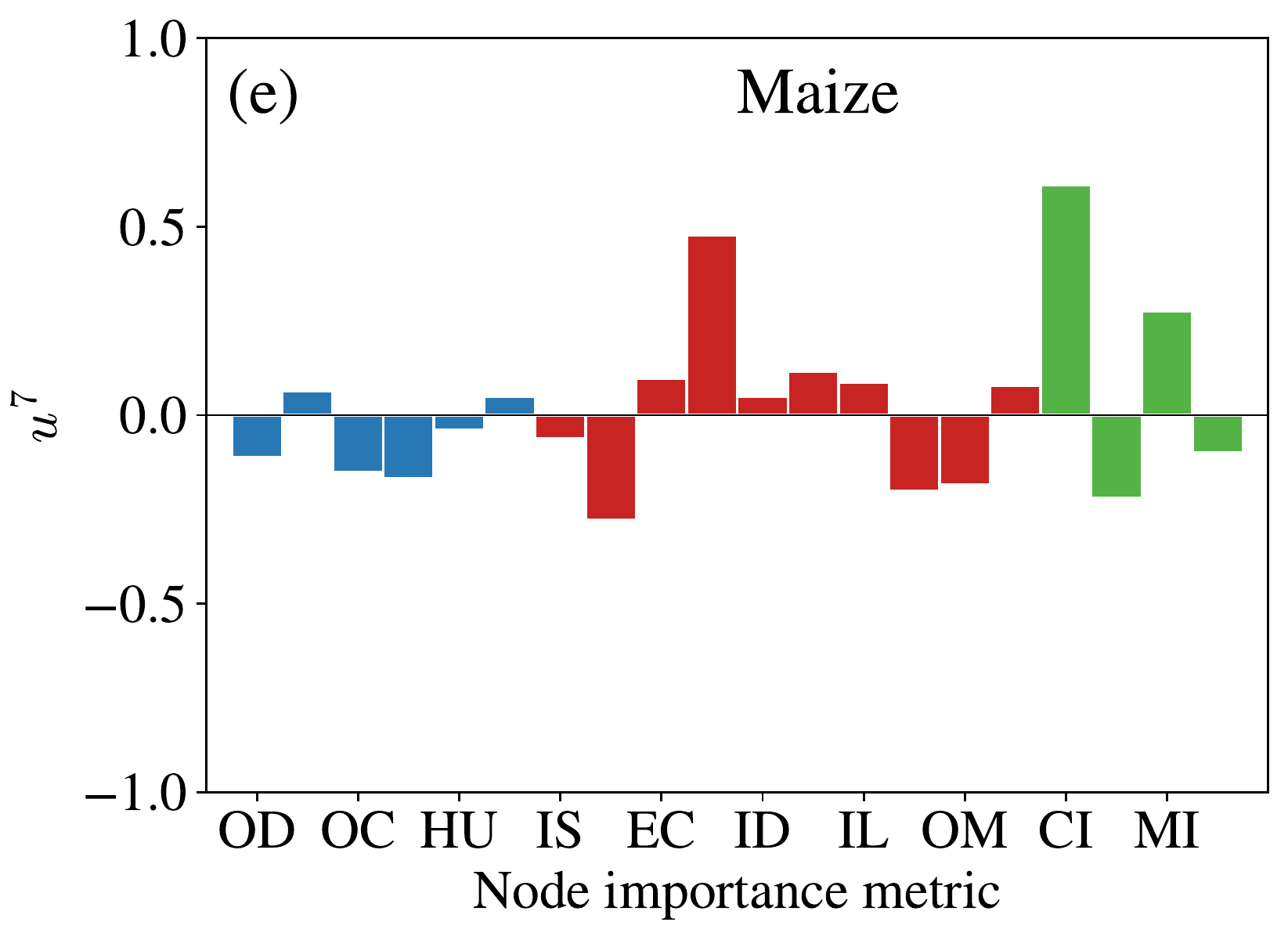}
      \includegraphics[width=0.233\linewidth]{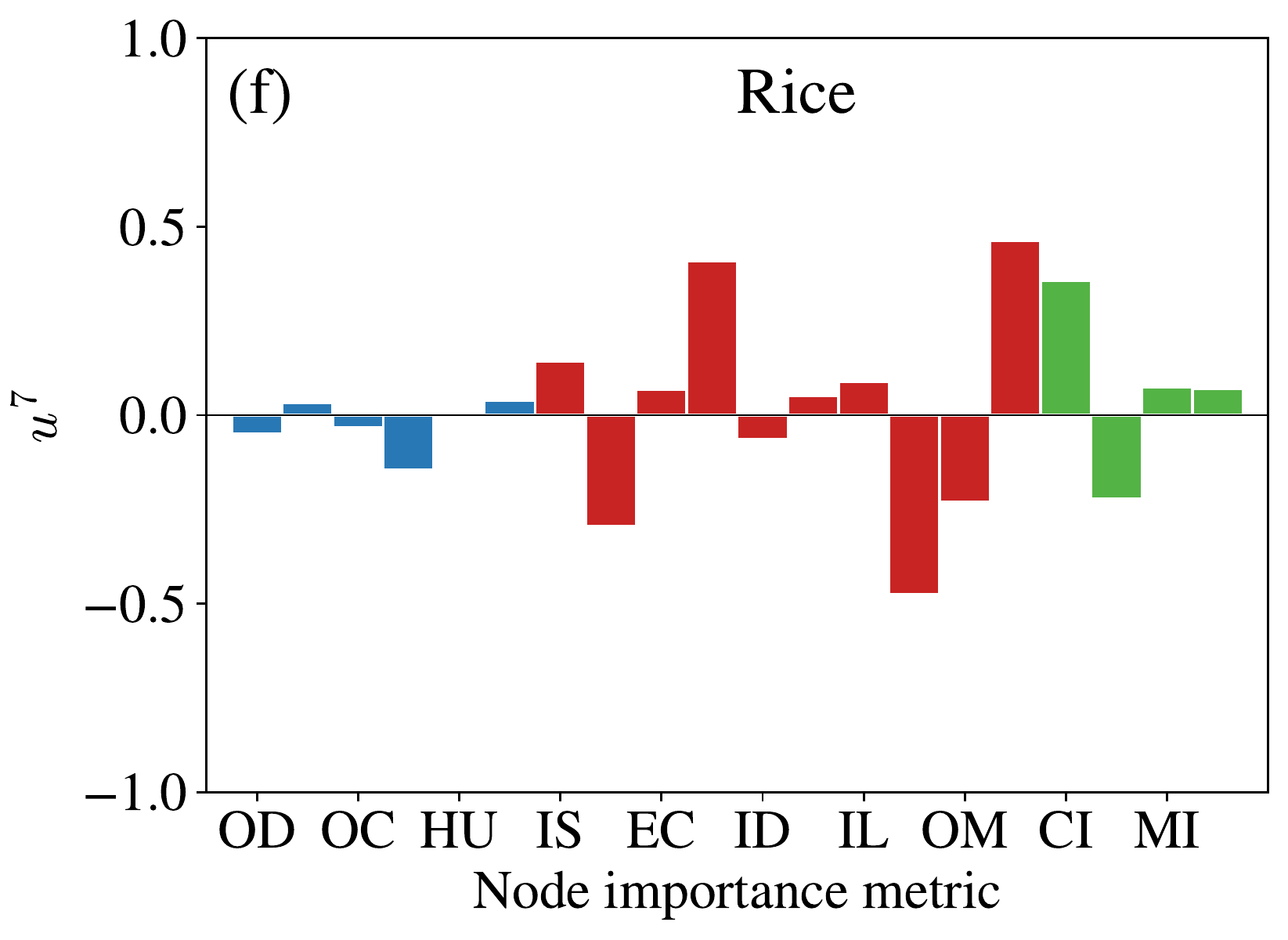}
      \includegraphics[width=0.233\linewidth]{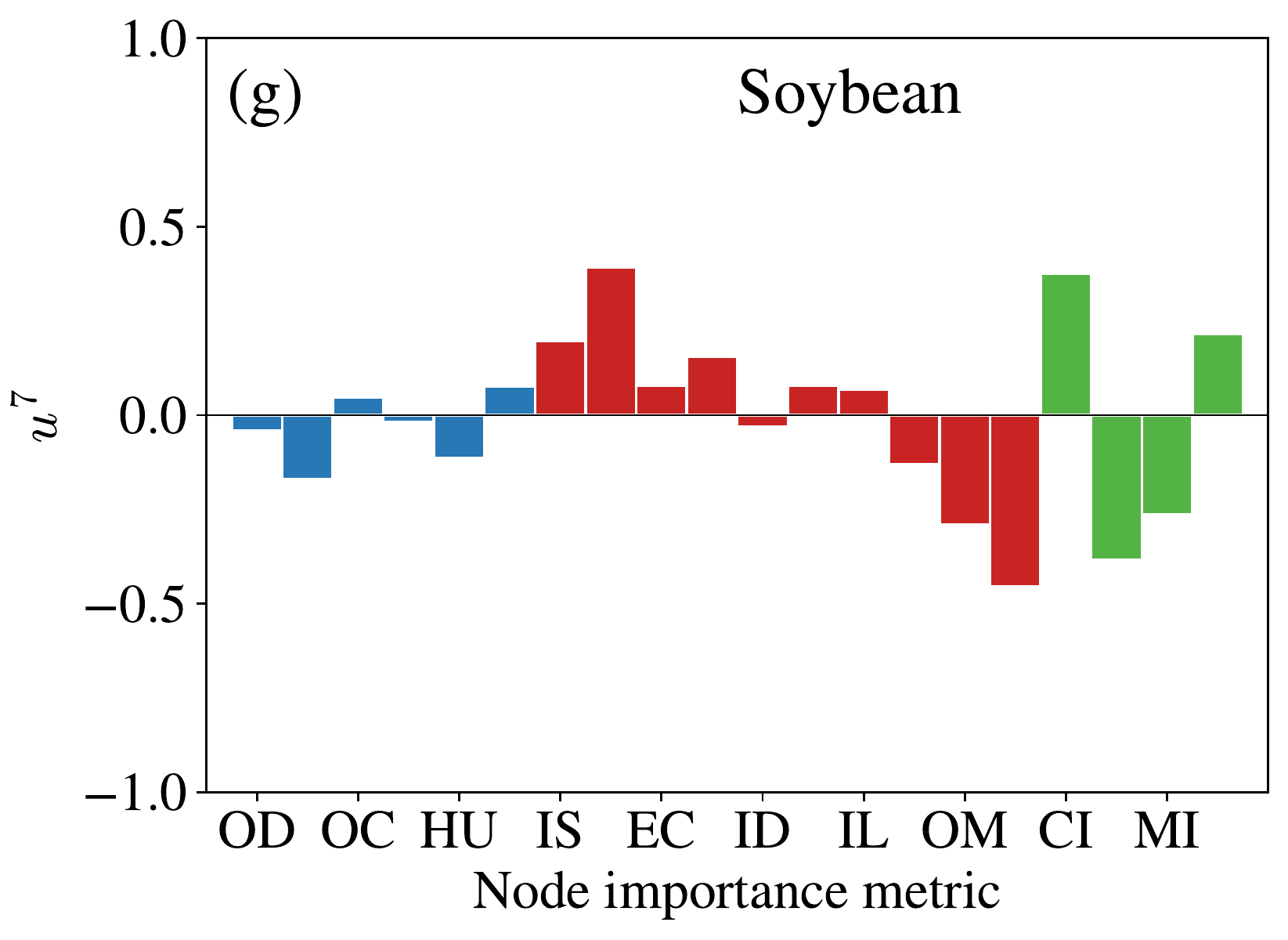}
      \includegraphics[width=0.233\linewidth]{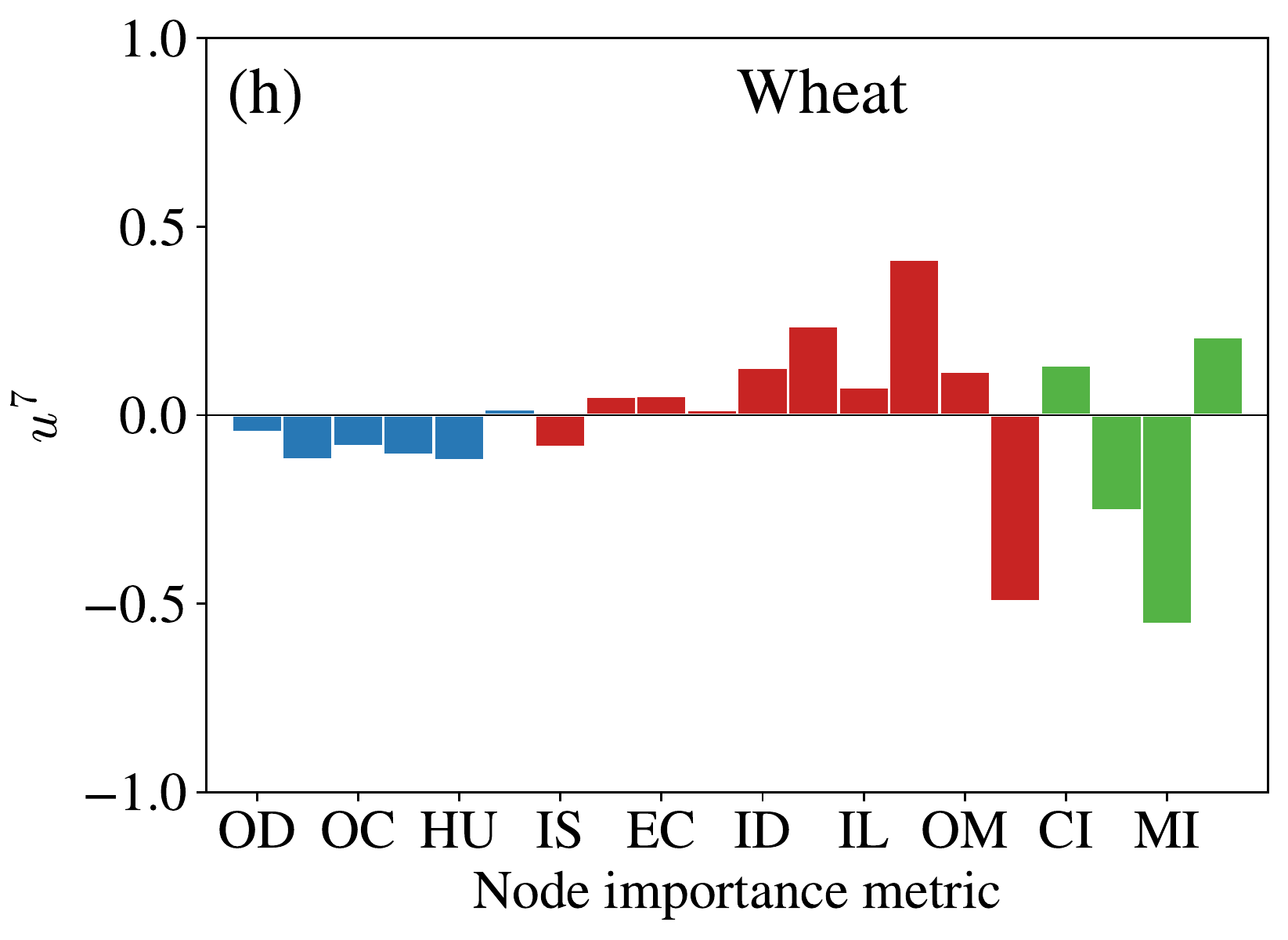}
      \includegraphics[width=0.233\linewidth]{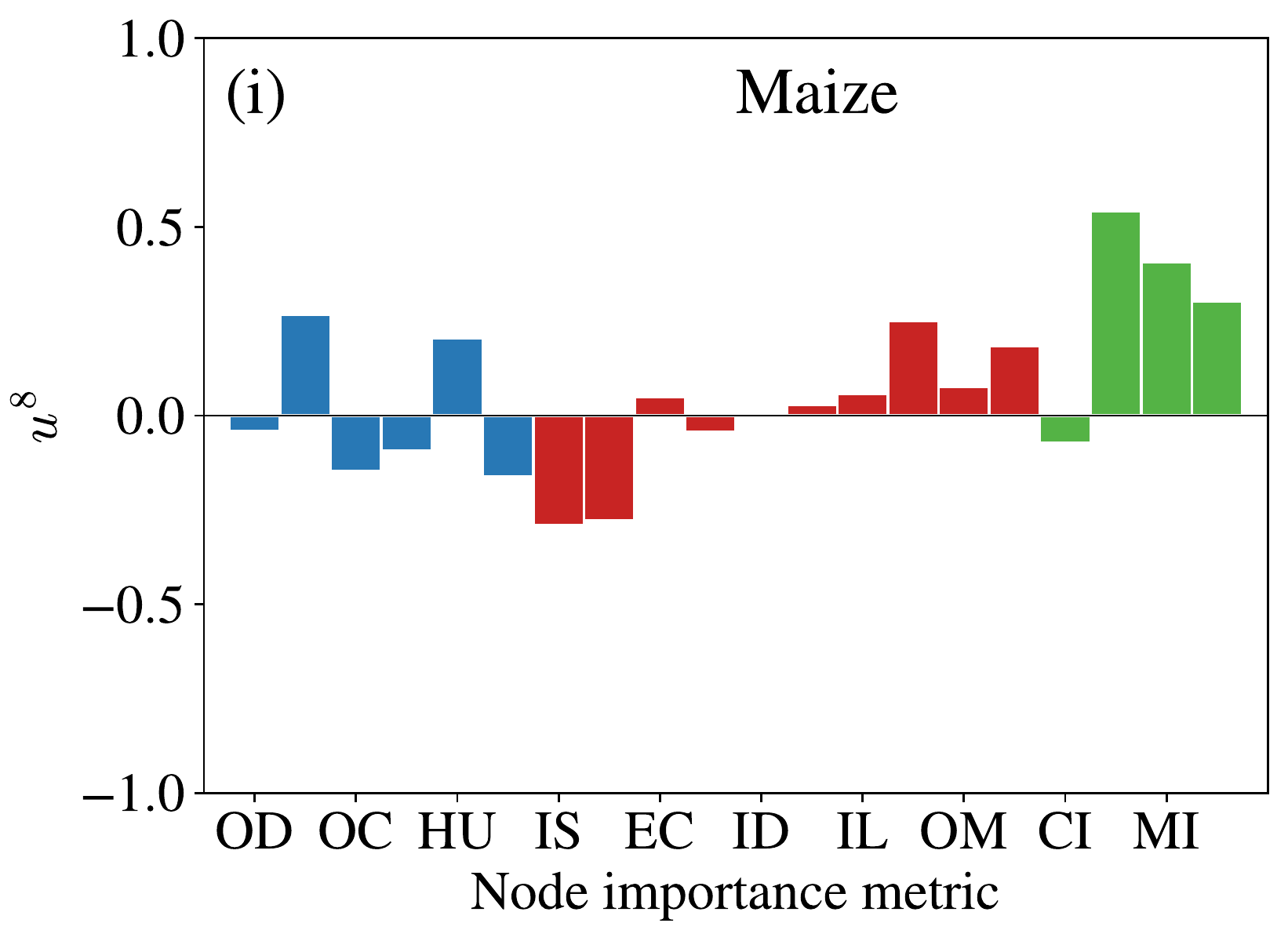}
      \includegraphics[width=0.233\linewidth]{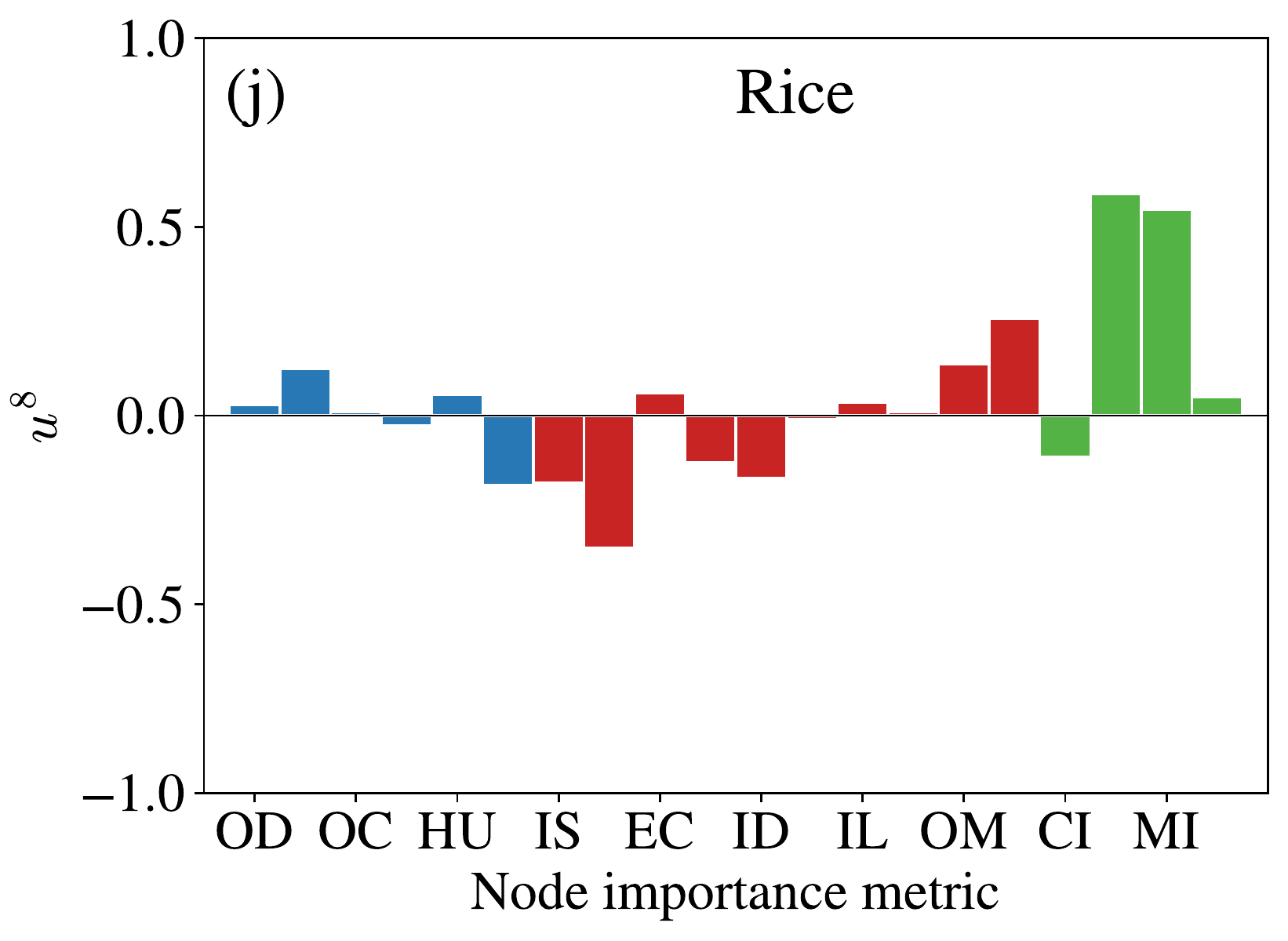}
      \includegraphics[width=0.233\linewidth]{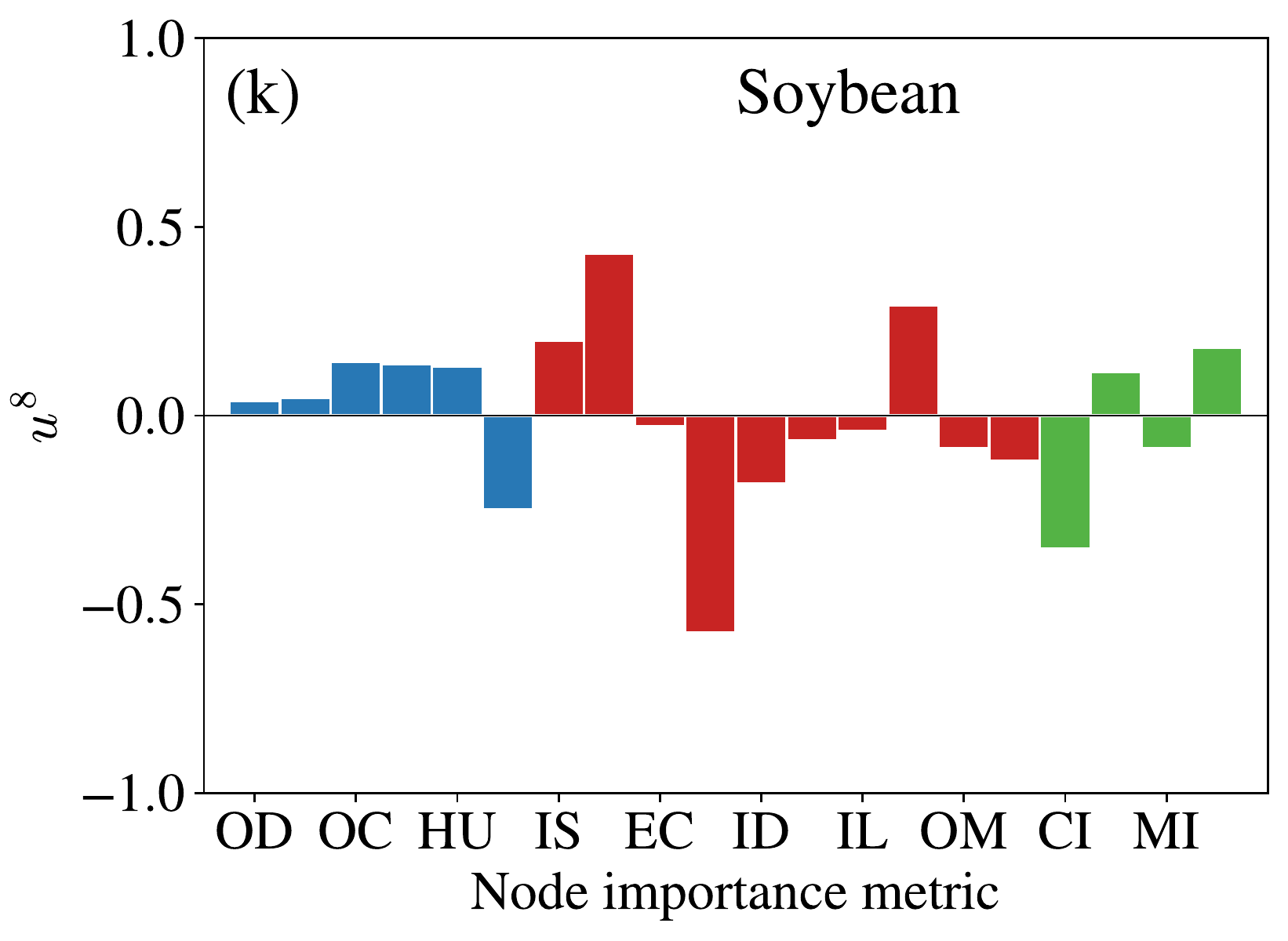}
      \includegraphics[width=0.233\linewidth]{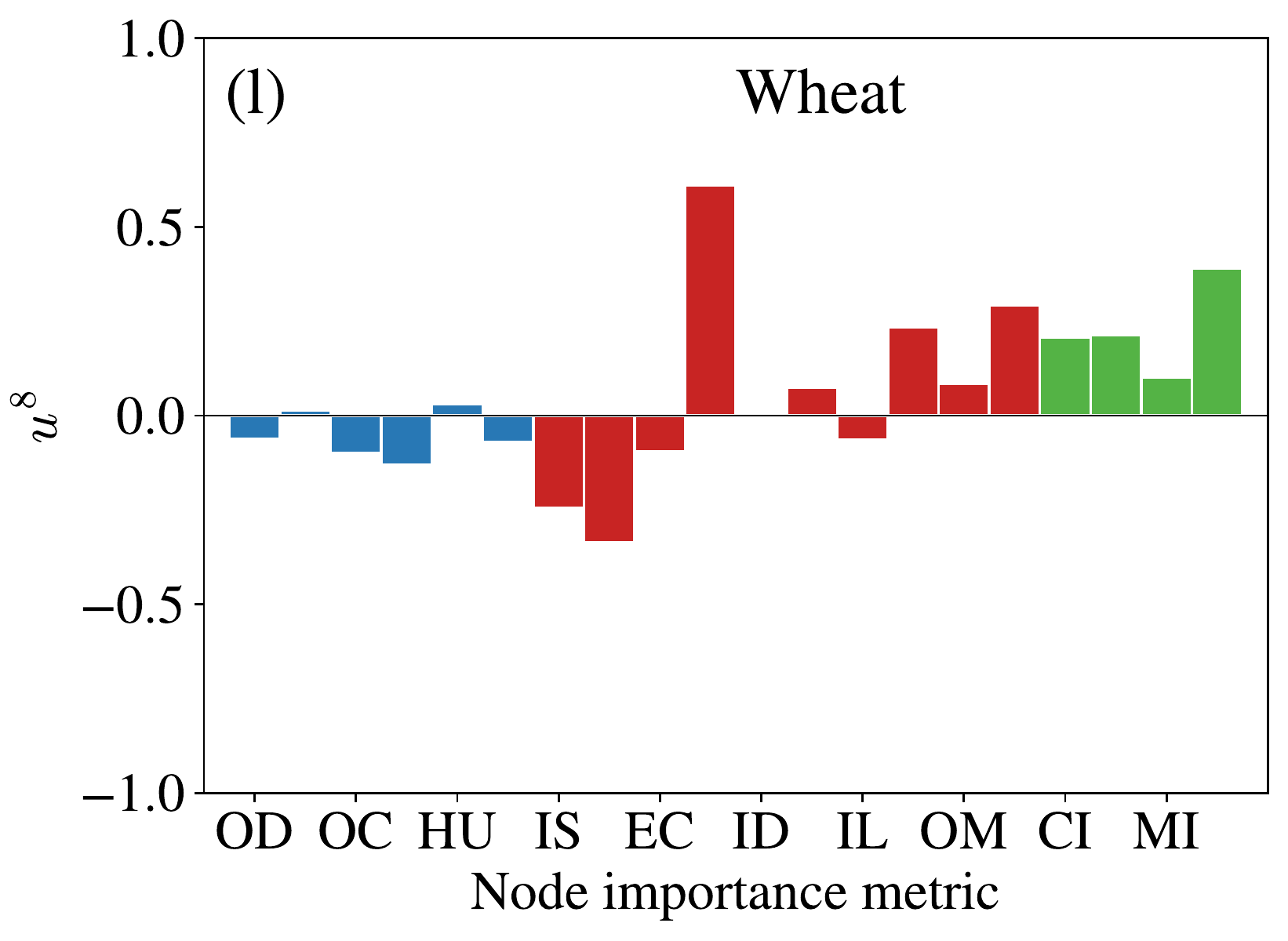}
      \includegraphics[width=0.233\linewidth]{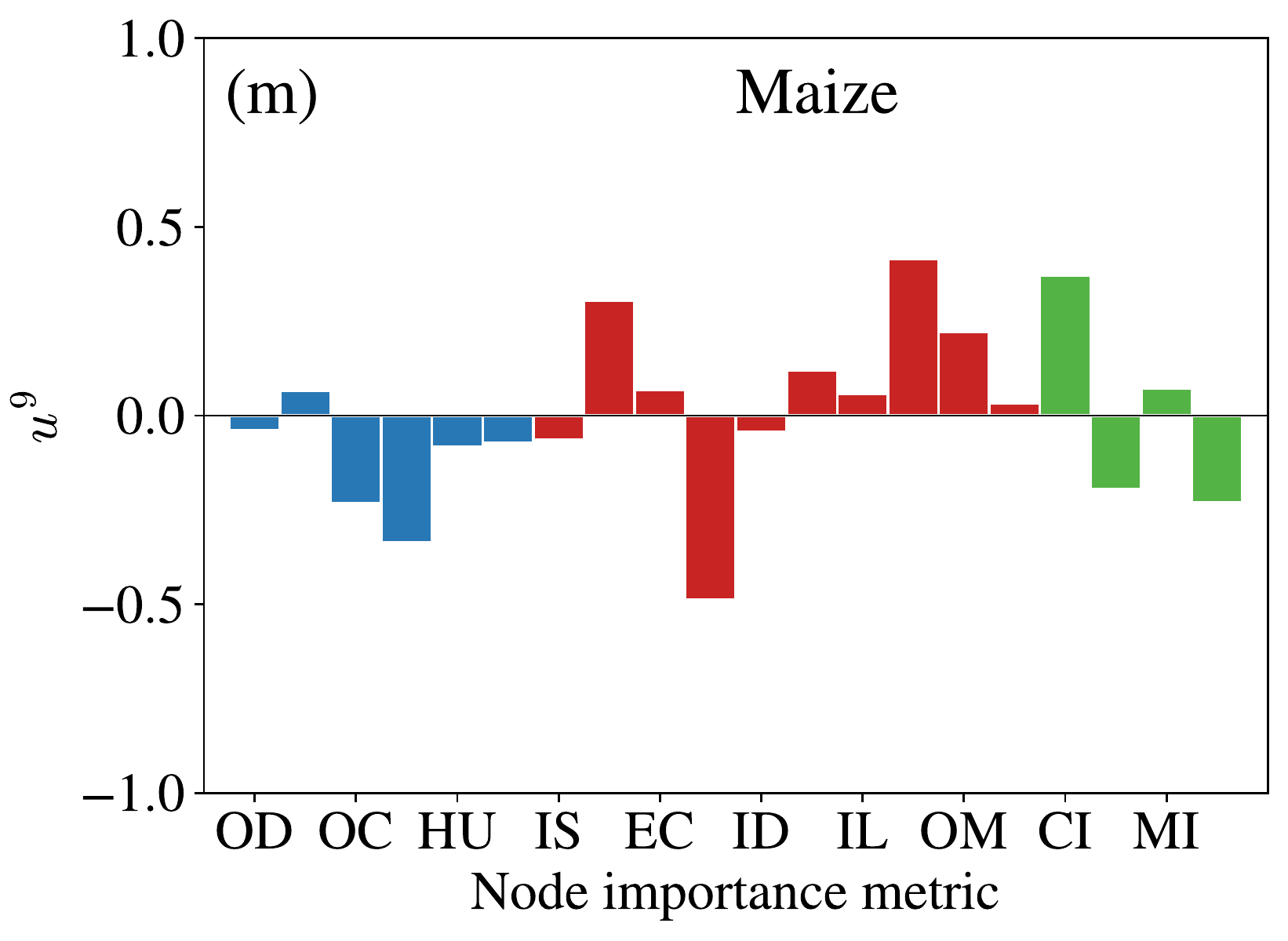}
      \includegraphics[width=0.233\linewidth]{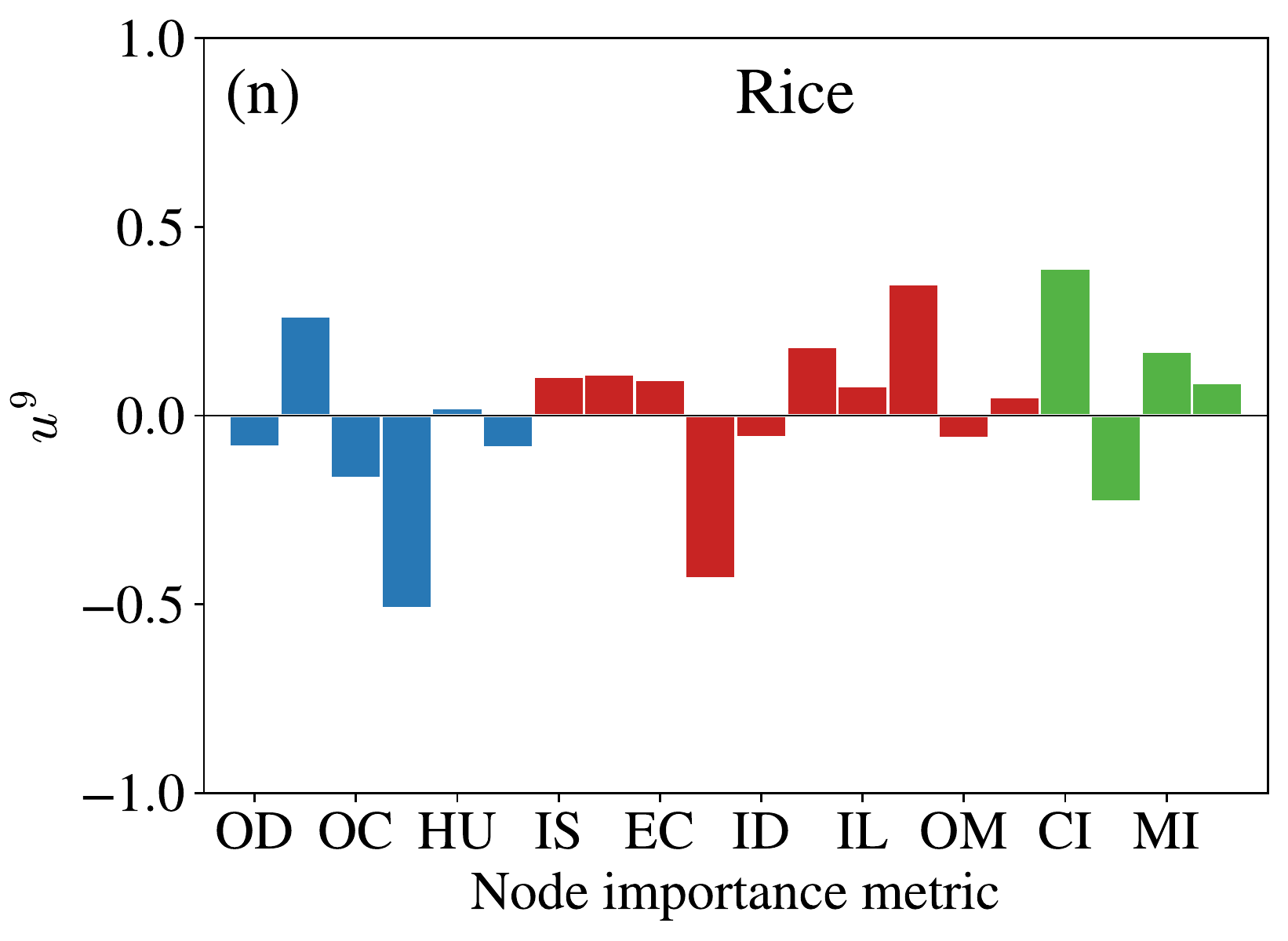}
      \includegraphics[width=0.233\linewidth]{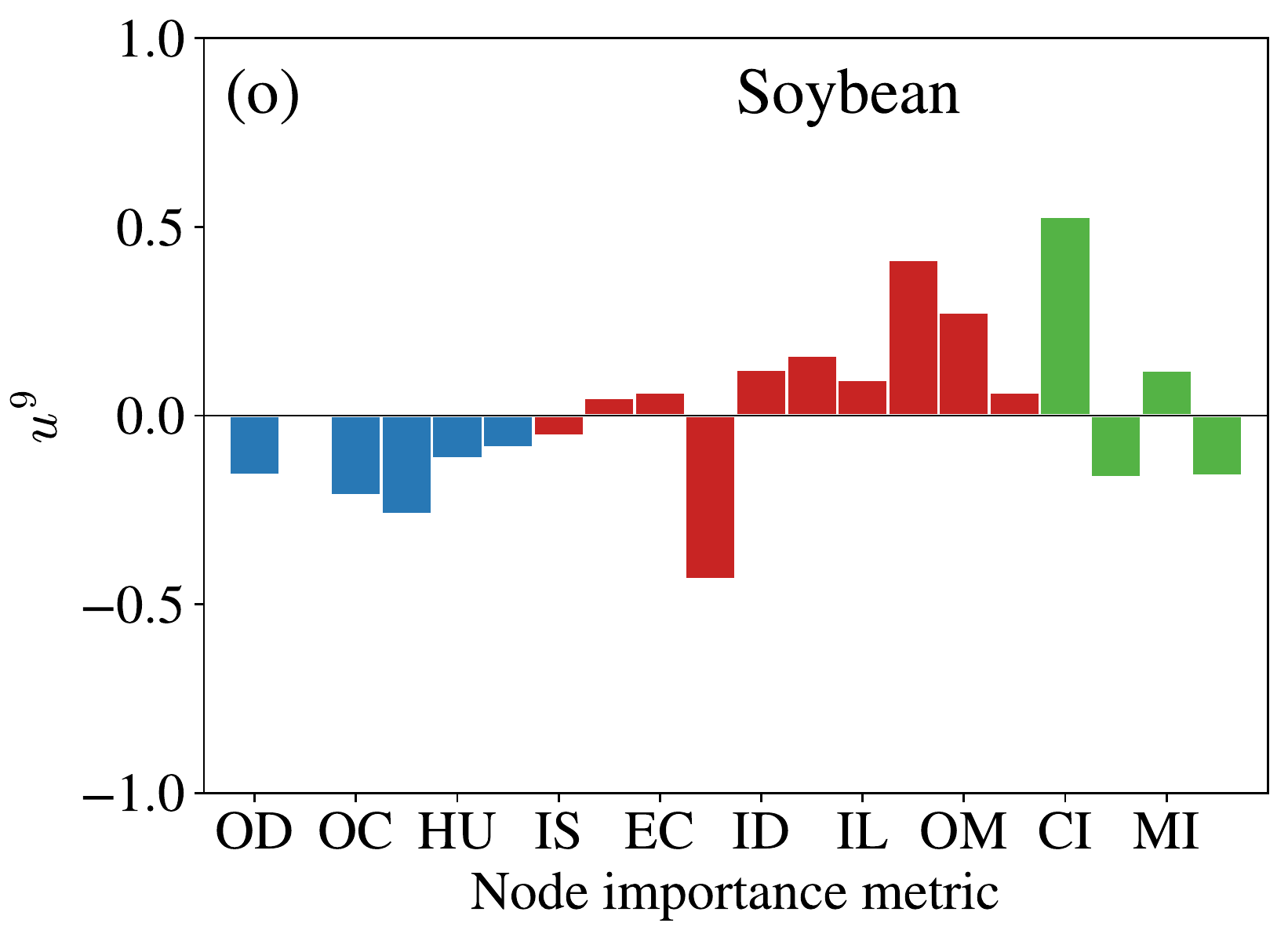}
      \includegraphics[width=0.233\linewidth]{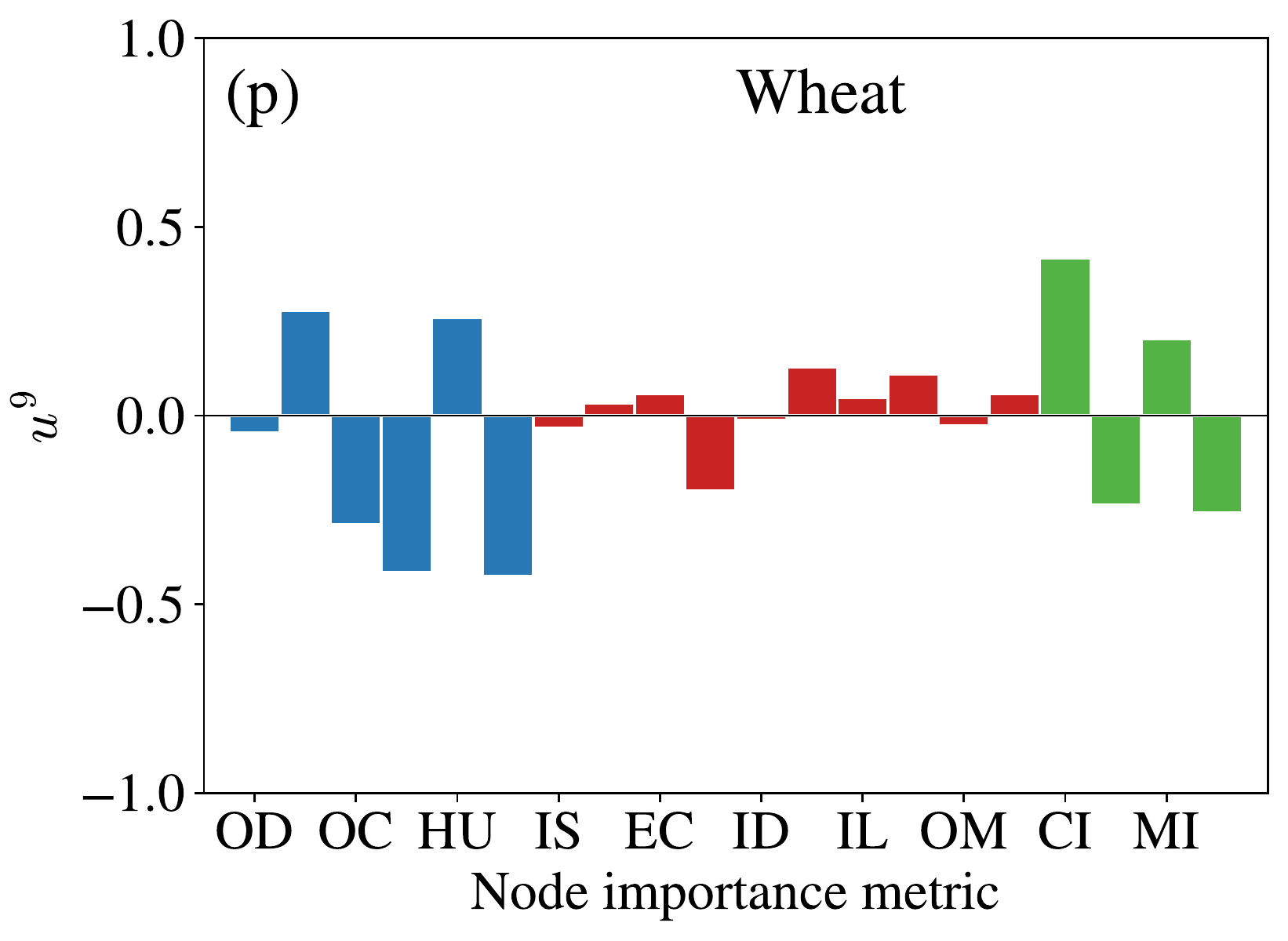}
      \includegraphics[width=0.233\linewidth]{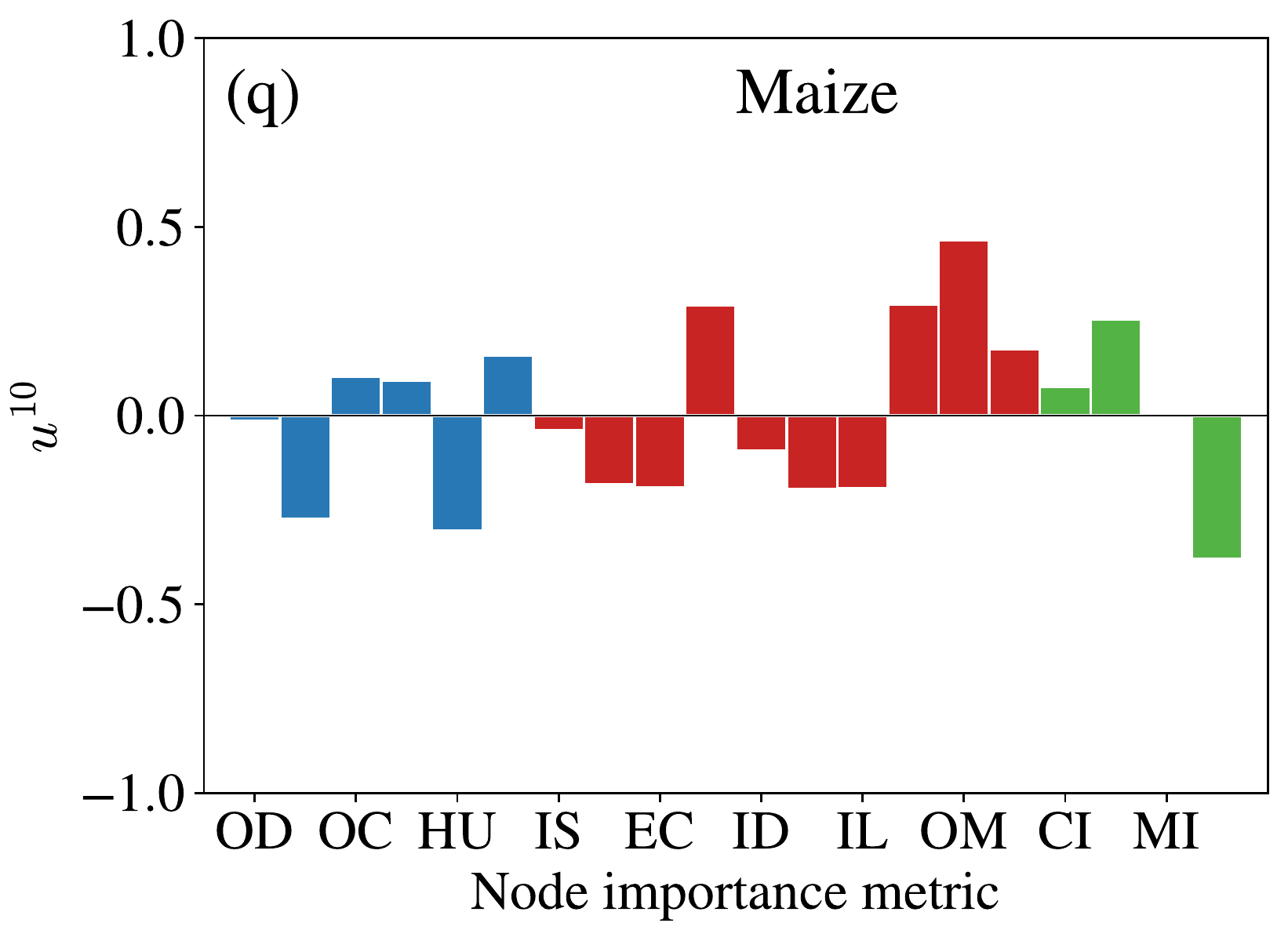}
      \includegraphics[width=0.233\linewidth]{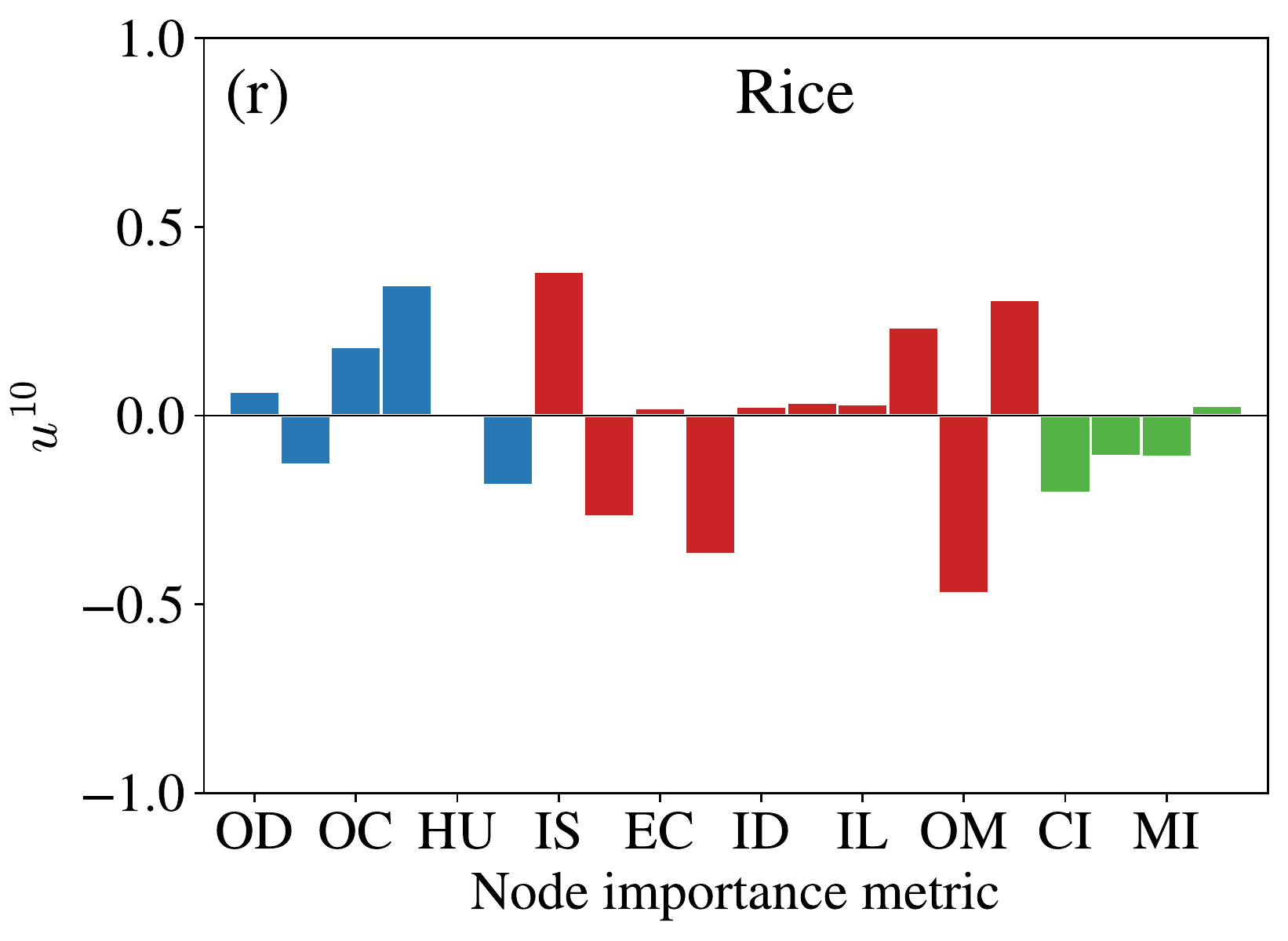}
      \includegraphics[width=0.233\linewidth]{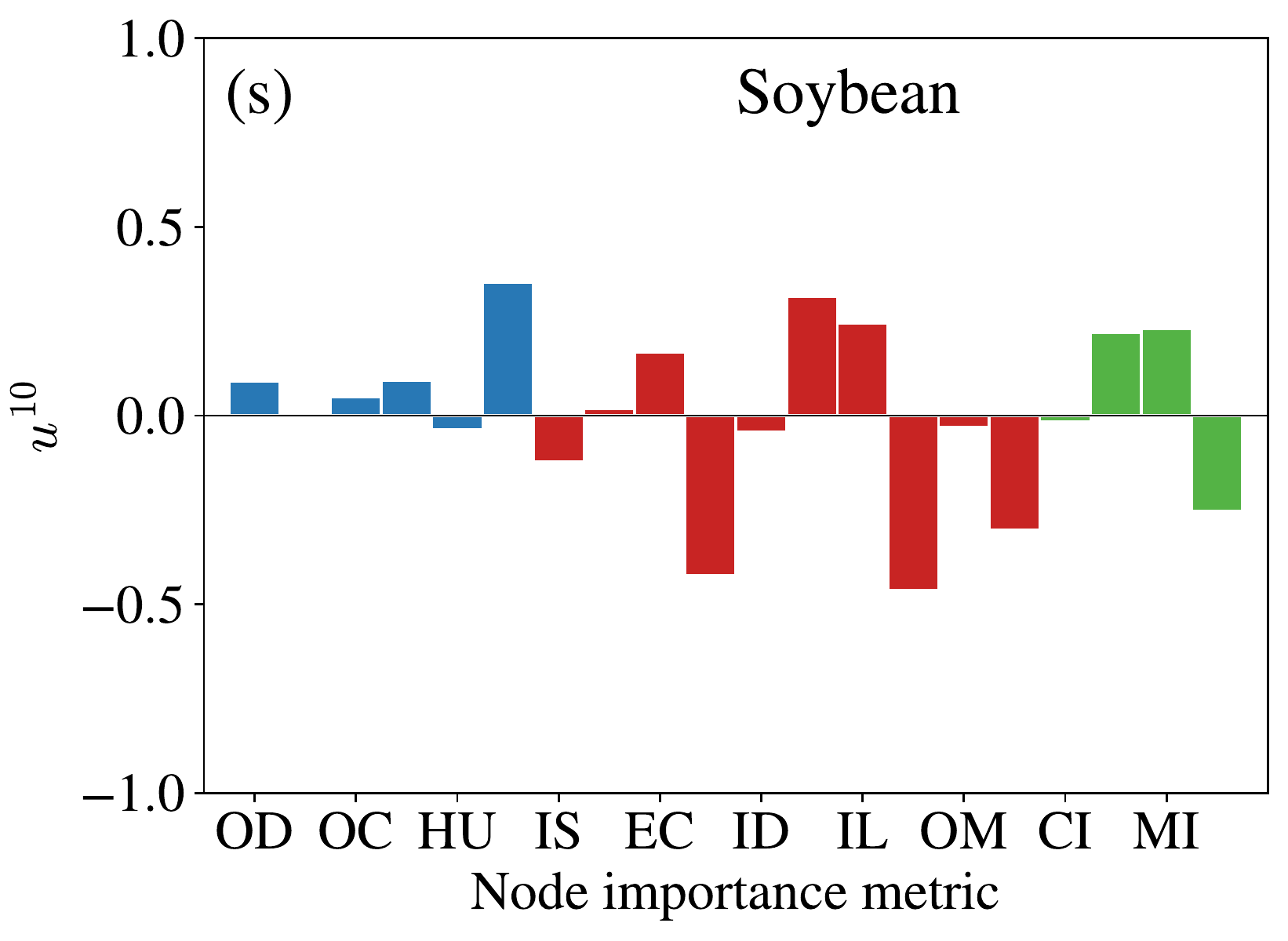}
      \includegraphics[width=0.233\linewidth]{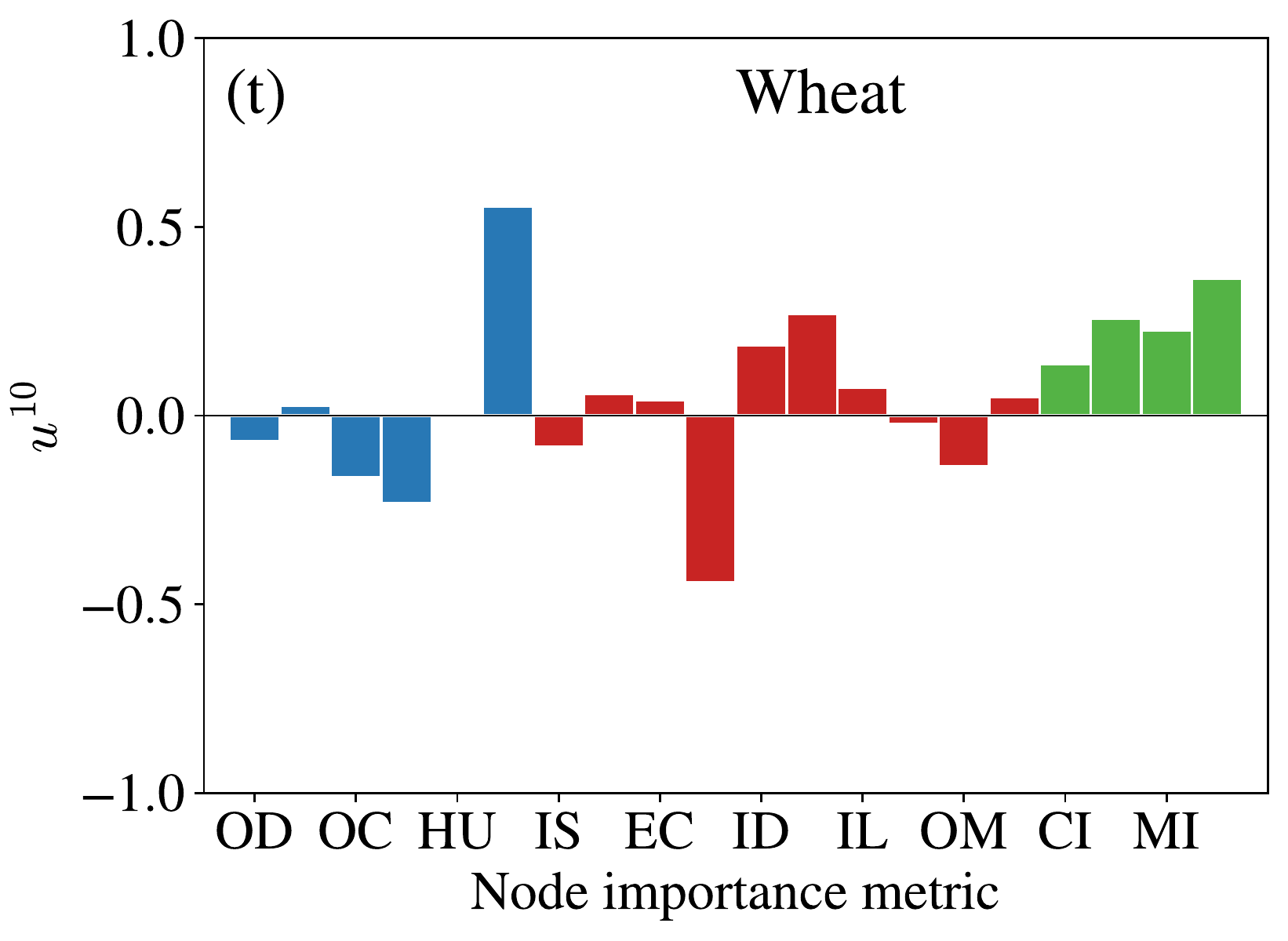}
      \caption{Components of the eigenvector $u_6-u_{10}$ of the eigenvalues $\lambda_6-\lambda_{10}$ given by Eq.~(\ref{Eq:RMT:PDF:eigenvalue}) of random matrix theory (RMT) in 2020.}
      \label{Fig:iCTN:PDF:eigenvalue:6-10:2020}
\end{figure}

 \begin{figure}[h!]
      \centering
      \includegraphics[width=0.233\linewidth]{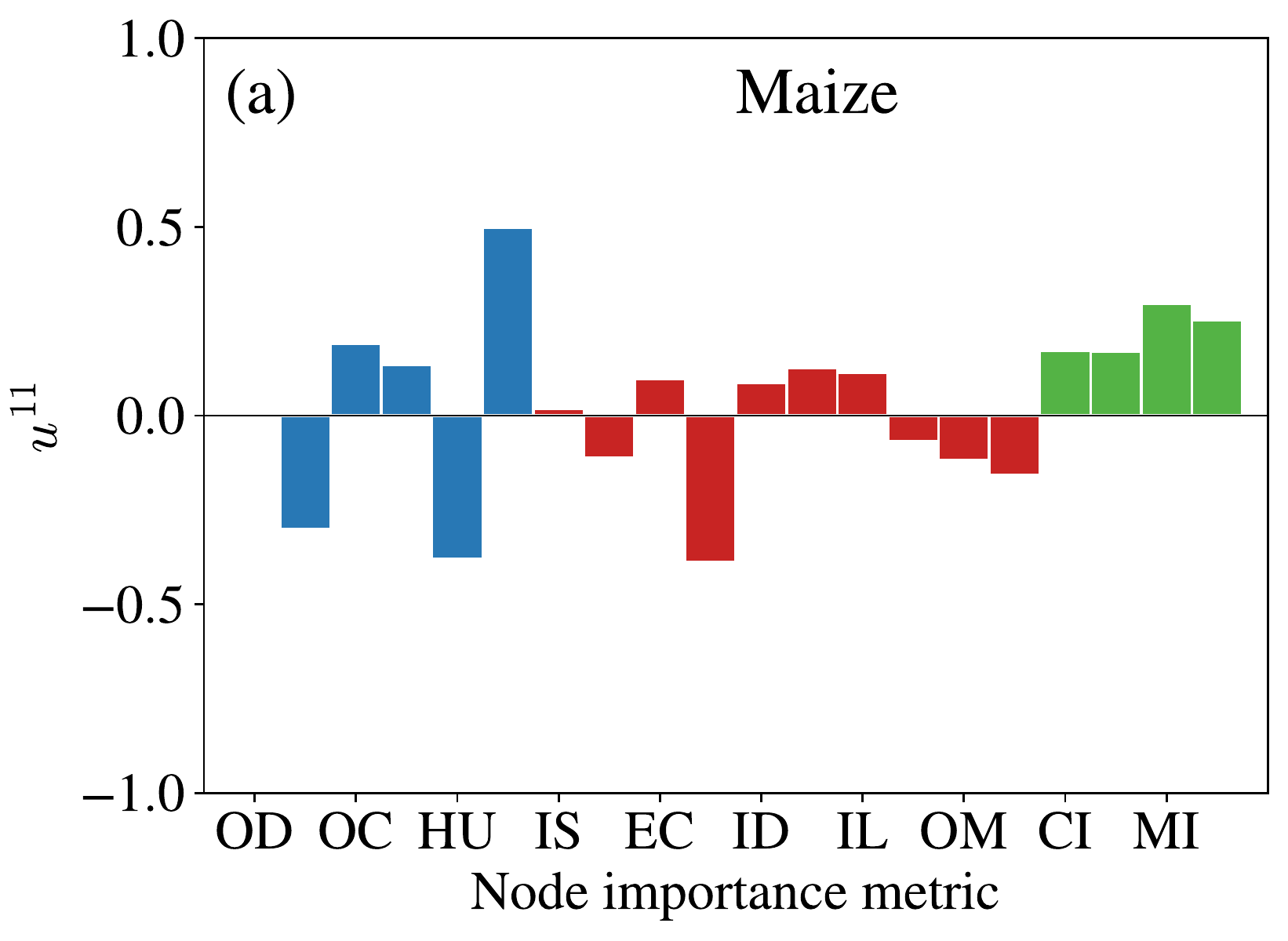}
      \includegraphics[width=0.233\linewidth]{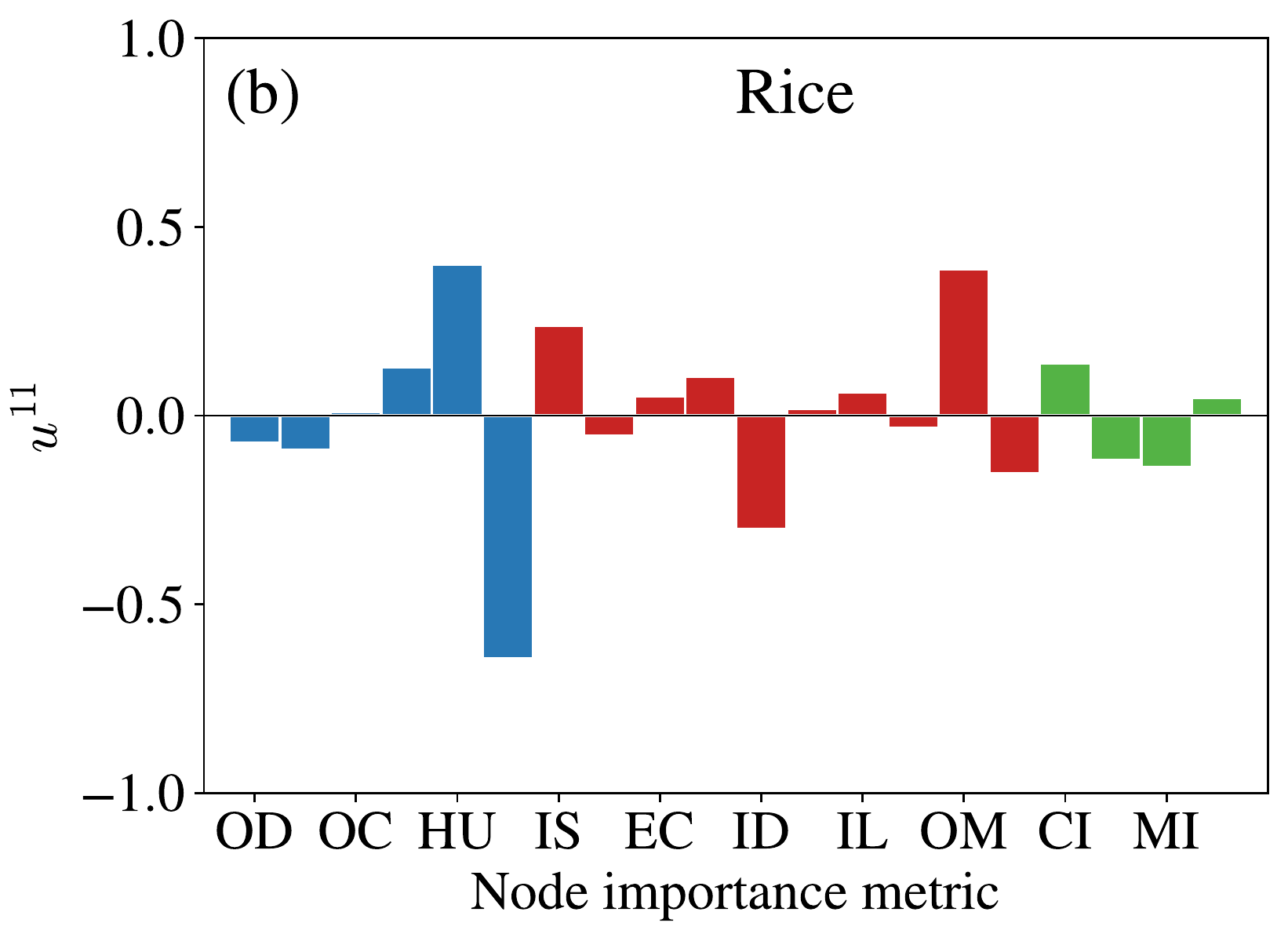}
      \includegraphics[width=0.233\linewidth]{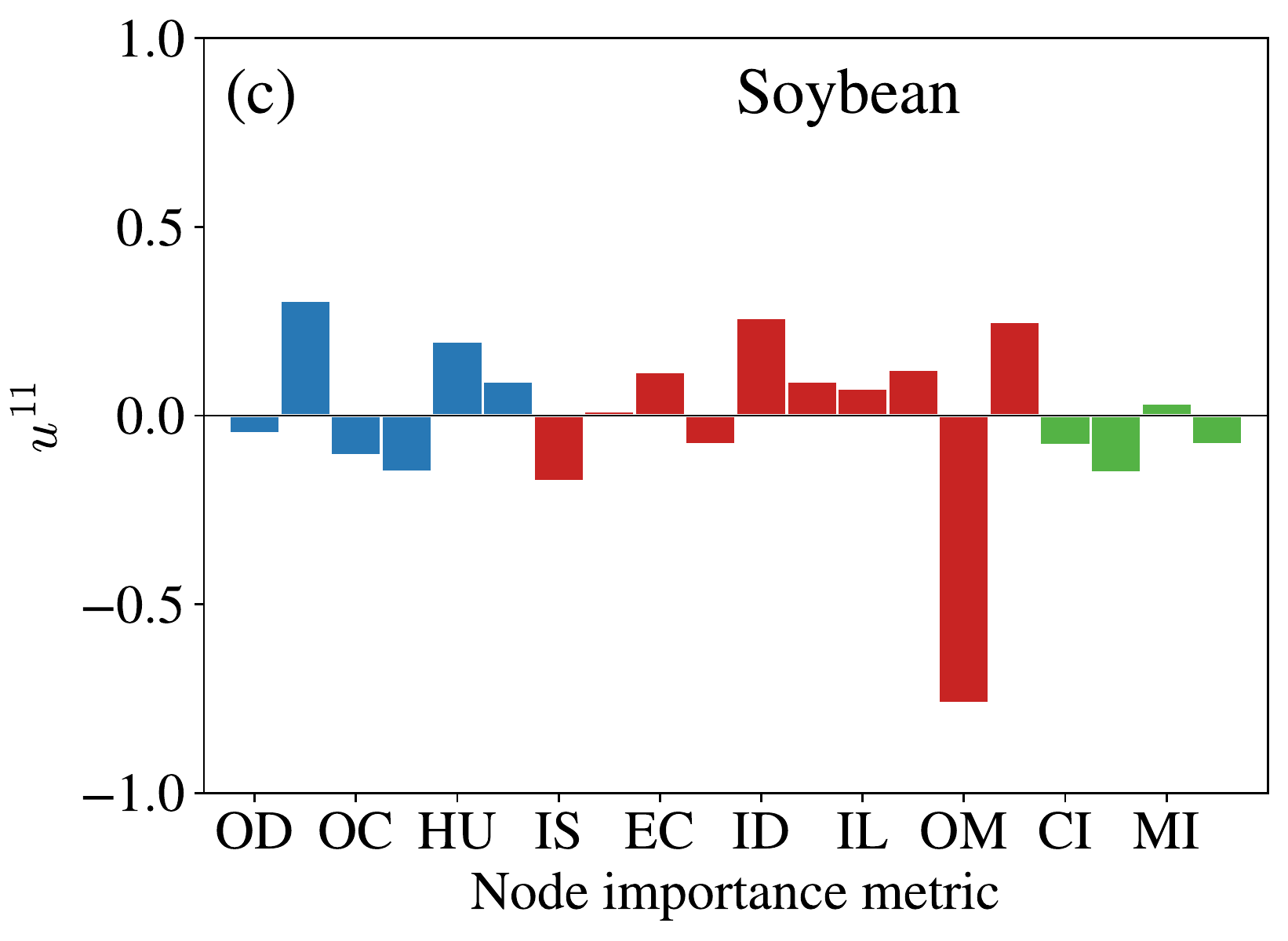}
      \includegraphics[width=0.233\linewidth]{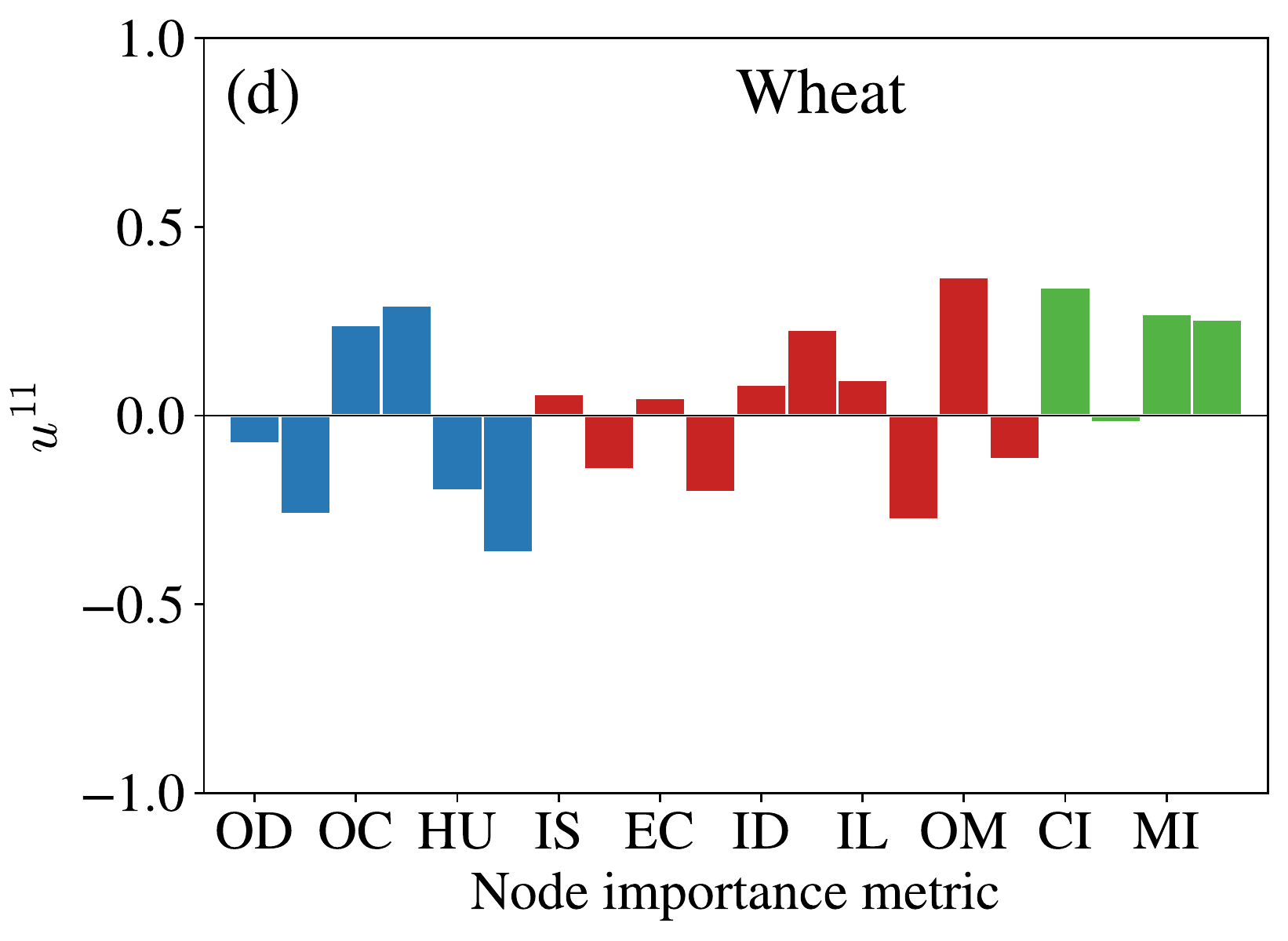}
      \includegraphics[width=0.233\linewidth]{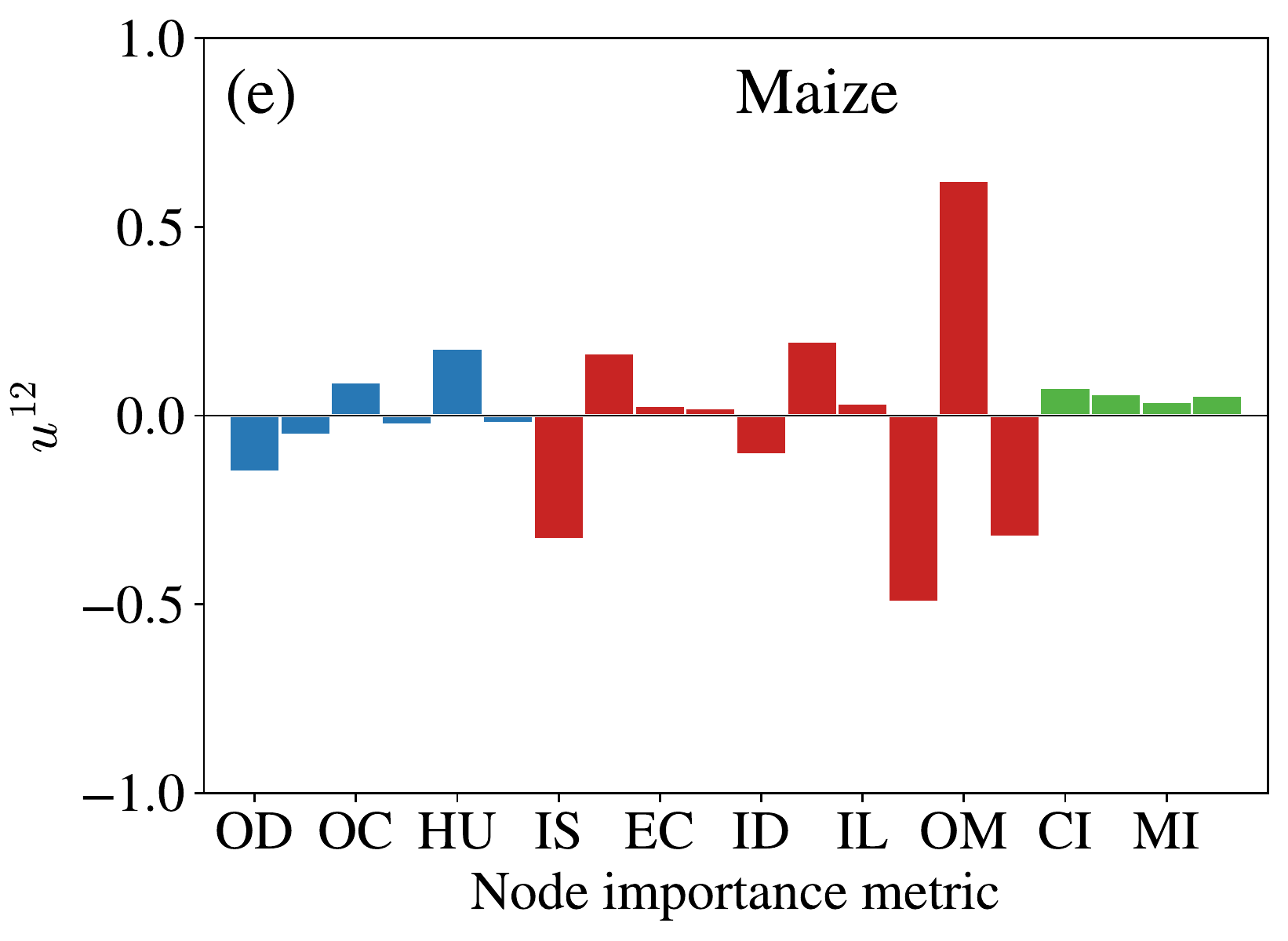}
      \includegraphics[width=0.233\linewidth]{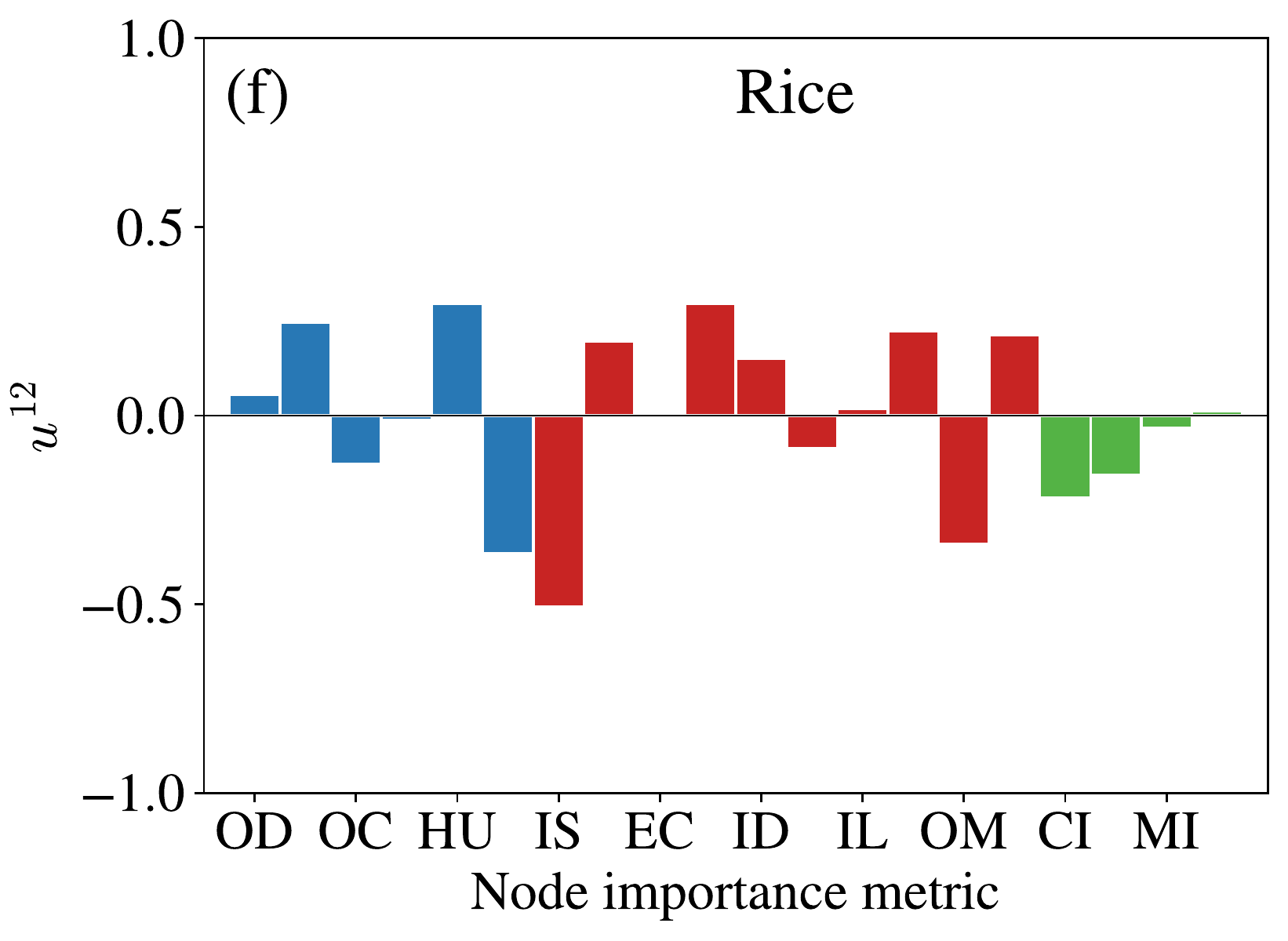}
      \includegraphics[width=0.233\linewidth]{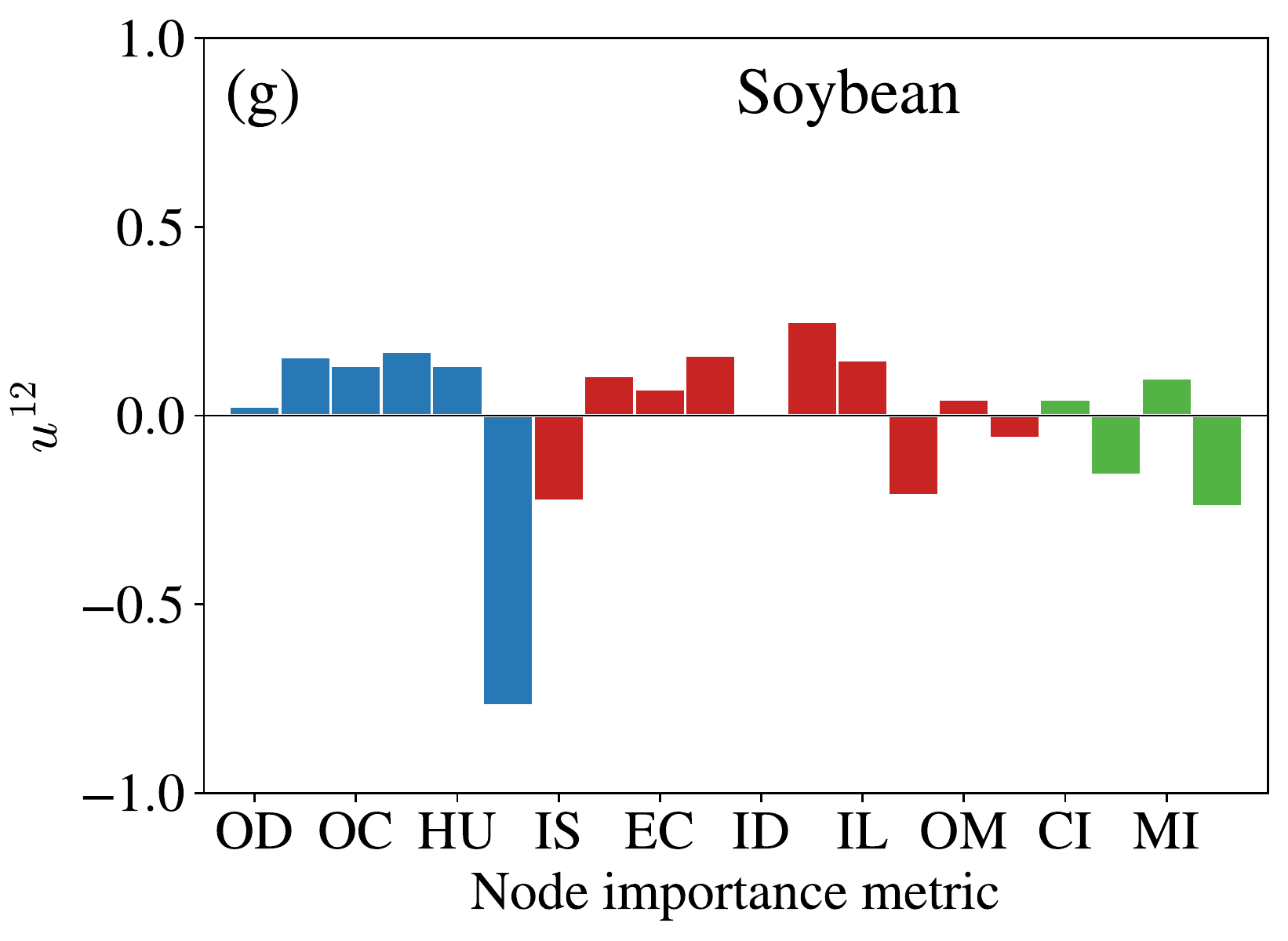}
      \includegraphics[width=0.233\linewidth]{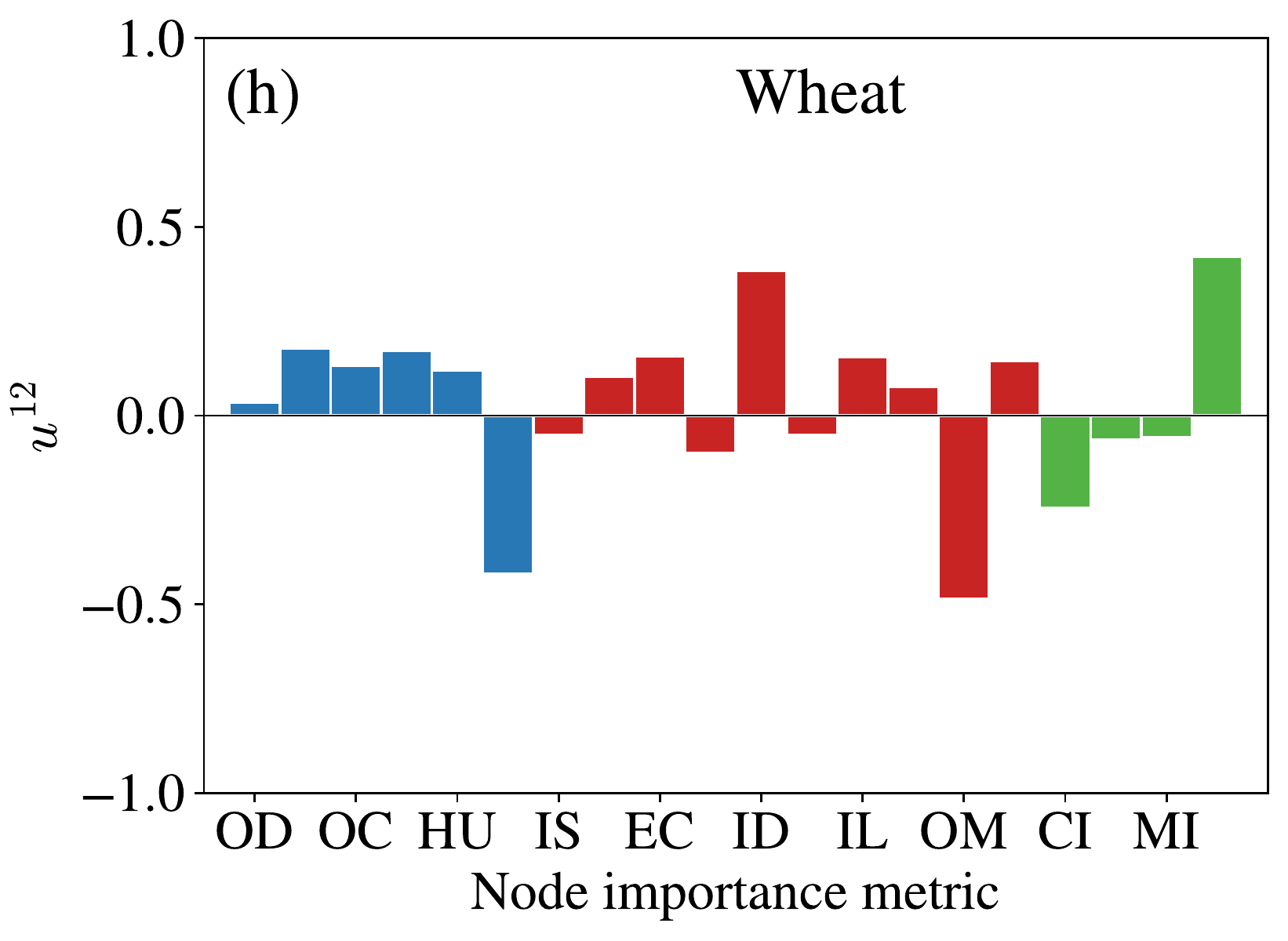}
      \includegraphics[width=0.233\linewidth]{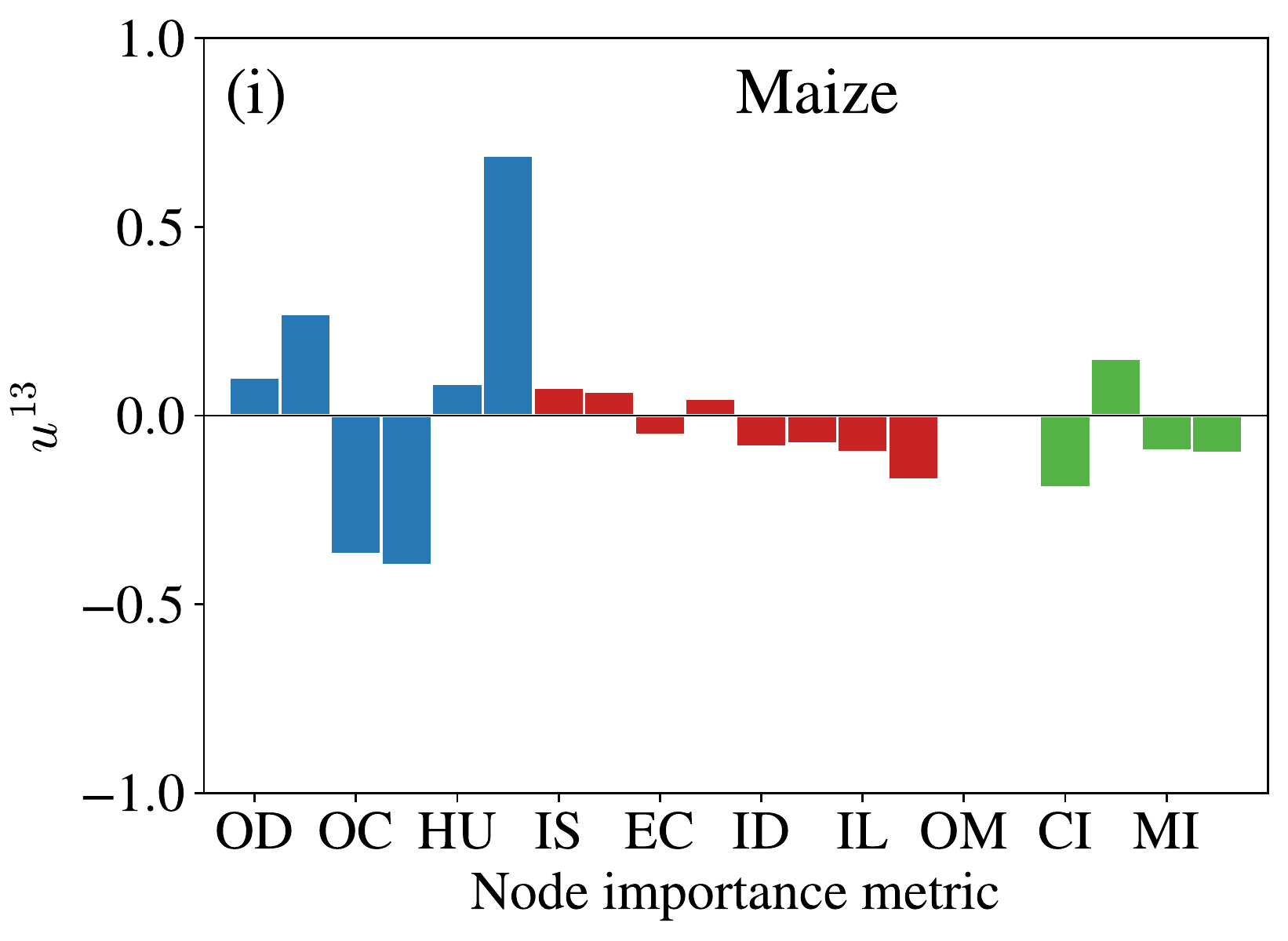}
      \includegraphics[width=0.233\linewidth]{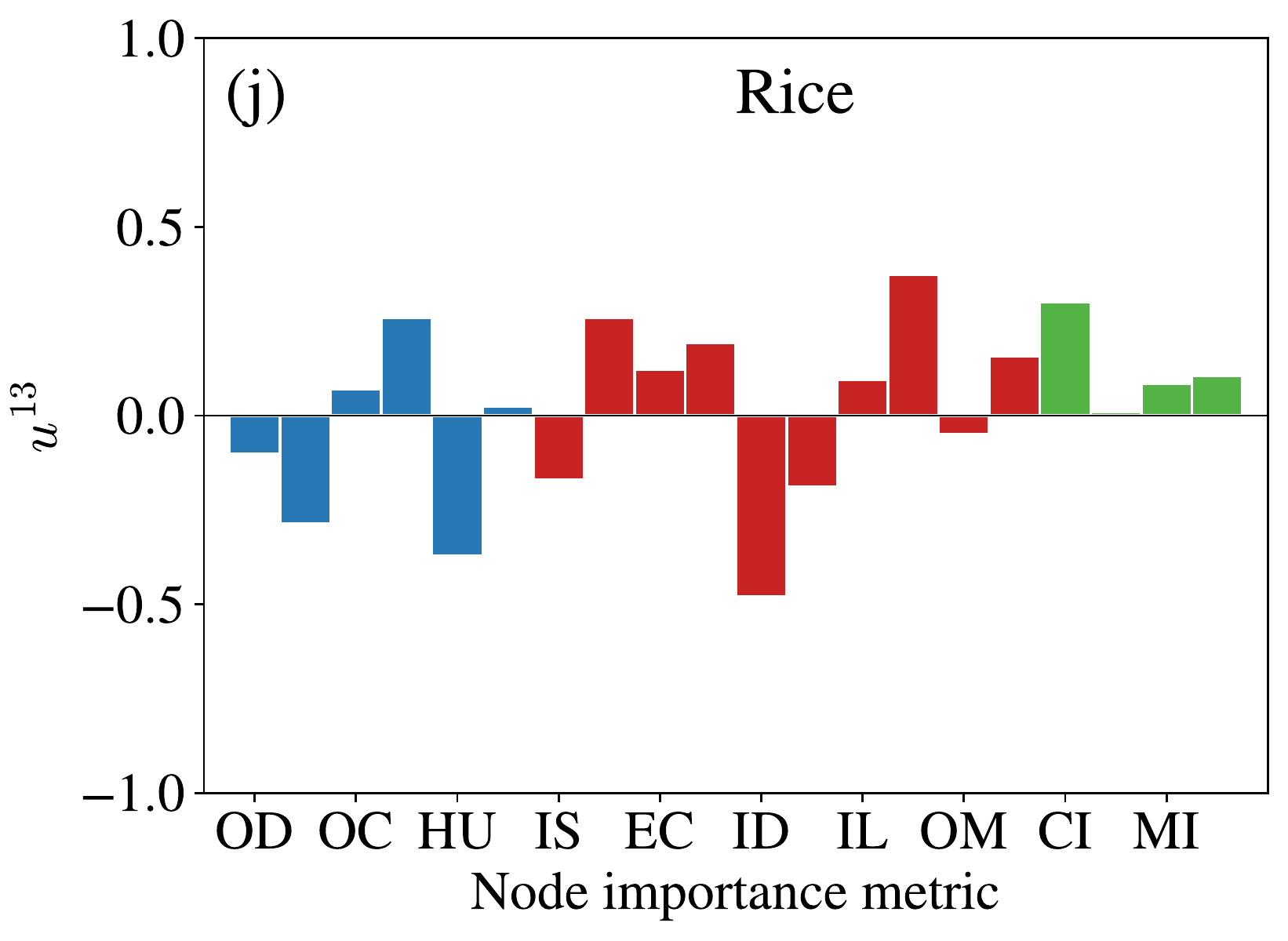}
      \includegraphics[width=0.233\linewidth]{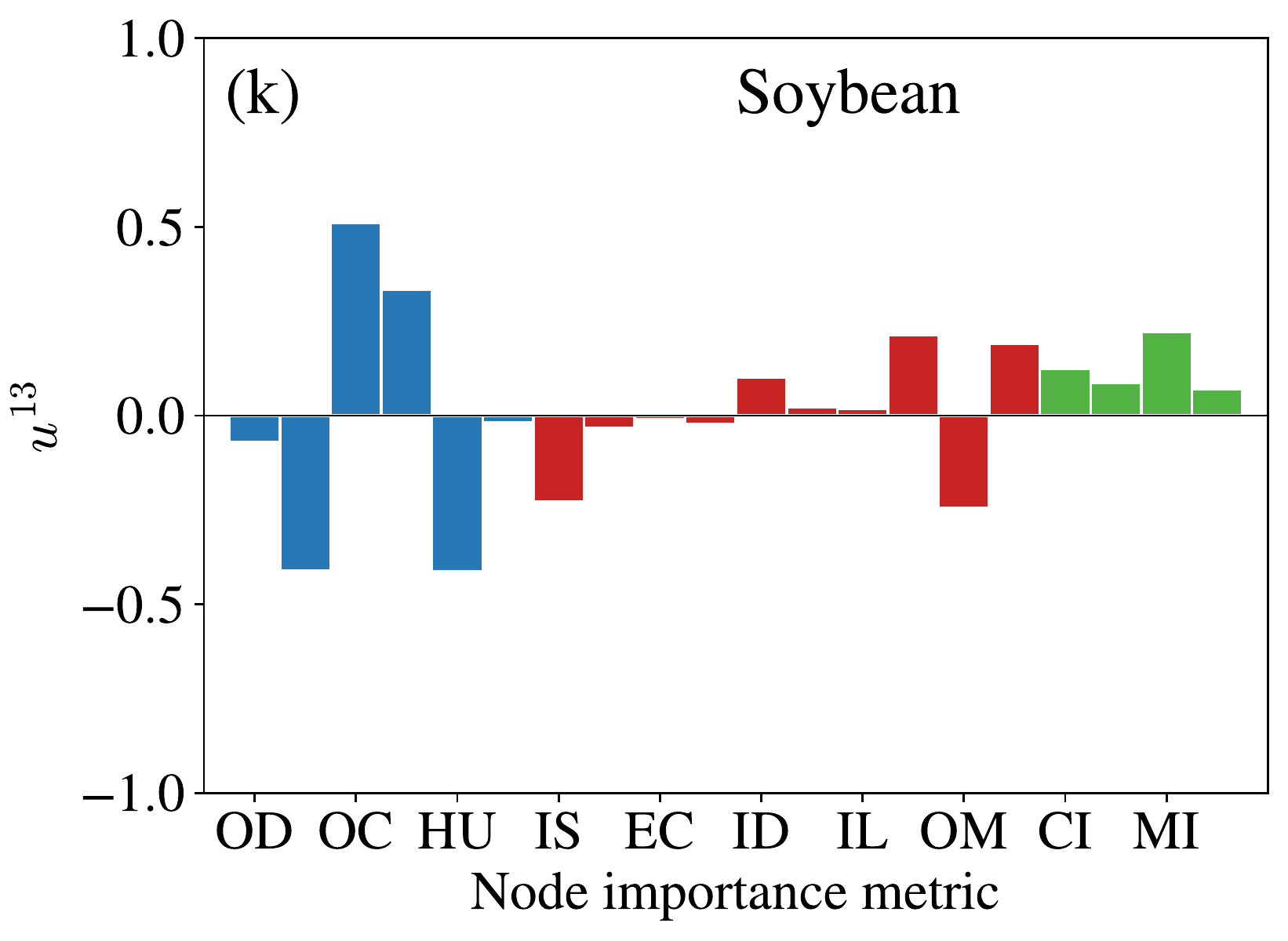}
      \includegraphics[width=0.233\linewidth]{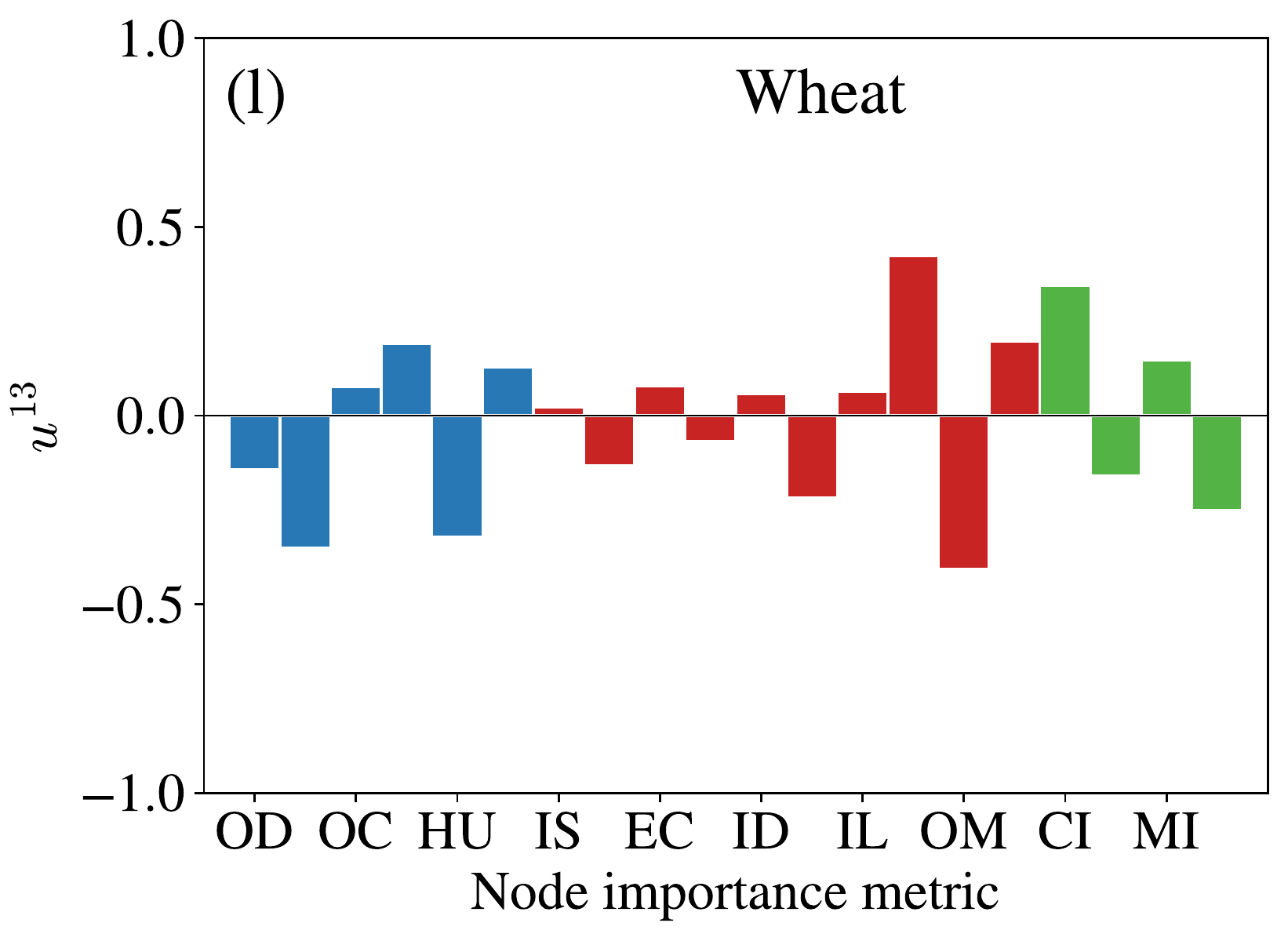}
      \includegraphics[width=0.233\linewidth]{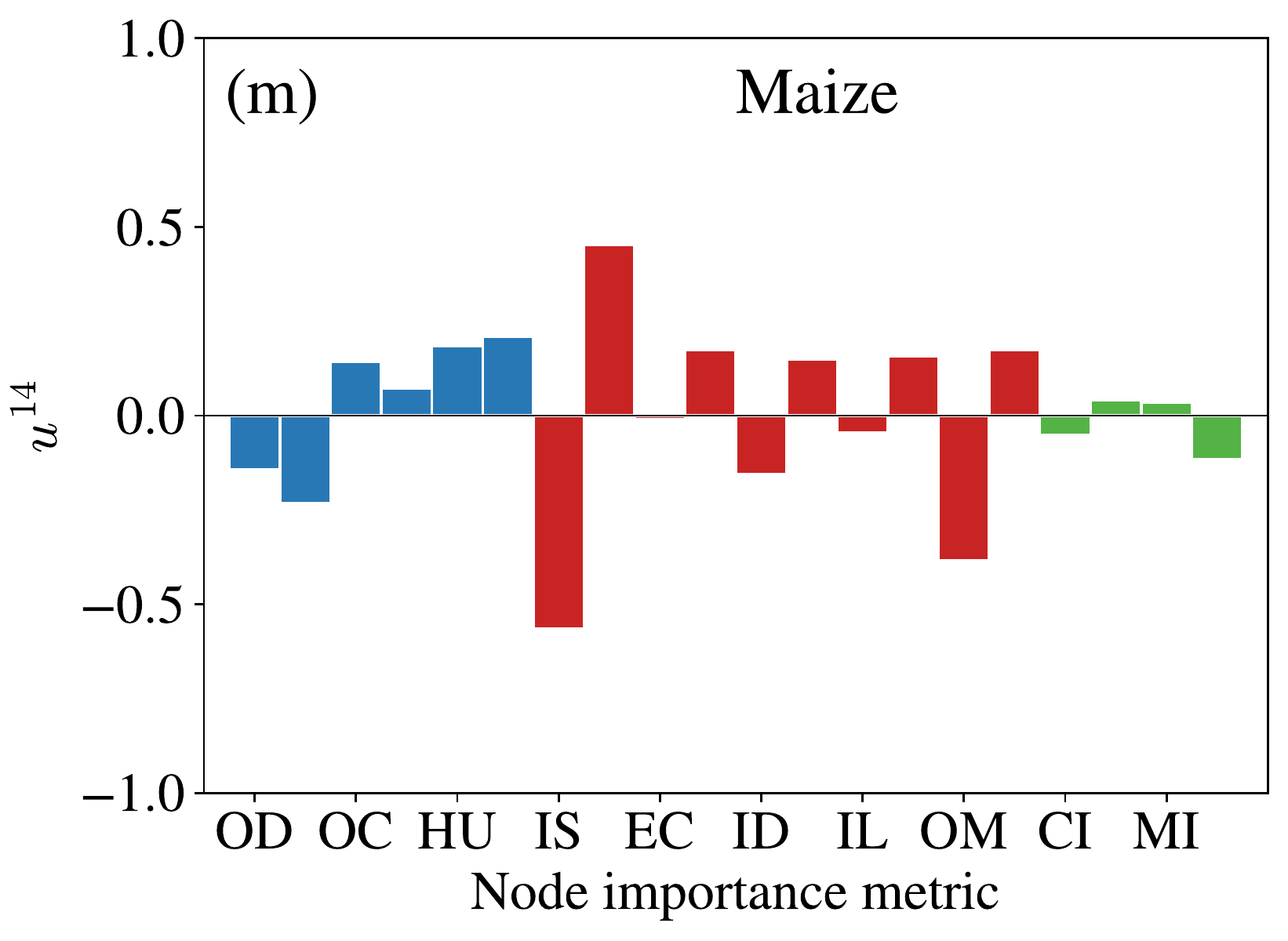}
      \includegraphics[width=0.233\linewidth]{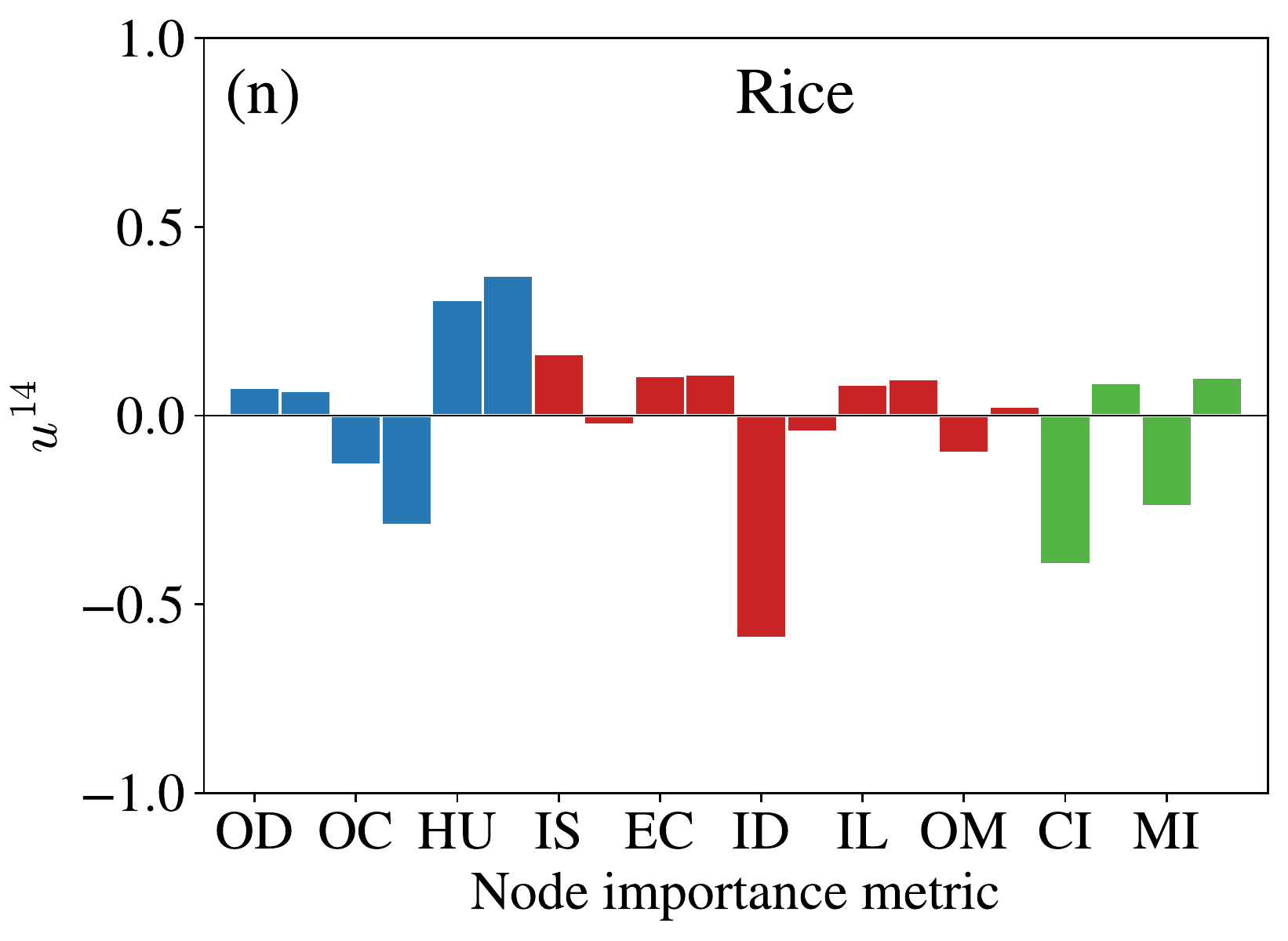}
      \includegraphics[width=0.233\linewidth]{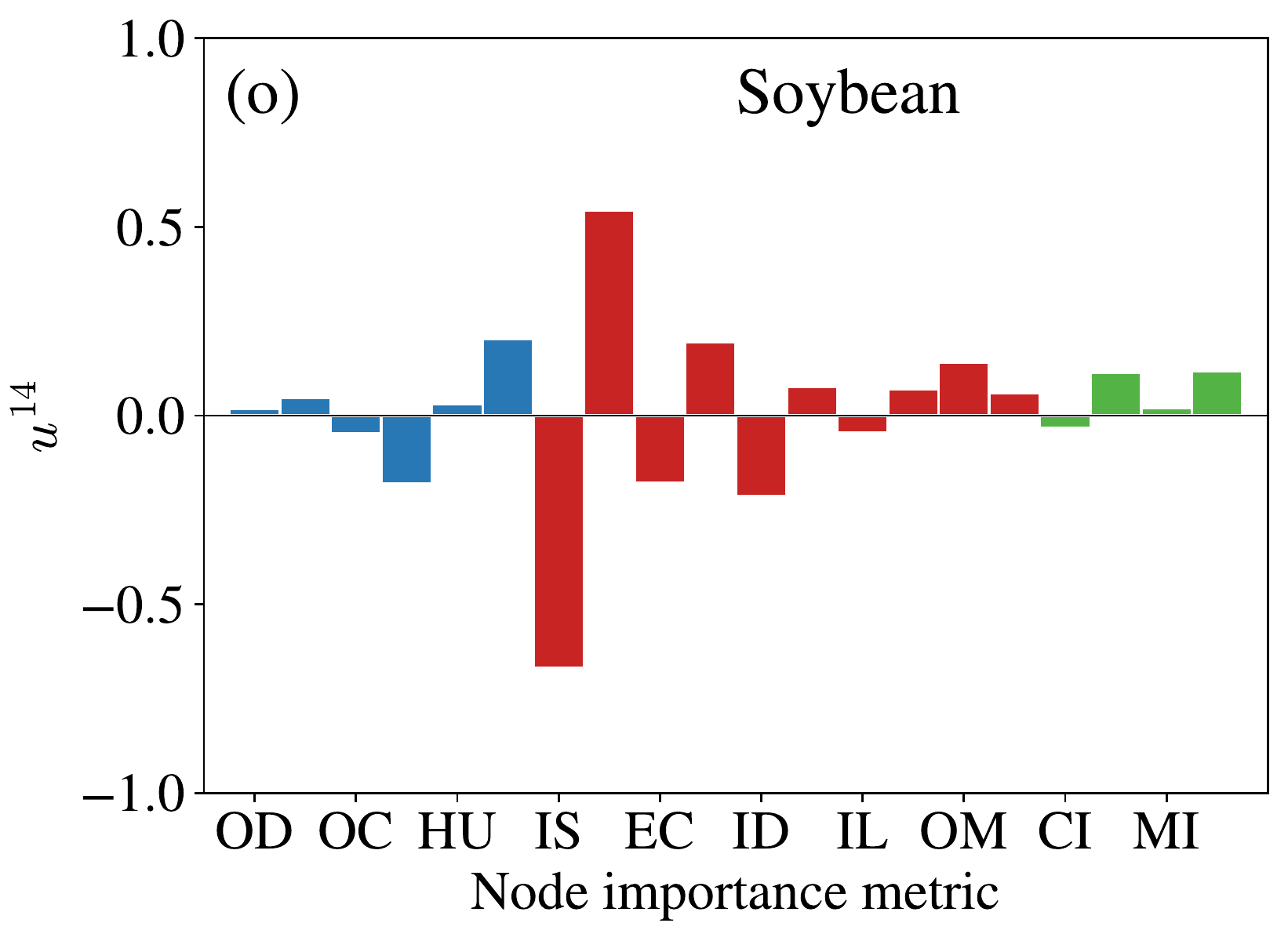}
      \includegraphics[width=0.233\linewidth]{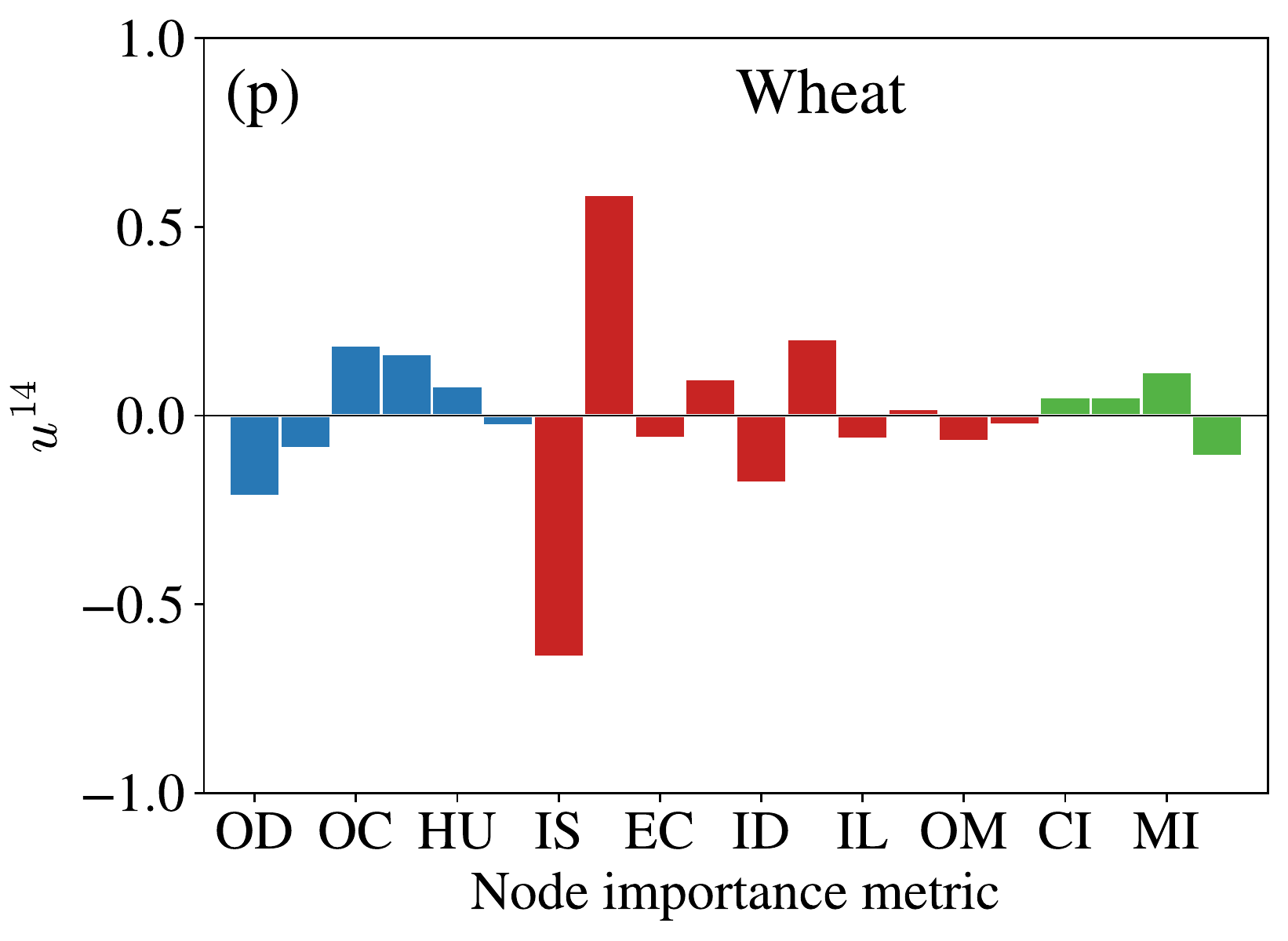}
      \includegraphics[width=0.233\linewidth]{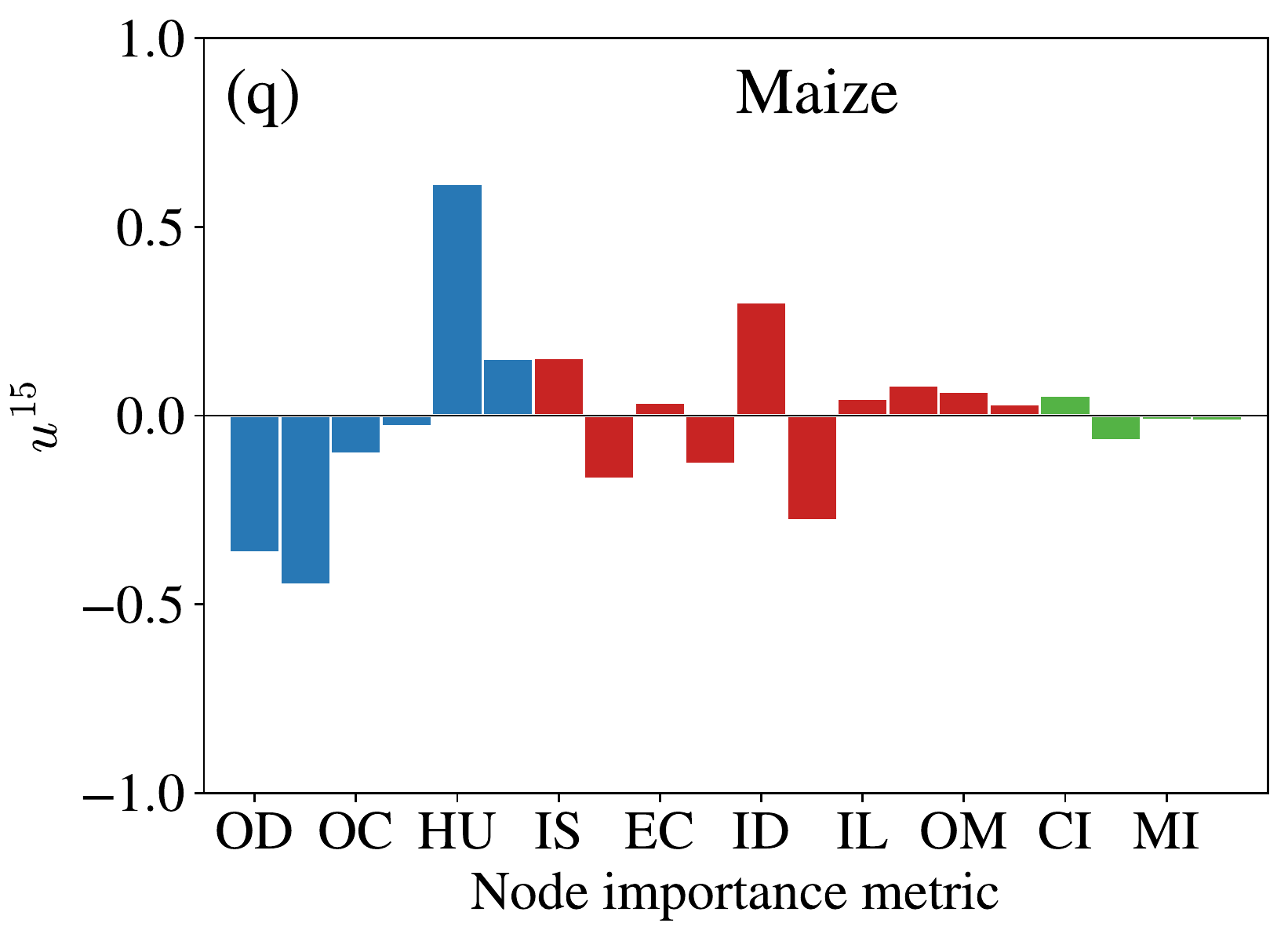}
      \includegraphics[width=0.233\linewidth]{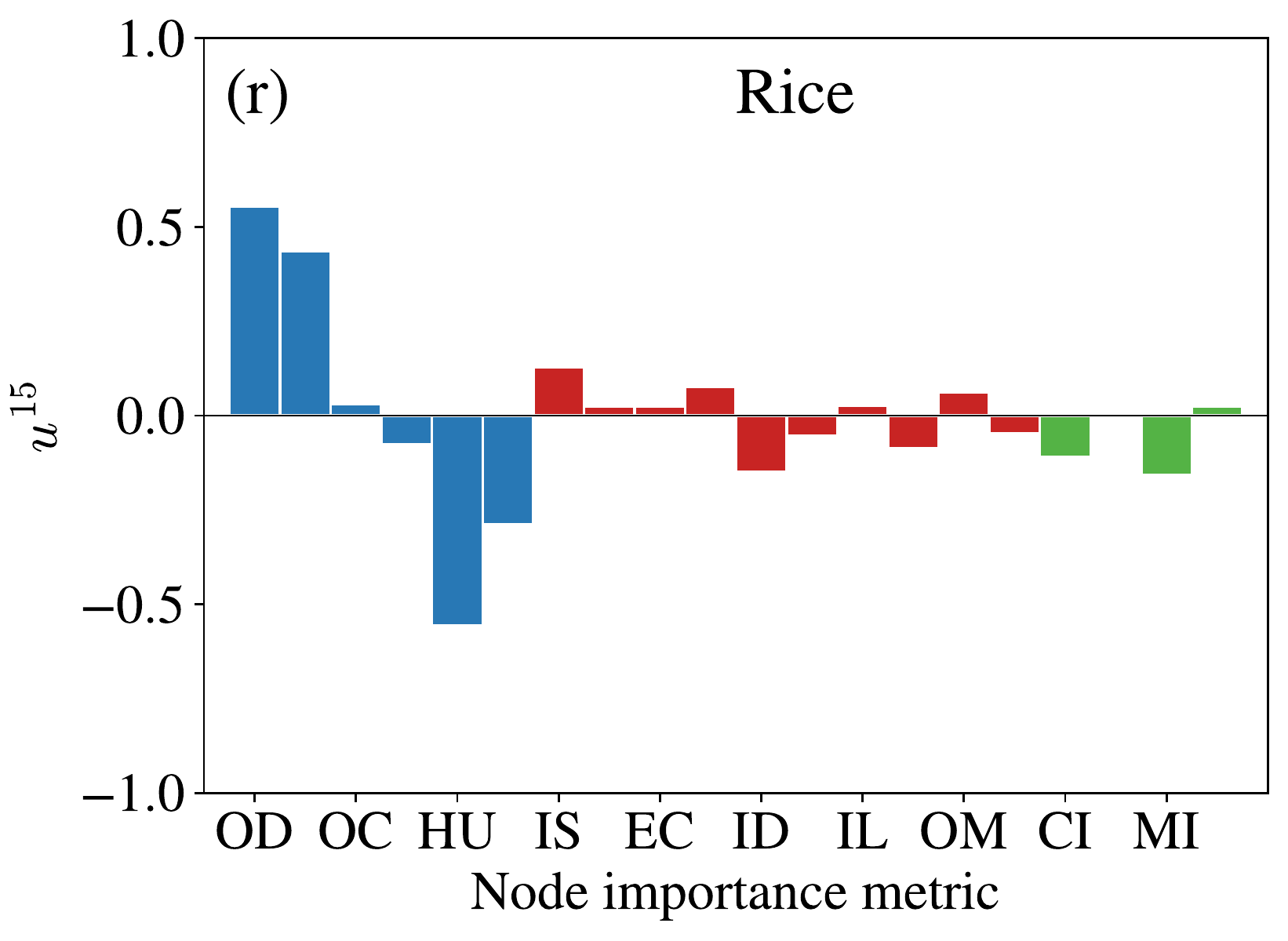}
      \includegraphics[width=0.233\linewidth]{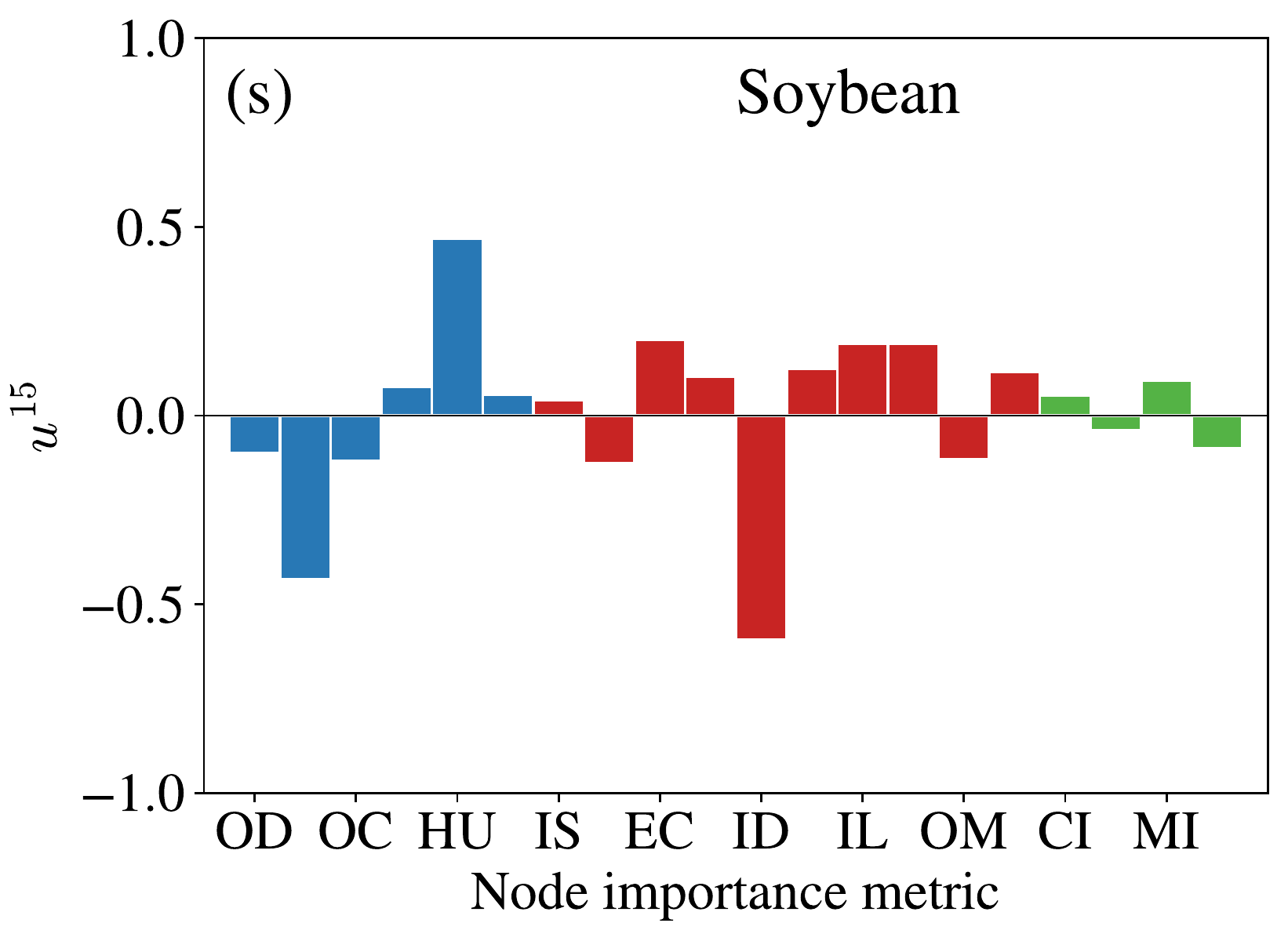}
      \includegraphics[width=0.233\linewidth]{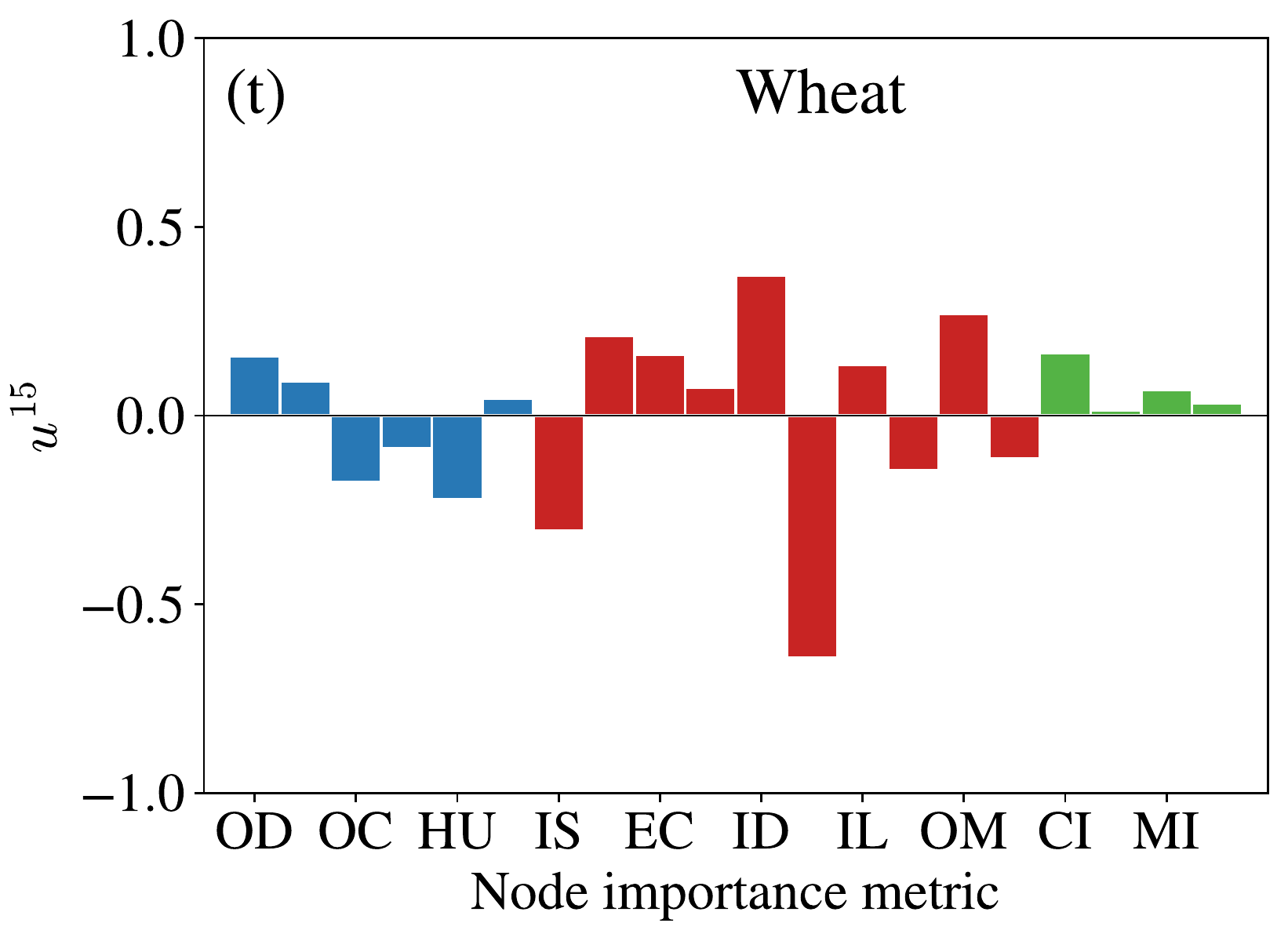}
      \caption{Components of the eigenvector $u_{11}-u_{15}$ of the eigenvalues $\lambda_{11}-\lambda_{15}$ given by Eq.~(\ref{Eq:RMT:PDF:eigenvalue}) of random matrix theory (RMT) in 2020.}
      \label{Fig:iCTN:PDF:eigenvalue:11-15:2020}
\end{figure}

 \begin{figure}[h!]
      \centering
      \includegraphics[width=0.233\linewidth]{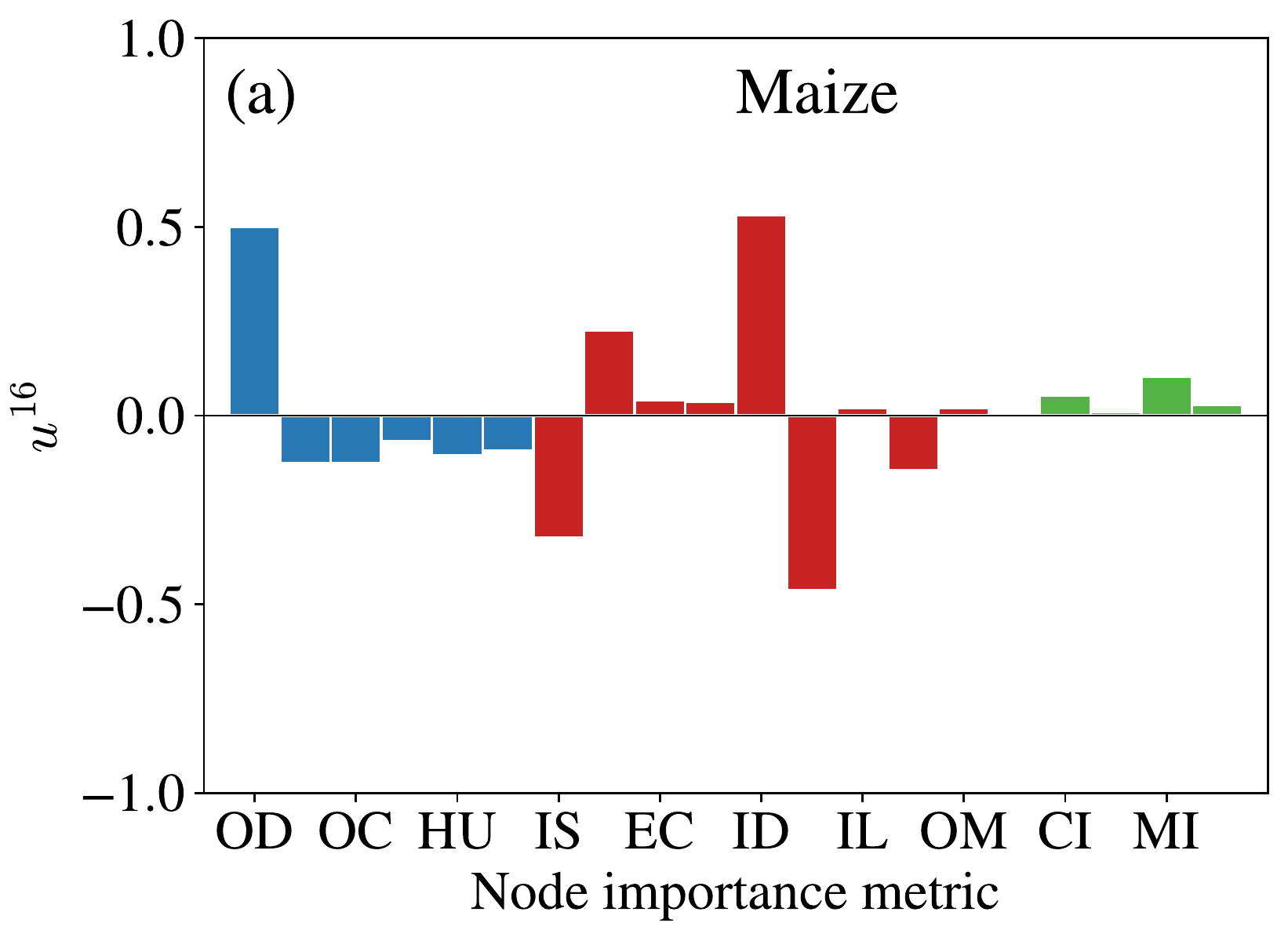}
      \includegraphics[width=0.233\linewidth]{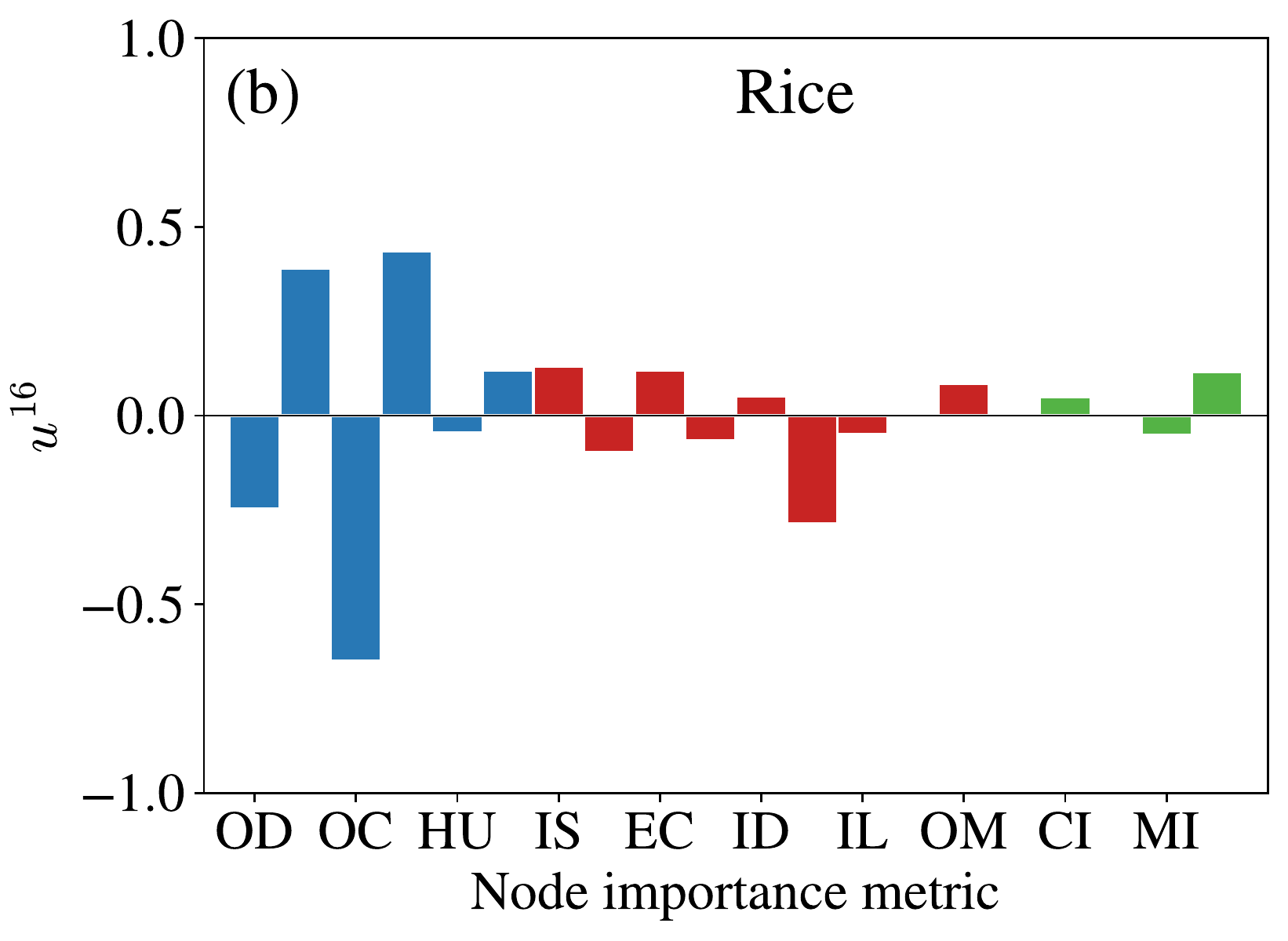}
      \includegraphics[width=0.233\linewidth]{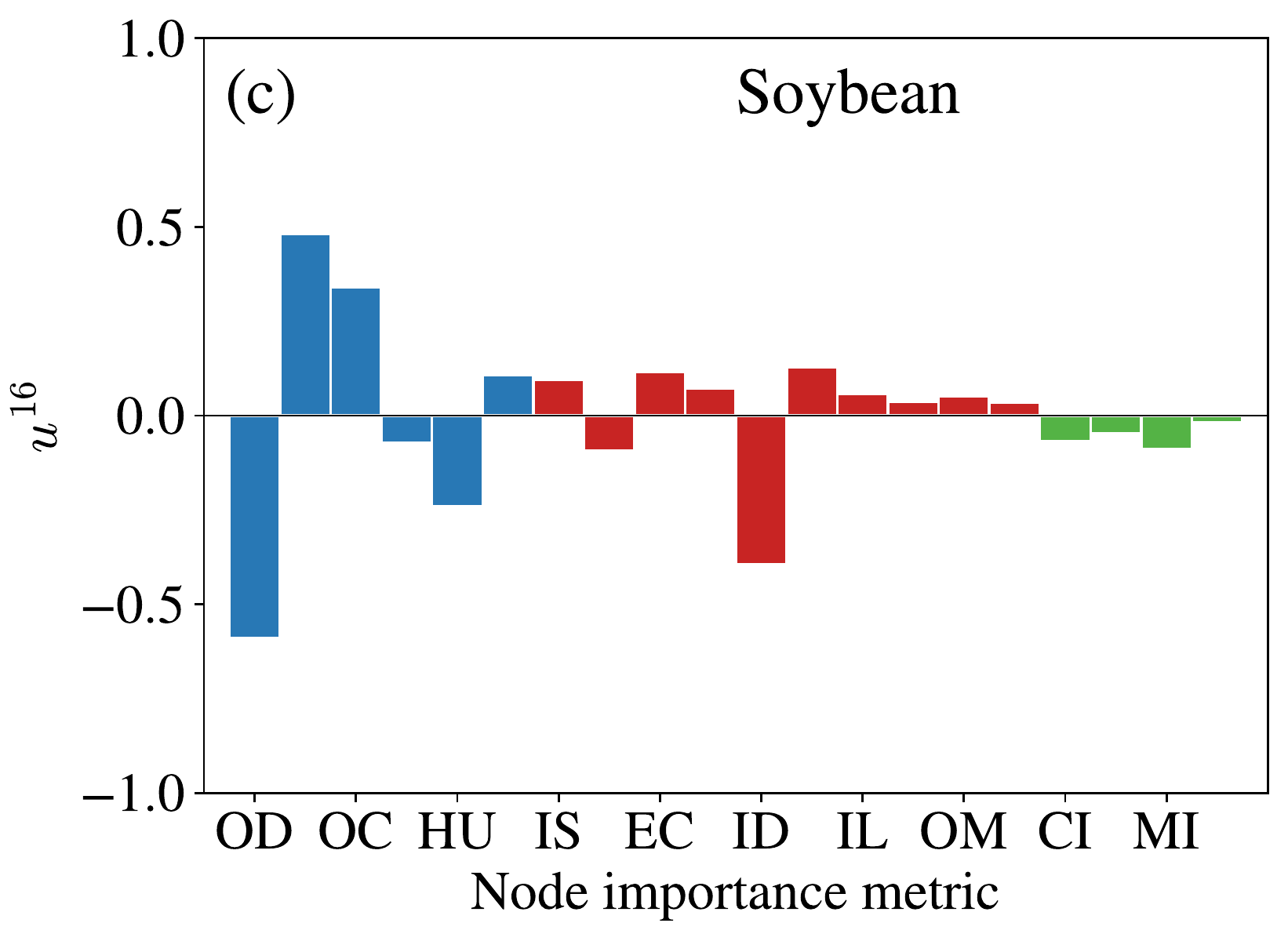}
      \includegraphics[width=0.233\linewidth]{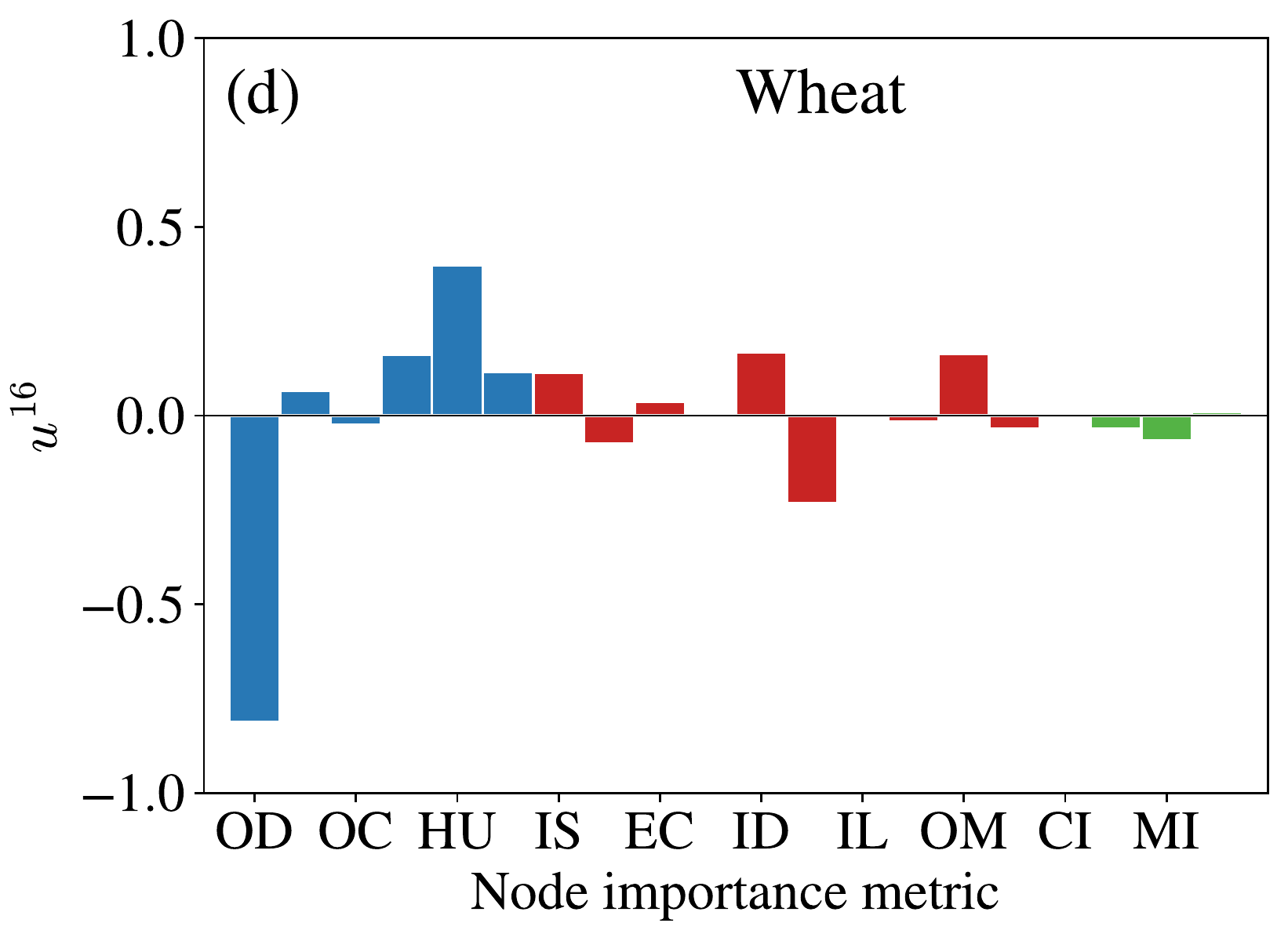}
      \\
      \includegraphics[width=0.233\linewidth]{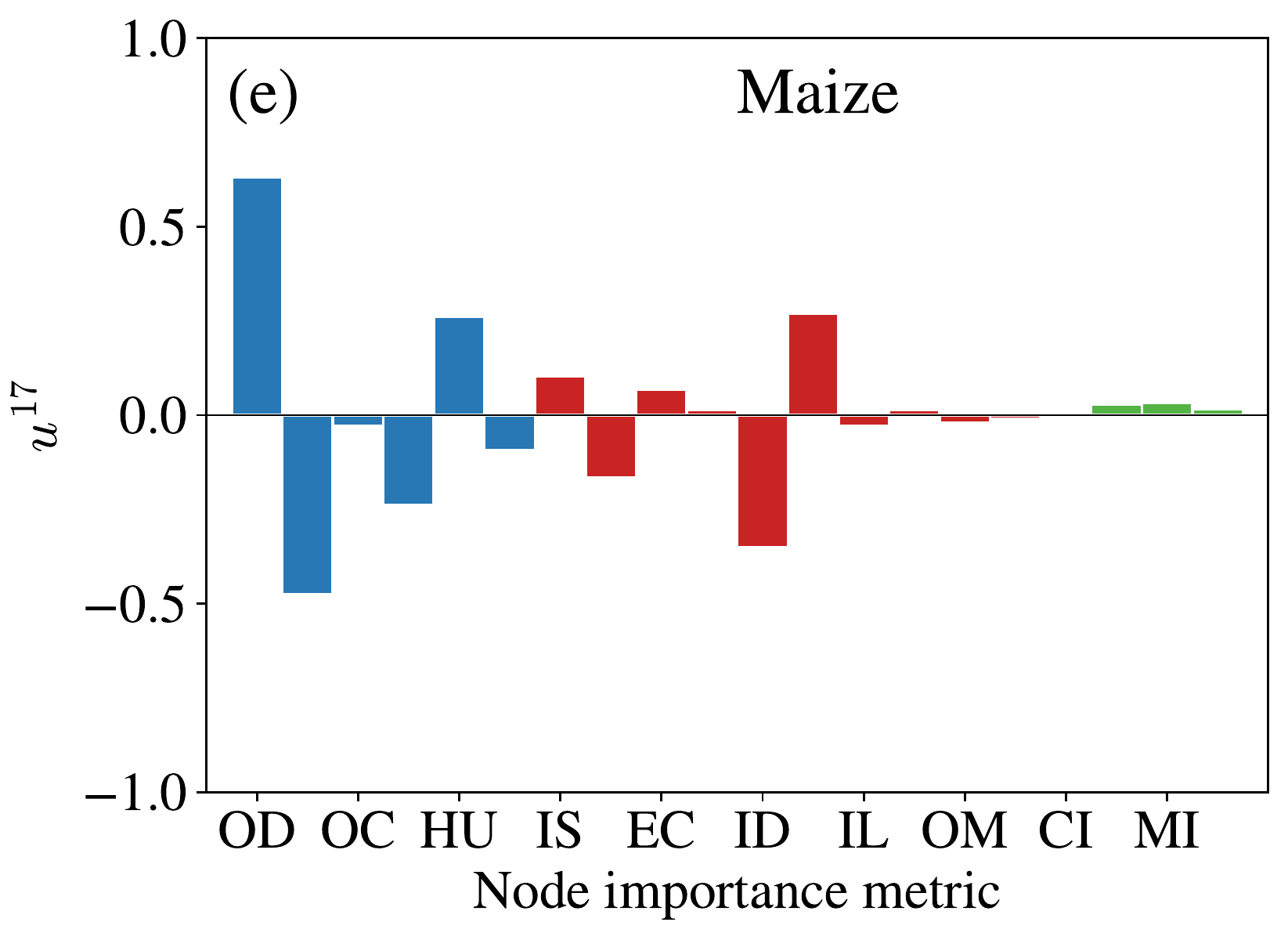}
      \includegraphics[width=0.233\linewidth]{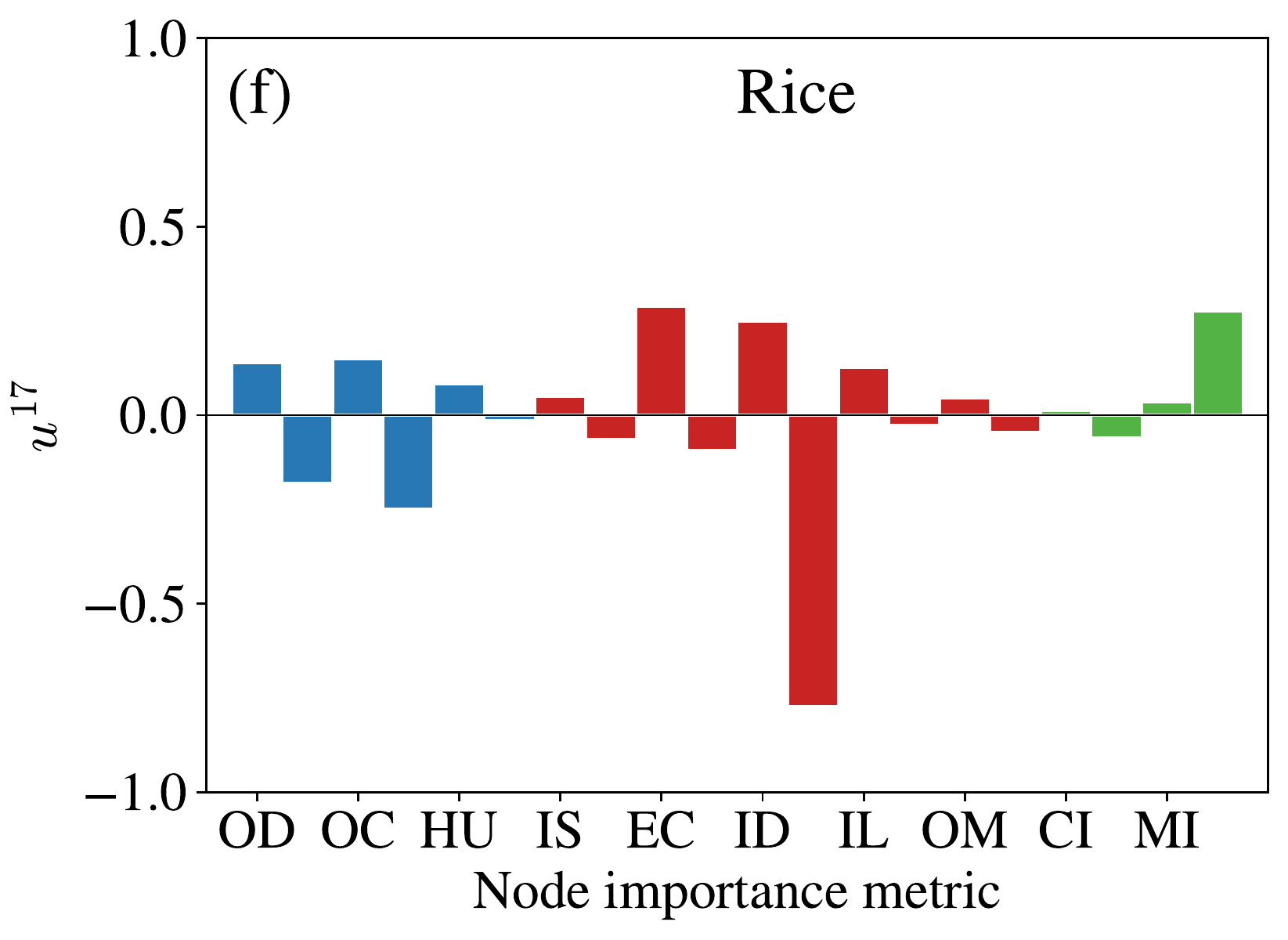}
      \includegraphics[width=0.233\linewidth]{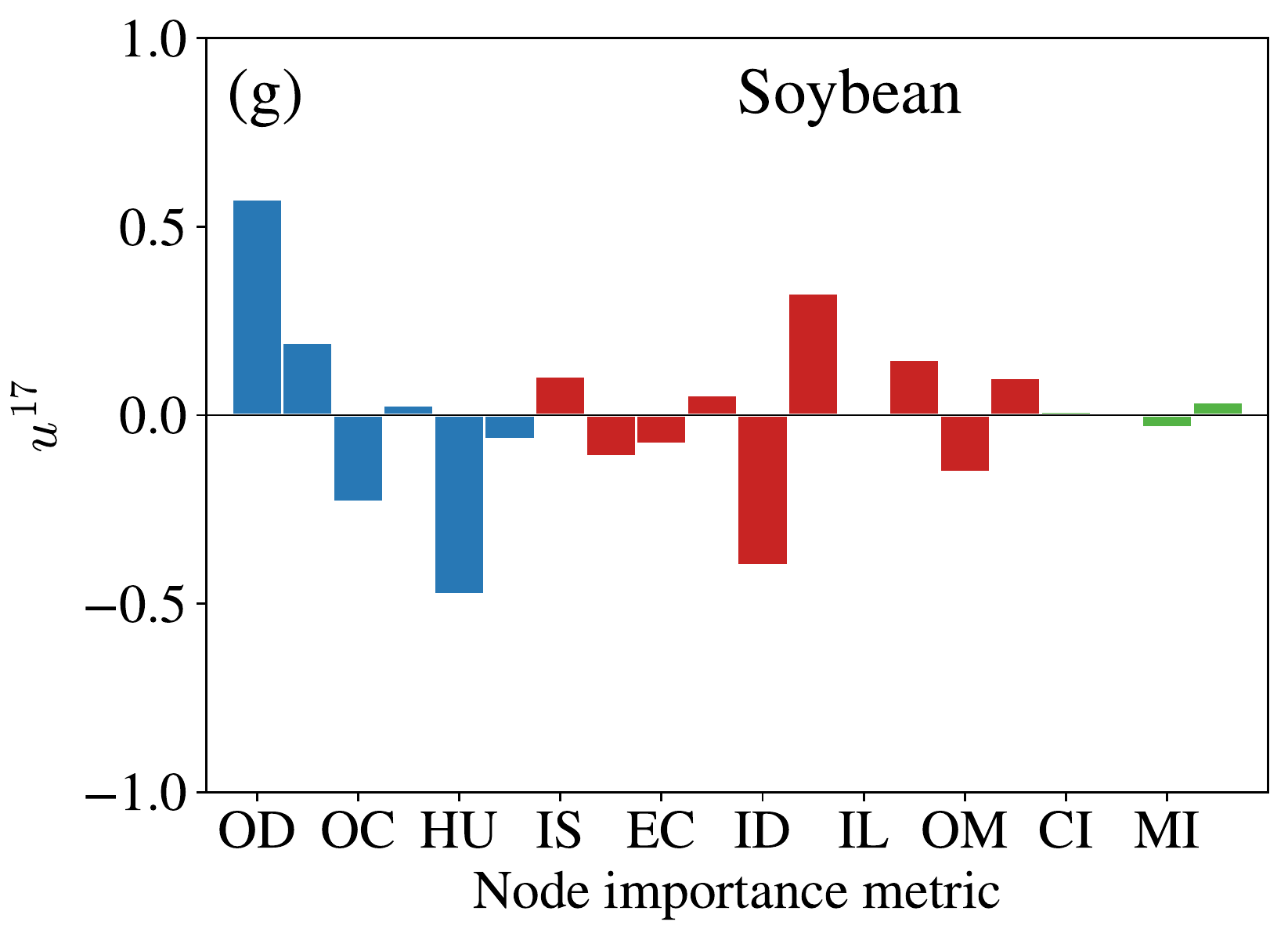}
      \includegraphics[width=0.233\linewidth]{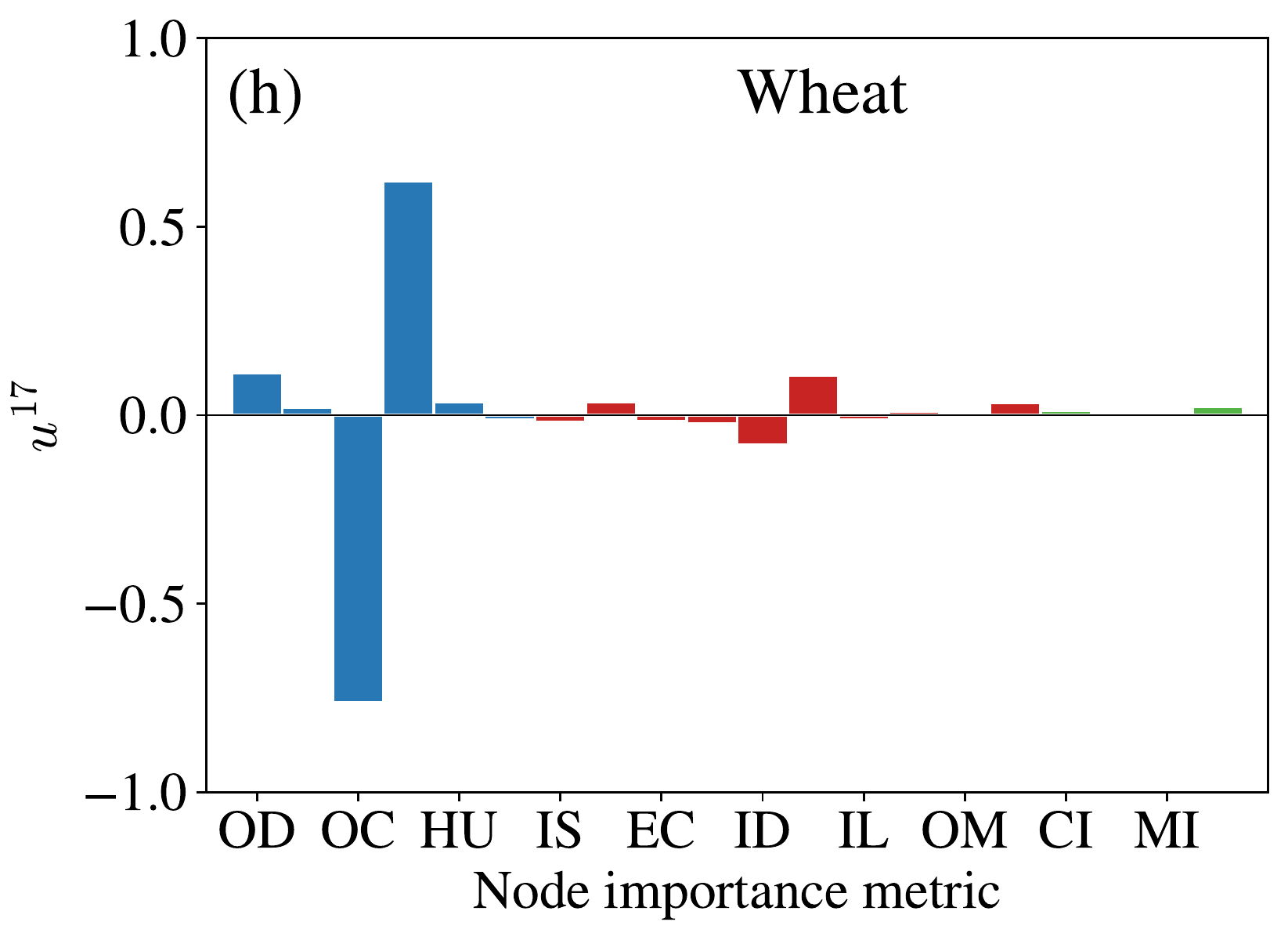}
      \\
      \includegraphics[width=0.233\linewidth]{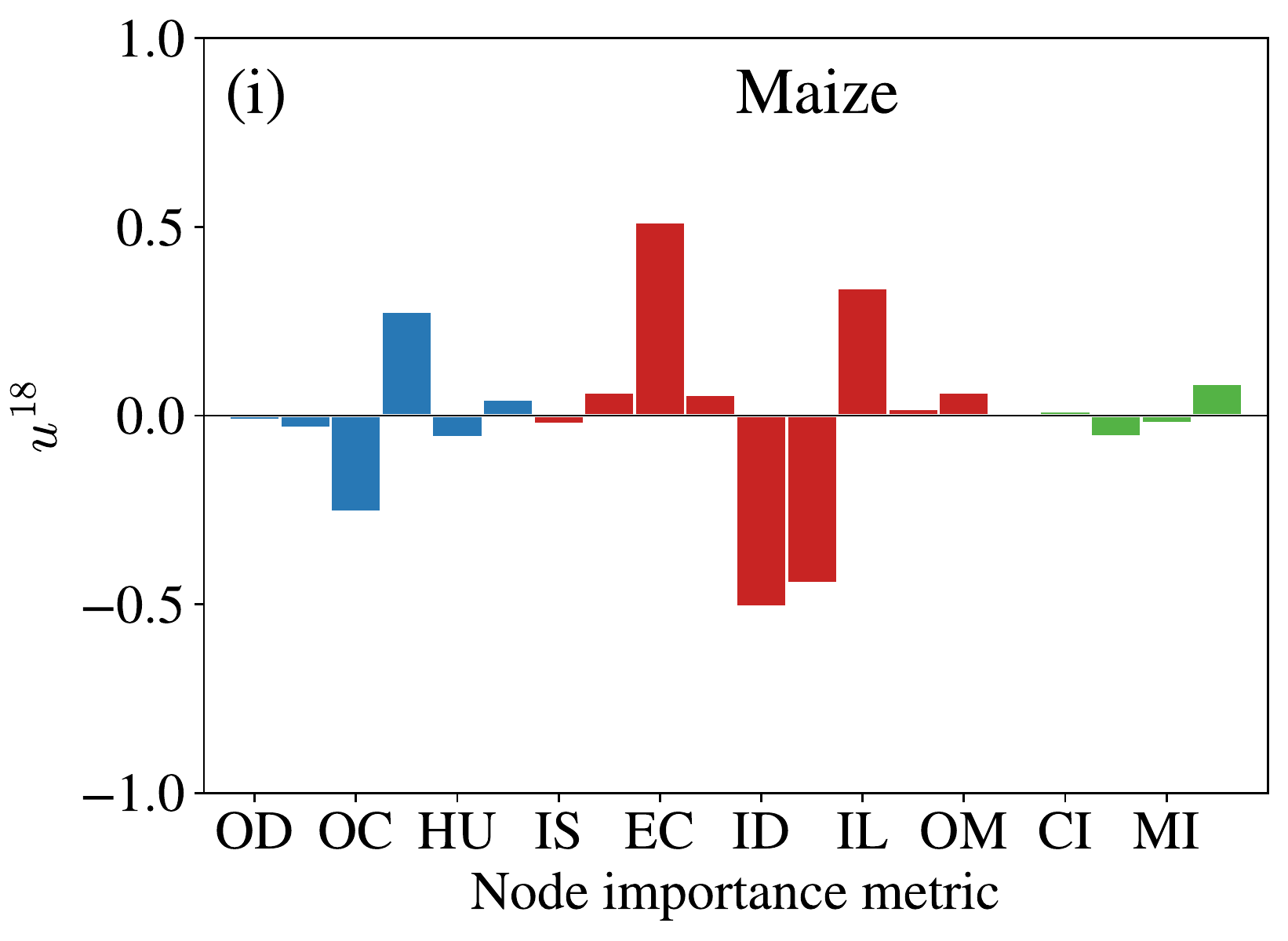}
      \includegraphics[width=0.233\linewidth]{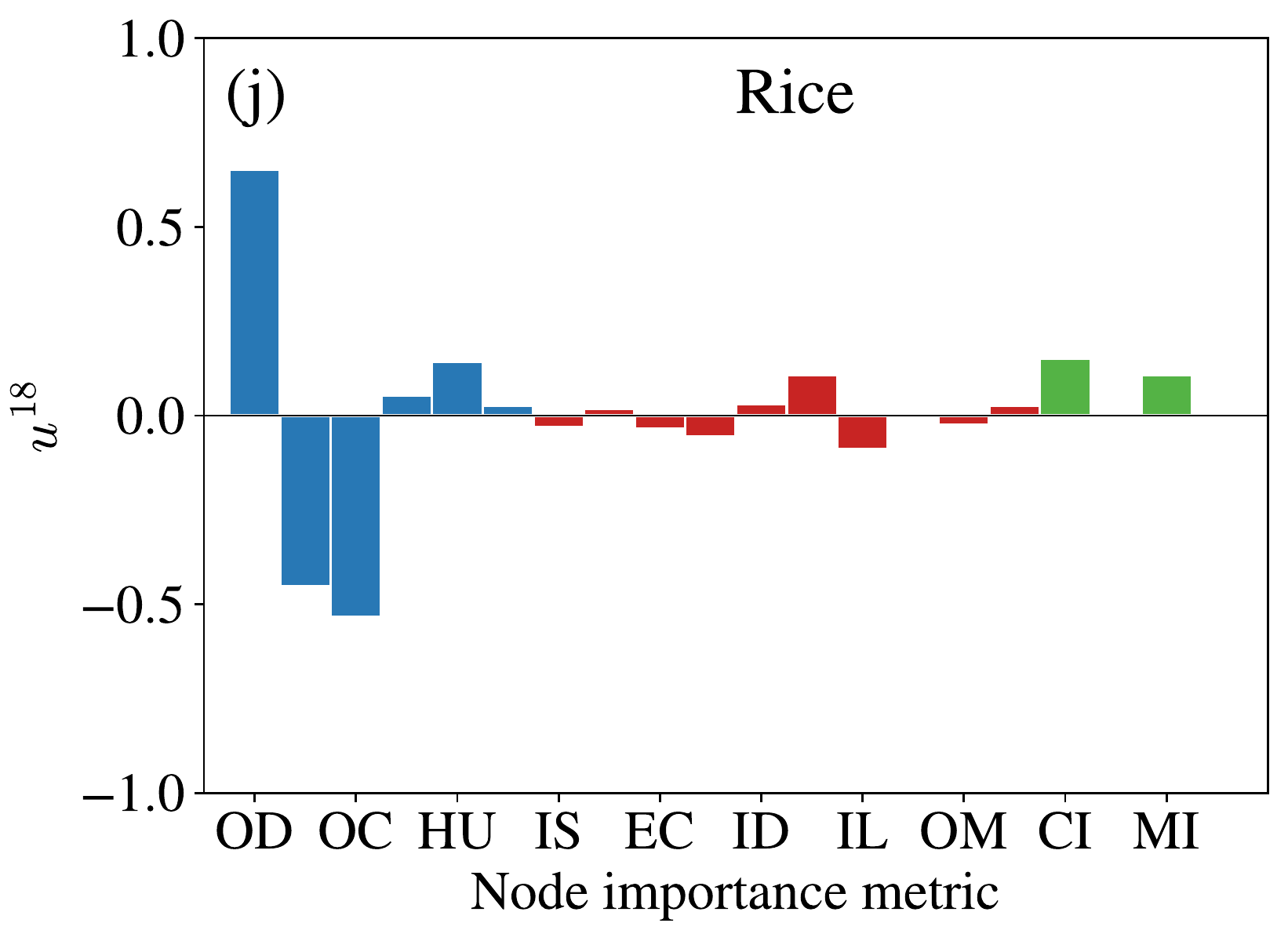}
      \includegraphics[width=0.233\linewidth]{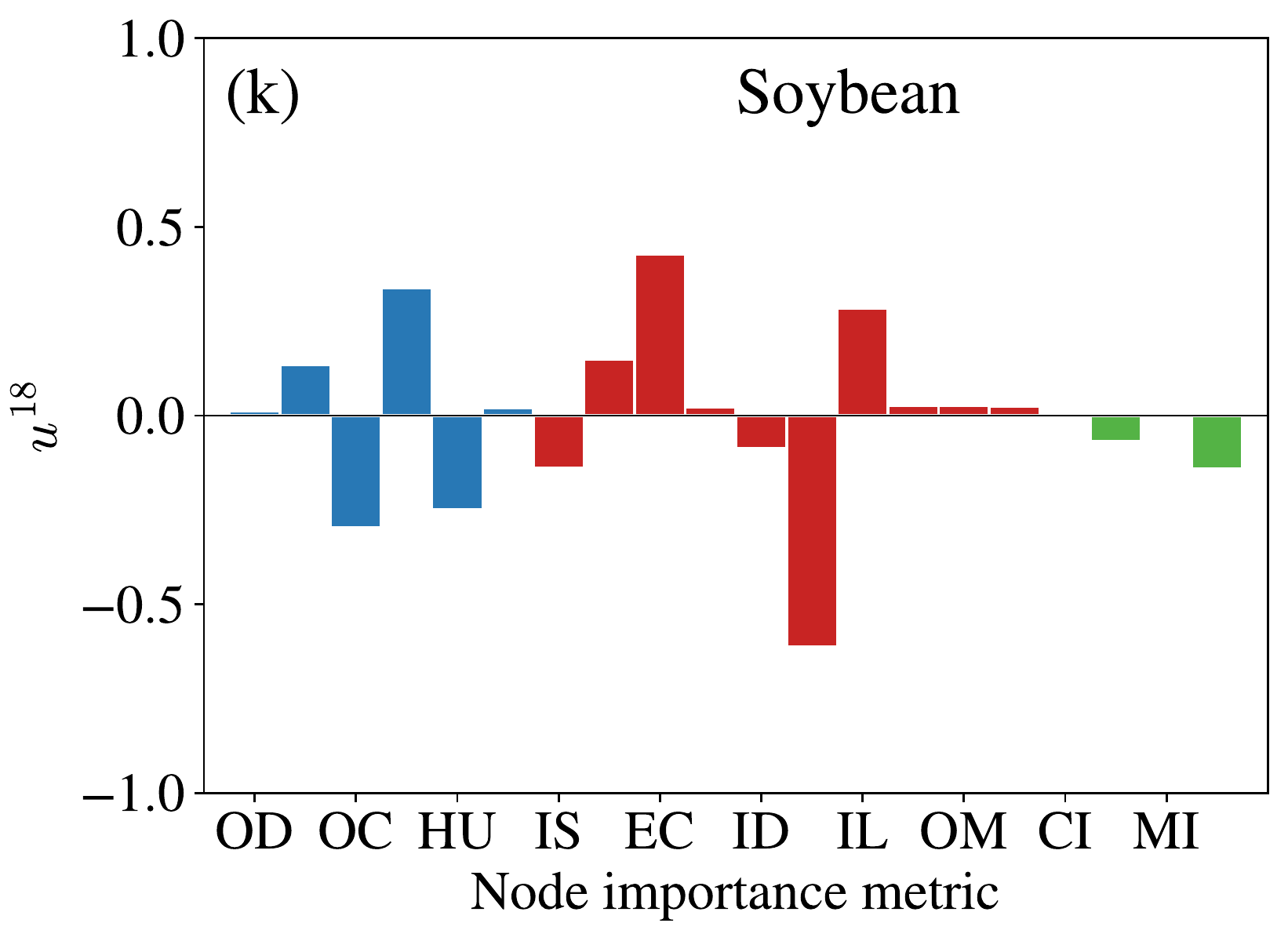}
      \includegraphics[width=0.233\linewidth]{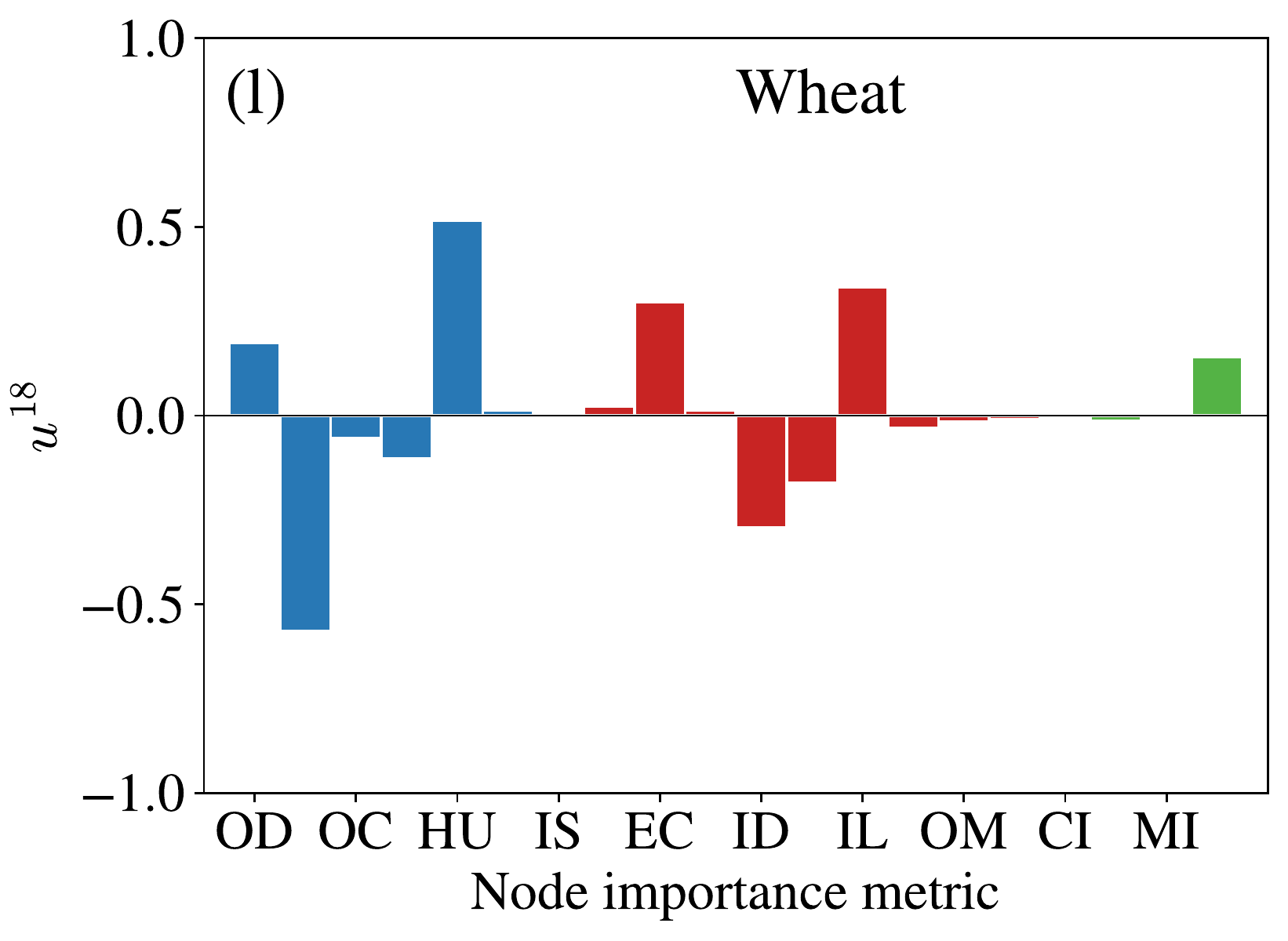}
      \\
      \includegraphics[width=0.233\linewidth]{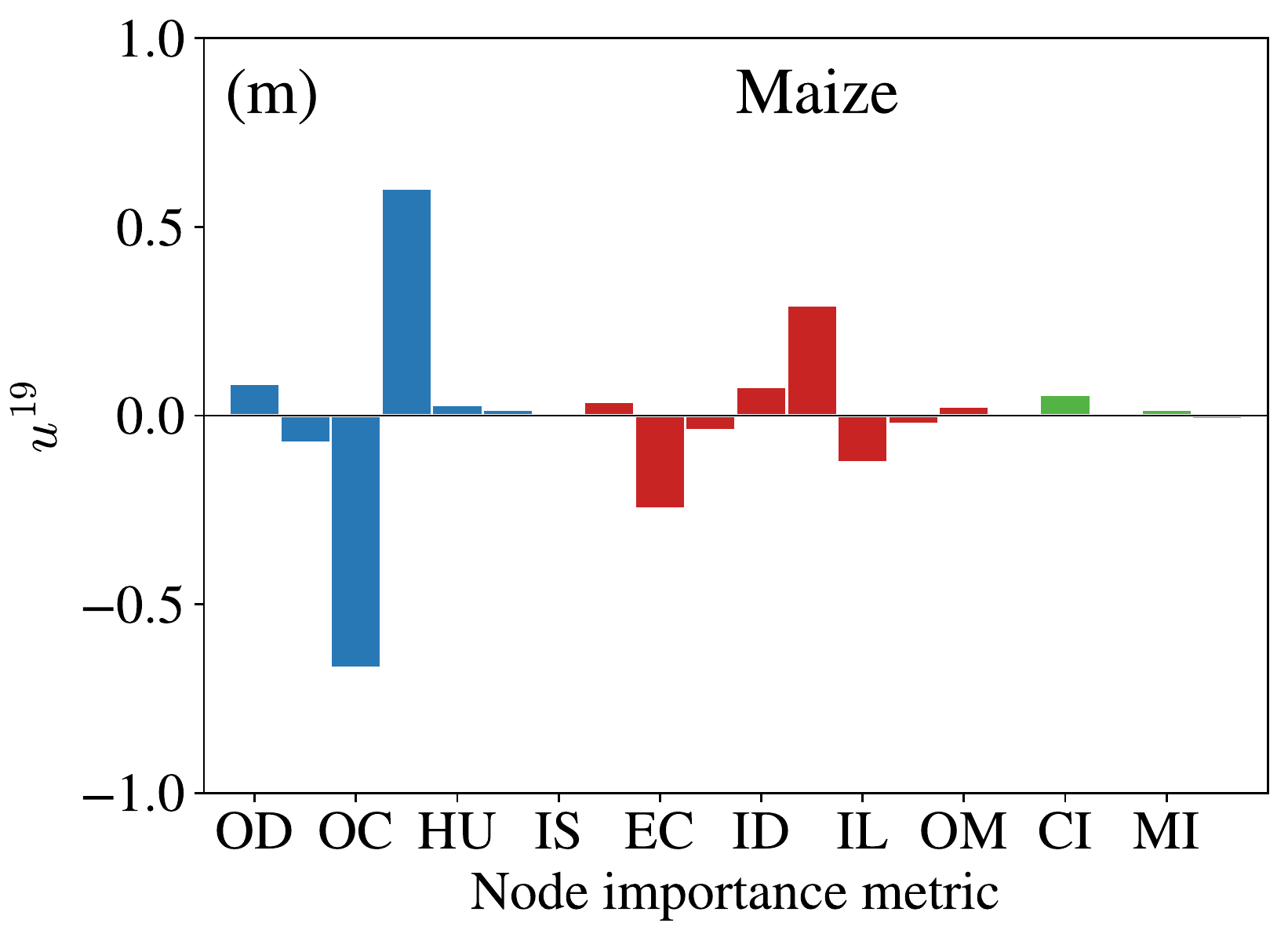}
      \includegraphics[width=0.233\linewidth]{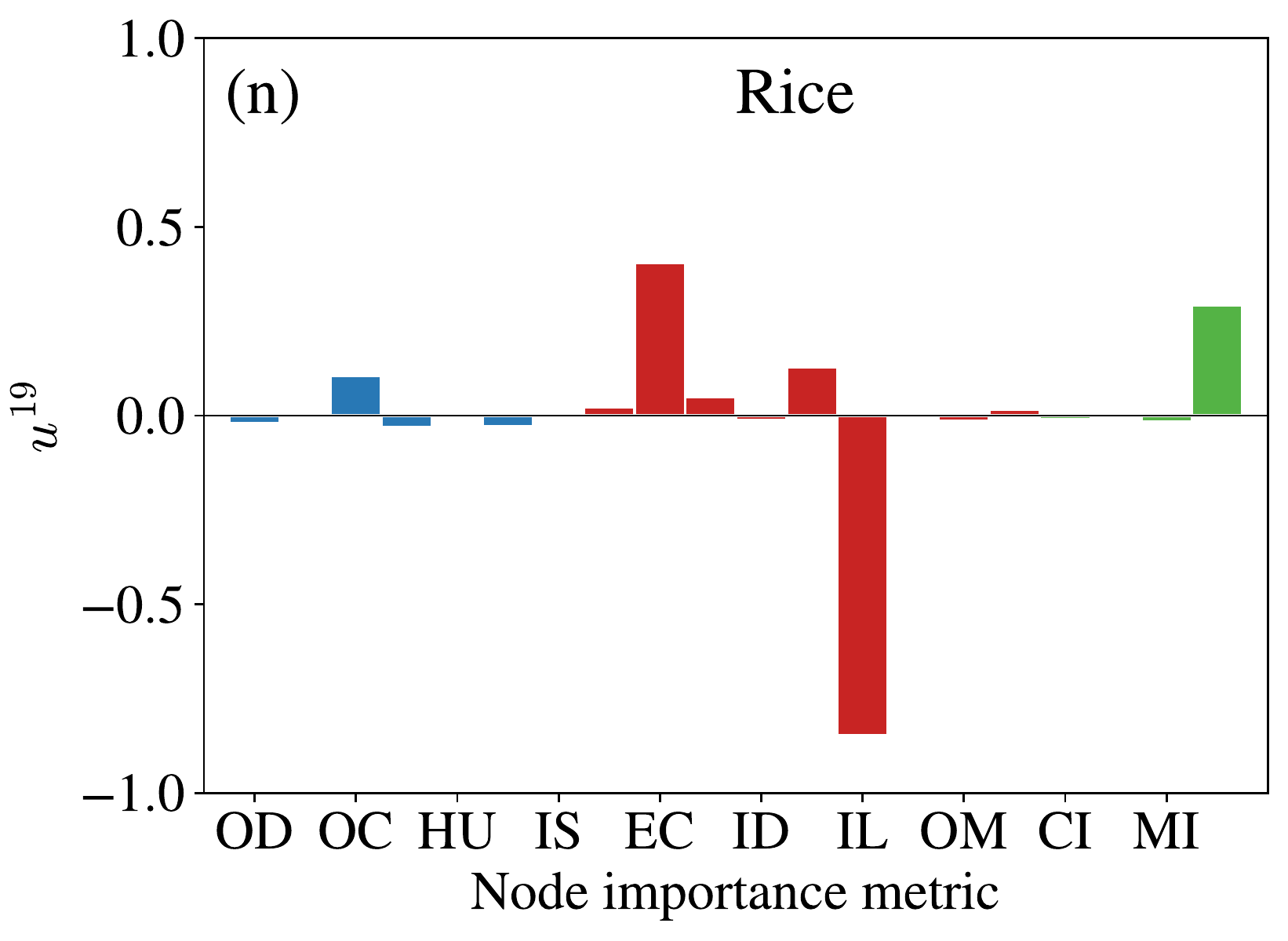}
      \includegraphics[width=0.233\linewidth]{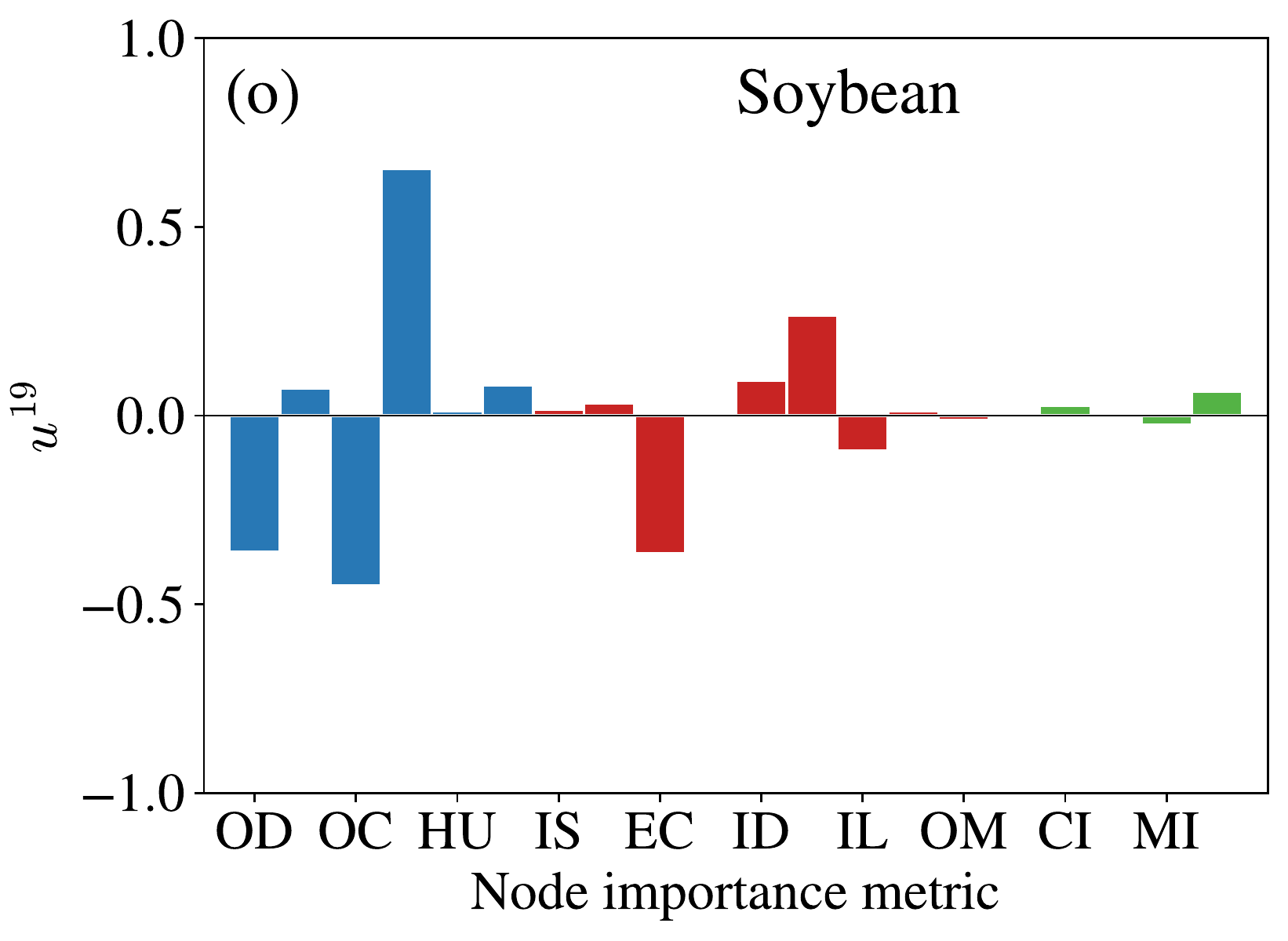}
      \includegraphics[width=0.233\linewidth]{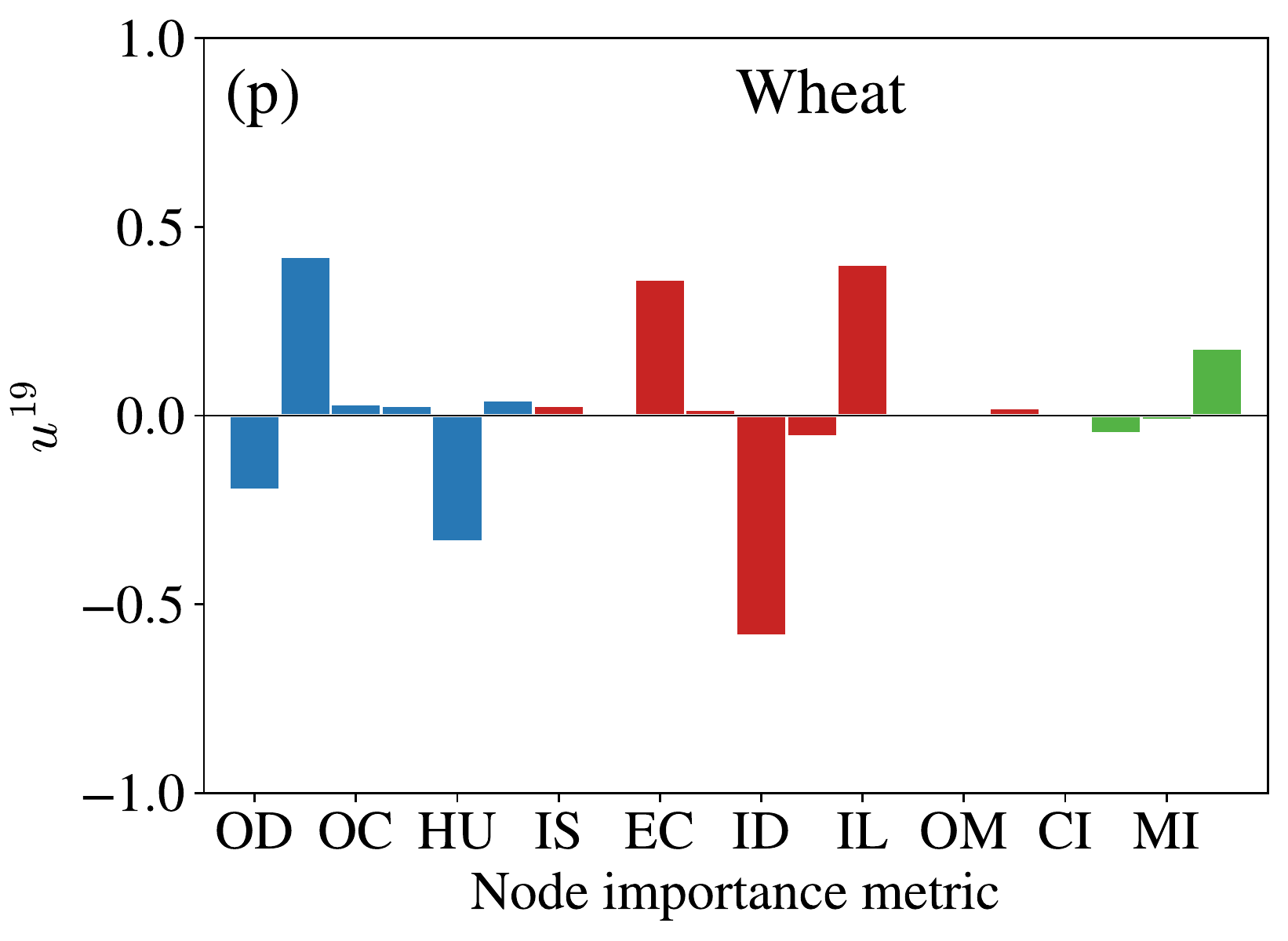}
      \\
      \includegraphics[width=0.233\linewidth]{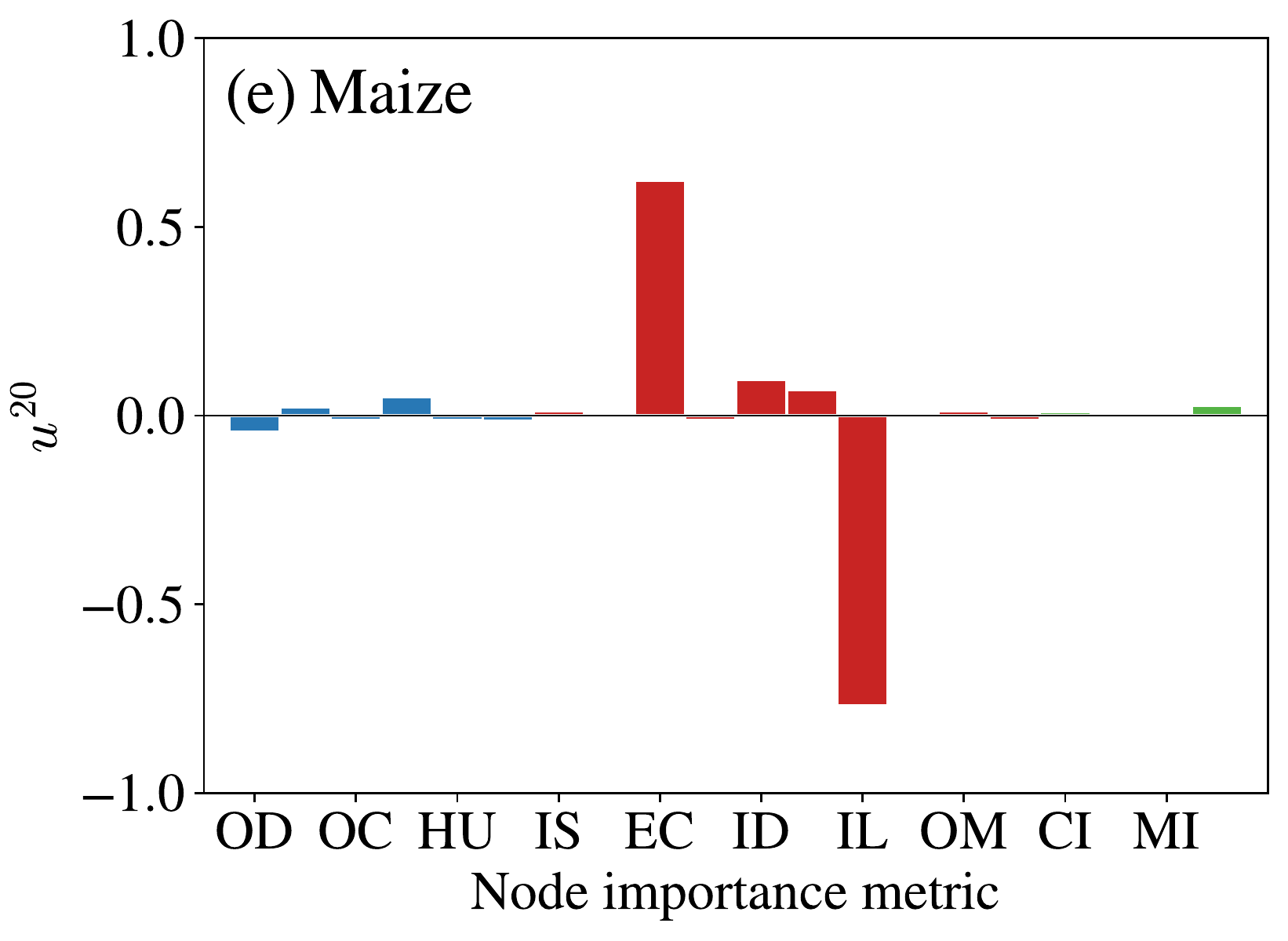}
      \includegraphics[width=0.233\linewidth]{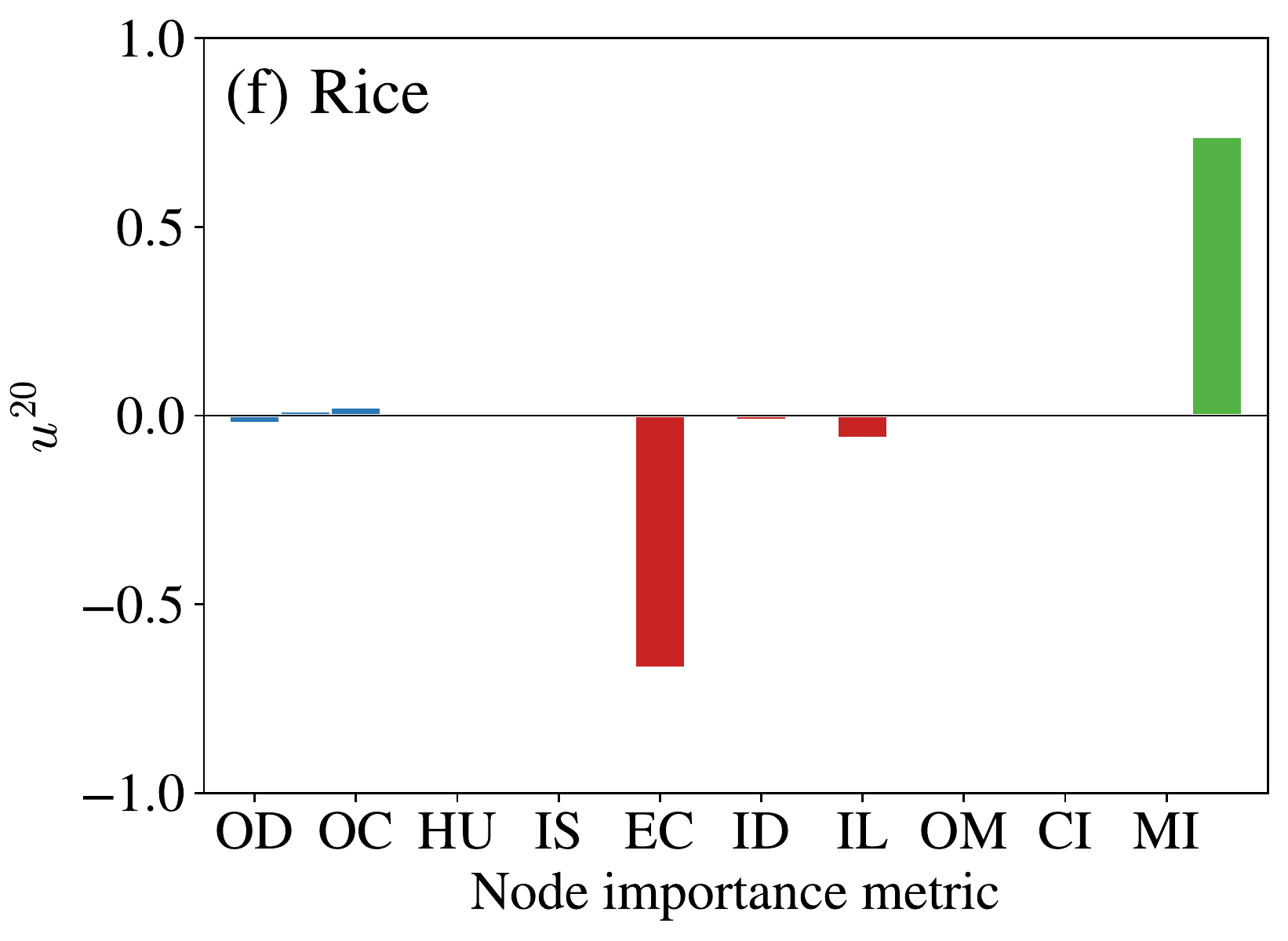}
      \includegraphics[width=0.233\linewidth]{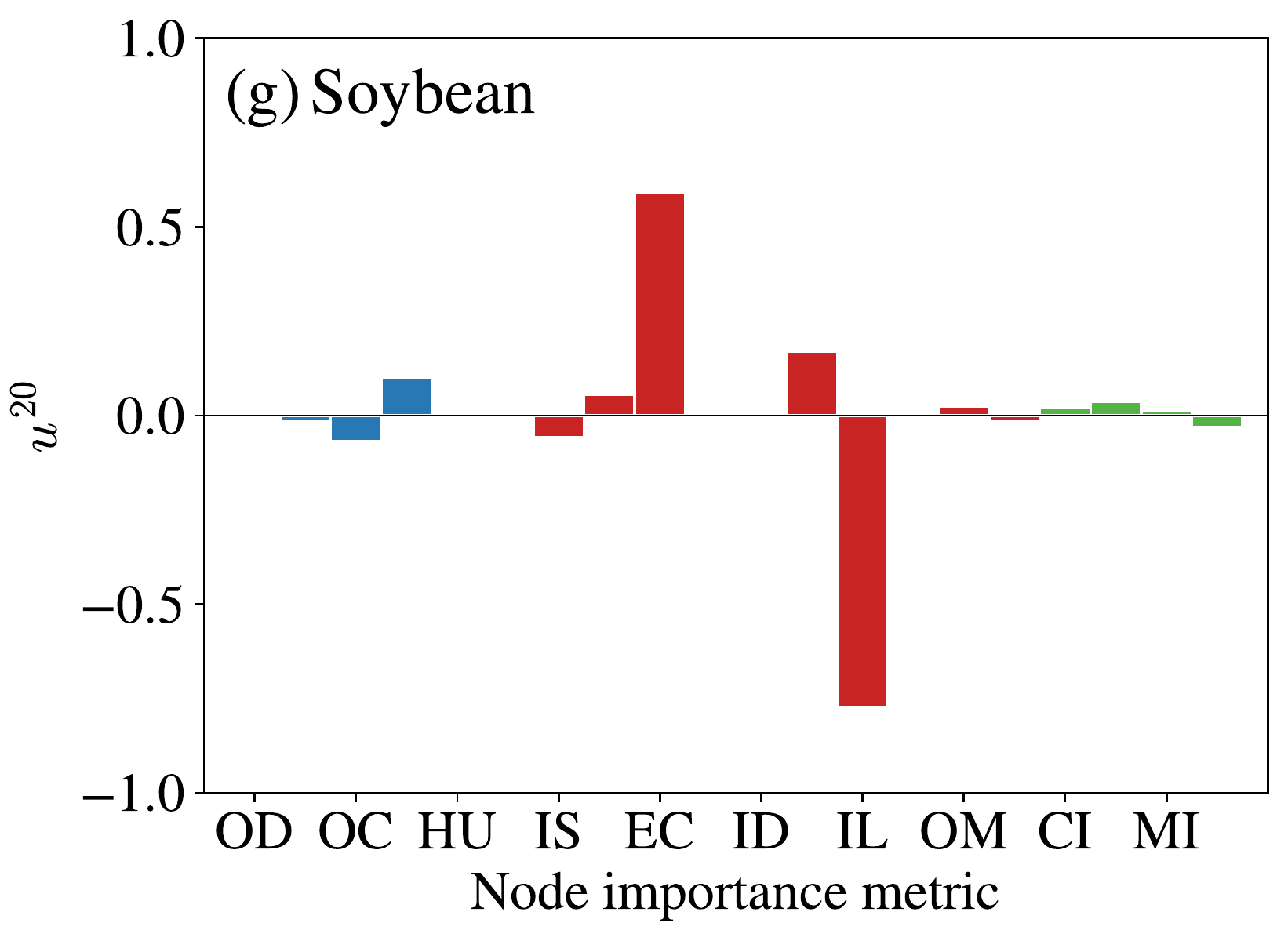}
      \includegraphics[width=0.233\linewidth]{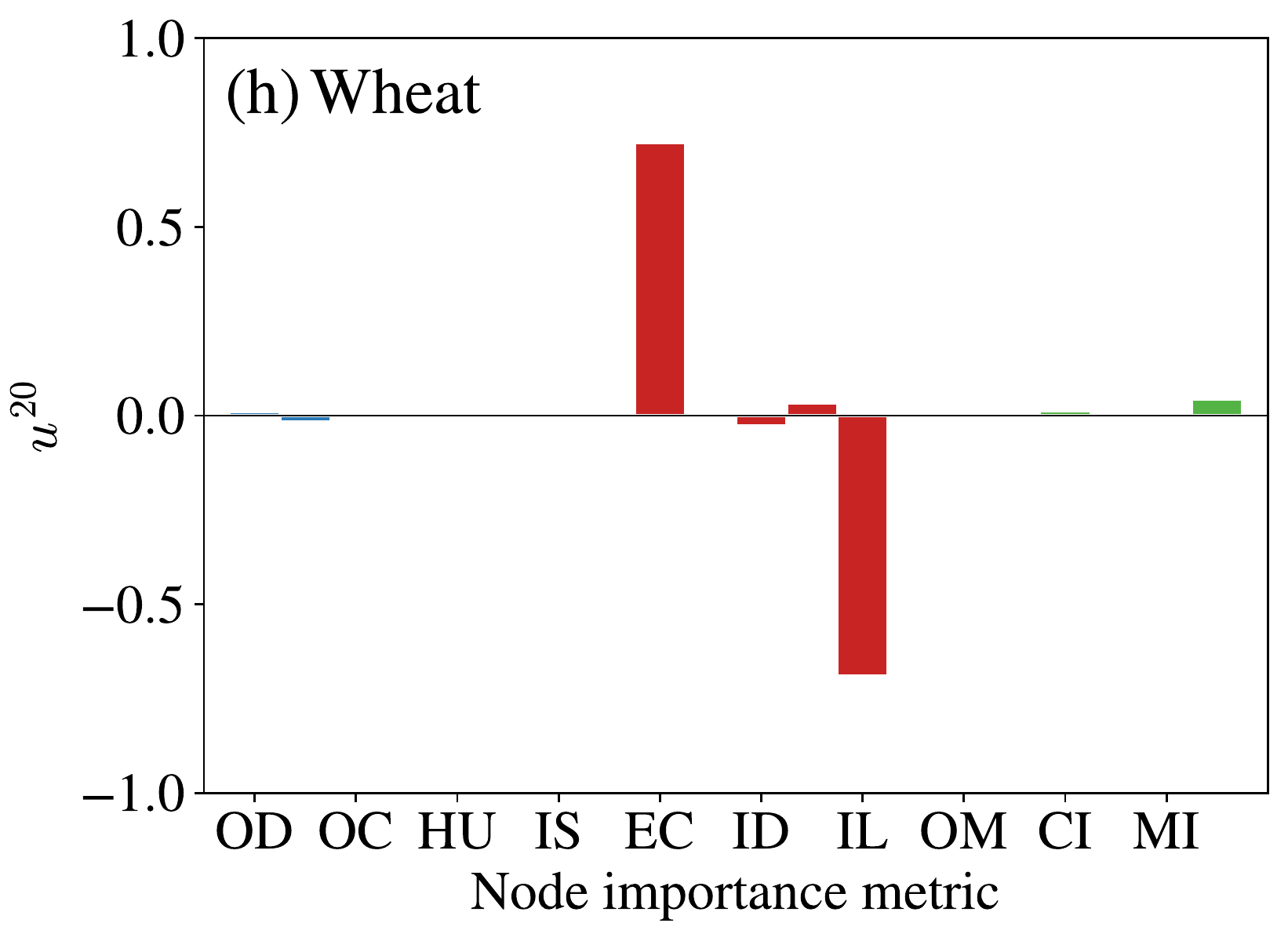}
      \caption{Components of the eigenvector $u_{16}-u_{20}$ of the eigenvalues $\lambda_{16}-\lambda_{20}$ given by Eq.~(\ref{Eq:RMT:PDF:eigenvalue}) of random matrix theory (RMT) in 2020.}
    \label{Fig:iCTN:PDF:eigenvalue:16-20:2020}
\end{figure}

\end{document}